\documentclass[11pt,a4paper,oneside]{report}
\usepackage[T1]{fontenc}
\usepackage{lmodern}
\usepackage[a4paper,margin=28mm,headheight=14pt,headsep=8mm,footskip=12mm]{geometry}
\usepackage{amsmath,amssymb,bm}
\usepackage{booktabs,tabularx,array,longtable}
\usepackage{graphicx,xcolor,tikz,pgfplots}
\usetikzlibrary{arrows.meta,positioning,calc,fit,decorations.pathmorphing}
\usepgfplotslibrary{groupplots}
\pgfplotsset{compat=1.18}
\usepackage{microtype,setspace,fancyhdr,etoolbox}
\usepackage[section]{placeins}
\usepackage{xurl}
\usepackage{cite}
\usepackage{hyperref}
\usepackage{bookmark}
\hypersetup{unicode=true,pdfencoding=auto,
  colorlinks=true,linkcolor=blue!35!black,citecolor=blue!35!black,urlcolor=blue!35!black,
  pdftitle={New Developments in Diagrammatic Monte Carlo for Interacting Lattice Fermions: Efficient Algorithms and Applications},
  pdfauthor={Boyuan Shi},
  pdfsubject={PhD thesis in Physics},
  pdfkeywords={diagrammatic Monte Carlo, Hubbard model, quantum simulation, large flavour expansion, combinatorial summation, Renyi entropy},
  pdflang={en-GB},bookmarksnumbered=true}
\AtBeginEnvironment{figure}{\onehalfspacing}
\AtBeginEnvironment{table}{\onehalfspacing}
\fancypagestyle{plain}{\fancyhf{}\fancyfoot[C]{\thepage}}
\newcommand{\Tr}{\operatorname{Tr}}
\newcommand{\dd}{\mathrm d}
\newcommand{\ii}{\mathrm i}

\newcommand{\Nf}{N_f}
\newcommand{\Ns}{N_s}

\newcommand{\Z}{\mathcal Z}

\newcolumntype{Y}{>{\raggedright\arraybackslash}X}
\newcommand{\frontchapter}[1]{\chapter*{#1}\phantomsection\addcontentsline{toc}{chapter}{#1}\markboth{#1}{}}
\begin{document}
\pagenumbering{arabic}
\thispagestyle{plain}
\begin{center}
\vspace*{21mm}
{\LARGE\bfseries New Developments in Diagrammatic Monte Carlo for Interacting Lattice Fermions: Efficient Algorithms and Applications\par}
\vspace{22mm}
{\Large Boyuan Shi\par}
\vspace{22mm}
{\large Department of Physics\par}
{\large Imperial College London\par}
\vspace{18mm}
{\large Thesis for the degree of\par}
{\large Doctor of Philosophy\par}
\vfill
{\normalsize October 2026\par}
\vspace{12mm}
\end{center}
\clearpage

\frontchapter{Abstract}
Calculating thermodynamic and subsystem quantities for interacting fermions
supports comparisons with cold-atom experiments, tests of other quantum
simulators and assessments of many-body approximations. This thesis develops
diagrammatic methods for these calculations.

Starting from a collective $G$--$\Sigma$ functional integral, I construct
an inverse-flavour expansion about the Hartree saddle. Its Gaussian
fluctuations generate a screened interaction, and enumerating higher-order
diagrams gives the non-Gaussian corrections. A discrete Lehmann
representation of the dynamical interaction allows selected internal
vertices to be integrated analytically. The density calculation covers
31 state points of the six-flavour Hubbard model on a $70\times70$
lattice at $T/t=0.15$. At every point, the central value of the first
non-Gaussian correction is positive and smaller than the Gaussian term.

The generalization of combinatorial summation (CoS) developed here
accommodates flavour-dependent propagators and expansion shifts.
Loop-head labels specify the allowed contractions in CoS++, while shared
partial computations give connected sums for interaction subsets.
A determinant-polynomial sieve evaluates the same sums through local
coefficient filters. At order nine, it is $24$--$25$ times faster
than the specified CoS baselines in three-flavour benchmarks on dense
random matrices. In the thermodynamic applications, adding the eighth
order narrows the spread of Pad\'e approximants for double occupancy on
a $2048$-site honeycomb lattice with two spin components. Three-flavour
calculations on $1024$ square-lattice sites give a continuation close to
the inverse-flavour truncation.

For the second R\'enyi entropy, I derive a fermionic replica path integral
and a stable evaluation of its two-time Green functions. The
connected-determinant calculation reaches tenth order on a $3\times3$
cluster, with low-order coefficients checked against deterministic
integration and exact diagonalization, and ninth order on larger lattices.
On $50\times50$ sites at $(U/t,\mu/t,\beta t)=(3,1,8)$, the ninth-order
continuation at shift $\alpha=1.3$ gives $S_2/N_s=0.06695$ with a
propagated Monte Carlo uncertainty of $3.0\times10^{-5}$. A comparison
of two expansion shifts at their respective retained orders gives an
endpoint difference of $2.6\times10^{-4}$. The replica formulation makes subsystem-purity
calculations possible on these lattices within the connected-diagram
framework used for the thermodynamic observables.
\clearpage

\clearpage
\phantomsection\addcontentsline{toc}{chapter}{Contents}
\tableofcontents
\clearpage\phantomsection\addcontentsline{toc}{chapter}{List of Figures}
\listoffigures
\clearpage\phantomsection\addcontentsline{toc}{chapter}{List of Tables}
\listoftables
\frontchapter{Publications and Research Contributions}
\label{front:contributions}
The research chapters develop thermodynamic and subsystem calculations
for interacting lattice fermions. Chapters~\ref{c2:chapter}--\ref{c4:chapter}
expand the research reported in my three single-author papers:
\begin{enumerate}
\item B. Shi, \emph{Semideterministic and stochastic sampling of Feynman
diagrams with $1/N_f$ expansions}, Physical Review Research
\textbf{7}, 023177 (2025) \cite{ShiLargeNf2025}.
\item B. Shi, \emph{CoS++: Towards More General and Explicit Implementations
for Sampling High-Order Feynman Diagrammatic Series},
arXiv:2507.12349 (2025) \cite{ShiCoS2025}.
\item B. Shi, \emph{Diagrammatic Monte Carlo for Fermionic R\'enyi
Entanglement Entropy}, arXiv:2605.21632 (2026) \cite{ShiRenyi2026}.
\end{enumerate}

Section~\ref{c3:fast:section} extends this work with a determinant-polynomial
sieve for connected sums. I derive its local coefficient filters and
compare its C++ implementation with direct CoS evaluation at fixed
configurations. The timing results in Fig.~\ref{c3:fast:speedup} and
Table~\ref{c3:fast:timings} belong to this additional study.

In Chapter~\ref{c2:chapter}, I derive the inverse-flavour expansion from
the collective action through to the integrals used in the calculation.
This includes the normal-ordering prescription, diagram weights and
analytical integration of internal vertices. The six-flavour densities in
Section~\ref{sec:author-density-results} are those I reported in
Ref.~\cite{ShiLargeNf2025}. The chapter resolves their saddle, Gaussian and first non-Gaussian
contributions.

In Chapter~\ref{c3:chapter}, I present the CoS++ recurrences and their proofs,
including typed-head summation, circuit compression and subset readout.
These constructions extend the CoS method
of Kozik \cite{KozikCoS2024}. I also derive the integration measures and estimators used in the lattice examples. The lattice figures reproduce the numerical results reported in
Ref.~\cite{ShiCoS2025}.

Chapter~\ref{c4:chapter} gives the fermionic replica trace and path integral
in detail. I derive a stable evaluation of its Gaussian propagator and
use it to calculate connected coefficients for the second R\'enyi entropy.
The numerical analysis compares low-order coefficients with deterministic integration and exact diagonalization, and examines continuation on larger lattices using the results reported in Ref.~\cite{ShiRenyi2026}.
\clearpage

\frontchapter{Notation and Abbreviations}
The following symbols are used throughout the thesis. Additional notation
is defined in the relevant derivations.

\begin{longtable}{@{}p{29mm}p{117mm}@{}}
\toprule
Symbol & Meaning\\
\midrule\endhead
$H$, $K$ & Physical Hamiltonian and grand Hamiltonian
$K=H-\mu\widehat N$ for a common chemical potential. For unequal
flavour potentials,\newline\mbox{$K=H-\sum_a\mu_a\widehat N_a$}.\\
$\widehat N$, $N_{\mathrm{at}}$ & Total particle-number operator and its mean.\\
$N_f$ & Number of fermion flavours. The notation $\mathrm{SU}(N)$ in
quoted publication titles uses $N=N_f$.\\
$N_s$, $N_c$, $\nu$ & Number of sites, number of unit cells and sites per
unit cell, with $N_s=\nu N_c$.\\
$\beta$, $T$ & Inverse temperature and temperature; $k_{\mathrm B}=1$.\\
$t$, $U$, $\mu$ & Hopping scale, physical onsite pair interaction and
chemical potential, unless a rescaled coupling is stated.\\
$Z$, $\psi$, $\omega$ & Grand partition function,
$\psi=\ln Z/N_s$, and $\omega=-T\psi$.\\
$n$, $\rho$ & $n=\langle\widehat N\rangle/N_s$ is the total density per site.
In the Hartree equations of Chapter~\ref{c2:chapter}, $\rho$ denotes
density per flavour; in Chapter~\ref{c4:chapter}, $\rho$ is a density
operator. \\
$\xi$, $\alpha$ & Auxiliary series parameter and expansion shift;
see the definitions in Chapters~\ref{c2:chapter}--\ref{c4:chapter}.\\
$A$, $\bar A$ & Spatial subsystem and its complement.\\
$S_n(A)$ & R\'enyi entropy of index $n$ of subsystem $A$.\\
$\widetilde Z_n(A)$ & Unnormalized replicated partition function,
$\Tr\rho_A^n=\widetilde Z_n(A)/Z^n$.\\
$f_A$ & In Chapter~\ref{c4:chapter},
$f_A=[\ln\widetilde Z_2(A)-2\ln Z]/N_s=-S_2(A)/N_s$.\\
$C$, $\Sigma$ & Covariance matrices in statistical discussions; the
collective field $\Sigma$ of Chapter~\ref{c2:chapter} is defined as a bilocal field.\\
\bottomrule
\end{longtable}

The research chapters use $g=\langle\mathcal T_\tau cc^\dagger\rangle$,
without a leading minus sign. Equal-time prescriptions and counterterms
are stated where each expansion is defined. The equilibrium Green function $g$ depends on the time difference; the replica Green function for a spatial cut generally depends on both absolute times. 

\begin{longtable}{@{}p{29mm}p{117mm}@{}}
\toprule
Abbreviation & Meaning\\
\midrule\endhead
AFQMC & Auxiliary-field quantum Monte Carlo.\\
BCS & Bardeen--Cooper--Schrieffer.\\
CDet & Connected-determinant diagram summation.\\
CoS, CoS++ & Combinatorial summation and the generalizations developed in this thesis.\\
DAG & Directed acyclic graph; here often an arithmetic circuit.\\
DiagMC & Diagrammatic Monte Carlo.\\
DLR & Discrete Lehmann representation.\\
DMFT & Dynamical mean-field theory.\\
DQMC & Determinant quantum Monte Carlo.\\
ED & Exact diagonalization.\\
FFT & Fast Fourier transform.\\
MC, MCMC & Monte Carlo and Markov-chain Monte Carlo.\\
MCMCMC & Many-configuration Markov-chain Monte Carlo.\\
NLCE & Numerical linked-cluster expansion.\\
QMC & Quantum Monte Carlo.\\
RG-MCMC & Renormalization-group Markov-chain Monte Carlo.\\
RPA & Random-phase approximation.\\
SB-CoS & Symmetry-broken combinatorial summation.\\
VEGAS & Adaptive importance-sampling integration.\\
\bottomrule
\end{longtable}
\clearpage

\chapter[Precision Many-Body Calculations for Interacting Fermions]{Precision Many-Body Calculations\\for Interacting Fermions}
\label{ch:benchmarks}

Precise predictions for interacting fermions enter experimental thermometry,
tests of approximate many-body methods and assessments of states prepared
on quantum processors. In each application, the comparison concerns a
specified Hamiltonian, ensemble and observable, with an error scale set
by the physical question. This thesis develops diagrammatic calculations
of lattice-fermion thermodynamics and subsystem R\'enyi entropy. Cold-atom
experiments provide direct applications, and the same quantities can
serve as references for other quantum implementations and classical
many-body methods.

The information needed from a calculation depends on how it will be
used. Fitting a measured density profile involves both the equation of
state and its parameter derivatives, which determine the sensitivity
of the inferred temperature or coupling. Absolute free energies enter
thermodynamic integration and variational thermal-state calculations.
Measurements of subsystem purity extend these comparisons to nonlinear
properties of a reduced state beyond its density and energy. This chapter
introduces these tasks and the numerical methods available for them,
providing the setting for the diagrammatic constructions in the research
chapters.

\section{Interacting fermions and quantitative reference problems}
\label{c1:v7:fermions}

The Hubbard model is a common test of many-body methods because a local
interaction and single-particle hopping already produce demanding
correlation problems. Numerical comparisons have examined its energy,
occupancy and self-energy across coupling and filling, while finite-temperature
studies compare the evolution of local and magnetic observables
\cite{LeBlancBenchmark2015,Schafer2021Footprints,QinComputational2022}.
For two fermion components, each site has four local occupation states;
the unrestricted many-body basis of $N_s$ sites has dimension $4^{N_s}$.
At a fixed filling between the empty and fully occupied limits, even
the fixed-particle-number sector grows exponentially with system size. Many-body methods consequently target
selected quantities through a representation suited to the problem.

A flavour denotes an internal fermion state, such as a spin component.
Its multiplicity is independent of both particle number and spatial size.
In the calculations below, it determines the interaction combinatorics
and supplies the parameter of an inverse-flavour expansion. Calculations
on the same lattice may target a local density, a much smaller fluctuation
correction or a subsystem entropy, each with a different precision target.

\label{sec:model}

\subsection{A common Hamiltonian and ensemble}

I use the repulsive Fermi--Hubbard model to specify the Hamiltonian and ensemble in these comparisons. For $\Nf$ equivalent fermion flavours,
consider
\begin{equation}
 H=-\sum_{\langle ij\rangle,\alpha}
 t_{ij}\bigl(c^\dagger_{i\alpha}c_{j\alpha}
 +c^\dagger_{j\alpha}c_{i\alpha}\bigr)
 +U\sum_i\sum_{\alpha<\gamma}n_{i\alpha}n_{i\gamma}
 +\sum_i v_i n_i,
 \label{eq:hubbard}
\end{equation}
where each hopping bond is counted once, the hopping amplitudes are real
in the present notation, $n_i=\sum_\alpha n_{i\alpha}$ and
$\alpha=1,\ldots,\Nf$. The potential $v_i$ allows for confinement or a
specified spatial modulation. Throughout this chapter $k_{\mathrm B}=1$,
$\beta=1/T$, and the chemical potential is introduced through the grand
Hamiltonian $K=H-\mu\widehat N$, with $\widehat N=\sum_i n_i$.
The grand partition function is
\begin{equation}
 \Z(T,\mu,U)=\Tr\exp[-\beta(H-\mu\widehat N)],
 \qquad \Omega=-T\ln\Z.
 \label{eq:grand-partition}
\end{equation}
The interaction parameter $U$ in Eq.~\eqref{eq:hubbard} is the interaction
energy of one pair of different flavours. A large-$\Nf$ formulation may
rescale this parameter. Such a rescaling is a change of convention that
must be stated when calculations are compared at different $\Nf$ or with
an experimental value of $U/t$.

Optical lattices realize this class of models by restricting the atomic
motion to lattice bands and relating tunnelling and interactions to
matrix elements of localized orbitals. The microscopic assumptions and
their regimes of validity are well established
\cite{Bloch2008ManyBody,Bloch2012QuantumGases,Gross2017QuantumSimulations}. For alkaline-earth
and alkaline-earth-like atoms, the nuclear-spin structure provides access
to enlarged flavour symmetries, giving $\mathrm{SU}(\Nf)$ Hubbard models
a direct experimental motivation \cite{Gorshkov2010TwoOrbital,Taie2012Pomeranchuk}.
For computation, the additional flavours enlarge both the physical
parameter space and the combinatorial content of the many-body problem.

Corrections to the simplest Hubbard description become relevant when
their effects approach the target accuracy. Further-neighbour hopping
changes the one-body matrix, an intersite density coupling adds a term
such as $V\sum_{\langle ij\rangle}n_i n_j$, and flavour-dependent fields
can give different propagators to different components. Specifying these
terms connects the experimental realization to the Hamiltonian used in
the calculation. The broader range of such corrections is reviewed in
Ref.~\cite{Dutta2015NonstandardHubbard}.

\subsection{Particle number, system size and correlation length}

A lattice reference calculation distinguishes several sizes. The number of lattice
sites is $\Ns$, the mean particle number is
$N_{\mathrm{at}}=\langle\widehat N\rangle$, and the mean filling is
$n=N_{\mathrm{at}}/\Ns$. The number of flavours $\Nf$ counts internal states and sets no spatial scale. A measurement may further refer to a subregion $A$
containing $|A|$ sites.  A cloud containing thousands of atoms can sample a range of
fillings and local chemical potentials, and a correlation measurement may use
only a selected part of that cloud.

The correlation length $\ell_{\mathrm{corr}}$ determines whether a finite
sample reproduces the bulk observable of interest. If all relevant
correlations decay well inside the measured region, a uniform
thermodynamic-limit calculation may provide a suitable reference. If
correlations extend across that region, boundaries and the shape of the
sample become part of the problem. The relevant size depends on the observable. A density may be close to its bulk value while a
long-distance correlation or a subsystem entropy remains sensitive to
geometry.

In an inhomogeneous system a local-density approximation identifies
$\mu_i=\mu-v_i$ and evaluates a uniform equation of state at the local
chemical potential. This approximation can be applied when the potential varies slowly over the relevant correlation scale. Near a sharp boundary or where the correlation length becomes comparable
to the trap scale, a finite inhomogeneous calculation can test the
local-density approximation.

The equilibrium calculations developed in this thesis provide thermal
reference states and static responses. In the discussion of dynamical
experiments, these correspond to a specified initial state or endpoint
of the preparation.

\section{What makes a useful classical reference?}
\label{c1:v7:reference}
\label{sec:programme}

A classical reference and the result being assessed are compared at a
common Hamiltonian, ensemble, geometry and observable. This specification
is needed for experiments and for comparisons between numerical methods.
Agreement in a tractable regime establishes reference points from which
to explore more difficult parameters, accompanied by the corresponding
size and error analysis \cite{Altman2021Architectures,Daley2022PracticalAdvantage}.

The intended comparison sets the accuracy scale. Distinguishing predictions
separated by $\Delta O$ calls for an uncertainty smaller than that
separation; inferring a parameter also depends on the observable's
response to it. Derivatives and entropies formed from several estimates
further depend on their covariance. A parameter grid of jointly calculated
observables can supply this information in a form that a value at one
isolated point cannot.

For extensive thermodynamics I use
\begin{equation}
 \psi=\frac{\ln\mathcal Z}{N_s},\qquad
 \omega=\frac{\Omega}{N_s}=-T\psi .
 \label{eq:psi}
\end{equation}
I denote the thermal entropy per site by $s=S/N_s$.
An uncertainty $\delta\psi$ corresponds to $N_s\delta\psi$ in the total
partition logarithm. Accuracy per site and accuracy of the complete
system must therefore be stated separately. A Gibbs-state comparison
that depends on a total free-energy difference can be more demanding
than a comparison of intensive observables.

\subsection{The observation model}

A comparison with experimental data includes the observation procedure as
well as the many-body observable. Image analysis can average several sites
within a pixel, select particular occupancies, or combine measurements
from a finite region. Applying the same spatial averaging, detection
response and conditioning to the prediction gives the quantity to compare
with the reported data.

The observation model can be written as
\begin{equation}
 y_a=\mathcal M_a\!\left[O(H,\theta)\right]+\epsilon_a,
 \label{eq:measurement-map}
\end{equation}
where $O$ is a theoretical observable, $\theta$ denotes state and model
parameters, $\mathcal M_a$ specifies the measurement map and $\epsilon_a$
is the measurement error. Numerical accuracy concerns the evaluation of
$O$ for a specified input. The accuracy of an experimental comparison
also depends on $\theta$ and $\mathcal M_a$. Below, $\bm O(\theta)$ denotes the predicted measurements after this map has
been applied, and $J_{a\lambda}=\partial O_a/\partial\theta_\lambda$ is their
response matrix. The parameter dependence of this predicted signal
includes the measurement map as well as the many-body state.

\subsection{Error budget of a benchmark}

The error budget refers to a specified observable and normalization.
The sampling error of a Monte Carlo estimate depends on autocorrelation as well as on the number of measurements. A stationary scalar estimator with finite variance, summable autocorrelations
and integrated autocorrelation time
$\tau_{\mathrm{int}}=\tfrac12+\sum_{\ell\geq1}\rho(\ell)$ has large-sample variance
\begin{equation}
 \operatorname{Var}\overline O
 \simeq \frac{2\tau_{\mathrm{int}}}{M}
 \operatorname{Var}O_{\mathrm{inst}},
 \label{eq:mc-variance}
\end{equation}
where $M$ is the number of measured steps. Consequently, if a step
costs $C_{\mathrm{step}}$, reaching a specified statistical precision
depends on the product
\begin{equation}
 C_{\mathrm{step}}\,
 2\tau_{\mathrm{int}}\,
 \operatorname{Var}O_{\mathrm{inst}},
 \label{eq:cost-variance}
\end{equation}
as well as on the target error. Ratio estimators and several jointly
measured observables require their corresponding covariance treatment.
An algorithm can gain by making each integrand cheaper to evaluate, reducing its variance or exploring configurations more efficiently \cite{ShiLargeNf2025,MCMCMC2021}.

Alongside sampling uncertainty, the error budget includes the effects of
finite size, time discretization, cluster or diagram order, and numerical
representation. Their relative importance depends on the method and the
observable. For a continued series, variation across reconstruction choices
measures sensitivity to that choice, and successive truncations show how
the estimate responds to additional coefficients. Further orders and independent reference values help assess the omitted
contributions. A set of reference cases can also be used to examine whether
proposed reconstruction intervals attain their stated coverage.

Calibrated parameters contribute a further covariance
$J C_{\mathrm{cal}}J^{\mathsf T}$ to the predicted measurements, to leading
order in their uncertainty. Here $C_{\mathrm{cal}}$ is the parameter
covariance and $J$ is the response matrix defined above. For independent
calibration and numerical errors this term adds to the numerical
covariance; shared fits or data produce cross terms. The balance of these
contributions determines how much improving the solver changes the final
comparison, as illustrated by the tasks in Table~\ref{tab:requirements}.

\begin{table}[htbp]
\centering
\small
\renewcommand{\arraystretch}{1.2}
\begin{tabularx}{\textwidth}{@{}>{\raggedright\arraybackslash}p{39mm}YY@{}}
\toprule
Use of reference data & Required numerical information & Error assessment\\
\midrule
Temperature inference & Observable and temperature dependence, including correlated nuisance parameters & Statistical covariance and numerical bias assessed on the scale of the inferred temperature.\\
Model discrimination & Several observables and their response to candidate Hamiltonian changes & Independent predictions at fixed fitted parameters; sensitivity to the proposed correction resolved.\\
Entropy or pressure reconstruction & A consistent thermodynamic surface and a known reference & Controlled integration, derivatives and cancellation errors.\\
Large-system comparison & Bulk data or the actual finite geometry, as appropriate & Size and observation-region effects distinguished from solver errors.\\
Nonlocal correlations & The same subsystem observable and ensemble as the measurement & Geometry, normalization and thermal contributions included.\\
\bottomrule
\end{tabularx}
\caption[Numerical information required for experimental inference]{Numerical information and error assessment for experimental inference.
}
\label{tab:requirements}
\end{table}

\subsection{Correlated errors in differences and derivatives}

Thermodynamic derivatives and differences can amplify input uncertainty. A central difference,
\begin{equation}
 \partial_\mu\psi\simeq
 \frac{\psi(\mu+h)-\psi(\mu-h)}{2h},
 \label{eq:finite-difference}
\end{equation}
has a truncation error that decreases with $h$ for a sufficiently smooth
function. The noise on each input enters with a coefficient of magnitude $1/(2h)$.
For independent inputs with equal standard deviation $\sigma$, the
resulting derivative has standard deviation $\sigma/(\sqrt{2}h)$.
With shared sampling, a positive covariance reduces this statistical
error, while a negative covariance increases it. Explicitly,
\begin{equation}
 \operatorname{Var}\!\left[
 \frac{\widehat\psi_+-\widehat\psi_-}{2h}\right]
 =\frac{\operatorname{Var}\widehat\psi_+
 +\operatorname{Var}\widehat\psi_-
 -2\operatorname{Cov}(\widehat\psi_+,\widehat\psi_-)}{4h^2}.
 \label{eq:fd-covariance}
\end{equation}
The derivative error estimate has to balance truncation against the $1/h$ growth of noise; reducing the difference step alone can lower the accuracy.

For several jointly estimated inputs, the
general linearized propagation rule is
\begin{equation}
 \operatorname{Var}F(\bm x)
 \simeq \bm\nabla F^{\mathsf T}C_{\bm x}\bm\nabla F.
 \label{eq:covariance-propagation}
\end{equation}
For a linear combination $F(\bm x)=\bm a^{\mathsf T}\bm x$, this
relation is exact, with $\bm\nabla F=\bm a$. For nonlinear quantities,
it is the leading small-error approximation about the input means.
Off-diagonal covariances enter with coefficients fixed by this gradient,
so shared sampling can either increase or decrease the final variance.
This formulation applies to differences of predictions, quadrature sums
and combinations of partition logarithms using their jointly estimated
inputs.

Thermodynamic identities provide additional diagnostics. For example,
the Maxwell relation
\begin{equation}
 \left.\partial_\mu s\right|_T
 =\left.\partial_T n\right|_\mu
 \label{eq:maxwell}
\end{equation}
tests derivatives extracted through different routes. Such a comparison must account for numerical differentiation errors and statistical inputs shared between the two routes. It is especially useful when the benchmark is intended to
support an entropy estimate or an isentropic trajectory, for which a
small inconsistency can propagate over a range of parameters.

\subsection{Parameter inference and model response}
\label{sec:reference-inference}

Error propagation also determines the precision with which parameters can
be inferred from observables. In thermometry, for example, a measured
value $y$ is compared with a calculated function $O(T)$, whose temperature
response converts the observable uncertainty into a temperature uncertainty. In a local
linear approximation, with other parameters fixed,
\begin{equation}
 \delta T\simeq\frac{\delta O}{|\partial_T O|}.
 \label{eq:thermometry-sensitivity}
\end{equation}
Here $\delta O$ represents the uncertainty relevant to the comparison;
independent statistical contributions from experiment and computation
can be combined in quadrature, while systematic shifts must be assessed
separately. Reducing numerical error matters most when it dominates the uncertainty of the inferred temperature.

Equation~\eqref{eq:thermometry-sensitivity} also identifies a limitation
of precision alone. If an observable is weakly dependent on temperature,
even an accurate value may be a poor thermometer. The numerical task is
then to calculate an observable with greater sensitivity, or to combine
several observables whose parameter dependences are complementary. A calculation over a range of temperatures can thus help select a thermometer as well as calibrate it.

For several unknown parameters, the linearized observation model is
$\bm y\simeq\bm O(\theta_0)+J\,\delta\theta+\bm\epsilon$.
Using the same response matrix $J$ to infer parameters from the data gives
the local Gaussian parameter covariance
\begin{equation}
 C_\theta\simeq (J^{\mathsf T}C^{-1}J)^{-1},
 \label{eq:parameter-covariance}
\end{equation}
where $C=\operatorname{Cov}(\bm\epsilon)$ is the covariance of the combined errors entering the comparison. Equation~\eqref{eq:parameter-covariance} is the inverse curvature of the Gaussian negative log-likelihood $\tfrac12(\bm y-\bm O)^{\mathsf T}C^{-1}(\bm y-\bm O)$ for the linearized model. This expression assumes
a locally identifiable linearized model, a fixed covariance $C$,
and a fit without an informative prior. A Gaussian prior adds its
precision matrix to $J^{\mathsf T}C^{-1}J$. A discrepancy between the physical model and the experiment would require an additional error term. Numerical noise in $\bm O$ and uncertainty in
its derivatives can degrade this inference. A systematic numerical bias can shift the inferred parameters. Define
$\bm b_{\mathrm{num}}=\bm O_{\mathrm{num}}-\bm O_{\mathrm{exact}}$
at the true parameters. To the same linear order, fitting the numerical
model to the measurements produces the parameter shift
\begin{equation}
 \delta\theta_{\mathrm{bias}}
 \simeq -(J^{\mathsf T}C^{-1}J)^{-1}
 J^{\mathsf T}C^{-1}\bm b_{\mathrm{num}}.
 \label{eq:parameter-bias}
\end{equation}
An apparently precise fitted temperature can therefore inherit a bias
from its numerical reference.

Observables excluded from a fit can also test candidate corrections to the
Hamiltonian. Suppose a small perturbation $\delta H=\lambda\widehat X$ changes an
observable by
\begin{equation}
 \delta O=\lambda\chi_{OX}+O(\lambda^2),
 \qquad
 \chi_{OX}=\left.\frac{\partial\langle\widehat O\rangle}
 {\partial\lambda}\right|_{\lambda=0}.
 \label{eq:model-response}
\end{equation}
For a parameter-independent $\widehat O$, its equilibrium response can
be expressed through an imaginary-time connected correlation,
\begin{equation}
 \chi_{OX}=-\int_0^\beta\dd\tau\,
 \langle\delta\widehat X(\tau)\,
 \delta\widehat O(0)\rangle,
 \label{eq:kubo-static}
\end{equation}
with the usual imaginary-time ordering understood. A direct parameter
dependence of $\widehat O$ adds its explicit derivative. Accurate
calculations of such responses quantify whether an omitted Hamiltonian
term could explain an observed discrepancy. 

The observable's susceptibility sets the required tolerance on a Hamiltonian parameter and must be calculated at the relevant temperature and filling. Work on error
propagation in quantum simulation provides a broader setting for this
question, although bounds for one class of dynamical tasks cannot be
transferred unchanged to equilibrium thermodynamics
\cite{Flannigan2022Errors}.

\section{Cold atoms as quantitative many-body experiments}
\label{c1:v7:cold-atoms}
\label{sec:experimental-benchmark-evidence}
Optical lattices give access to measurements ranging from local occupation
to spatially resolved correlations. Numerical references play several
roles in these experiments: inferring temperature from an equation of
state or a fluctuation, testing the magnetic environment of a dopant, and
comparing thermodynamics across flavour numbers. The following account
groups the experimental developments by these physical tasks and by the
observables and parameter ranges that enter each comparison.

\subsection{Optical-lattice realizations and quantum gas microscopy}

Jaksch \emph{et~al.}\@ derived a tunable Bose--Hubbard model for atoms in
optical lattices, and Greiner \emph{et~al.}\@ observed the superfluid-to-Mott
transition \cite{CA_Jaksch1998,CA_Greiner2002}. In the fermionic case,
J\"ordens \emph{et~al.}\@ identified a Mott-insulating regime through suppressed
double occupancy, a reduced compressibility inferred from its atom-number dependence, and an excitation gap
\cite{CA_Jordens2008}. The relation between experimental controls and the
Fermi--Hubbard Hamiltonian is reviewed in Ref.~\cite{CA_Esslinger2010}.
Identifying an insulating regime is only the first step towards determining
its temperature. A joint prediction of density, double occupancy and their
response to particle number constrains that inference more strongly than
one observable alone.

Tunable honeycomb lattices allowed Dirac points to be moved and merged,
while periodic driving realized the Haldane band structure
\cite{CA_Tarruell2012,CA_Jotzu2014}. These experiments establish
single-particle dispersions that can subsequently enter an interacting
calculation. A fermionic optical superlattice has also been used to
engineer local offsets and spin couplings \cite{CA_Chalopin2025}.
This range of controls, reviewed by Sch\"afer \emph{et~al.}\@, makes the
hopping matrix, sublattice structure and orbital content important inputs
to a many-body solver \cite{CA_Schafer2020}.

Site-resolved imaging made the occupation distribution itself accessible.
The first quantum gas microscopes resolved individual site occupancies
\cite{CA_Bakr2009,CA_Sherson2010}; in the bosonic Mott-insulator experiment
of Ref.~\cite{CA_Sherson2010}, the density and its variance already gave
temperature and entropy information through comparison with theory.
Resolving these quantities spatially distinguishes a cold incompressible
region from the more entropic part of a trapped cloud. Fermionic
microscopes subsequently extended this access to potassium and lithium
\cite{CA_Cheuk2015,CA_Parsons2015,CA_Haller2015}. Comparisons of the measured
correlators include missing atoms, spin-readout errors and total-number
fluctuations in the observation model.

Local fluctuations provide a natural progression from the ideal gas to
interacting phases. Suppressed number fluctuations revealed Pauli blocking
in a spin-polarized lattice gas \cite{CA_Omran2015}, giving a noninteracting
reference with which to assess detection. In the repulsive two-dimensional
model, local-moment measurements distinguished metallic, Mott-insulating
and band-insulating regimes \cite{CA_Cheuk2016}. The occupation variance
and local moment therefore add experimentally accessible tests of the
state beyond its mean density.

\subsection{Equation-of-state thermometry and trap thermodynamics}
\label{sec:indispensable}

A trapped density profile contains thermodynamic information through the
spatial variation of chemical potential. Ho and Zhou developed methods
to extract homogeneous properties from that profile under equilibrium
and a known confinement potential \cite{CA_Ho2010}. A numerical equation
of state supplies the temperature and interaction dependence, allowing
different radial regions to test a common description of the cloud.
Fluctuation measurements provide a complementary route: M\"uller
\emph{et~al.}\@ used the fluctuation--dissipation relation with a
local-density approximation, and Sanner \emph{et~al.}\@ measured suppressed
noise relative to an ideal-gas reference \cite{CA_Muller2010,CA_Sanner2010}.
Because these procedures have different systematic uncertainties from an
interacting equation-of-state fit, their comparison tests the temperature
inference as well as the underlying thermodynamics.

In continuum Fermi gases, precise thermodynamics has already constrained
many-body descriptions. Nascimb\`ene \emph{et~al.}\@ extracted a homogeneous
equation of state, and Horikoshi \emph{et~al.}\@ measured universal
thermodynamic functions \cite{CA_Nascimbene2010,CA_Horikoshi2010}.
Ku \emph{et~al.}\@ identified the superfluid transition through density,
pressure and compressibility \cite{CA_Ku2012}. In these continuum experiments, thermodynamic derivatives revealed features that are hard to resolve in the density alone. A direct equation-of-state comparison with bold DiagMC provided an
independent numerical benchmark for this thermodynamics
\cite{VanHoucke2012Emulator}.

\label{sec:benchmark-inference}

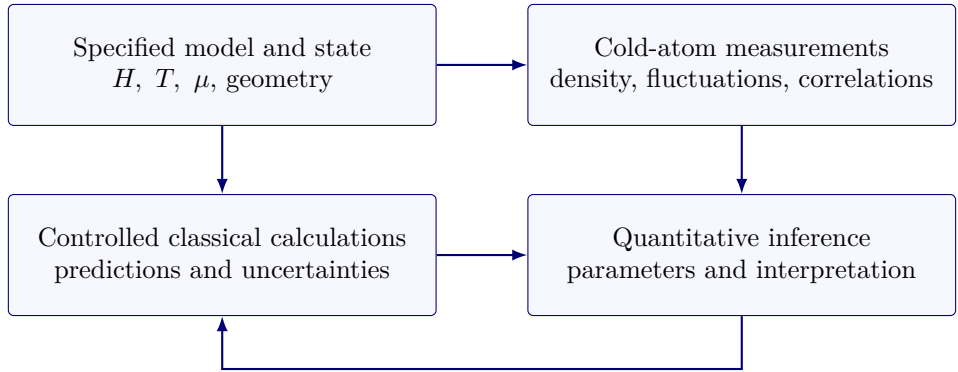
\begin{figure}[!htbp]
\centering
\begin{tikzpicture}[node distance=9mm and 12mm,
  box/.style={draw=blue!45!black,rounded corners=2pt,fill=blue!3,
  text width=54mm,minimum height=16mm,align=center,font=\small},
  arr/.style={-{Latex[length=2mm]},thick,draw=blue!45!black}]
\node[box] (model) {Specified model and state\\$H,\ T,\ \mu,$ geometry};
\node[box,right=of model] (experiment) {Cold-atom measurements\\density, fluctuations, correlations};
\node[box,below=of model] (numerics) {Controlled classical calculations\\predictions and uncertainties};
\node[box,below=of experiment] (inference) {Quantitative inference\\parameters and interpretation};
\draw[arr] (model) -- (experiment);
\draw[arr] (model) -- (numerics);
\draw[arr] (experiment) -- (inference);
\draw[arr] (numerics) -- (inference);
\draw[arr] (inference.south) -- ++(0,-7mm) -| (numerics.south);
\end{tikzpicture}
\caption[Precision calculations in quantum simulation]{Calculation and measurement in an equilibrium quantum-simulation benchmark. Both use the specified Hamiltonian, state and geometry. The return arrow represents a further calculation prompted by the comparison, such as a more temperature-sensitive observable or a refined parameter grid. }
\label{fig:benchmark-cycle}
\end{figure}

In a trapped gas, temperature and chemical potential are usually inferred
from measurements rather than set independently. An equation of state
must be evaluated over the possible parameter range, and the prediction
must include confinement and the observation region. Diagrammatic
thermodynamics has been compared with the equation of state of a unitary
Fermi gas \cite{VanHoucke2012Emulator}. In lattice systems, dynamical
mean-field calculations helped interpret metallic and insulating profiles
\cite{Schneider2008Fermions}, while later two-dimensional measurements
used numerical linked-cluster references
\cite{CocchiEoS2016,Cocchi2017}. Figure~\ref{fig:benchmark-cycle} shows the
calculation and measurement steps in such an equilibrium comparison.

Fluctuation relations provide another route to thermometry. For a
Hamiltonian satisfying $[H,\widehat N]=0$, the grand-canonical ensemble
gives
\begin{equation}
 \frac{\partial\langle\widehat N\rangle}{\partial\mu}
 =\beta\bigl(\langle\widehat N^2\rangle
 -\langle\widehat N\rangle^2\bigr).
 \label{eq:fdt-number}
\end{equation}
Here particle number is conserved under Hamiltonian evolution, while the
grand-canonical ensemble includes states with different particle numbers.
Fluctuation-based thermometry that requires no numerical equation of state has been proposed within a local-density approximation
\cite{Zhou2011Thermometry} and implemented using density correlations
in a Fermi--Hubbard gas \cite{Hartke2020Thermometry}.
Using this relation in a finite measured region requires attention to
which fluctuations are observed and which susceptibility is inferred.
The variance of a site occupancy, the variance of a regional particle
number and the total-number variance are different objects. Calculations
of spatial correlations help connect these objects and determine the
validity of a local approximation. 

At lower temperatures, spin correlations become powerful thermometers.
McKay and DeMarco reviewed the difficulty of cooling strongly correlated
gases in optical lattices \cite{CA_McKay2011}.
Xu \emph{et~al.}\@ reported $T/t=0.05_{-0.05}^{+0.06}$ at half filling by comparing measured spin correlations with numerically exact calculations \cite{Xu2025Cryogenic}. In the doped setting, they compared short-range spin correlations with constrained-path auxiliary-field results. The additional constraint bias must be considered when using that comparison for thermometry.

\subsection{Spin, charge and higher-order correlations}

Magnetic correlations extend thermometry beyond the charge equation of
state. Entropy redistribution made nearest-neighbour spin correlations
observable in optical lattices \cite{CA_Greif2013}; spin-sensitive Bragg
scattering in three dimensions provided a complementary thermometer
through comparison with DQMC and NLCE \cite{CA_Hart2015}.

Microscopy resolved the two-dimensional spin correlator \cite{CA_Parsons2016}, charge and spin correlations as functions of doping \cite{CA_Cheuk2016Correlations}, and spin- and density-resolved antiferromagnetic correlations in Hubbard chains \cite{CA_Boll2016}.
The measured dependence on separation, density and temperature can be compared with calculated correlators using the experimental normalization of the spin operator.
Uniform-susceptibility measurements also probed the suppression of
antiferromagnetic correlations away from half filling
\cite{CA_Drewes2017}.

The spatial scale of magnetic correlations also determines the appropriate
reference. Mazurenko \emph{et~al.}\@ observed correlations extending across
an approximately eighty-site two-dimensional system
\cite{CA_Mazurenko2017}. Since the infinite isotropic two-dimensional
model has no antiferromagnetic long-range order at finite temperature,
the comparison concerns this finite sample and its boundaries. By
contrast, the three-dimensional experiment of Shao \emph{et~al.}\@
studied a bulk antiferromagnetic transition in a uniform lattice of
approximately $800\,000$ sites \cite{CA_Shao2024}, for which the numerical
transition and its finite-size extrapolation are the relevant targets.

In one-dimensional chains, string correlators revealed magnetic order
hidden by mobile charge defects, and the magnetic wavevector was tracked
as doping and spin imbalance changed \cite{CA_Hilker2017,CA_Salomon2019}.
Vijayan \emph{et~al.}\@ followed the dynamical counterpart of this spin--charge
structure, observing their real-time separation after particle removal
\cite{CA_Vijayan2020}.

In two dimensions, conditional imaging tests how a dopant changes its
magnetic environment. Patterns consistent with geometric strings were
found in the doped Hubbard model \cite{CA_Chiu2019}, and measurements
around individual magnetic polarons followed the evolution towards
Fermi-liquid behaviour with increasing doping
\cite{CA_Koepsell2019,CA_Koepsell2021}. These joint spin--charge moments resolve the magnetic environment of a
dopant beyond its local density or energy. Chalopin \emph{et~al.}\@ measured
spin--charge correlators up to fifth order and compared their temperature
and doping dependence with DQMC and minimally entangled typical
thermal-state calculations, relating the resulting correlation scales
to the pseudogap regime \cite{CA_Chalopin2026}.

Triangular-lattice Mott experiments used NLCE and Monte Carlo results
for thermometry \cite{CA_Mongkolkiattichai2023}. Continuously changing
from square to triangular geometry revealed different magnetic
responses on the particle- and hole-doped sides \cite{CA_Xu2023}.
The observation of Nagaoka polarons resolved ferromagnetic regions
around particle dopants and contrasting correlations near hole dopants
\cite{CA_Lebrat2024}. Modelling these results requires the actual hopping signs and lattice geometry. A nearest-neighbour bipartite model is particle--hole symmetric and would miss the difference between the particle- and hole-doped sides.

Ladders and mixed-dimensional couplings give another set of test cases.
Hubbard ladders were used to probe a finite-temperature Haldane regime
with nonlocal edge and bulk diagnostics \cite{CA_Sompet2022}.
Experiments with engineered mixed-dimensional couplings observed magnetically mediated hole pairing \cite{CA_Hirthe2023} and, subsequently, individual stripe structures \cite{CA_Bourgund2025}. These anisotropic couplings and geometries enter the one-body matrix of a diagrammatic calculation. Changing that matrix is straightforward, although the convergence of the resulting series may change substantially.

Charge transport in a doped lattice gas and spin transport in a Mott
insulator probe the motion of the corresponding conserved quantities
\cite{CA_Brown2019,CA_Nichols2019}. A thermodynamic calculation contributes
the static susceptibility, such as the compressibility that relates
charge diffusion to conductivity.

For the attractive model, doublon-density measurements and Monte Carlo
comparisons constrained pairing correlations \cite{CA_Mitra2018}.
Spin- and density-resolved imaging later measured nonlocal fermion
pairing directly \cite{CA_Hartke2023}. Local double occupancy, pair
extent and long-range phase coherence test different properties of
the paired state. The distinction between fitted observables and independent predictions follows the procedure discussed in Section~\ref{sec:indispensable}.

\subsection{Enlarged \texorpdfstring{$\mathrm{SU}(N_f)$}{SU(Nf)} flavour symmetries}
\label{sec:requirements}

Flavour multiplicity and orbital couplings are distinct controls in
experiments with alkaline-earth-like atoms. A degenerate mixture with
$\mathrm{SU}(2)\times\mathrm{SU}(6)$ symmetry was realized by Taie
\emph{et~al.}\@ \cite{CA_Taie2010}. Nuclear-spin-independent interactions
in strontium probe the interaction symmetry \cite{CA_Zhang2014}, while
the two-orbital ytterbium experiment accesses spin exchange with
$\mathrm{SU}(6)$ symmetry \cite{CA_Scazza2014}.

The effect of varying spin multiplicity was measured in a repulsive
one-dimensional liquid \cite{CA_Pagano2014}. Comparisons of two and
four components in an optical superlattice found enhanced magnetic
correlations associated with Pomeranchuk cooling \cite{CA_Ozawa2018}.
Together with the two-dimensional equation-of-state measurements
\cite{Pasqualetti2024EoS}, these studies call for calculations of entropy
and correlations as functions of flavour number. Chapters~\ref{c2:chapter} and~\ref{c3:chapter} develop inverse-flavour expansion and flavour-dependent summation methods towards these calculations, applying them here to density and $\ln Z$ at fixed $N_f$. Spin-resolved strontium microscopy with access to all ten spin-state
occupations has been reported in a preprint \cite{CA_GasFerrer2026},
opening a route to flavour-resolved correlation measurements.

Hazzard \emph{et~al.}\@ calculated trapped SU($N_f$) Hubbard thermodynamics
with a high-temperature expansion in a regime where temperature exceeds
the tunnelling scale. They used it to predict final temperatures after
adiabatic loading from a noninteracting gas and compared the effect of
flavour number at specified initial conditions
\cite{Hazzard2012HighTemperature}.

The two-dimensional $\mathrm{SU}(N_f)$ experiment of Pasqualetti
\emph{et~al.}\@ provides a particularly relevant example. It measured
equations of state and occupation probabilities for $N_f=3,4,6$ in a
single-layer square lattice, with approximately two thousand atoms,
and compared the observations with determinant Monte Carlo and
numerical linked-cluster calculations. Density fluctuations provided a second estimate of temperature \cite{Pasqualetti2024EoS}. Numerical results were used both to determine the state and as predictions tested against the measurements. 

A central objective for new methods is to enlarge the region in which
independent comparisons are possible. For the $\mathrm{SU}(N_f)$ lattice
problem, this means obtaining thermodynamic information where increasing
flavour number and decreasing temperature restrict established solvers.
Experimental studies of the Mott crossover and antiferromagnetic correlations show that these questions are accessible in atomic systems \cite{Hofrichter2016Mott,Taie2022SUNCorrelations}.
In the lowest-entropy two- and three-dimensional regimes of
Ref.~\cite{Taie2022SUNCorrelations}, the numerical comparisons were
limited by the convergence range of the methods used.

The connection between the equation-of-state measurements of
Ref.~\cite{Pasqualetti2024EoS} and the subsequent CoS calculation in
Ref.~\cite{KozikCoS2024} illustrates this strategy. The numerical method
was compared in an overlapping regime before being used to investigate
lower-temperature $\mathrm{SU}(6)$ thermodynamics. The strong-coupling results also required series continuation, as discussed in Section~\ref{sec:series-control}.

Within an overlap region, calculations can be compared over a range of interaction strengths, densities and temperatures. Agreement tests the reported errors and gives reference points for later calculations, including expansions formulated with a different reference action.

\section{Reference calculations for quantum hardware}
\label{c1:v7:hardware}

Fermionic models can be implemented by encoding occupations in qubits
or by engineering the physical couplings of an analogue simulator.
Both approaches produce many-body predictions through a specified state
preparation and measurement protocol. Their native interactions, control
and readout differ across the architectures surveyed by Altman
\emph{et~al.}\@ \cite{Altman2021Architectures}. Translating these features
into a common Hamiltonian and observable defines the reference problem
against which their outputs can be compared.

\subsection{Application benchmarks for interacting fermions}

Device characterization measures properties such as gate errors, leakage
or measurement fidelity. An application benchmark asks how accurately a
specified calculation has been performed. The distinction matters when
many operations are combined, errors are correlated, or the reported
observable is insensitive to some departures from the target state.
Certification protocols differ in the information they require and in
the conclusions they support \cite{Eisert2020Certification}. For the
applications considered here, a concrete target is an energy, density,
correlation function or subsystem purity at specified physical parameters.

Ground-state preparation illustrates how a physical application defines
its own accuracy criterion. Dallaire-Demers \emph{et~al.}\@ proposed a
one-dimensional Fermi--Hubbard benchmark in which the Bethe-ansatz energy
sets an effective fermionic length: the longest chain satisfying their
criterion \cite{DallaireDemers2020Benchmark}. On superconducting hardware,
Stanisic \emph{et~al.}\@ used a Hamiltonian variational ansatz and error
mitigation to study $1\times8$ and $2\times4$ lattices, comparing energies,
densities and spin correlations with classical calculations
\cite{Stanisic2022HubbardHardware}. Their occupation encoding uses two
qubits per spinful site, so the 16-qubit calculation represents eight
sites. These examples benchmark ground states. Changing the dimension,
observable or temperature changes the reference problem; in particular,
a thermal comparison specifies the ensemble and temperature as well as
the Hamiltonian.

For a small lattice, exact diagonalization can calculate the entire
spectrum and the required thermal trace. Beyond that range, a classical
method may still evaluate a selected observable without storing all
many-body eigenstates. This is the setting in which the thermodynamic
calculations of Chapters~\ref{c2:chapter} and~\ref{c3:chapter} can provide
reference data. Their usefulness depends on the temperature, coupling,
flavour number and accuracy of the particular series. A reference grid
should state these parameters, the lattice and boundary conditions, and
the uncertainty of each observable. A larger grid can then be compared
with new devices or improved preparations without repeating an exact
state-vector calculation.

\subsection{Native-electron simulators and the effective model}

Semiconductor quantum dots offer another experimental use of fermionic
reference calculations. Hensgens \emph{et~al.}\@ independently tuned
filling and tunnel coupling in a triple-dot array and characterized the
crossover from individual to collective Coulomb blockade
\cite{Hensgens2017QuantumDots}. Dehollain \emph{et~al.}\@ studied three
electrons on a four-dot plaquette, using pairwise spin readout to follow
the change in spin-state probabilities as the geometry was tuned towards an open chain
\cite{Dehollain2020Nagaoka}. In these devices the mobile particles are
electrons, and gate control changes the terms of their effective lattice
Hamiltonian.

Charge-addition thresholds, local occupations and spin probabilities
probe the confinement and charging interactions of these arrays under
their preparation conditions. The charging model of
Ref.~\cite{Hensgens2017QuantumDots}, for example, contains both onsite
and intersite Coulomb terms. Their relative importance sets the range
in which an onsite Hubbard description is adequate, independently of
the numerical precision used to solve that reduced model.

A calibrated finite-array calculation can first be compared with charging
or correlation measurements on the same geometry. Varying tunnel
coupling, filling and local offsets then tests whether one parameter set
describes several measurements. For larger arrays, the corresponding
reference calculation retains the device's interaction terms and targets
the chosen observables at its experimental temperature. Extending the
diagrammatic methods below to such a device entails deriving the
contraction rules for those additional interactions.

\subsection{Calibrated parameters and predictions outside the fit}

Calibration establishes the relation between a device's nominal controls
and the parameters of its effective Hamiltonian. In an encoded simulation, calibration and
compilation determine the implemented hopping and interaction gates. In
an analogue realization, confinement, coupling inhomogeneity and unwanted
terms can affect the mapping to the model. The resulting discrepancy in
a measured observable depends on its parameter response. Flannigan
\emph{et~al.}\@ studied the propagation of several error sources in
quantitative quantum simulation, including analogue and digital
implementations of many-body dynamics \cite{Flannigan2022Errors}.
Their observable-dependent analysis motivates specifying an error target
for the physical quantity being compared.

The observation model of Section~\ref{sec:reference-inference} includes
this hardware calibration. In a thermal preparation, density and double
occupancy can constrain temperature, chemical potential and interaction
strength, after which spin correlations test the same calibrated model.
Repeating that comparison at larger sizes probes whether one effective
Hamiltonian continues to describe the preparation. Parameter errors and
correlations shared with the calibration enter through the joint inference
framework already established in Section~\ref{sec:reference-inference}.

Particle-number postselection illustrates how the observation procedure
enters this test. It conditions the ensemble on a fixed sector and removes
the number fluctuations present in a grand-canonical calculation, which
can alter finite-system correlations even at the same mean density. The
reference can account for this by using the selected sector or calculating
the conditioning explicitly. Spatial averaging, readout corrections and
the normalization of density enter the same observation map before fitting.

\subsection{Testing mitigation and comparing platforms}

An independent reference allows an error-mitigation procedure to be
assessed through its effect on the observable, its bias and the measurements
needed for a given precision. Postselection, for example, can suppress
some errors at the cost of discarding outcomes, so a reduction in bias
may increase the number of raw shots. In the Fermi--Hubbard experiment of
Ref.~\cite{Stanisic2022HubbardHardware}, number-sector postselection and
symmetry averaging were combined with training based on classically
tractable fermionic linear-optics circuits.

For a finite-depth preparation, two classical comparisons distinguish
errors of the ideal preparation algorithm from errors introduced by the
hardware. Evaluating the ideal trial circuit gives its predicted energy;
the equilibrium energy of the specified Hamiltonian sets the physical
target. The methods in this thesis provide that equilibrium target once
the state and ensemble are specified. Assessing the trial circuit itself
also involves its state representation or a suitable circuit simulation.

Cross-platform comparisons add another source of evidence. Elben
\emph{et~al.}\@ developed randomized-measurement protocols for overlaps
and purities of states prepared on separate devices, and demonstrated
experiment--theory comparisons for entangled ten-qubit states in an ion
simulator \cite{Elben2020CrossPlatform}. Comparing a few thermodynamic
observables imposes fewer constraints than such state-overlap measurements.
It can nevertheless be performed for the observables that a given
many-body calculation resolves reliably. A subsystem purity adds a
nonlinear constraint to energy and density, as discussed in
Section~\ref{c1:v7:renyi}.

Cross-device comparisons and classical references expose different error
sources. Two platforms can share an approximation or an incorrectly
specified target, while a numerical reference can carry truncation or
representation error. Varying the preparation and numerical approximations
across several sizes and parameter points helps locate the source of a
common discrepancy.

\subsection{Cost at a specified physical accuracy}

Resource comparisons require the same final error target. The runtime of
a classical calculation includes preparation of propagators and structural
data, repeated integration and reconstruction of the physical answer.
Equations~\eqref{eq:mc-variance} and~\eqref{eq:cost-variance} express the
sampling cost at a specified statistical precision in terms of the
per-update cost, variance and integrated autocorrelation time. A faster connected evaluator
can reduce that cost, but its gain in a complete calculation depends on
how much time is spent in the other operations and whether the estimator
changes. Section~\ref{c3:fast:section} measures a fixed-configuration
contraction cost; using that evaluator inside a sampler is a further
comparison.

Daley \emph{et~al.}\@ discuss practical quantum advantage in terms of
physical problems, relevant observables and the classical methods
available for their calculation \cite{Daley2022PracticalAdvantage}.
The reference regime can expand as either hardware or classical algorithms
improve. For equilibrium fermions, systematic variation of diagram order,
reference shift and system size is needed alongside the sampling error.
If the dominant uncertainty is in continuation, increasing the number of
measurements in Eq.~\eqref{eq:mc-variance} will improve the
coefficient estimate without determining the omitted orders.

An equilibrium reference may also test the beginning or endpoint of a
dynamical experiment. One can compare late-time local observables with a
thermal prediction at energy and particle number fixed by the conserved
quantities, provided thermalization is a justified hypothesis. Integrable
systems, additional conserved charges and finite-size ensemble differences
require their corresponding descriptions. 

\section{Partition functions and free energies}
\label{c1:v7:free-energy}

The absolute logarithm of the partition function is needed to normalize
a Gibbs state or to determine the minimum of a thermal variational
objective. Its parameter derivatives predict observables, and differences
between parameter points enter thermodynamic transformations. The connected
vacuum diagrams evaluated in Chapters~\ref{c2:chapter} and~\ref{c3:chapter}
provide a route to this logarithm.

\subsection{Observable predictions from a common thermodynamic potential}

Use the grand-canonical convention $K=H-\mu\widehat N$ and
$\Omega=-\beta^{-1}\ln Z$, with $k_{\rm B}=1$. When $H$ has no explicit
temperature dependence, its first derivatives give
\begin{equation}
 \left.\frac{\partial\Omega}{\partial\mu}\right|_{T}=-\langle\widehat N\rangle,
 \qquad
 \left.\frac{\partial\Omega}{\partial T}\right|_{\mu}=-S_{\rm th},
 \qquad
 \frac{\partial\Omega}{\partial g}
 =\left\langle\frac{\partial H}{\partial g}\right\rangle,
 \label{c1:v7:hf:potential-derivatives}
\end{equation}
where $g$ is a physical Hamiltonian parameter, and the other parameters
are fixed in the final derivative. For the onsite coupling convention
$H_{\rm int}=U\sum_i\sum_{a<b}n_{ia}n_{ib}$, this derivative counts
onsite pairs. Its conversion to a quoted double occupancy depends on
whether the pairs are summed over flavours or normalized per pair.
The thermodynamic entropy is
\begin{equation}
 S_{\rm th}=\ln Z+\beta\bigl(E-\mu\langle\widehat N\rangle\bigr),
 \qquad E=\langle H\rangle.
 \label{c1:v7:hf:thermodynamic-entropy}
\end{equation}
Thus a reference based on $\ln Z$ can be compared with measurements even
when the thermodynamic potential itself is not directly measured.

Derivatives demand control of parameter dependence as well as accuracy
at isolated points. For example, a nearly constant offset in an
approximation to $\Omega$ may have little effect on a density derivative,
while a small rapidly varying error can have a large effect. Conversely,
accurate derivatives determine the potential only up to a constant on
a connected parameter domain. An equation of state, an absolute free energy and a response function
therefore probe different aspects of the same thermodynamic surface:
its derivatives, its additive normalization and its local curvature.

\subsection{An absolute reference for thermodynamic integration}

At fixed temperature and interaction, a measured or calculated density
can reconstruct a grand-potential difference:
\begin{equation}
 \Omega(T,\mu)-\Omega(T,\mu_0)
 =-\int_{\mu_0}^{\mu}\langle\widehat N\rangle(T,\mu')\,\mathrm d\mu'.
 \label{c1:v7:hf:mu-integration}
\end{equation}
Cocchi \emph{et~al.}\@ obtained pressure and entropy from thermodynamic
measurements of a two-dimensional Hubbard gas, using the density profile
to reconstruct pressure \cite{Cocchi2017}. This illustrates why the
integration range and its normalization enter an experimental error
budget. A reference value of $\Omega$ at one endpoint fixes the constant;
values along the path can test interpolation and the contribution of
unmeasured low-density tails.

An analogous relation holds when energy is integrated over temperature.
For fixed $\mu$ and Hamiltonian parameters,
\begin{equation}
 \ln Z(\beta_2)-\ln Z(\beta_1)
 =-\int_{\beta_1}^{\beta_2}
 \langle K\rangle_{\beta}\,\mathrm d\beta.
 \label{c1:v7:hf:beta-integration}
\end{equation}
For a finite lattice with the full fermionic Fock space, the infinite-
temperature value is $\ln Z(0)=N_sN_f\ln2$. Restricting the trace to a
fixed number sector changes this reference to the logarithm of that
sector's dimension. In either convention, obtaining a low-temperature
potential from integration requires information over the intervening
range. Direct diagrammatic values at selected temperatures can check
that reconstruction or shorten the interval that must be integrated.

The physical interaction strength provides a further integration path.
At fixed $T$ and $\mu$, write $H(U)=H_0+UV$ with
$V=\sum_i\sum_{a<b}n_{ia}n_{ib}$. Then
\begin{equation}
 \Omega(U)-\Omega(0)=\int_0^U\langle V\rangle_{U'}\,\mathrm dU'.
 \label{c1:v7:hf:coupling-integration}
\end{equation}
The noninteracting endpoint is known from single-particle energies.
This relation can compare direct connected-vacuum summation with an
integration of pair occupancies. At fixed chemical potential, the density
is allowed to change along the path. A fixed-density comparison must
instead include the required chemical-potential adjustment and use the
appropriate thermodynamic potential. In a finite system, matching the
mean particle number of a grand-canonical state does not exactly reproduce
a canonical number projection. The chosen ensemble therefore belongs to
the definition of the reference data.

A density offset accumulates over the full chemical-potential interval,
while a sampled integration path combines values that may share
calibration or Monte Carlo inputs. For a discretized quadrature, the
integral is a weighted linear combination of those values, so its
statistical error follows directly from Eq.~\eqref{eq:covariance-propagation}.

\subsection{The minimum free energy as a target for thermal-state preparation}

A Gibbs state minimizes a thermal free-energy functional over normalized
density operators. Keeping the grand-canonical notation, define
\begin{equation}
 \mathcal G_\beta[\sigma]
 =\operatorname{Tr}(\sigma K)-\beta^{-1}S_{\rm vN}(\sigma),
 \qquad S_{\rm vN}(\sigma)=-\operatorname{Tr}(\sigma\ln\sigma).
 \label{c1:v7:hf:gibbs-objective}
\end{equation}
For $\rho_\beta=e^{-\beta K}/Z$, substituting
$\ln\rho_\beta=-\beta K-\ln Z$ into the quantum relative entropy gives
\begin{align}
 D(\sigma\Vert\rho_\beta)
 &=\operatorname{Tr}\sigma(\ln\sigma-\ln\rho_\beta)\notag\\
 &=-S_{\rm vN}(\sigma)+\beta\operatorname{Tr}(\sigma K)+\ln Z\notag\\
 &=\beta\bigl(\mathcal G_\beta[\sigma]-\Omega\bigr)\geq0.
 \label{c1:v7:hf:free-energy-relative-entropy}
\end{align}
The exact $\Omega$ is therefore an absolute target for a variational
thermal calculation. Setting $K=H$ gives the usual free-energy functional. A canonical
fixed-$N$ calculation additionally restricts the trace and the candidate
states to that particle-number sector.

Chowdhury, Low and Wiebe proposed a variational Gibbs-state algorithm
based on free-energy minimization, with entropy estimation through
Fourier approximations to the logarithm \cite{Chowdhury2020Gibbs}.
Consiglio \emph{et~al.}\@ used two coupled registers to parameterize the
probabilities and basis of a thermal state. They benchmarked Ising and
XXZ models in simulations and tested the Ising preparation on IBM devices
\cite{Consiglio2024Gibbs}. These are concrete settings in which a
variational objective can be monitored during optimization, while an
independent equilibrium calculation determines how far the final value
lies above the true minimum.

For a family
$\sigma_\theta=U_\theta\operatorname{diag}(p_\theta)U_\theta^\dagger$,
unitary invariance gives
$S_{\rm vN}(\sigma_\theta)=-\sum_jp_{\theta,j}\ln p_{\theta,j}$.
The chosen parameterization can therefore make entropy accessible even
though generic density-matrix reconstruction is costly. This also
introduces a specific source of uncertainty: the entropy estimated from
finite counts is a nonlinear function of the inferred probabilities.
The energy and entropy errors both enter the variational free energy.
In a noisy implementation, the entropy in the objective must describe
the actual prepared state. Noise-induced mixing can alter it relative to
the entropy obtained from the ideal circuit probabilities.

Equation~\eqref{c1:v7:hf:free-energy-relative-entropy} identifies precisely
which extra information is needed to interpret a free-energy comparison
as a state-quality statement. The candidate must be a normalized positive
density operator, and its energy and von Neumann entropy must be known
with controlled errors. The second R\'enyi entropy studied in
Chapter~\ref{c4:chapter} is a different functional. For an extensive system the same free-energy-density difference
$\delta g$ gives $D=\beta N_s\delta g$, so its effect on the total relative
entropy grows with system size. A global state comparison can consequently
demand much tighter per-site precision than a local-observable benchmark.

\subsection{Learning physical couplings from thermal data}

Suppose the inverse temperature is independently known and the candidate
Hamiltonian family is linear in its parameters,
\begin{equation}
 H_{\boldsymbol\theta}=\sum_a\theta_a O_a,
 \qquad
 \rho_{\boldsymbol\theta}
 =\frac{e^{-\beta H_{\boldsymbol\theta}}}{Z(\boldsymbol\theta)}.
 \label{c1:v7:hf:learning-family}
\end{equation}
The $O_a$ are specified Hermitian operators, for example hopping,
onsite-pair or number terms with a fixed sign convention. The generator
can equally be a grand Hamiltonian with physical chemical potentials
included among the parameters. If measured averages are
$m_a=\operatorname{Tr}(\rho_{\rm data}O_a)$, the quantum cross-entropy
objective is
\begin{equation}
 \mathcal L(\boldsymbol\theta)
 =-\operatorname{Tr}(\rho_{\rm data}\ln\rho_{\boldsymbol\theta})
 =\beta\sum_a\theta_a m_a+\ln Z(\boldsymbol\theta).
 \label{c1:v7:hf:learning-loss}
\end{equation}
The entropy of $\rho_{\rm data}$ is constant with respect to the model
parameters, so minimizing this objective is equivalent to minimizing
$D(\rho_{\rm data}\Vert\rho_{\boldsymbol\theta})$ within the chosen
family. The linear term is obtained from the specified measured
expectations; evaluating the model term requires $\ln Z$.

The trace derivative is valid even when the terms in $H$ do not commute.
Differentiating the exponential and using cyclicity under the trace yields
\begin{equation}
 \partial_{\theta_a}\ln Z=-\beta\langle O_a\rangle_{\boldsymbol\theta},
 \qquad
 \partial_{\theta_a}\mathcal L
 =\beta\bigl(m_a-\langle O_a\rangle_{\boldsymbol\theta}\bigr).
 \label{c1:v7:hf:learning-gradient}
\end{equation}
Thus a thermodynamic solver can be called at successive trial parameters
until its predicted moments agree with the measured ones within their
uncertainties. This requires a parameter-dependent reference surface and
consistent derivatives. Numerical bias in these moments can otherwise
be absorbed into the inferred couplings.

The log-partition function also enters the statistical theory of learning
Hamiltonians. Anshu \emph{et~al.}\@ established sample-efficiency results
for local Hamiltonians using maximum-entropy estimation from Gibbs states
and convexity of the log-partition function
\cite{Anshu2021HamiltonianLearning}. Implementing the resulting optimization
calls for model expectations at successive trial couplings, linking the
inference problem to the thermodynamic calculations above.

For noncommuting operators, the Hessian involves an imaginary-time
response. With $\delta O_a=O_a-\langle O_a\rangle_{\boldsymbol\theta}$,
\begin{equation}
 \partial_{\theta_a}\partial_{\theta_b}\ln Z
 =\beta^2\int_0^1\mathrm ds\,
 \operatorname{Tr}\!\left(
 \rho_{\boldsymbol\theta}^{\,s}\delta O_a
 \rho_{\boldsymbol\theta}^{\,1-s}\delta O_b\right).
 \label{c1:v7:hf:learning-hessian}
\end{equation}
This Kubo--Mori covariance reduces to the ordinary covariance when the
relevant operators commute with the Gibbs generator. Directions with
small curvature are weakly constrained by the thermal data. The energy
zero must be fixed because an additive multiple of the identity leaves
the normalized Gibbs state unchanged. If temperature and every energy
scale are both unknown, only their products with $\beta$ are identifiable
without additional calibration. An auxiliary expansion shift is instead held within the numerical
representation of this fixed Hamiltonian; the fitted parameters are its
physical couplings.

\subsection{Entropy budgets for state preparation}

The thermal entropy in Eq.~\eqref{c1:v7:hf:thermodynamic-entropy} can guide
the preparation of a target state. For a quasistatic, thermally isolated
process at fixed particle number, a reversible endpoint satisfies
\begin{equation}
 S(T_f,g_f,N)=S(T_i,g_i,N),
 \label{c1:v7:hf:isentropic-endpoints}
\end{equation}
where $g$ denotes the controlled Hamiltonian parameters. A computed
entropy surface determines the ideal final temperature or, inversely,
the maximum initial entropy compatible with a desired endpoint. A
finite-system canonical calculation uses the fixed-$N$ partition
function. A grand-canonical calculation must account for number
fluctuations or justify the ensemble conversion when used for that
prediction.

Chiu \emph{et~al.}\@ prepared a low-entropy band-insulating region by
entropy redistribution and expanded it towards a half-filled Hubbard
system; their analysis also identified entropy increase during the
preparation \cite{Chiu2018StateEngineering}. For multicomponent gases,
Ibarra-Garc\'ia-Padilla \emph{et~al.}\@ calculated SU($N_f$) Hubbard
thermodynamics and examined how entropy and cooling depend on flavour
number and temperature \cite{Ibarra2021UniversalThermodynamics}.
These studies require comparisons at specified filling and entropy.
Increasing the number of flavours changes the local state count and the
energy scales, so its cooling benefit must be evaluated in the temperature
range of interest.

For a continuously varied Hamiltonian parameter $g$, the direction of an
ideal isentropic ramp follows from a thermodynamic response. In a smooth
phase at fixed $N$, use
$\partial_g F=\langle\partial_gH\rangle$ and
$S=-\partial_TF$. For nonzero heat capacity $C_{g,N}=T(\partial_TS)_{g,N}$,
\begin{equation}
 \left.\frac{\mathrm dT}{\mathrm dg}\right|_{S,N}
 =\frac{T}{C_{g,N}}
 \left.\frac{\partial}{\partial T}
 \left\langle\frac{\partial H}{\partial g}\right\rangle
 \right|_{g,N}.
 \label{c1:v7:hf:isentropic-slope}
\end{equation}
For a ramp of the onsite coupling, the right-hand side contains the
temperature derivative of the onsite-pair expectation. Flavour number
enters the entropy surfaces themselves, so comparisons between different
$N_f$ use their respective entropy curves.

Comparing an observed endpoint with Eq.~\eqref{c1:v7:hf:isentropic-endpoints}
can reveal an excess thermal load relative to ideal preparation. Particle
loss, entropy exchange with a reservoir and deviations from equilibrium
must be included in that comparison. A thermodynamic entropy inferred
by fitting an equilibrium state is distinct from the entropy of the full
isolated quantum state, which is conserved under unitary evolution.

\subsection{Equilibrium free energies in nonequilibrium work experiments}

An equilibrium calculation can also provide an independent reference for
a work-distribution experiment. Consider an initial canonical Gibbs
state of $H_i$ at inverse temperature $\beta$, an isolated unitary drive
$U$, and projective energy measurements at the initial and final times.
The two-measurement definition assigns work $W=E_m^f-E_n^i$ to a pair
of outcomes \cite{Talkner2007Work}. Averaging over their probabilities gives
\begin{align}
 \left\langle e^{-\beta W}\right\rangle
 &=\sum_{n,m}\frac{e^{-\beta E_n^i}}{Z_i}
 |\langle m_f|U|n_i\rangle|^2
 e^{-\beta(E_m^f-E_n^i)}\notag\\
 &=\frac{Z_f}{Z_i}=e^{-\beta\Delta F}.
 \label{c1:v7:hf:jarzynski}
\end{align}
The last equality uses unitarity and the two equilibrium partition
functions at the same $\beta$. The final state of the driven system can
be far from equilibrium; $Z_f$ refers to the Gibbs state of the final
Hamiltonian. Jensen's inequality then gives
$\langle W\rangle\geq\Delta F$ for this protocol.

Hahn \emph{et~al.}\@ tested quantum Jarzynski and Crooks relations with
up to 16 interacting degrees of freedom on trapped-ion and superconducting
quantum processors, and analysed platform-dependent dissipative errors
\cite{Hahn2023Jarzynski}. A classical calculation of $\Delta F$ fixes the
right-hand side independently of the sampled work distribution. For a
fermionic lattice realization, this would require both endpoint
Hamiltonians to lie in a regime where their equilibrium expansions can
be evaluated reliably. Particle exchange or an open-system drive requires
the corresponding fluctuation relation and ensemble convention.

\section{Reference data for approximate many-body methods}
\label{c1:v7:approximation}

Researchers choosing a cluster size, bath discretization or variational
space need to know how the remaining approximation changes an observable.
Increasing these resources can improve the internal solution while leaving
another source of bias almost unchanged. An independent reference gives a
way to measure the resulting physical error. For example, solving an
impurity problem more accurately within a fixed embedding improves the
impurity calculation but leaves the embedding approximation in place.

\subsection{What independent comparisons have established}

LeBlanc \emph{et~al.}\@ compared energies, double occupancies and
imaginary-frequency self-energies of the square-lattice Hubbard model
using auxiliary-field, diagrammatic, embedding, tensor and other methods
\cite{LeBlancBenchmark2015}. Their study emphasized extrapolation to a
common thermodynamic limit and used agreement between independent methods
to establish reference values. Its observables and state points define the
scope of that agreement. A new approximation can be tested against those
values before it is used in a regime with fewer independent checks.

A finite-temperature example is the weak-coupling, half-filled Hubbard
comparison of Sch\"afer \emph{et~al.}\@ \cite{Schafer2021Footprints}.
Diagrammatic and determinantal Monte Carlo supplied references for
quasiparticle and magnetic observables across the metallic and
pseudogap crossover regimes. The comparison identified where local
approximations remained accurate and where spatial spin fluctuations
became important. Comparing several observables allowed the authors to
relate a discrepancy to missing correlations in an approximation.

The reference problem must be specified before attributing a difference
to the solver. A cluster embedded in an infinite medium, a cylinder with
open boundaries and a finite periodic torus can have different answers
at the same nominal $U/t$, temperature and filling. One can either compare
matching finite systems or extrapolate the relevant size and boundary
limits. For an observable $O$, the quantity to report is
\begin{equation}
 \Delta O_M=O_M-O_{\rm ref},
 \label{c1:v7:ar:method-error}
\end{equation}
with the approximation level of method $M$ and the reference uncertainty
stated alongside it. Its statistical error follows from
Eq.~\eqref{eq:covariance-propagation} with gradient $(1,-1)$; the input
covariance includes any shared samples or normalization data. The
comparison also retains the size and truncation errors of the chosen
physical calculations.

\subsection{Cluster, bath and variational-space choices}

Embedding calculations replace part of a lattice problem by an impurity
coupled to a bath. Single-site dynamical mean-field theory and its cluster
extensions differ in the spatial correlations treated explicitly
\cite{Georges1996DMFT,Maier2005Clusters}. A reference calculation can be
used to decide how much a larger cluster changes the physical error,
separately from the numerical accuracy of the impurity solver. A cluster large enough for the local density can still leave appreciable
errors in double occupancy or the momentum-dependent self-energy.

Finite-temperature density matrix embedding theory provides a concrete
example. Sun \emph{et~al.}\@ constructed thermal baths by matching
successive orders of a mean-field density matrix or Hamiltonian
\cite{Sun2020FTDMET}. They compared one-dimensional Hubbard results with
Bethe-ansatz data and studied the half-filled two-dimensional model
against DMFT and dynamical-cluster results. Bath construction and bath
size were thus choices to be tested against independently obtained
thermodynamic quantities. A reference data set for another lattice or
flavour number would allow the same assessment in that new setting.

Tensor methods have their own resource choices. In the finite-temperature
Hubbard study of Wietek \emph{et~al.}\@, minimally entangled typical
thermal states were used on width-four cylinders; comparisons in the
small-system, zero-temperature and half-filled limits checked different
parts of the calculation \cite{Wietek2021METTS}. For a new comparison,
the bond dimension, thermal sampling error and cylinder geometry should
be varied or matched according to the quantity being tested. A discrepancy
with a periodic-lattice result can contain a boundary effect as well as
a tensor truncation error.

Variational density matrices introduce a further distinction between
optimization and expressiveness. Liu \emph{et~al.}\@ combined an
autoregressive probability distribution with a unitary quantum circuit
to represent a mixture of orthogonal pure states, and demonstrated
thermal calculations for the quantum Ising model
\cite{Liu2021VariationalThermal}. The construction permits evaluation of
a variational free energy and entropy. For such a normalized, positive
trial state, the equilibrium free energy of
Section~\ref{c1:v7:free-energy} gives an absolute comparison target.
Increasing network capacity or circuit depth can then be assessed by the
remaining free-energy gap and by additional observables. The comparison propagates the errors of both the candidate energy and its
von Neumann entropy into this gap.

The energy and entropy should also be examined separately. Errors in the
two terms of $F=E-TS$ can compensate, and the cancellation changes with
temperature. A comparison can therefore use the free energy as an overall
objective and reserve density, double occupancy or subsystem purity for
additional tests. The variational upper bound applies when the candidate
is a legitimate density operator and its free energy is evaluated
consistently. For an embedding calculation reported as a set of approximate observables,
the upper-bound interpretation is justified only if those observables
come from such a density operator and a consistent entropy evaluation.

A study aimed at selecting a practical approximation would hold a state
point fixed, increase the approximation parameter, and compare the full
sequence with a common reference. Repeating this procedure over a
specified temperature and coupling range produces an error map. That
map can distinguish a low-cost approximation adequate for a local density
from one adequate for a small response or correlation. Its resolution
should be chosen according to the intended use: a thermometer needs the
local temperature dependence, while a comparison of nearby states needs
the difference between their predictions.

\subsection{Testing an expansion through its successive corrections}

The inverse-flavour calculation in Chapter~\ref{c2:chapter} provides
references within a systematic organization of the interaction. The
Hartree saddle, Gaussian contribution and first non-Gaussian term resolve
successive changes in the density. At the 31 six-flavour state points
analysed there, the central non-Gaussian increment is positive and smaller
than the Gaussian increment \cite{ShiLargeNf2025}.

At $U/t=2.3$ and $\mu/t=-2.3$, for example, the saddle density is
$0.39798$ and the Gaussian increment is $0.08224$, giving $0.48022$.
Subtracting these two contributions from the cumulative density $0.487$
gives the non-Gaussian increment $0.00678$
(Table~\ref{tab:table3-author-data}). Resolving this smaller correction
calls for greater precision than resolving the Gaussian change.
With the deterministic saddle and RPA values fixed, the difference has
the same Monte Carlo statistical uncertainty as the cumulative density.
Numerical representation, discretization and series-truncation effects
are treated separately in Section~\ref{c2:results:precision}.

The reference state and the expansion parameter also matter. A
$1/N_f$ expansion resums density fluctuations into screened propagators
before the higher corrections are evaluated. A bare-interaction expansion
organizes corrections by the number of interaction insertions. The two
truncations contain different sets of contributions. The three-flavour
partition-function comparison in Chapter~\ref{c3:chapter} therefore
checks more than a second implementation of the same integral. It compares
two ways of approximating one thermodynamic quantity.

Independent comparisons are needed even when a sequence appears stable.
Kozik, Ferrero and Georges found cases in Hubbard-like models where a
skeleton self-energy series converged to an unphysical branch
\cite{KozikLuttingerWard2015}. That result concerns the specified
skeleton construction and models. It illustrates why stability under
adding terms must be combined with checks of the physical solution and
of the assumptions entering the representation.

\subsection{References for reconstruction and data reuse}

For a series evaluated beyond its direct convergence domain, the object
being tested includes the continuation procedure. Rossi \emph{et~al.}\@
used a conformal-Borel reconstruction for a fermionic diagrammatic series
with zero convergence radius \cite{RossiResummation2018}. Such
reconstructions use information or assumptions about the analytic
structure in addition to the finite coefficient list. An uncertainty
obtained by propagating coefficient noise through a fixed reconstruction
is conditional on that choice.

A useful benchmark for a reconstruction consists of a truncated
coefficient vector, its covariance and an independently known physical
value. The choice of reconstruction parameters can be made without
consulting that value and tested afterwards. One can then compare the
bias at successive orders, the response to coefficient noise and the
coverage of proposed uncertainty intervals. The small-cluster calculation
in Chapter~\ref{c4:chapter}, where exact diagonalization is available,
provides ingredients for this test. The different continuation branches
there also show why an apparently stable parameter window needs an
external check.

Reference results can subsequently be used more cheaply through
interpolation. An experimental fit may require repeated evaluations of
$n(T,\mu,U)$, while a dense many-body calculation at every fitting step
would be expensive. A tabulated thermodynamic potential with its
uncertainties can be fitted by a smooth surrogate and checked at parameter
points excluded from training. If the surrogate is an intensive grand
potential $\omega=\Omega/N_s$, its derivatives define
\begin{equation}
 n=-\left.\frac{\partial\omega}{\partial\mu}\right|_T,
 \qquad
 s=-\left.\frac{\partial\omega}{\partial T}\right|_\mu,
 \label{c1:v7:ar:surrogate-consistency}
\end{equation}
where the other Hamiltonian parameters and system size are fixed. A
smooth surrogate satisfies the Maxwell relation, Eq.~\eqref{eq:maxwell},
by construction. Its accuracy can be assessed at withheld reference
points for both the potential and its derivatives. Regions with large
prediction uncertainty or strong experimental sensitivity then guide
where additional thermodynamic calculations are most valuable.

\section{Subsystem purity and R\'enyi entropy}
\label{c1:v7:renyi}

A local equation of state constrains only some properties of a many-body
state. Measurements of subsystem purity test information contained in a
reduced density matrix beyond a chosen set of local expectation values.
This provides another use for precise many-body calculations: predicting
a nonlinear observable that is now accessible through interference and
randomized-measurement protocols.

\subsection{Information beyond local densities}

For a normalized state $\rho$, a spatial region $A$ and a fixed
identification of its fermionic modes, define
\begin{equation}
 \rho_A=\Tr_{\bar A}\rho,
 \qquad p_A=\Tr\rho_A^2,
 \qquad S_2(A)=-\ln p_A.
 \label{c1:v7:ar:purity}
\end{equation}
Because purity depends quadratically on the eigenvalues of $\rho_A$,
it contains information beyond a selected set of linear expectation
values. On two fermionic modes, for example, both the coherent state
$|\psi\rangle=(|10\rangle+|01\rangle)/\sqrt2$ and the mixture
$(|10\rangle\langle10|+|01\rangle\langle01|)/2$ give occupation
$1/2$ on each mode. Their two-mode purities, one and $1/2$ respectively,
distinguish the coherence that the occupations do not resolve.

In the grand-canonical infinite-temperature limit at fixed Hamiltonian
parameters and chemical potential, each of the $N_f|A|$ modes in $A$
is maximally mixed on the full fermionic Fock space. Its
purity is $2^{-N_f|A|}$ and $S_2(A)=N_f|A|\ln2$. This extensive
entropy is present even in a product of maximally mixed site states.
The empty region has purity one and zero entropy. These elementary
limits test a numerical or measurement normalization and also show why
a finite-temperature entropy must be interpreted with its state
preparation. A fixed-total-particle-number ensemble has a different
infinite-temperature density matrix because of the global constraint.

For a pure state of the full system, $S_2(A)$ characterizes entanglement
between $A$ and its complement. For a thermal mixed state, it also
contains thermal mixing. Pichler \emph{et~al.}\@ analysed this distinction
for a finite-temperature one-dimensional Fermi--Hubbard model while
proposing a two-copy measurement scheme for spinful fermionic atoms
\cite{Pichler2013}. A numerical reference must specify the temperature,
particle-number convention, region and boundary conditions. Comparing
several regions or temperatures can constrain more of the state than a
single purity does.

The possible information in spatial entropy is well established in
condensed-matter theory. Ground-state scaling contains boundary terms
and, in appropriate systems, universal subleading contributions
\cite{Eisert2010Area,Laflorencie2016Entanglement}. Free-fermion Fermi
surfaces can produce logarithmic enhancements
\cite{Wolf2006FermionArea,GioevKlich2006}. These examples motivate
calculations over region sizes and shapes. Extracting a small subleading
term requires accuracy in the difference of entropies after larger
contributions have been removed.

\subsection{Interference and randomized measurements}

Copy interference makes a nonlinear state function accessible through
measurements on a larger system. Daley \emph{et~al.}\@ proposed coupling
copies of lattice bosons and measuring occupation parity to extract
R\'enyi entropies \cite{Daley2012Entanglement}; Islam \emph{et~al.}\@
realized many-body copy interference with ultracold bosonic atoms and
measured purity and spatial R\'enyi entropy \cite{Islam2015}.
The corresponding numerical prediction represents the prepared copies,
the spatial partition and the detection operations used to measure the
observable.

In the fermionic implementation, the mode ordering determines the signs
in a copy exchange. Pichler \emph{et~al.}\@ proposed tunnelling between
corresponding sites of two copies followed by site- and spin-resolved
occupation measurements \cite{Pichler2013}. The ordinary swap of Fock
configurations and the exchange of fermionic creation operators have
different signs, a distinction shared by the measurement protocol and
the replicated numerical trace. Chapter~\ref{c4:chapter} derives this
trace and its boundary conditions as the starting point for the
Gaussian propagator.

Randomized measurements offer a different experimental route. Elben
\emph{et~al.}\@ proposed random quenches in Hubbard and spin systems,
followed by projective measurements, to recover R\'enyi moments from
statistical correlations \cite{Elben2018}. Brydges \emph{et~al.}\@
measured subsystem entropies on a trapped-ion simulator using randomized
measurements and followed entanglement growth in disordered and clean
dynamics \cite{Brydges2019Randomized}. Local random rotations and their
correlations were developed further in Ref.~\cite{Elben2019Correlations};
Ref.~\cite{Elben2023Toolbox} reviews the broader measurement framework.
The reference needed for an equilibrium fermionic experiment is the
moment of its specified thermal state and region. The measurement
protocol may be shared with a dynamical experiment even though the
physical state being compared is different.

Copy preparation enters the measured quantity itself. For two possibly
different reduced states and an ordinary swap operator $\mathcal S_A$,
\begin{equation}
 \Tr\!\left[\mathcal S_A(\rho_A\otimes\sigma_A)\right]
 =\Tr(\rho_A\sigma_A).
 \label{c1:v7:ar:copy-overlap}
\end{equation}
This becomes the purity when the two states coincide. Unequal loading,
temperatures or preparation errors can therefore change an interference
result even if the swap and detection operations are ideal. A predicted purity benchmarks the complete protocol. Independent checks
of each copy and the detection map can then locate preparation or readout
errors when the measured overlap differs from that prediction.

\subsection{Whole-system thermal purity and a spatial replica numerator}

The ordinary partition function already determines the purity of a full
Gibbs state. With $\rho_\beta=e^{-\beta K}/Z(\beta)$ and fixed grand
Hamiltonian $K$, multiplication of the two thermal operators gives
\begin{equation}
 \Tr\rho_\beta^2=\frac{Z(2\beta)}{Z(\beta)^2},
 \qquad
 S_2(\rho_\beta)=2\ln Z(\beta)-\ln Z(2\beta).
 \label{c1:v7:ar:global-purity}
\end{equation}
The chemical potential and all other parameters in $K$ are the same at
both temperatures. This identity makes the calculation at $2\beta$ part
of the reference task; it can be substantially harder than the one at
$\beta$. For a closed system evolving from a pure state, global purity
remains one under unitary evolution. Local thermalization in such an
experiment does not replace its global state by the Gibbs state in
Eq.~\eqref{c1:v7:ar:global-purity}.

Tracing out a spatial complement changes the problem. Define
\begin{equation}
 R_A=\Tr_{\bar A}e^{-\beta K},
 \qquad \widetilde Z_2(A)=\Tr_A R_A^2.
 \label{c1:v7:ar:replica-numerator}
\end{equation}
Then
\begin{equation}
 p_A=\frac{\widetilde Z_2(A)}{Z(\beta)^2},
 \qquad S_2(A)=2\ln Z(\beta)-\ln\widetilde Z_2(A).
 \label{c1:v7:ar:replica-ratio}
\end{equation}
The partial trace retains the influence of the complement on the state
in $A$. The replica numerator joins two copies along $A$, changing the
Gaussian boundary condition and producing a propagator that generally
depends on two absolute imaginary times.

For $L=\ln Z$ and $R=\ln\widetilde Z_2$, the statistical uncertainty of
$S_2=2L-R$ follows from Eq.~\eqref{eq:covariance-propagation} with gradient
$(2,-1)$. This combination is exact at the level of the logarithms;
finite-order estimates also carry the representation and continuation
errors of both components.

\subsection{Choosing regions and measurement resources}

A theoretical purity can inform the design of a measurement before the
experiment is run. Consider an ideal two-copy measurement of a correctly
defined subsystem swap observable with outcomes $X=\pm1$ and mean
$p_A$. For $M$ independent repetitions,
\begin{equation}
 \operatorname{Var}\overline X=\frac{1-p_A^2}{M},
 \qquad
 M\simeq\frac{1-p_A^2}{\varepsilon^2p_A^2}
 \label{c1:v7:ar:swap-budget}
\end{equation}
gives a relative standard error $\varepsilon$ in the purity. Each
repetition uses two copies. The estimate describes this ideal binary
measurement; randomized protocols have their own allocation of shots
between random operations and repeated readouts
\cite{Elben2018,Elben2019Correlations}. Readout errors and uncertainty in
the implemented random operations must also be included.

Because $p_A=e^{-S_2(A)}$, a large entropy implies a small purity signal.
A reference calculation can identify regions and temperatures for which
the proposed measurement budget is adequate. In the small-error regime,
$\delta S_2\simeq-\delta p_A/p_A$, so a desired absolute entropy error
translates into a relative purity precision. Finite-sample bias from
taking the logarithm matters when the purity estimate is broad. Comparing
analysis procedures on data generated from a known reference state can
test that bias before interpreting an experimental discrepancy.

With temperature and filling fixed as in Section~\ref{c1:v7:hardware},
purities for regions excluded from the fit test additional properties of
the prepared state. Reference curves over several region sizes and
temperatures provide more constraints than one purity, although these
moments do not determine the reduced-state spectra.

\subsection{Independent entropy calculations and spatial information}

Numerical R\'enyi calculations have several independent formulations.
Grover expressed the reduced density matrix of interacting fermions as a
sum of free-fermion operators in determinantal Monte Carlo, allowing
R\'enyi moments in ground and thermal states to be evaluated
\cite{Grover2013}. Wang and Troyer developed a continuous-time
interaction-expansion calculation that samples the interaction correction
to the entropy \cite{WangTroyer2014}. Their approach is a useful point
of comparison for an interaction expansion of the replica partition
function. Earlier swap estimators and subsequent improved estimators
also provide references in spin and related lattice systems
\cite{Hastings2010,Luitz2014Improving,Zhao2022Increment}.

Matching the subsystem and thermal or ground-state limit allows these
algorithms to be compared first on small systems, where fermionic signs
and replica normalization can be checked, and subsequently in statistical
efficiency. The half-filled honeycomb Hubbard calculations of D'Emidio
\emph{et~al.}\@ resolve subleading entropy contributions across ground-state
regimes \cite{DEmidio2024Honeycomb}. Their low-temperature and size limits
set the corresponding target for a finite-temperature expansion.

Spatial combinations can reduce an extensive thermal contribution. For
example,
\begin{equation}
 I_2(A:B)=S_2(A)+S_2(B)-S_2(A\cup B)
 \label{c1:v7:ar:renyi-combination}
\end{equation}
has been investigated through finite-size scaling in an interacting spin
model \cite{Melko2010Mutual}. Its interpretation differs from that of
von Neumann mutual information. For example, the separable mixture
$(|00\rangle\langle00|+|11\rangle\langle11|)/2$ has $I_2=\ln2$ purely from
classical correlations. Thus $I_2$ includes classical as well as quantum
correlations in a mixed state; region and temperature scans probe their
spatial structure through the specified R\'enyi index.

Chapter~\ref{c4:chapter} computes absolute subsystem entropies at specified
parameters and examines their numerical reconstruction. Further scans
could compare regions of equal area and different perimeter, or track a
chosen geometry as the thermal entropy decreases. Those calculations would
test whether a measured spatial contribution is consistent with the target
thermal model. Extending the replica construction to additional moments
or fermionic negativity would require the corresponding trace operators
and boundary conditions to be derived separately
\cite{WangXu2025Negativity}.

\section{Capabilities and limitations of numerical approaches}
\label{sec:limitations}

Different numerical representations are tractable in different parts of
the Hubbard parameter space. Temperature, doping, flavour number and
geometry change their costs, and a method that resolves one observable
may still be inaccurate for another. Cross-method benchmark studies compare the accuracy and range of applicability of these methods
\cite{LeBlancBenchmark2015,QinComputational2022}.

\subsection{Exact diagonalization and the growth of the state space}

Exact diagonalization is a valuable reference because it solves the
specified finite problem without a perturbative truncation. It can
test signs, normalization factors and observable definitions, and it
allows direct access to partition functions when enough of the spectrum
is available.  Its limitation is the growth of the state
space. A site with $\Nf$ fermionic flavours has $2^{\Nf}$ local Fock
states before symmetry restrictions, so the unrestricted lattice space
has dimension $2^{\Nf\Ns}$.

Symmetries, sparse methods and restrictions to selected sectors greatly
extend what can be done, but do not remove this growth. Finite-temperature
thermodynamics can require contributions from many states, and a
reduced-density-matrix observable can require information beyond a small
set of low-energy eigenvalues. 

A new method that reproduces exact small-system results still needs tests of its large-system and high-order behaviour: a small cluster may not contain the long correlations responsible for the phenomenon of interest. Finite-size comparisons and bulk methods then
extend the checks available from exact diagonalization.

\subsection{Determinant quantum Monte Carlo}

Determinant quantum Monte Carlo (DQMC) is a central source of
finite-temperature information for fermionic lattice models. Introducing auxiliary fields allows the fermions to be integrated out;
the remaining field configurations carry determinant weights and are
sampled stochastically. Under suitable conditions the statistical weight
is nonnegative. The repulsive two-flavour Hubbard model on a bipartite
lattice with the appropriate nearest-neighbour hopping and
particle--hole-symmetric half filling is an important example. In such
regimes DQMC can provide highly accurate benchmarks, subject to control
of statistical, finite-size and time-discretization errors
\cite{LeBlancBenchmark2015}.

The difficulty changes when the positivity condition is lost. Writing a
real weight as $w=s|w|$ gives
\begin{equation}
 \langle O\rangle
 =\frac{\langle Os\rangle_{|w|}}{\langle s\rangle_{|w|}}.
 \label{eq:sign-reweighting}
\end{equation}
The average sign can become small as temperature decreases or system
size grows. Its relation to the free-energy difference follows from
the ratio of partition sums \cite{TroyerWiese2005}. If $\Z_{|w|}$ denotes
the partition sum of absolute weights, then
\begin{equation}
 \langle s\rangle_{|w|}
 =\frac{\Z}{\Z_{|w|}}
 =\exp[-\beta\Ns\Delta\omega],
 \label{eq:average-sign}
\end{equation}
where $\Delta\omega=\omega-\omega_{|w|}$ is the difference between
the corresponding grand-potential densities; $\Delta\omega\geq0$ because $\Z\leq\Z_{|w|}$. When
that difference is nonzero, the ratio estimator can require rapidly
increasing sampling effort to maintain a fixed relative accuracy.
This behaviour is a principal reason that calculations in a doped
regime can be much more difficult than those at a nearby sign-free
point.
For $M$ independent samples, the sign itself has relative standard error
$\sqrt{1-\langle s\rangle^2}/(\sqrt M\,|\langle s\rangle|)$,
which grows as $e^{\beta N_s\Delta\omega}/\sqrt M$ when the mean
sign is small. Correlated sampling replaces $M$ by the appropriate
effective sample size. An observable ratio additionally depends on
its numerator--denominator covariance.

Long correlations impose a finite-size limitation even when the sampling
weight is nonnegative: the lattice must be large enough to resolve the
bulk observable. Conversely, enlarging a sign-problematic
lattice can make the reweighting problem harder at the same time that
it reduces a finite-size bias. With a conventional imaginary-time
discretization, the time step adds another extrapolation. Continuous-time
formulations remove that particular discretization but retain their own
sampling and model-dependent limitations \cite{RubtsovCTINT2005}.

Blankenbecler, Scalapino and Sugar developed an auxiliary-field formulation
in which the fermions are integrated out and local field updates are
computed from determinant information \cite{BSS1981}. Hirsch's discrete Hubbard--Stratonovich transformation made this approach particularly useful for lattice
simulations \cite{HirschHS1983,HirschHubbard1985}. The determinant
represents propagation through a complete field configuration. Its size
and stabilization requirements differ from those of the interaction-vertex determinants of continuous-time and connected-determinant methods (Section~\ref{sec:diagmc-route}).

Loh \emph{et~al.}\@ examined the exponential degradation of the average sign
and showed that neglecting the sign can distort different observables to different degrees, so that a reasonable-looking energy need not imply an accurate correlation function
\cite{LohSign1990}. Numerical stabilization controls roundoff when imaginary-time propagation produces widely separated matrix scales \cite{BauerStable2020}. The resulting improvement in propagator accuracy leaves the sign reweighting unchanged.

The continuous-time formulations mentioned above expand the trace, as in the methods of Rombouts, Heyde and Jachowicz
and of Rubtsov \emph{et~al.}\@ \cite{RomboutsCT1999,RubtsovCTINT2005}.
Impurity solvers can instead expand in impurity--bath hybridization
\cite{WernerCTHYB2006}, or combine a weak-coupling expansion with
auxiliary fields \cite{GullCTAUX2008}. The review in
Ref.~\cite{GullCTReview2011} describes these constructions and their
applicable regimes. The configuration spaces and sign structures of impurity solvers differ from those of finite-density lattice calculations, so their accessible temperature ranges need separate assessment.

DQMC gives precise reference data in its favourable regimes. Doping,
frustrated hopping or changes in flavour structure can make sign
reweighting too costly at the desired temperature. These regimes motivate alternative representations, including the diagrammatic expansions of Section~\ref{sec:diagmc-route}.

\subsection{Numerical linked-cluster expansions}

Numerical linked-cluster expansions (NLCEs) use a cluster representation
of the thermodynamic problem. An extensive property per
site is expressed as a sum of connected-cluster weights,
\begin{equation}
 \frac{P}{\Ns}=\sum_c L(c)W_P(c),\qquad
 W_P(c)=P(c)-\sum_{s\subsetneq c}W_P(s),
 \label{eq:nlce}
\end{equation}
where embeddings and subcluster multiplicities are included in the
corresponding sums. Each finite cluster is solved numerically, while
the linked-cluster construction yields the infinite-lattice observable
when the series converges \cite{RigolNLCE2006}. 

The limiting issue is whether the retained clusters capture the
correlations at the chosen temperature and parameters. As longer
correlations become important, increasing the maximum cluster size
becomes necessary, while solving those clusters grows more expensive.
Resummation can extend the convergence range. Stability over a few truncations is a convergence diagnostic, but gives no strict error bound.

NLCEs provide finite-temperature thermodynamic reference data for two-dimensional optical-lattice models \cite{KhatamiThermodynamics2011}. They can converge in regimes where the Monte Carlo sign is unfavourable. Enlarged flavour spaces make the individual cluster calculations more costly; comparisons for $\mathrm{SU}(N_f)$ models show the different temperature and parameter ranges accessible to DQMC and NLCE \cite{Ibarra2021UniversalThermodynamics}.

For cold-atom interpretation, an important difficulty arises when
the desired inference follows a path out of the converged domain.
A density reference available at higher temperature may no longer
support a low-temperature fit; a thermodynamic integration may
require values in a region where successive cluster orders do not
stabilize. Beyond that point, the cluster series can no longer serve as the reference along the integration path.

\subsection{Auxiliary-field, tensor and embedding methods}

Projector auxiliary-field quantum Monte Carlo (AFQMC) gives access to
ground-state observables. In favourable cases it gives precise results after projection, time-step and size limits are controlled.
For sign-problematic systems, constrained-path or related approximations
can extend the accessible regime, but introduce a bias that must be
distinguished from the statistical error \cite{Zhang1997CP}.
Accuracy for a ground-state energy, moreover, cannot simply be transferred to an entropy or temperature-dependent response
\cite{LeBlancBenchmark2015}. Finite-temperature constrained AFQMC
extends this strategy to a grand-canonical ensemble, with its
constraint remaining a distinct approximation
\cite{He2019FTAFQMC}.

Tensor-network methods offer a different route by restricting the
representation of the state or density operator. Their success depends
on the entanglement structure, geometry and resources needed to control
the bond dimension. Accuracy on a chain or cylinder has to be reassessed for a two-dimensional bulk system at finite temperature. For example, minimally entangled
typical thermal states have resolved finite-temperature correlations
of the doped Hubbard model on width-four cylinders
\cite{Wietek2021METTS}. Extrapolating those results to the two-dimensional bulk requires additional control of the cylinder width. Dynamical
mean-field theory (DMFT) and cluster embedding methods treat selected local or cluster correlations nonperturbatively, with
an approximation to the remaining spatial structure
\cite{Georges1996DMFT,Maier2005Clusters,QinComputational2022}.

Where the methods have overlapping regimes of accuracy, their results
provide cross-checks of sampling, extrapolation and truncation errors.
Table~\ref{tab:methods} summarizes these methods. Diagrammatic Monte Carlo
(DiagMC), including connected-determinant (CDet) and combinatorial-summation
(CoS) evaluators, is discussed in Section~\ref{sec:diagmc-route}.

\begin{table}[htbp]
\centering
\small
\renewcommand{\arraystretch}{1.24}
\begin{tabularx}{\textwidth}{@{}>{\raggedright\arraybackslash}p{29mm}YY@{}}
\toprule
Method & Main output & Main limitation\\
\midrule
Exact diagonalization & Exact finite-system spectrum and observables within the solved space & Exponential state-space growth; finite-system information may not resolve bulk correlations.\\
DQMC & Finite-temperature lattice observables; particularly strong in sign-free regimes & Sign reweighting, lattice size, time step where applicable, and sampling correlations require separate control.\\
NLCE & Thermodynamic-limit quantities from linked finite-cluster contributions & Accessible cluster order and series convergence deteriorate when the required correlations exceed the retained clusters.\\
AFQMC & Accurate projected-state results in favourable regimes; wider reach with constraints & Projection and finite-size limits; additional constraint bias in constrained-path calculations.\\
Tensor networks and embeddings & Complementary access to spatial correlations and selected regimes & Tensor or cluster truncation becomes harder to converge for wider geometries and lower temperatures.\\
DiagMC, CDet and CoS & High-order series and, where appropriate, direct bulk observables & Diagram order, sampling, series convergence or continuation, and observable complexity remain limiting.\\
\bottomrule
\end{tabularx}
\caption[Numerical methods and their principal limitations]{Numerical methods and their principal limitations for the benchmark
problems discussed in this section.
See Refs.~\cite{RigolNLCE2006,LeBlancBenchmark2015,QinComputational2022,RossiCDet2017,KozikCoS2024}.}
\label{tab:methods}
\end{table}

\subsection{Coverage in observables}

The observable being measured determines the estimator needed from a
many-body calculation. Density and energy alone do not provide the
subsystem entropy or a frequency-dependent response. For example, the
observation of a pseudogap in a Fermi--Hubbard quantum simulator
combined compressibility measurements with lattice-modulation spectroscopy
\cite{Kendrick2026Pseudogap}. A quantitative comparison with the latter
uses the response to the applied modulation at the measured temperature
and parameters. The single-particle spectral function supplies related
physical information, but corresponds to a different probe.

Experimental comparisons usually require calculations over a parameter range. A thermometer needs the temperature dependence of its observable, while a response function probes variation of a specified Hamiltonian parameter. An isentropic trajectory requires several thermodynamic quantities over a surface. Reusing configurations for related measurements and controlling the variance of their differences can reduce the cost relative to a collection of independent calculations.

\section{Diagrammatic Monte Carlo: history and open problems}
\label{sec:diagmc-route}

\label{c1:expanded-diagmc-history}

Diagrammatic Monte Carlo evaluates the terms of a many-body expansion
by stochastic integration. In its early formulation, diagram topology
and continuous internal variables were sampled together, as in the
polaron problem \cite{Prokofev1998Polaron}. Applications to correlated
fermions developed this strategy into a tool for thermodynamic-limit
calculations \cite{Kozik2010CorrelatedFermions}. This representation replaces the finite-system sampling problem with the evaluation and reconstruction of a high-order series.

For an observable $F$, an auxiliary expansion parameter $\xi$ gives
the schematic representation
\begin{equation}
 F(\xi)=\sum_{n=0}^{\infty}a_n\xi^n,
 \qquad F_{\mathrm{physical}}=F(1).
 \label{eq:diagram-series}
\end{equation}
The coefficients contain sums over diagrams and integrations over
internal variables. The choice of reference theory determines their
form and the analytic structure of the series. A shifted action or
a partially dressed expansion can incorporate physical information into the starting point
\cite{ShiftedAction2016,OneLoop2020}. The reference is chosen to permit efficient reconstruction of the physical result with a reliable error estimate.

The infinite lattice can often be built into the propagators used
for the expansion. This avoids a conventional finite-size
extrapolation for bulk observables. The computational difficulty
then depends strongly on perturbation order and the required
cancellations. The limitations differ from those of auxiliary-field simulations whose average sign decays with configuration size. The resulting series must converge at the physical point or admit a reliable continuation to it.

The developments below changed both the configurations sampled by Monte
Carlo and the algebra performed within each configuration. Their combination
sets the orders and physical regimes accessible to diagrammatic methods.

\subsection{From continuous time and worm updates to diagram sampling}

An important precursor was the continuous-time formulation of quantum Monte
Carlo by Prokof'ev, Svistunov and Tupitsyn. Their 1996 construction sampled
virtual transitions at continuous imaginary times instead of introducing a
small time step \cite{Prokofev1996ContinuousTime}. The subsequent worldline
formulation developed this idea together with updates involving worldline
discontinuities \cite{Prokofev1998Worldline}. Continuous time removes the time-step extrapolation, although autocorrelation and finite-size convergence still need to be controlled.

The worm algorithm enlarges the sampled configuration space to include configurations with two open ends. Motion of these ends connects closed-worldline configurations
while also providing access to a Green function. The original worm algorithm illustrated this strategy for bosonic lattice systems, including
disorder, grand-canonical sampling and nonzero winding numbers
\cite{Prokofev1998Worm}. The term \emph{worm} describes this use of an enlarged
configuration space and its updates. Worldline and diagrammatic formulations can use these updates with their respective weights and estimators.
Boninsegni, Prokof'ev and Svistunov later combined continuous-space worm
sampling with a diagrammatic treatment of contributions to the pair
potential \cite{Boninsegni2006Worm}. 

The founding polaron work of Prokof'ev and Svistunov instead made a Green
function's diagrammatic expansion the sampled object
\cite{Prokofev1998Polaron}. A configuration then specifies an order, a
connectivity and the continuous or discrete variables carried by that graph.
The sum over orders and integrals is explored numerically instead of being truncated after a few hand-evaluated diagrams.
Mishchenko \emph{et~al.}\@ developed improved estimators and spectral
analysis for the Fr\"ohlich polaron, obtaining information about the energy,
effective mass and phonon cloud \cite{MishchenkoPolaron2000}. The later
methodological account by Van Houcke \emph{et~al.}\@ formulated the
approach for interacting fermions and illustrated it through Hubbard-model
equations of state \cite{VanHoucke2010DiagMC}. 

A single impurity simplifies the testing of diagrammatic expansions while retaining its dressing by a many-body environment. The Fermi-polaron calculations of
Prokof'ev and Svistunov showed how stochastic coefficient evaluation could
be combined with the summation of a divergent, sign-alternating series
\cite{Prokofev2008FermiPolaron}. Their longer methodological treatment
explained self-energy sampling and self-consistency in terms of dressed
propagators \cite{Prokofev2008Bold}. The statistically resolved coefficients can be resummed even when their ordinary partial sums fail to approach the physical answer.

Vlietinck, Ryckebusch and
Van Houcke compared propagator choices and resummations in the
three-dimensional Fermi-polaron problem and found that dressing as many
lines as possible was not always the most effective strategy
\cite{Vlietinck2013FermiPolaron}. Their two-dimensional study established a
separate benchmark for polaron and molecule properties and the transition
between them \cite{Vlietinck2014TwoDimensional}. More recently, Van Houcke, F. Werner and Rossi reached order thirty with determinant diagrammatics,
identified exponential large-order growth and used a conformal mapping to
recover the polaron energy \cite{VanHoucke2020Polaron}. 

Finite-density gases brought thermodynamic observables into reach of
several diagrammatic formulations. Burovski \emph{et~al.}\@ obtained
finite-temperature thermodynamics and the unitary transition temperature
of a balanced attractive gas using determinant diagrammatic simulations
and a dilute-lattice extrapolation \cite{Burovski2006Critical}.
Bold DiagMC then supplied an independent normal-state equation of state
for direct comparison with a continuum-gas experiment
\cite{VanHoucke2012Emulator}. The resonant-gas algorithm described by
Van Houcke \emph{et~al.}\@ treats the ultraviolet structure in both
self-consistent and non-self-consistent schemes
\cite{VanHoucke2019ResonantGas}. The range of calculated observables expanded when Rossi \emph{et~al.}\@ combined bold sampling and
Borel resummation to obtain normal-phase contact and momentum-distribution
benchmarks \cite{Rossi2018Contact}, resolving short-distance properties
of the gas.

The single-impurity setting also permits a sign-free lattice expansion.
Pascual, Boronat and Van Houcke constructed it for one minority particle
in an otherwise ideal majority gas on the square lattice, reaching orders
above two hundred \cite{Pascual2025Polaron}. This construction exploits
the absence of exchange among minority particles; a finite minority
density introduces such exchanges into the diagrammatic sum.

\subsection{Reference problems, dressed expansions and continuation}

Bold-line sampling reorganizes the expansion around self-consistently
dressed propagators \cite{Prokofev2007Bold}. By incorporating repeated
processes into the reference lines, it can improve the remaining series.
However, convergence of these equations also involves selecting the
physical solution among their possible branches.

Kozik, Ferrero and Georges demonstrated this problem in Hubbard-like
systems: the skeleton self-energy functional can have distinct branches,
and a convergent skeleton sequence can select an unphysical one
\cite{KozikLuttingerWard2015}. The shifted-action construction of Rossi \emph{et~al.}\@ provides a way to formulate dressed schemes through an
auxiliary expansion and to state conditions for their applicability
\cite{ShiftedAction2016}. The auxiliary parameter specifies the path from the reference problem to the target action.

Van Houcke \emph{et~al.}\@ analysed the
sequence of self-consistent propagators in two models with known exact solutions,
showing both correct and misleading convergence and testing a criterion
that distinguishes them without prior knowledge of the exact propagator
\cite{VanHoucke2024Branches}. A related preprint by Kim, P. Werner and Kozik found that a strange-metal-like solution of a bold Hubbard theory
can lie on the unphysical branch in a weakly doped regime
\cite{Kim2020Strange}.  These examples distinguish convergence of the self-consistency iteration
from selection of the physical branch, which can be tested by the proposed
diagnostics or an independent solution.

Other developments change the reference problem from which the diagrams are constructed. Rossi, \v{S}imkovic and Ferrero formulated
high-order renormalized expansions in spacetime and demonstrated the
inclusion of nonperturbative information such as a gap or a dynamical
mean-field reference \cite{Renormalized2020}. Their one-loop-renormalized
vertex expansions provide another construction in which selected
processes are resummed before the remaining connected determinant
calculation \cite{OneLoop2020}. The homotopic action of Kim \emph{et~al.}\@ places several convergence-improving transformations within a common
framework of action deformation \cite{HomotopicAction2021}. 

Expansions around dynamical mean-field theory treat local correlations nonperturbatively and expand in the remaining nonlocal correlations. Iskakov, Antipov and Gull
stochastically sampled all topologies built from two-particle dual vertices
\cite{Iskakov2016Dual}. Gukelberger, Kozik and Hafermann subsequently
examined the convergence of this dual series and the consequences of
neglecting higher-order vertices \cite{Gukelberger2017Dual}. Increasing diagram order within a fixed truncated effective action leaves its omitted higher-order vertices absent. The uncertainty consequently includes a vertex-truncation error as well as the summation error.

Strong-coupling formulations treat the local interaction nonperturbatively from the outset.
Carlstr\"om's spin-charge transformation allowed diagrammatic calculations
including the infinite-on-site-repulsion limit
\cite{Carlstrom2018SpinCharge}. His subsequent strong-coupling framework
uses connected local vertices that are nonperturbative in the local
interaction \cite{Carlstrom2021Strong}. Applications to the attractive
Hubbard model revealed nontrivial connectivity of excitation spectra
\cite{Carlstrom2023Spectral}. A later preprint proposed integrating this
organization with a dynamical mean-field solution so that the remaining
expansion isolates nonlocal corrections \cite{Carlstrom2023DMFT}.
A distinct variational expansion built on dynamical mean-field theory
was developed by Wang and Haule and applied to the three-dimensional
Hubbard N\'eel transition and critical behaviour \cite{WangHaule2025DMFT}. 

Variational reference problems have also been used beyond the contact Hubbard interaction.
Chen and Haule combined a variational starting point with grouped
diagrammatic sampling for the long-range Coulomb electron gas
\cite{ChenHaule2019}. Haule and Chen then obtained quasiparticle and Landau
parameters for that system \cite{Haule2022Excitations}. For quantum
impurities, Li, Wallerberger and Gull used recursive determinants with general interactions and hybridization functions, identifying many orbitals with moderate correlations as a favourable regime for their implementation \cite{Li2020Impurity}.

\label{sec:series-control}

The accuracy at $\xi=1$ depends on both the coefficients and the continuation. The physical point may lie outside the direct
convergence disk, and different choices of the reference theory can
produce very different finite-order behaviour. Analytic
continuation or resummation is then part of the calculation, with
its own assumptions and diagnostics. Conformal-Borel methods have
been used to extract physical information even from fermionic
series with zero convergence radius
\cite{RossiResummation2018}. These reconstructions depend on specific assumptions about the analytic structure of the series.

A concrete limitation appears in the six-flavour CoS equation of state
of Ref.~\cite{KozikCoS2024}. At $T/t=0.15$ and $U/t=12$, its
estimated resummation systematic uncertainty becomes comparable to the
propagated coefficient error. The work also reports divergent direct
partial sums in the studied regimes, requiring reconstruction before
a physical value can be assigned. For these state points, continuation is a substantial source of uncertainty even after the coefficient errors have been reduced.

The reconstruction can be tested by increasing the maximum order and comparing continuation procedures with justified assumptions. Alternative expansions of the same physical result provide further checks. The resulting reconstruction uncertainty can exceed the statistical error propagated from the coefficient covariance.

\subsection{Connected summation, complexity and integration}

For fermions, a large set of diagrams may contribute with opposing
signs. If the individual contributions are sampled separately,
the cancellation is recovered only statistically. Summing diagrams
at fixed interaction vertices can perform much of that cancellation
before Monte Carlo integration. CDet uses determinants to combine contractions and a subset
recursion to remove disconnected contributions
\cite{RossiCDet2017}. This changes the algebraic task from explicit
enumeration of a factorial number of diagrams to a structured
calculation with exponential order dependence.

Equation~\eqref{eq:cost-variance} combines the work per step with the variance and autocorrelation factors. A more expensive integrand can reduce the total cost when deterministic cancellation lowers its variance enough to compensate for the extra work per sample.

Fast principal-minor algorithms improve the building blocks of
determinant-based calculations \cite{FastPrincipalMinors2022}.
Many-configuration Markov-chain Monte Carlo can exploit the
additional subconfiguration information already obtained in an
evaluation \cite{MCMCMC2021}. CoS offers a further way to encode connected diagrams in a dynamic
program and an arithmetic circuit \cite{KozikCoS2024}. \mbox{Combining} these evaluations and samplers can increase the accessible order and provide additional estimators for the chosen observable.

For dynamical observables, the required subtraction depends on reducibility.
Moutenet, Wu and Ferrero compared determinant algorithms for quantities
such as the self-energy, including recursive evaluation of
one-particle-irreducible contributions \cite{Moutenet2018Dynamical}.
\v{S}imkovic and Kozik developed determinant summation of irreducible
self-energy diagrams and used the resulting high-order information to
reconstruct momentum-dependent quantities in correlated regimes
\cite{Irreducible2019}. The vacuum functional, connected Green function and irreducible self-energy require different subtraction formulae. In this thesis the relevant objects are the connected contributions to $\ln Z$ and to the replica free-energy difference.

Rossi \emph{et~al.}\@ related exponential convergence in diagram order to
polynomial complexity in inverse accuracy \cite{PolynomialComplexity2017}.
The evaluator-cost part of their argument can be written as follows.
At fixed physical parameters, suppose the truncation error satisfies
$E_n\leq C_{\mathrm{tr}}r^n$, with $C_{\mathrm{tr}}>0$ and $0<r<1$.
For $0<\epsilon<C_{\mathrm{tr}}$, it is sufficient to retain terms through order
$n_\epsilon=\lceil\ln(C_{\mathrm{tr}}/\epsilon)/\ln(1/r)\rceil$.
If one fixed-coordinate evaluation at order $k$ costs $w_k=O(a^k)$,
with $a>1$ independent of $\epsilon$, the work for one evaluation of
each retained order obeys
\begin{equation}
 W_{\mathrm{eval}}(\epsilon)
 =\sum_{k=0}^{n_\epsilon}w_k
 =O\!\left(\epsilon^{-p}\right),
 \qquad p=\frac{\ln a}{\ln(1/r)}>0.
 \label{eq:conditional-complexity}
\end{equation}
Indeed, $a^{n_\epsilon}\leq a(C_{\mathrm{tr}}/\epsilon)^p$, and the
geometric sum of evaluation costs is $O(a^{n_\epsilon})$. Polynomial
prefactors in order add powers of $\ln(1/\epsilon)$ to this scaling.
This estimate concerns the evaluator work at a given truncation tolerance. The full Monte Carlo runtime also depends on the samples needed for the statistical tolerance, their variance and autocorrelation. Applying the estimate requires a geometric remainder bound for the model and expansion under study.

Sturt and Kozik's
preprint parallelizes CoS arithmetic on graphics processors to increase the throughput of fixed-vertex evaluation \cite{Sturt2024Parallel}. Leoni and Franchini developed a normalizing-flow
proposal for global sampling of diagrams and demonstrated reduced
correlation and statistical error for the single-site Holstein polaron
\cite{Leoni2024Flow}. Extending this gain to finite-density fermionic sampling requires tests with the corresponding measure.

Efficiently summed diagrammatic integrands can also be integrated deterministically. N\'u\~nez Fern\'andez \emph{et~al.}\@ used
tensor-cross interpolation to obtain a compressed representation of
high-order diagram sums and demonstrated accurate real-time impurity
calculations \cite{Nunez2022TensorTrain}. This separates algebraic summation from the choice of integration:
the same fixed-coordinate quantity can be integrated by Monte Carlo,
adaptive quadrature or a low-rank approximation. Direct low-order sums
provide a common check of the evaluators in
Chapters~\ref{c2:chapter} and~\ref{c3:chapter} before integration errors
are introduced.

\subsection{Physical applications and cross-method comparisons}

Deng \emph{et~al.}\@
combined bold sampling with a Cooper-channel treatment to determine
competing pairing states in an emergent Bardeen--Cooper--Schrieffer regime
\cite{Deng2015}. \v{S}imkovic, Deng and Kozik extended this analysis to
next-nearest-neighbour hopping \cite{Simkovic2021BCS}. 

At stronger correlations and lower doping, Wu \emph{et~al.}\@ used
controlled diagrammatic summation to examine the pseudogap and the
nodal--antinodal distinction \cite{Wu2017}. Subsequent studies of the
half-filled model resolved an extended crossover from Fermi-liquid to
quasi-antiferromagnetic behaviour \cite{Simkovic2020Crossover} and showed
that characteristic changes in spin and charge correlations need not
occur at the same temperature \cite{Kim2020SpinCharge}. The corresponding local observables consequently have different temperature sensitivities in thermometry. Lenihan \emph{et~al.}\@ examined entropy and specific heat in
the doped model, linking thermodynamic signatures to its non-Fermi-liquid
crossover \cite{Lenihan2021Entropy}. 

Thermodynamic-limit momentum resolution has connected the low-temperature
evolution of the doped pseudogap with stripe results from ground-state
methods, in the calculations of \v{S}imkovic, Rossi, Georges and Ferrero
\cite{Simkovic2024}. The pseudogap was found in regimes with both long and short magnetic correlation lengths, so the correlation length alone cannot identify its mechanism. Rossi \emph{et~al.}\@
also compared interaction-enhanced nesting in spin-fermion and Hubbard
models \cite{Rossi2024Nesting}. These calculations resolve momentum-dependent features averaged out in local density and energy data.

Cross-method studies connect this microscopic information to the accuracy
of approximate descriptions. Sch\"afer \emph{et~al.}\@ compared methods
for the half-filled two-dimensional model, using diagrammatic and
determinant Monte Carlo as quantitative references for local quantities,
quasiparticles and magnetic correlations \cite{Schafer2021Footprints}.
Comparisons with experiment test the same observables against a physical
realization. A preprint by Currie, Sturt and Kozik compared
thermodynamic-limit DiagMC results with the low-temperature Hubbard
simulator of Xu \emph{et~al.}\@ and reported agreement over the observables
and temperatures accessible in that experiment
\cite{Currie2025Validation,Xu2025Cryogenic}.

At a phase transition, a perturbative expansion must represent the singularity of the infinite-system observable. Lenihan \emph{et~al.}\@ used
high-order coefficients to locate the N\'eel transition of the doped
three-dimensional Hubbard model \cite{Lenihan2022}. Their strategy extracts
the position of a singularity without requiring the series to evaluate a
regular observable exactly at that singularity. In an ordered phase,
expanding about an appropriate broken-symmetry reference offers a different
route. A preprint by Spada \emph{et~al.}\@ developed high-order connected
expansions around BCS theory, including polarized superfluid states
\cite{Spada2021BCS}. Garioud \emph{et~al.}\@ introduced spin-symmetry-broken
CDet calculations within the antiferromagnetic phase
\cite{Garioud2024}. 

The choice of microscopic input also extends diagrammatic calculations
beyond lattice-model benchmarks. Molecular correlation energies have been
computed by sampling Hugenholtz graphs \cite{Bighin2023Molecules}, and a
zero-temperature formulation for finite systems has been demonstrated on
the Richardson pairing model \cite{Brolli2025Finite}. For material-specific
polarons, first-principles electron--phonon matrix elements can enter the
diagrammatic expansion directly \cite{Luo2025FirstPrinciples}. These
applications retain the stochastic diagram sum while changing the
one-particle basis, interaction vertices and reference problem.

Currie and Kozik's lowest-Landau-level
calculation uses CoS to investigate the emergence of an incompressible
one-third-filled fractional quantum Hall state
\cite{Currie2024FQH}. The same work also examines the long-time Green function at half
filling and reports behaviour consistent with a pseudogap. For the
attractive Lieb lattice, pairing-response calculations by Tupitsyn
\emph{et~al.}\@ yield a characteristic temperature that provides a
controlled \emph{upper bound} on the superconducting transition temperature
\cite{FlatBand2026}.

\section{Questions and contributions of this thesis}
\label{sec:thesis-contributions}
\label{c1:v7:contributions}

The research chapters examine three stages of a diagrammatic calculation:
the choice of expansion, evaluation of its connected coefficients, and
construction of an observable requiring a replicated trace. Their
numerical examples concern finite-temperature lattice fermions. The
reference tasks discussed in this chapter specify potential uses of the
results, while the following chapters establish the methods and test them
on stated models and parameters.

\begin{table}[!htbp]
\centering\small
\renewcommand{\arraystretch}{1.2}
\begin{tabularx}{\textwidth}{@{}p{18mm}YY@{}}
\toprule
Chapter & Question and construction & Numerical evidence\\
\midrule
2 & Organizing fluctuations: collective fields, $1/N_f$ counting and analytic internal integration. & Low-order graph and integral checks; successive six-flavour density contributions.\\
3 & Evaluating connected sums: flavour-resolved CoS, shared subset outputs and determinant-polynomial evaluation. & Exact small-order comparisons, C++ kernel timings and lattice thermodynamic examples.\\
4 & Extending the observable: fermionic replica trace and stable two-time Green functions. & Deterministic and ED comparisons, coefficient covariance and continuation tests.\\
\bottomrule
\end{tabularx}
\caption[Research questions and their tests]{The three research questions and the calculations used to examine them.}
\label{tab:thesis-map}
\end{table}

\subsection{An inverse-flavour expansion for thermodynamics}

In Chapter~\ref{c2:chapter}, I derive the collective $G$--$\Sigma$ action
and expand about its Hartree saddle. Gaussian fluctuations generate a
screened density interaction, and flavour counting organizes the remaining
connected diagrams by powers of $1/N_f$ \cite{ShiLargeNf2025}.
I derive the graph weights, momentum routing and contact prescription,
and integrate internal vertices analytically through a discrete Lehmann
representation. The remaining integrals are estimated with adaptive
sampling and reference-normalization chains. The six-flavour density
example resolves the saddle, Gaussian and first non-Gaussian contributions,
providing the kind of comparison with low-order approximations discussed
in Section~\ref{c1:v7:approximation}.

\subsection{Connected evaluation with general flavour structure}

Chapter~\ref{c3:chapter} concerns the connected sum at fixed interaction
coordinates. Building on CoS \cite{KozikCoS2024}, I develop
flavour-resolved recurrences, arithmetic-circuit reductions and terminal
outputs for interaction subsets. These outputs can be used by the
many-configuration sampler to evaluate several candidate configurations
from shared intermediate work. Section~\ref{c3:fast:section} develops
a determinant-polynomial representation of the same contraction problem
and a coefficient sieve that extracts the interaction moments
from principal-minor tables and applies connected subtraction. Its
arithmetic analysis and fixed-configuration C++ timings compare this
route with direct CoS evaluation.

The chapter includes two-spin honeycomb and three-flavour square-lattice
thermodynamic calculations reported in Ref.~\cite{ShiCoS2025}, with
their respective implementations stated explicitly. The three-flavour
partition-logarithm comparison uses an inverse-flavour result as an
independent reference. The accuracy and cost of a complete sampled
coefficient also depend on the integration and continuation stages.

\subsection{A replicated calculation of subsystem entropy}

Chapter~\ref{c4:chapter} develops the fermionic replica formulation for
$S_2(A)$ introduced in Ref.~\cite{ShiRenyi2026}. I derive the trace with
its fermionic ordering signs and the corresponding Gaussian boundary
conditions. The subsystem cut produces a Green function depending on two
absolute imaginary times. Scaled matrix formulas evaluate this kernel,
after which connected-determinant summation and sampling give the
interaction coefficients of the replicated quantity.

Deterministic low-order integration and exact diagonalization test the
small-cluster calculation. Larger lattices require continuation of finite
series, which I examine through changes of order, shift and reconstruction
parameters. These tests supply numerical examples for the independent
entropy comparisons discussed in Section~\ref{c1:v7:renyi}; they also
identify requirements for improving the final continued estimates.

Chapter~\ref{ch:conclusions} brings these results back to the reference
problems of this introduction and identifies the additional calculations
needed for wider parameter coverage and experimental use.

\chapter[Diagrammatic Monte Carlo with \texorpdfstring{$1/N_f$}{1/Nf} Expansions]{Diagrammatic Monte Carlo\\with $1/N_f$ Expansions}
\label{c2:chapter}

An expansion in the inverse flavour number $1/N_f$ organizes collective
fluctuations around a self-consistent reference. In the $G$--$\Sigma$
functional integral developed here, the Hartree saddle supplies that
reference and fixes the propagators entering the successive thermodynamic
corrections.
Its quadratic fluctuations sum repeated density bubbles into a screened
interaction, while higher vertices generate the non-Gaussian corrections.
Starting from this action, I derive the diagram weights and use the discrete
Lehmann representation (DLR) to integrate selected internal vertices
analytically. Monte Carlo sampling then evaluates the remaining integral.
The formulation and density calculations were first presented in my
work in Ref.~\cite{ShiLargeNf2025}.

The numerical application is the onsite $\mathrm{SU}(6)$ Hubbard model
on a $70\times70$ square lattice at $T/t=0.15$. The densities calculated
in Ref.~\cite{ShiLargeNf2025} resolve the saddle, Gaussian/RPA and first
non-Gaussian contributions as the chemical potential varies
(Figs.~\ref{fig:table3-author-cumulative} and~\ref{fig:table3-author-increments}).
Their diagrammatic origin is illustrated by the first non-Gaussian
topologies in Fig.~\ref{c2:gr:first-graphs}. A three-flavour partition-logarithm
calculation later provides a comparison with CoS++ in
Chapter~\ref{c3:chapter}.

\section{Conventions and model parameters}
\label{c2:scope}

The flavour number $N_f$ is a model parameter. A vacuum graph contains
$E$ screened interaction lines and contributes at non-Gaussian order
$k$ in $1/N_f$; several values of $E$ generally occur at the same $k$.
The numbers of unit cells and physical sites are denoted by $N_c$ and
$N_s=\nu N_c$, respectively, with $\nu=1$ on the square lattice and
$\nu=2$ on the honeycomb lattice.

The physical Hamiltonian $H$ excludes the chemical-potential term,
which is introduced in $K=H-\mu\widehat N$. Units are chosen with
$k_{\mathrm B}=1$ and $\beta=1/T$. The conventional onsite pair coupling
is $U_{\mathrm{phys}}$. In taking the formal large-$N_f$ limit, however,
I hold the scaled density kernel fixed, with the conversion given by
Eq.~\eqref{c2:th:coupling-map}.
The label $U/t$ in the numerical figures denotes $U_{\mathrm{phys}}$,
as in Table~III of Ref.~\cite{ShiLargeNf2025}.

The matrix derivation accommodates more than one site per cell, but the
high-order contact rules and numerical results below concern onsite
interactions. An instantaneous interaction between different sites
identifies the two times while retaining distinct spatial or sublattice
labels, and therefore requires a different contact-contraction rule.

In Ref.~\cite{ShiLargeNf2025}, I solved for the screened interaction on a
time grid and then fitted its DLR coefficients. Here I also describe a
direct coefficient-space solution of the same equation. Both routes use
the contact prescription and diagrammatic expansion derived below.

\subsection{Relation to functional and screened-interaction methods}

Matsubara formulated equilibrium statistical mechanics in terms of
finite-temperature propagators \cite{Matsubara1955}. Auxiliary-field
transformations express interactions as integrals over additional fields
\cite{HubbardHS1959}. The $G$--$\Sigma$ construction below imposes a constraint on a fermion
bilinear, with the measure, multiplier contour and equal-time prescription
derived from the regulated fermionic integral of the present Hamiltonian.

Thermodynamic functionals of the dressed propagator go back to Luttinger
and Ward \cite{LuttingerWard1960}. Baym and Kadanoff, and subsequently
Baym, related self-consistent approximations to conservation laws and
thermodynamic consistency \cite{BaymKadanoff1961,BaymConserving1962}.
Deriving the energy, density and response functions from one functional
treats these observables within the same approximation. Whether a particular
truncation preserves the relevant conservation laws or converges still
depends on the chosen functional and the diagrams retained. Assessing the
finite-order accuracy of the $1/N_f$ truncations used here requires higher
orders and numerical benchmarks.

Bohm and Pines separated collective density motion from screened electronic
interactions in a Coulomb system \cite{BohmPines1953}. Hedin's equations
express the one-particle Green function and self-energy in terms of a
screened interaction and a vertex function \cite{Hedin1965}. The onsite
Hubbard expansion developed here likewise incorporates repeated density
fluctuations into a screened line. Its density channel, flavour factors
and instantaneous term, as well as the subtraction in the Gaussian free
energy, follow from the collective action.

\section{The flavour-scaled model and its collective fields}
\label{c2:th:model}

The expansion is defined by specifying both the physical Hamiltonian and the
quantity held fixed as the flavour number is varied. Let $i=(\mathbf R,a)$
label a physical lattice site, where $\mathbf R$ labels one of $N_c$ unit
cells and $a=1,\ldots,\nu$ labels sites within a cell. The flavour index
is $m=1,\ldots,N_f$. A flavour-independent one-body Hamiltonian and an
instantaneous, real, symmetric density kernel $v_{ij}=v_{ji}$ define
\begin{align}
 H&=\sum_{ij,m}h^0_{ij}c^\dagger_{im}c_{jm}+H_{\mathrm{int}},
 &K&=H-\mu\widehat N,
 \label{c2:th:hamiltonian}\\
 H_{\mathrm{int}}
 &=\frac{1}{2N_f}\sum_{ij}\sum_{m,m'}v_{ij}
 c^\dagger_{im}c^\dagger_{jm'}c_{jm'}c_{im}.
 \label{c2:th:normal-ordered-interaction}
\end{align}
The double sum runs over ordered pairs $(i,j)$, so the factor $1/2$
prevents double counting.
The creation and annihilation operators in
Eq.~\eqref{c2:th:normal-ordered-interaction} are normal ordered. Normal
ordering also fixes the equal-time prescription used at Gaussian order.

Writing $n_i=\sum_m c^\dagger_{im}c_{im}$ gives the exact operator identity
\begin{equation}
 H_{\mathrm{int}}=\frac{1}{2N_f}\sum_{ij}v_{ij}
 \left(n_i n_j-\delta_{ij}n_i\right).
 \label{c2:th:normal-order-identity}
\end{equation}
For the onsite model, $v_{ij}=v_0\delta_{ij}$, this becomes
$H_{\mathrm{int}}=(v_0/N_f)\sum_i\sum_{m<m'}n_{im}n_{im'}$.
Consequently the conventional physical Hubbard coupling obeys
\begin{equation}
 v_0=N_fU_{\mathrm{phys}}=2u,
 \qquad u=\frac{N_f}{2}U_{\mathrm{phys}}.
 \label{c2:th:coupling-map}
\end{equation}
The formal large-$N_f$ expansion holds $v$ fixed, so the physical onsite
coupling $U_{\mathrm{phys}}=v_0/N_f$ varies with $N_f$.
At a specified physical flavour number the two
descriptions represent the same model after this conversion. For the $N_f=6$ benchmarks in Table~III of Ref.~\cite{ShiLargeNf2025},
$U$ denotes the physical coupling, and Eq.~\eqref{c2:th:coupling-map}
gives $u=3U_{\mathrm{phys}}$.

The conversion also determines derivatives with respect to the coupling. At fixed
$N_f$, differentiating a result with respect to $U_{\mathrm{phys}}$ brings
the factor $\partial v_0/\partial U_{\mathrm{phys}}=N_f$. Differentiation
with respect to $u$ instead brings a factor of two. These physical response derivatives keep $N_f$ fixed. They therefore
vary a different parameter from the formal expansion, which varies
$N_f$ at fixed $v$.

\subsection{Green functions, point splitting and the multiplier contour}
\label{c2:th:conventions}

With the compact label $1=(\tau_1,i_1)$, an integral over this label
denotes a time integral and a site sum. The fermionic kernel is
$A_0=\partial_\tau+h^0-\mu$, and its inverse has the convention
\begin{equation}
 g(1,2)=\langle\mathcal T_\tau c(1)c^\dagger(2)\rangle,
 \qquad
 g(\tau;E)=\frac{e^{-E\tau}}{1+e^{-\beta E}}\quad(0<\tau<\beta).
 \label{c2:th:green-convention}
\end{equation}
Thus $g(0^-;E)=-f(E)$, where $f(E)=(e^{\beta E}+1)^{-1}$.
This $g$ has the opposite sign to the Green function defined with a minus
sign before the time-ordered expectation value. The density, polarization
and screened interaction below use the same convention.

The path integral follows from the operator trace with an explicit
time regulator. Divide $\beta$ into $M$ intervals of width
$\Delta\tau=\beta/M$ and insert fermionic coherent-state resolutions of
the identity between successive factors of $e^{-\Delta\tau K}$:
\begin{equation}
 I=\int d\bar\psi\,d\psi\,e^{-\bar\psi\psi}
 |\psi\rangle\langle\bar\psi|.
 \label{c2:th:coherent-completeness}
\end{equation}
All physical sites and flavours are included in this notation. The trace
closes the Grassmann fields antiperiodically,
$\psi_0=-\psi_M$ and $\bar\psi_0=-\bar\psi_M$. Coherent-state overlaps
generate the discrete time derivative. Normal ordering makes the
short-time Hamiltonian symbol a polynomial with creation operators
replaced by $\bar\psi_\ell$ and annihilation operators by
$\psi_{\ell-1}$. To first order in the time step this gives
\begin{align}
 S_M=\sum_{\ell=1}^M\left[
 \sum_{i,m}\bar\psi_{im,\ell}
       (\psi_{im,\ell}-\psi_{im,\ell-1})
 +\Delta\tau K(\bar\psi_\ell,\psi_{\ell-1})\right],\notag\\
 Z=\lim_{M\to\infty}\int D(\bar\psi,\psi)\,e^{-S_M}.
 \label{c2:th:regulated-action}
\end{align}
The omitted short-time terms vanish in this limit for a finite lattice
and fixed $N_f$. The interaction symbol is
\begin{equation}
 K_{\mathrm{int}}(\bar\psi_\ell,\psi_{\ell-1})
 =\frac{1}{2N_f}\sum_{ij,m,m'}v_{ij}
 \bar\psi_{im,\ell}\bar\psi_{jm',\ell}
 \psi_{jm',\ell-1}\psi_{im,\ell-1}.
 \label{c2:th:coherent-interaction}
\end{equation}
Keeping the creation and annihilation fields on adjacent time slices
preserves normal ordering in the continuum limit. In particular, terms
with identical site and flavour vanish by Grassmann nilpotency before
any collective-field approximation. The same regulator determines the
one-sided contact prescription: occupation insertions use $0^-$, with
the anticommutator fixing the jump to $0^+$. Contact contractions below
are therefore limits of the time-sliced identity, taken before multiplying
the discontinuous propagator factors.

For the quadratic theory, the regulated Grassmann integral is a finite
determinant of its discrete kernel $A_{0,M}$. Antiperiodic closure fixes
its continuum limit to the physical free partition function
\begin{equation}
 Z_0=\prod_{\mathbf k,\ell}
 [1+e^{-\beta\epsilon_\ell(\mathbf k)}]^{N_f},
 \label{c2:th:free-normalization}
\end{equation}
where $\epsilon_\ell$ are eigenvalues of $h^0-\mu I$. Recovering this
partition function sets the absolute normalization of the trace logarithm
and the common reference for adding saddle and Gaussian contributions.
The same time regulator determines their contact terms.

Introduce a flavour-averaged bilocal field
$G(1,2)=N_f^{-1}\sum_m\psi_m(1)\bar\psi_m(2)$ and a multiplier $\Sigma$.
With the imaginary-time regulator in place, the bilinear constraint on
$G$ is enforced by a functional delta integral with exponent
$iN_f\Tr[\Sigma(G-N_f^{-1}\sum_m\psi_m\bar\psi_m)]$.
Integrating the fermions then gives the collective action
\begin{equation}
 \frac{S[G,\Sigma]}{N_f}
 =-\Tr\ln(A_0-i\Sigma)-i\Tr(\Sigma G)
 +\frac12\int_{1,2,3,4}
 G(1,2)v(2,3)\delta_{12}\delta_{34}G(3,4).
 \label{c2:th:collective-action}
\end{equation}
Here $v(1,2)=v_{i_1i_2}\delta_\beta(\tau_1-\tau_2)$, where
$\delta_\beta$ is the periodic bosonic delta distribution. The trace and
coincident-time symbols inherit the original normal-ordered regulator.
For example, density insertions use a creation field infinitesimally later
than its annihilation field, giving $G(\tau,\tau+0^+)=-\rho$.

The fields in Eq.~\eqref{c2:th:collective-action} are integrated along
the contours that implement the bilinear constraint. For a
repulsive density interaction the stationary multiplier is generally
imaginary, while $i\Sigma_*$ is real. Expanding around that point assumes
an admissible deformation to its local steepest-descent contour. The
expansion is local to the stable branch continuously connected to the
chosen reference. Its physical interpretation for the uniform solution
is discussed with the density results in Section~\ref{c2:results:precision}.

The origin of the first two terms in the action can be checked before
taking any stationary approximation. With the trace pairing used here,
$\Tr(\Sigma\psi\bar\psi)=-\bar\psi\Sigma\psi$ after Grassmann
reordering. Integrating the constrained quadratic form
$\bar\psi(A_0-i\Sigma)\psi$ produces $\det(A_0-i\Sigma)^{N_f}$, whose
negative logarithm is the first term of Eq.~\eqref{c2:th:collective-action}.
The remaining multiplier coupling yields the second term and the density
product the last.
The normalization of the auxiliary delta identity is fixed by recovering
the original fermion integral.

\section{The Hartree saddle and its matrix structure}
\label{c2:th:saddle}

Varying the two collective fields independently gives
\begin{align}
 G_*&=(A_0-i\Sigma_*)^{-1}\equiv g_H,
 \label{c2:th:dyson}\\
 s(1,2)\equiv i\Sigma_*(1,2)
 &=\delta_{12}\int_3v(1,3)g_H(3,3^+).
 \label{c2:th:saddle-equation}
\end{align}
Let $\rho_i=\langle c^\dagger_{im}c_{im}\rangle_H$ denote the density per
flavour in a static flavour-uniform saddle.
Then
\begin{equation}
 s_i=-\sum_jv_{ij}\rho_j,
 \qquad h_H=h^0-\mu I-\operatorname{diag}s.
 \label{c2:th:hartree-shift}
\end{equation}
The physical Hartree energy shift $-s_i$ is positive for a repulsive
onsite kernel and positive density.
The direct density field is retained at leading order, with exchange
and the cancellation of same-flavour onsite contributions supplied by
subsequent terms in the expansion. This order assignment determines the
factor $N_f$ in the saddle and the matching contact correction at
Gaussian order.

For a translation-invariant state, choose
\begin{equation}
 c_{\mathbf k a m}=\frac{1}{\sqrt{N_c}}
 \sum_{\mathbf R}e^{i\mathbf k\cdot\mathbf R}c_{\mathbf R a m},
 \qquad
 c_{\mathbf R a m}=\frac{1}{\sqrt{N_c}}
 \sum_{\mathbf k}e^{-i\mathbf k\cdot\mathbf R}c_{\mathbf k a m}.
 \label{c2:th:fermion-fourier}
\end{equation}
No internal-position phase is attached to $a$. Diagonalize the $\nu\times\nu$
Hartree matrix as
\begin{equation}
 h_H(\mathbf k)Q(\mathbf k)=Q(\mathbf k)E(\mathbf k),
 \qquad Q^\dagger Q=I.
 \label{c2:th:bands}
\end{equation}
The background is determined self-consistently by
\begin{align}
 \rho_a&=\frac{1}{N_c}\sum_{\mathbf k,\ell}
 |Q_{a\ell}(\mathbf k)|^2 f(E_\ell(\mathbf k)),
 &s_a&=-\sum_b v_{ab}(\mathbf 0)\rho_b,
 \label{c2:th:hartree-density}\\
 g_{H,ab}(\tau,\mathbf k)
 &=\sum_\ell Q_{a\ell}(\mathbf k)
 \frac{e^{-E_\ell(\mathbf k)\tau}}{1+e^{-\beta E_\ell(\mathbf k)}}
 Q^*_{b\ell}(\mathbf k),\quad 0<\tau<\beta.
 \label{c2:th:band-green}
\end{align}
Here $v_{ab}(\mathbf0)$ is the zero-momentum component in cell coordinates;
the indices $a,b$ retain the sublattice structure.
The total density per physical site is
$n=N_f\nu^{-1}\sum_a\rho_a$. A one-site square-lattice cell has $\nu=1$;
a larger cell enlarges the Hartree matrices. The $\rho_a$ may vary between
sublattices while remaining equal for all flavours.

The first saddle variation uses
$\delta\Tr\ln A=\Tr(A^{-1}\delta A)$, with
$\delta A=-i\delta\Sigma$. The resulting $+i\Tr(g_H\delta\Sigma)$
cancels the variation of the multiplier term only when $G_*=g_H$.
Variation with respect to $G$ then yields the local density field in
Eq.~\eqref{c2:th:saddle-equation}. In this derivation, the negative
equal-time value of $g_H$ fixes the sign of the physical Hartree shift.

Substitution into the stationary action gives the extensive saddle result
\begin{equation}
 \ln Z_{\mathrm{sp}}=N_f\left[
 \sum_{\mathbf k,\ell}\ln(1+e^{-\beta E_\ell(\mathbf k)})
 +\frac{\beta N_c}{2}\sum_{ab}\rho_a v_{ab}(\mathbf0)\rho_b
 \right].
 \label{c2:th:saddle-lnz}
\end{equation}
The second term removes the double counting already present in the
Hartree-shifted one-particle energies. Equation~\eqref{c2:th:saddle-lnz} is the total $\ln Z$; division by $N_s$
or $\beta N_s$ gives the corresponding intensive quantities. An accurate
saddle is needed because the fluctuation kernel and every subsequent
diagram use this background. For the numerical example below, this
background is the spatially uniform, flavour-symmetric scalar saddle.

\section{Gaussian fluctuations and screened interactions}
\label{c2:th:gaussian}

Write the fluctuations in rescaled form,
\begin{equation}
 G=g_H+N_f^{-1/2}\eta,\qquad
 \Sigma=\Sigma_*+N_f^{-1/2}\sigma.
 \label{c2:th:field-rescaling}
\end{equation}
This removes the overall flavour factor from the quadratic action.
For bilocal kernels use composition $(AB)(1,2)=\int_3A(1,3)B(3,2)$
and the bilinear pairing $(A,B)=\Tr(AB)=\int_{1,2}A(1,2)B(2,1)$.
Define operators on bilocal variations by
\begin{align}
 (\mathcal B A)(1,2)&=\int_{3,4}g_H(1,3)A(3,4)g_H(4,2),
 \label{c2:th:bubble-operator}\\
 (\mathcal V A)(1,2)&=\delta_{12}\int_3 v(1,3)A(3,3^+).
 \label{c2:th:interaction-operator}
\end{align}
The transpose of a bilocal operator $\mathcal M$ with respect to this
pairing is defined by $(A,\mathcal M B)=(\mathcal M^\top A,B)$.
In components, $\mathcal M^\top(1,2;3,4)=\mathcal M(4,3;2,1)$:
interchanging the paired slots also reverses the indices within each slot.

Since the linear terms vanish at the saddle, the quadratic term is
\begin{align}
 S_2&=\frac12\left[-(\sigma,\mathcal B\sigma)
 -2i(\sigma,\eta)+(\eta,\mathcal V\eta)\right],
 \label{c2:th:quadratic-action}\\
 \mathcal K&=
 \begin{pmatrix}-\mathcal B&-iI\\-iI&\mathcal V\end{pmatrix}.
 \label{c2:th:hessian}
\end{align}
The minus sign in the upper-left block comes from $i^2$.
The mixed block follows directly from $-i\Tr(\Sigma G)$.
The mixed block is imaginary, so the Gaussian integral over $\sigma$
and $\eta$ follows the constraint contours specified in
Section~\ref{c2:th:conventions}.

In components the two Hessian kernels are
$\mathcal B(1,2;3,4)=g_H(1,3)g_H(4,2)$ and
$\mathcal V(1,2;3,4)=\delta_{12}v(1,3)\delta_{34}$, with the
coincident arguments understood as above. Although $\mathcal B$ consists
of two fermion propagators, it acts on a bosonic bilocal variation.
Its projection onto density insertions therefore carries bosonic
frequencies. The full four-point notation is needed to derive the
block inverse; after projection the screened interaction has only two
density endpoints. 

\subsection{Resolvent form of the block inverse}
\label{c2:th:schur}

The density operator $\mathcal V$ annihilates many off-diagonal bilocal
variations, so it is generally singular on the full bilocal space.
The screened resolvents $R=(I-\mathcal V\mathcal B)^{-1}$ and
$\overline R=(I-\mathcal B\mathcal V)^{-1}$ nevertheless give the
block inverse whenever they exist. Direct multiplication yields
\begin{equation}
 \mathcal K^{-1}=
 \begin{pmatrix}
 R\mathcal V&iR\\
 i\overline R&-\overline R\mathcal B
 \end{pmatrix}.
 \label{c2:th:block-inverse}
\end{equation}
Since $R\mathcal V=\mathcal V\overline R$, the covariance may be written
with either of these equivalent product orders. Equation~\eqref{c2:th:block-inverse}
requires the screened resolvents to exist and also applies when a component
or eigenvalue of the bare interaction vanishes.

The rescaled multiplier covariance is $D_\Sigma=R\mathcal V$. Its density
part is local within each bilocal pair,
\begin{equation}
 D_\Sigma(1,2;3,4)=\delta_{12}D(1,3)\delta_{34},
 \qquad D=v+v\Pi D,
 \label{c2:th:positive-covariance}
\end{equation}
where $\Pi(1,2)=g_H(1,2)g_H(2,1)$ with the regulated endpoint convention.
 For diagram
generation it is convenient to use the rescaled field $X=i\sigma$. Its
covariance is instead
\begin{equation}
 W\equiv\langle XX\rangle=-D,
 \qquad W=-v+v\Pi W.
 \label{c2:th:negative-w}
\end{equation}
The conventions with bare terms $+v$ in $D$ and $-v$ in $W$ are thus
equivalent. The unrescaled fluctuation covariance contains
an additional $1/N_f$; that factor is absent from $D$ because of
Eq.~\eqref{c2:th:field-rescaling}.

\subsection{Fourier normalization and the polarization matrix}
\label{c2:th:fourier}

A translation-invariant two-point kernel uses the inverse-transform
factor $1/N_c$, whereas the fermion operators in
Eq.~\eqref{c2:th:fermion-fourier} use symmetric factors $1/\sqrt{N_c}$.
Thus take $F(\mathbf q)=\sum_{\mathbf R}e^{i\mathbf q\cdot\mathbf R}
F(\mathbf R)$ with inverse prefactor $1/N_c$. Time transforms use
$F(i\Omega_n)=\int_0^\beta d\tau\,e^{i\Omega_n\tau}F(\tau)$ and inverse
$\beta^{-1}\sum_n$, where $\Omega_n=2\pi n/\beta$. Then
\begin{align}
 g_H(\mathbf k,i\omega_l)&=[h_H(\mathbf k)-i\omega_l I]^{-1},
 &\omega_l&=(2l+1)\pi/\beta,
 \label{c2:th:matsubara-green}\\
 \Pi_{ab}(\tau,\mathbf q)
 &=\frac1{N_c}\sum_{\mathbf k}
 g_{H,ab}(\tau,\mathbf k+\mathbf q)g_{H,ba}(-\tau,\mathbf k),
 \label{c2:th:polarization-time}\\
 W(\mathbf q,i\Omega_n)
 &=-[I-v(\mathbf q)\Pi(\mathbf q,i\Omega_n)]^{-1}v(\mathbf q).
 \label{c2:th:screened-matrix}
\end{align}
Here $\Pi$ is the single-flavour polarization bubble: the flavour
multiplicity has cancelled against the $1/N_f$ in the scaled interaction.
With chemical potential as the source, the static density response of
the Hartree reference is $-\Pi(\mathbf q,0)$. A single level has $\Pi(i\Omega_n)=-\beta f(1-f)\delta_{n0}$, which provides a simple
check on the denominator of Eq.~\eqref{c2:th:screened-matrix}.

Separating the instantaneous part gives
\begin{align}
 W(\tau,\mathbf q)&=-v(\mathbf q)\delta_\beta(\tau)
 +\Delta W(\tau,\mathbf q),
 \label{c2:th:contact-split}\\
 \Delta W&=-v\Pi v+v\Pi\circ\Delta W.
 \label{c2:th:dynamic-equation}
\end{align}
Here $\circ$ denotes the time convolution, including matrix multiplication.
All factors of $N_c$ have already been included in
Eq.~\eqref{c2:th:polarization-time}. With a unitary Fourier normalization, the screened-interaction coefficients instead
represent $\widetilde{\Delta W}=\Delta W/\sqrt{N_c}$. Evaluating the physical
propagator then restores $\sqrt{N_c}$, as described in
Section~\ref{c2:dlr:representation}.

On a finite periodic lattice, these momentum sums contain exactly $N_c$
discrete points. In a bulk calculation each
normalized sum $N_c^{-1}\sum_{\mathbf k}$ is replaced by a Brillouin-zone
integral normalized to one. The internal sublattice dimension $\nu$
enters through the matrix indices and the conversion $N_s=\nu N_c$. Bosonic periodicity, fermionic antiperiodicity and
the reversal of both endpoints of the second Green function determine
the momentum routing in Eq.~\eqref{c2:th:polarization-time}. The product $g_{H,ab}(\tau,\mathbf k+\mathbf q)g_{H,ba}(-\tau,\mathbf k)$
therefore reverses the matrix indices and time argument together.

\section{Gaussian free energy and equal-time limits}
\label{c2:th:gaussian-lnz}

The Gaussian determinant is of order $N_f^0$ at fixed $v$, whereas
Eq.~\eqref{c2:th:saddle-lnz} is of order $N_f$. Normalizing the Gaussian
constraint integral to its noninteracting value removes field-independent
factors. Its formal result is
\begin{equation}
 \ln Z_G=-\frac12\Tr_{\mathrm{reg}}\ln(I-\mathcal B\mathcal V).
 \label{c2:th:formal-determinant}
\end{equation}
The density projection reduces the nontrivial determinant to
$I-v\Pi$. The subscript ``reg'' denotes the equal-time prescription of the
normal-ordered Hamiltonian. In particular, the one-interaction closed
diagram must be evaluated with that prescription; an unrestricted
Matsubara trace of the continuum bubble gives a different first-order term.

The density-square interaction
$H_{\mathrm{sq}}=(2N_f)^{-1}\sum_{ij}v_{ij}n_i n_j$
provides an operator-level derivation of the normal-ordered first-order term.
Equation~\eqref{c2:th:normal-order-identity} shows that its chemical potential
must be shifted by $v_{ii}/(2N_f)$ to represent the same normal-ordered
grand Hamiltonian. Expanding the leading, order-$N_f$ thermodynamic
potential in this shift adds
\begin{equation}
 C_{\mathrm{no}}=\frac\beta2\sum_i v_{ii}\rho_i
 \label{c2:th:onebody-correction}
\end{equation}
at order $N_f^0$. The shift introduces a counterterm while keeping the
physical grand Hamiltonian unchanged.

There are thus two consistent ways to organize the calculation. One can
retain the point-split, normal-ordered coherent-state construction from
the outset and use its regulated diagram rules at every order. Alternatively,
one can work with the density-square operator and apply the exact
chemical-potential shift throughout. If the second route is itself
expanded in $1/N_f$, Taylor terms from that shift act on the saddle,
the Gaussian term and the higher corrections at their respective orders.
Equation~\eqref{c2:th:onebody-correction} gives the correction through
Gaussian order. Higher orders require the corresponding Taylor terms.
I use the point-split construction to define the expansion and the
density-square representation to check the Gaussian contact term.

Let $\rho_{ij}=\langle c^\dagger_{jm}c_{im}\rangle_H$ for one flavour.
Wick's theorem in this free Hartree background gives the equal-time
connected density correlation
$\langle\delta n_{im}\delta n_{jm}\rangle_H
=\delta_{ij}\rho_i-\rho_{ij}\rho_{ji}$. Since $\Pi$ has the opposite sign,
its complete frequency trace satisfies
\begin{equation}
 \sum_{\mathbf q,n}\operatorname{tr}[v(\mathbf q)\Pi(\mathbf q,i\Omega_n)]
 =\beta\left[\sum_{ij}v_{ij}\rho_{ij}\rho_{ji}
 -\sum_i v_{ii}\rho_i\right].
 \label{c2:th:equal-time-trace}
\end{equation}
The lowercase trace acts on the $\nu$ internal sites. The complete
momentum and bosonic-frequency sum in Eq.~\eqref{c2:th:equal-time-trace}
has the normalization appropriate to the dimensionless $\ln Z$.

Combining the term linear in $v\Pi$ in Eq.~\eqref{c2:th:equal-time-trace}
with Eq.~\eqref{c2:th:onebody-correction} yields
the exchange/contact contribution
\begin{align}
 C_{\mathrm{ex}}&=\frac\beta2\sum_{ij}v_{ij}\rho_{ij}\rho_{ji},
 \qquad Y_{\mathbf qn}=v(\mathbf q)\Pi(\mathbf q,i\Omega_n),\notag\\
 \ln Z_G&=C_{\mathrm{ex}}-\frac12\sum_{\mathbf q,n}
 \left\{\ln\det[I-Y_{\mathbf qn}]+
 \operatorname{tr}Y_{\mathbf qn}\right\}.
 \label{c2:th:regularized-gaussian}
\end{align}
For onsite interactions $C_{\mathrm{ex}}=\beta v_0\sum_i\rho_i^2/2$.
The expression in braces starts at second order in $Y$; its leading contribution to
$\ln Z_G$ is $\sum\operatorname{tr}Y^2/4$. This separation gives an
unambiguous first-order term and a rapidly decaying frequency tail.
The imaginary-time calculation therefore evaluates the contact term
separately, using Eq.~\eqref{c2:th:regularized-gaussian}.

\subsection{Resolvent integration and a normal-ordering check}
\label{c2:th:resolvent}

For a branch with $I-xY$ nonsingular on $0\leq x\leq1$, choose the
logarithm continuously from $x=0$. The exact finite-matrix identity is
\begin{equation}
 -\ln\det(I-Y)=\int_0^1dx\,
 \operatorname{tr}[Y(I-xY)^{-1}].
 \label{c2:th:log-resolvent}
\end{equation}
This identity follows by differentiating $\ln\det(I-xY)$ along the
specified nonsingular branch. In the region where the geometric series
converges, integrating each term $x^r$ gives the corresponding factor
$1/(r+1)$.
Subtracting the linear term gives
\begin{equation}
 -\frac12[\ln\det(I-Y)+\operatorname{tr}Y]
 =\frac12\int_0^1dx\,\operatorname{tr}
 \{Y[(I-xY)^{-1}-I]\}.
 \label{c2:th:ring-resolvent}
\end{equation}
Defining $\Delta W_x=-xv\Pi v+xv\Pi\Delta W_x$ in frequency space
rewrites the integrand as $-\operatorname{tr}(\Pi\Delta W_x)/2$.
The auxiliary $x$ integral is evaluated at fixed Hartree background
$g_H$.

For a multi-site cell the intermediate matrices can be complex. Conjugate
terms combine to a real $\ln Z$, with the logarithm branch fixed by
continuity in $x$ as in Eq.~\eqref{c2:th:log-resolvent}.

An elementary check is the isolated onsite level. Its exact partition
function is
\begin{equation}
 Z_{\mathrm{at}}=\sum_{r=0}^{N_f}\binom{N_f}{r}
 \exp\!\left[\beta\mu r-\frac{\beta v_0}{2N_f}r(r-1)\right].
 \label{c2:th:atomic-check}
\end{equation}
With $\rho_0=f(-\mu)$, to linear order in $v_0$,
$\ln Z_{\mathrm{at}}=N_f\ln(1+e^{\beta\mu})
-\beta v_0(N_f-1)\rho_0^2/2+O(v_0^2)$.
The saddle contributes $-\beta N_fv_0\rho_0^2/2$ and the contact term
contributes $+\beta v_0\rho_0^2/2$. Their sum recovers the atomic
expansion to $O(v_0)$ with the stated normal-ordering prescription,
flavour normalization and Gaussian sign.

Finally, all interactions beyond the quadratic collective action come
from the trace logarithm. In terms of $X=i\sigma$ they read
\begin{equation}
 S_{\geq3}=\sum_{p=3}^{\infty}\frac{N_f^{1-p/2}}{p}
 \Tr[(g_HX)^p].
 \label{c2:th:higher-vertices}
\end{equation}
The normal-ordering regulator remains attached to coincident-time
contractions throughout this expansion. The vertex coefficients in Eq.~\eqref{c2:th:higher-vertices}, together
with $W=\langle XX\rangle=-D$, determine the graph expansion below.

\section{Flavour counting in the connected expansion}
\label{c2:gr:counting}

\subsection{From the trace logarithm to a finite set of partitions}

For graph enumeration, denote the higher-order action by
$S_{\mathrm{int}}\equiv S_{\geq3}$ and introduce its trace vertices
$T_p[X]$. Equation~\eqref{c2:th:higher-vertices} then reads\begin{equation}
 S_{\mathrm{int}}[X]=\sum_{p\geq3}
 \frac{N_f^{1-p/2}}{p}\,T_p[X],
 \qquad T_p[X]=\Tr[(g_HX)^p].
 \label{c2:gr:vertices}
\end{equation}
The factor $1/p$ comes from the logarithm, and the factor $N_f$ before
the rescaling counts the flavours in a closed fermion loop. The factors of $i$ are absorbed into $X$ and hence into the negative
screened covariance used in the weights below.

The correction to the logarithm of the partition function is
\begin{equation}
 \Delta\ln Z=\ln\langle e^{-S_{\mathrm{int}}}\rangle_0
 =\sum_{m=1}^{\infty}\frac{(-1)^m}{m!}
 \langle S_{\mathrm{int}}^m\rangle_{0,c},
 \label{c2:gr:cumulant}
\end{equation}
where the subscript $c$ denotes a connected Gaussian cumulant, and the
average uses the quadratic collective-field theory. A choice of $m$
vertices is specified by an unordered partition
$p=(p_1,\ldots,p_m)$ with $p_i\geq3$. Gaussian pairing requires an
even total number of $X$ insertions. Define
\begin{equation}
 2E=\sum_{i=1}^m p_i,\qquad k=E-m.
 \label{c2:gr:orders}
\end{equation}
Multiplying the vertex powers gives
$\prod_iN_f^{1-p_i/2}=N_f^{-k}$. A fixed flavour order $k$ therefore collects graphs with different
numbers of screened lines $E$ and trace vertices $m$.

There are finitely many partitions at a fixed non-Gaussian order $k$.
Since $m\geq1$ and $2E\geq3m=3(E-k)$,
\begin{equation}
 k+1\leq E\leq3k.
 \label{c2:gr:order-window}
\end{equation}
For example, $k=1$ admits $(4)$ and $(3,3)$, with $E=2$ and $3$,
respectively.
For $k=2$ the complete set is
\begin{equation}
 (6),\quad(3,5),\quad(4,4),\quad(3,3,4),\quad(3,3,3,3),
 \label{c2:gr:k2-partitions}
\end{equation}
with $E$ from $3$ to $6$. A calculation at a single $E$ gives only part of the order-$k$ result:
at $E=4$, for example, only $(3,5)$ and $(4,4)$ occur. The complete result
sums over every $E$ allowed by Eq.~\eqref{c2:gr:order-window}.

\subsection{Wick pairing and the weight of a labelled contraction}

Let $\nu_s$ be the multiplicity of a vertex of size $s$ in a partition.
Selecting that multiset from the exponential gives $1/\prod_s\nu_s!$.
The trace logarithms contribute $\prod_i1/p_i$. Thus every labelled
Gaussian pairing for this partition has prefactor
\begin{equation}
 c_p=\frac{(-1)^m N_f^{-k}}
 {\displaystyle\prod_{i=1}^{m}p_i\prod_s\nu_s!}.
 \label{c2:gr:raw-weight}
\end{equation}
The remaining factor is the product of propagators, summed and integrated
over the internal labels. There are $(2E-1)!!$ perfect matchings of the
$2E$ insertions before connectedness is imposed. Pairings are generated
by selecting the first unpaired insertion, joining it to each possible
partner, and recursing on the unpaired set. Each unordered pair is
selected once, generating the $(2E-1)!!$ matchings with unit multiplicity.

The original bilocal notation explains why the resulting graph has an
ordinary density vertex at each insertion. A screened covariance has
the structure
\begin{equation}
 \langle X(1,2)X(3,4)\rangle_0
 =\mathcal W(1,3)\delta_{12}\delta_{34},
 \label{c2:gr:local-insertions}
\end{equation}
with the equal-time regulator understood. Each delta identifies the two indices of one insertion while leaving
the endpoints of $\mathcal W$ distinct. The $g_H$ factors therefore form the
directed cycles specified by the partition, and the screened lines join
pairs of their vertices. The additional identification made by an
onsite instantaneous line is introduced when constructing the $L$ cover
in Section~\ref{c2:gr:l-cover}.

Taking a logarithm removes disconnected vacuum contributions. In graph
language, the union of the directed fermion edges and the interaction
edges must be connected. Connectedness cannot be checked on the
interaction lines alone: a path can travel along a fermion cycle before
crossing a screened line to another cycle. Equivalently, one can test
whether the graph whose nodes are the fermion cycles is connected by
the inter-cycle screened lines. The two connectivity criteria can be
compared directly at low order.

\subsection{The complete first non-Gaussian correction}

Expanding Eq.~\eqref{c2:gr:cumulant} to order $1/N_f$ gives
\begin{equation}
 \Delta\ln Z_{k=1}
 =-\frac{1}{4N_f}\langle T_4\rangle_0
 +\frac{1}{18N_f}\langle T_3T_3\rangle_{0,c}.
 \label{c2:gr:first-correction}
\end{equation}
The first term contains three pairings on one directed four-vertex
cycle. Two are related by a cyclic relabelling and the third is distinct.
Their combined weights are $-1/(2N_f)$ and $-1/(4N_f)$.
The second term contains fifteen pairings on two directed triangles.
Nine have one inter-cycle screened line and an internal pairing on each
triangle. The other six have three inter-cycle lines and form two
orientation classes of three pairings each. Their combined weights
are $1/(2N_f)$ and two copies of $1/(6N_f)$.

\begin{figure}[htbp]
\centering
\begin{tikzpicture}[scale=0.92,
 g/.style={-{Latex[length=1.6mm]},thick},
 w/.style={draw=red!70!black,thick,decorate,decoration={snake,amplitude=0.6mm,segment length=2.1mm}},
 dot/.style={circle,fill=black,inner sep=1.2pt}]
\begin{scope}[shift={(-2.5,2.8)}]
\foreach \a/\ang in {1/45,2/135,3/225,4/315}{\node[dot] (a\a) at (\ang:0.75) {};}
\foreach \a/\b in {1/2,2/3,3/4,4/1}{\draw[g] (a\a) -- (a\b);}
\draw[w] (a1) to[bend right=50] (a2);\draw[w] (a3) to[bend right=50] (a4);
\node at (0,-1.25) {$-1/(2N_f)$};\node at (0,1.25) {$(4)$: adjacent pairs};
\end{scope}
\begin{scope}[shift={(2.5,2.8)}]
\foreach \a/\ang in {1/45,2/135,3/225,4/315}{\node[dot] (b\a) at (\ang:0.75) {};}
\foreach \a/\b in {1/2,2/3,3/4,4/1}{\draw[g] (b\a) -- (b\b);}
\draw[w] (b1) -- (b3);\draw[w] (b2) -- (b4);
\node at (0,-1.25) {$-1/(4N_f)$};\node at (0,1.25) {$(4)$: opposite pairs};
\end{scope}
\foreach \x/\name/\mode in {-4.5/c/1,0/d/2,4.5/e/3}{
\begin{scope}[shift={(\x,-1.2)}]
\foreach \a/\ang in {1/90,2/210,3/330}{
 \node[dot] (\name\a) at ({-0.9+0.55*cos(\ang)},{0.55*sin(\ang)}) {};
 \pgfmathtruncatemacro{\b}{\a+3}
 \node[dot] (\name\b) at ({0.9+0.55*cos(\ang)},{0.55*sin(\ang)}) {};
}
\foreach \a/\b in {1/2,2/3,3/1,4/5,5/6,6/4}{\draw[g] (\name\a) -- (\name\b);}
\ifnum\mode=1
 \draw[w] (\name1) -- (\name4);
 \draw[w] (\name2) to[bend right=45] (\name3);
 \draw[w] (\name5) to[bend right=45] (\name6);
 \node at (0,-1.4) {$1/(2N_f)$};
\else
 \draw[w] (\name1) -- (\name4);
 \ifnum\mode=2
  \draw[w] (\name2) .. controls (-1.38,-0.98) and (0.42,-0.98) .. (\name5);
  \draw[w] (\name3) .. controls (-0.42,-0.64) and (1.38,-0.64) .. (\name6);
 \else
  \draw[w] (\name2) .. controls (-1.38,-0.98) and (1.38,-0.98) .. (\name6);
  \draw[w] (\name3) .. controls (-0.42,-0.64) and (0.42,-0.64) .. (\name5);
 \fi
 \node at (0,-1.4) {$1/(6N_f)$};
\fi
\end{scope}}
\node at (0,0.1) {$(3,3)$: one bridge and two three-bridge orientation classes};
\end{tikzpicture}
\caption[Connected topologies at first non-Gaussian order]{The five connected topological representatives at first
non-Gaussian order. Arrows denote $g_H$ and wavy lines the negative
screened covariance $\mathcal W$. The displayed factors already include
the multiplicity of labelled Wick pairings. Line crossings without dots
are not additional vertices.}
\label{c2:gr:first-graphs}
\end{figure}
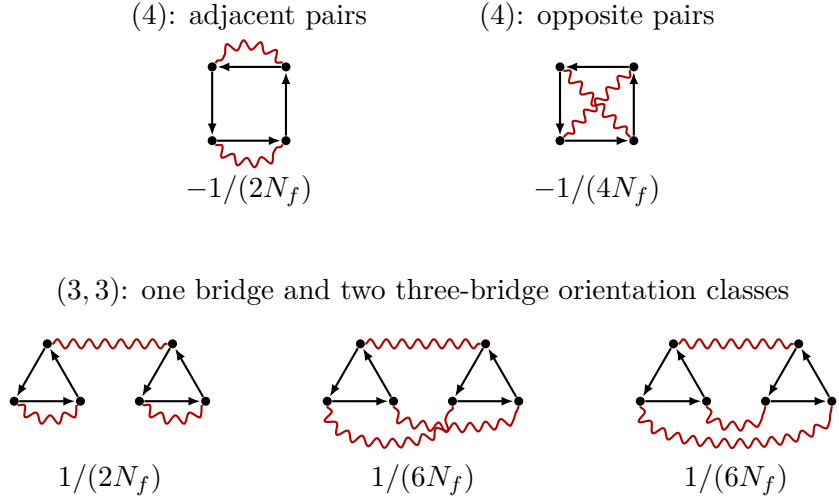

The weights above count labelled Wick pairings. The screened covariance,
fermion propagators and time ordering determine the value and sign of
each integral. Figure~\ref{c2:gr:first-graphs} shows the five representatives
with every insertion drawn.

The top row has two screened lines and one fermion cycle, giving
$k=E-m=2-1=1$. The bottom row has three screened lines and two cycles,
so $k=3-2=1$ as well. Grouping the equivalent pairings leaves five
distinct integrals with the weights shown below the graphs.

\section{Graph preparation and momentum routing}
\label{c2:gr:preparation}

\subsection{Graph-preparation sequence}

I prepare each graph structure once and reuse it during numerical
evaluation. For a selected pair $(E,k)$, the preparation sequence is:
\begin{enumerate}
 \item Generate nondecreasing integer partitions of $2E$ with all parts
 at least three and with $E-m=k$.
 \item Create a directed fermion cycle of length $p_i$ for each part,
 retaining distinct labels for its insertions.
 \item Enumerate the perfect matchings of the insertions and attach
 one screened edge to each pair.
 \item Remove disconnected graphs and assign the coefficient in
 Eq.~\eqref{c2:gr:raw-weight} to every surviving labelled pairing.
 \item Group isomorphic typed graphs and add their coefficients.
 \item Solve the momentum-conservation constraints for each representative.
 \item Select nonoverlapping $L$ blocks (Section~\ref{c2:dlr:l-function}),
 each integrating out one internal vertex, expand the remaining screened
 lines into instantaneous and dynamical pieces, and tabulate their
 connectivity, momentum combinations and combinatorial weights.
\end{enumerate}
Only steps 6 and 7 depend on the chosen integral representation: they
select momentum coordinates, internal integrations and the contact
expansion.
The topology and its combinatorial coefficient can be examined before
any Hartree calculation or Monte Carlo sampling is performed.

Each graph is specified by its screened edges, directed fermion edges,
coefficient and partition.
The graph for isomorphism testing is a directed multigraph: a fermion
edge has one directed orientation and one edge type; an interaction edge
is represented by two opposite directions of another type. I test
isomorphism with a VF2-family algorithm \cite{CordellaVF22004}, including
parallel edges and the type and direction of every edge in the comparison.

For a fixed partition, the natural relabelling group consists of cyclic
rotations of each directed fermion cycle and permutations of cycles of
equal length. Its order is $\prod_i p_i\prod_s\nu_s!$. If an orbit has
$M_\Gamma$ labelled pairings, the representative coefficient is
$M_\Gamma c_p$. By the orbit--stabilizer relation, $M_\Gamma c_p$ equals
$(-1)^mN_f^{-k}$ divided by the order of the residual automorphism group.
The orbit multiplicity and the denominator in $c_p$ thus combine into
the symmetry weight of the representative.

Table~\ref{c2:gr:counts} lists the resulting counts through $E=4$. The disconnected
pairings of $(4,4)$ are precisely the nine products of a four-vertex
pairing on each of the two separate cycles. Removing these nine products
leaves the 96 connected pairings contributing to $\ln Z$, out of the
105 pairings in the expansion of $Z$.

\begin{table}[htbp]
\centering
\begin{tabular}{@{}lrrrr@{}}
\toprule
Partition & $E$ & $k$ & Pairings (all / connected) & Topologies\\
\midrule
$(4)$ &2&1&$3/3$&2\\
$(6)$ &3&2&$15/15$&5\\
$(3,3)$ &3&1&$15/15$&3\\
$(8)$ &4&3&$105/105$&18\\
$(3,5)$ &4&2&$105/105$&7\\
$(4,4)$ &4&2&$105/96$&7\\
\bottomrule
\end{tabular}
\caption[Low-order graph enumeration counts]{Low-order combinatorial counts for the directed-cycle and
undirected-pairing rules. }
\label{c2:gr:counts}
\end{table}

\subsection{Physical weights and flavour-order sectors}

For a configuration $r$ descending from the representative $\Gamma(r)$
of partition $p(r)$, I use the physical coefficient
$c_r=M_{\Gamma(r)}c_{p(r)}$, with $c_p$ given by
Eq.~\eqref{c2:gr:raw-weight}. It includes the representative multiplicity
and the flavour power $N_f^{-k}$. Analytic $L$ integration and the
instantaneous/dynamical decomposition preserve this coefficient; the
interaction amplitudes remain in the propagator factors.

Each numerical integration covers a single $(E,k)$ sector. Its graphs
share the number of original time labels and free momenta, so their
reduced terms can be sampled on a common domain. Summing these integrals
over the allowed $E$ values gives the complete non-Gaussian order $k$,
to be added to the saddle and Gaussian contributions.

\subsection{Conservation equations and free loop momenta}

A graph in a translationally invariant system has one oriented momentum
per line. The direction of a screened edge is a bookkeeping choice;
reversing it also reverses its momentum and exchanges its matrix
indices. Let $A$ be the oriented incidence matrix and let $\bm p$ collect
all line momenta. Vertex conservation reads
\begin{equation}
 A\bm p=0\pmod{\mathbf G},
 \label{c2:gr:momentum-conservation}
\end{equation}
where $\mathbf G$ denotes reciprocal lattice vectors. The graph has
$2E$ insertion vertices, $2E$ fermion lines and $E$ screened lines.
For a connected graph the incidence matrix has rank $2E-1$, hence
\begin{equation}
 N_{\mathrm{free}}=3E-(2E-1)=E+1.
 \label{c2:gr:free-momenta}
\end{equation}
One may choose free edges by a spanning-tree construction or solve the
linear constraints by rank-revealing elimination. The remaining momenta
are integer linear combinations of the chosen free momenta,
$\bm p=B\bm q$, with $AB=0$. A cycle basis can be chosen with
coefficients in $\{-1,0,1\}$, so every routed momentum remains on the
finite periodic lattice's momentum grid.

The normalization follows from the Fourier convention: each of the
$3E$ lines carries an inverse-transform factor $1/N_c$. Summing
the $2E$ vertex positions produces $N_c^{2E}$ times the conservation
deltas. Dividing the extensive graph by $N_c$ for a result per cell
leaves $N_c^{-E-1}$, accompanying $E+1$ free momentum sums. A
result per physical site has the further factor $1/\nu$. Thus its
momentum measure is
\begin{equation}
 \frac{1}{\nu}\frac{1}{N_c^{E+1}}
 \sum_{\mathbf q_1,\ldots,\mathbf q_{E+1}}.
 \label{c2:gr:momentum-measure}
\end{equation}
Uniform discrete momentum sampling implements the normalized averages
in Eq.~\eqref{c2:gr:momentum-measure}.

\subsection{Compatible \texorpdfstring{$L$}{L} covers}
\label{c2:gr:l-cover}

A candidate $L$ block contains a screened line incident on the internal
vertex between two directed fermion lines. Replacing it by its defining
integral removes that internal time and its internal sublattice sum.
Two candidate blocks cannot both be chosen if they consume the same
propagator or internal vertex. Constructing a conflict graph of
candidates converts the selection into an independent-set problem.
I find a maximum independent set of this conflict graph to obtain a
largest compatible cover.

A compatible cover contains every original line exactly once, either
inside an $L$ block or as a remaining propagator. All such covers give
the same integral when the internal integrations are exact; maximizing
the cover reduces the numerical integration dimension.

For a remaining onsite instantaneous line,
$\mathcal W_{ab}^{\mathrm{inst}}(\tau,\mathbf q)
=-v_0\delta_{ab}\delta_\beta(\tau)$, the two endpoints are identified
in both time and sublattice index. An off-diagonal instantaneous matrix
element instead identifies the times while retaining distinct sublattice
labels. Expanding each remaining line into its instantaneous and dynamical
alternatives gives terms with the corresponding retained vertex labels.

Each $L$ block is specified by three external labels and two momentum
combinations. The remaining $G$ and $W$ factors are specified by their
endpoints and momenta. The resulting integral is evaluated through the
band propagators and the analytic time integrals derived next.

\section{Momentum representation and imaginary-time building blocks}
\label{c2:dlr:representation}

On a periodic $N_1\times N_2$ lattice with $N_c=N_1N_2$, the numerical
evaluator uses Eq.~\eqref{c2:th:fermion-fourier}. Suppressing the flavour
index, the same convention reads\begin{equation}
 c_{\mathbf k a}=\frac{1}{\sqrt{N_c}}\sum_{\mathbf R}
 e^{\mathrm i\mathbf k\cdot\mathbf R}c_{\mathbf R a},\qquad
 c_{\mathbf R a}=\frac{1}{\sqrt{N_c}}\sum_{\mathbf k}
 e^{-\mathrm i\mathbf k\cdot\mathbf R}c_{\mathbf k a}.
 \label{c2:dlr:fourier}
\end{equation}
The Fourier phase is attached to the unit-cell coordinate $\mathbf R$. 

\subsection{Band representation and the equal-time prescription}

In the convention of Eq.~\eqref{c2:dlr:fourier}, the nearest-neighbour
honeycomb model has off-diagonal hopping
$-t(1+e^{\mathrm i k_1}+e^{\mathrm i k_2})$, with $k_i=2\pi j_i/N_i$.
The Hermitian Hartree
one-body matrix, including the chemical potential, is diagonalized as
\begin{equation}
 h_H(\mathbf k)Q(\mathbf k)
 =Q(\mathbf k)\operatorname{diag}
 \bigl(\varepsilon_{0\mathbf k},\ldots,
       \varepsilon_{\nu-1,\mathbf k}\bigr),
 \qquad Q^\dagger Q=I.
 \label{c2:dlr:bands}
\end{equation}
The numerical propagator $g_H$ used below is the inverse of
$\partial_\tau+h_H$ on antiperiodic functions. Using the sign convention of Eq.~\eqref{c2:th:green-convention}, its
one-band expression on either side of zero is
\begin{equation}
 g_\varepsilon(\tau)=
 \begin{cases}
 \displaystyle\frac{e^{-\varepsilon\tau}}{1+e^{-\beta\varepsilon}},
       &0<\tau<\beta,\\[1mm]
 \displaystyle-\frac{e^{-\varepsilon\tau}}{1+e^{\beta\varepsilon}},
       &-\beta<\tau<0.
 \end{cases}
 \label{c2:dlr:band-green}
\end{equation}
Here $n_F\equiv f$ is the Fermi function defined in
Eq.~\eqref{c2:th:green-convention}. Thus $g_\varepsilon(0^- )=-n_F(\varepsilon)$ and
$g_\varepsilon(0^+)=1-n_F(\varepsilon)$: the jump is $+1$, as required
by its defining differential equation. The matrix propagator follows by
multiplying each band contribution by
$Q_{a\alpha}(\mathbf k)Q_{b\alpha}(\mathbf k)^*$ and summing over
$\alpha$. In particular,
\begin{equation}
 g_H^{ab}(0^-,\mathbf k)
 =-\sum_\alpha Q_{a\alpha}(\mathbf k)
 n_F(\varepsilon_{\alpha\mathbf k})Q_{b\alpha}(\mathbf k)^*.
 \label{c2:dlr:equal-time}
\end{equation}
The equal-time value in Eq.~\eqref{c2:dlr:equal-time} is used whenever
a contact contraction identifies two graph vertices.

For numerical evaluation define
\begin{equation}
 \overline\tau=\tau-\beta\lfloor\tau/\beta\rfloor,
 \qquad s_F(\tau)=(-1)^{\lfloor\tau/\beta\rfloor},
 \qquad F(x,\varepsilon)=\frac{e^{x\varepsilon}}{1+e^{\beta\varepsilon}}.
 \label{c2:dlr:wrapped}
\end{equation}
Away from equal times,
$g_\varepsilon(\tau)=s_F(\tau)F(\beta-\overline\tau,\varepsilon)$.
The factor $s_F$ accompanies the wrapped time and implements fermionic
antiperiodicity.

\subsection{Interaction normalization and conserved momenta}

Let $v_{ab}(\mathbf R)$ denote the density kernel entering the action
with prefactor $1/(2N_f)$, and define its unnormalized spatial
transform
\begin{equation}
 \mathcal V_{ab}(\mathbf q)=\sum_{\mathbf R}
 e^{\mathrm i\mathbf q\cdot\mathbf R}v_{ab}(\mathbf R).
 \label{c2:dlr:interaction-transform}
\end{equation}
For the onsite convention $v_{ii}=2u$, this is $2u\delta_{ab}$.
The calligraphic symbols $\mathcal V,\mathcal W$ are the field-theory
kernels $v,W$ written in the unnormalized momentum-space convention used
by the evaluator.
Comparison with the physical interaction between different flavours uses
$2u=N_fU_{\mathrm{phys}}$. The stored screened interaction, and hence its DLR coefficients in
Eq.~\eqref{c2:dlr:expansion}, use the symmetric transform
$W_s=\mathcal W/\sqrt{N_c}$. Restoring $\sqrt{N_c}$
gives the mixed-representation interaction
$\mathcal W$ used in the integrand. Its instantaneous part is
$-\mathcal V\delta_\beta(\tau)$, where $\delta_\beta$ is the periodic
delta distribution.

Momentum routing follows Section~\ref{c2:gr:preparation}. In matrix form,
reversing a density line exchanges its endpoints and sends
$(\tau,\mathbf q)$ to $(-\tau,-\mathbf q)$. The $E+1$ independent momenta carry the per-site measure of
Eq.~\eqref{c2:gr:momentum-measure}, written here as

\begin{equation}
 \frac{1}{\nu N_c^{E+1}}
 \sum_{\mathbf p_1,\ldots,\mathbf p_{E+1}}(\cdots).
 \label{c2:dlr:momentum-average}
\end{equation}
As in Eq.~\eqref{c2:gr:momentum-measure}, uniform momentum sampling
provides the factor $N_c^{-(E+1)}$ and $1/\nu$ converts to a per-site
quantity. The interaction lines in this expression are $\mathcal W$;
using $W_s$ requires restoring its factor $\sqrt{N_c}$.

\section{Discrete Lehmann representation}
\label{c2:dlr:screened}

Separating the instantaneous interaction is necessary both analytically
and numerically:
\begin{equation}
 \mathcal W(\tau,\mathbf q)
 =-\mathcal V(\mathbf q)\delta_\beta(\tau)
   +\delta\mathcal W(\tau,\mathbf q).
 \label{c2:dlr:contact-split}
\end{equation}
The delta distribution is handled exactly in graph contractions. Only the
regular dynamical part is compressed. A dense time grid could represent
that part, but repeated interpolation and storage become costly across
momenta and sublattice pairs. A short exponential expansion additionally
allows a time integral inside a diagram to be performed analytically.

The discrete Lehmann representation (DLR) constructs such an expansion
from a low-rank approximation of the imaginary-time spectral kernel.
The dimensionless spectral kernel determines its nodes for a given
cutoff $\Lambda=\beta\omega_{\max}$ and target tolerance
$\epsilon_{\mathrm{DLR}}$. In the resolved spectral
window, the required rank grows approximately as
$O(\log\Lambda\log(1/\epsilon_{\mathrm{DLR}}))$ for large $\Lambda$.
Imaginary-time interpolation nodes permit recovery of the coefficients
from a small set of values \cite{KayeDLR2022}. The \texttt{libdlr} library provides basis construction,
transforms and convolution operations \cite{KayeLibDLR2022}.

The construction uses dimensionless variables $x=\tau/\beta$ and
$y=\beta\omega$. A resolved sample of the kernel
on $x\in[0,1]$ and $y\in[-\Lambda,\Lambda]$ is subjected to a
rank-revealing decomposition. The selected frequency columns give the
exponentials and the selected time rows give the interpolation conditions.
The resulting dimensionless frequency nodes must be divided by $\beta$
before they are used with dimensional imaginary times. Both the decay rates and the resonance condition of the internal-vertex
integral (Section~\ref{c2:dlr:resonance-limits}) use these dimensional
frequencies. The frequency grid may be shared between matrix
components even though their coefficient vectors differ.

Cancellation between exponentials can make the coefficients sensitive
to small input perturbations even when the reconstructed function is
stable. Convergence must therefore be assessed from reconstructed values
and integrated quantities as well as from the coefficient vectors.

\subsection{Basis normalization and bosonic boundary conditions}

On the fundamental interval, introduce the normalized exponential
\begin{equation}
 k_l(\tau)=\frac{e^{-\omega_l\tau}}{1+e^{-\beta\omega_l}},
 \qquad 0\leq\tau\leq\beta,
 \label{c2:dlr:kernel}
\end{equation}
and approximate the stored interaction by
\begin{equation}
 \delta W_{s,ab}(\tau,\mathbf q)
 \simeq\sum_{l=1}^r p_{l,ab}(\mathbf q)k_l(\tau),
 \qquad
 \delta\mathcal W_{ab}=\sqrt{N_c}\,\delta W_{s,ab}.
 \label{c2:dlr:expansion}
\end{equation}
The fermionic denominator in Eq.~\eqref{c2:dlr:kernel} normalizes the basis
and is absorbed into its coefficients. The screened density interaction
is bosonic and is continued periodically: outside $[0,\beta)$ it is
evaluated at $\overline\tau$ with no fermionic sign. The bosonic endpoint relations are constraints on this reconstructed
expansion as a whole, satisfied within its representation error.

A fermionic
convolution introduces a minus sign when a time difference crosses the
boundary, whereas the interaction convolution below is periodic. The
zero-frequency kernel is simply $k_0(\tau)=1/2$ and can represent a
constant component with an appropriate coefficient. This normalization
keeps the basis finite at zero frequency. The selected frequencies
$\omega_l$ specify an interpolation basis, whose fitted coefficients
$p_{l,ab}$ reconstruct the imaginary-time interaction.

\subsection{From the RPA equation to a finite linear system}

Using the propagator convention above, define the unnormalized momentum
sum
\begin{equation}
 \mathcal P_{ab}(\tau,\mathbf q)=\sum_{\mathbf p}
 g_H^{ab}(\tau,\mathbf p)
 g_H^{ba}(-\tau,\mathbf p-\mathbf q).
 \label{c2:dlr:polarization}
\end{equation}
This auxiliary quantity satisfies $\mathcal P=N_c\Pi$, where $\Pi$ is
the normalized polarization used in the field-theory equations.
In the integral equations below, matrix products act on the sublattice
indices. The screened equation for $D=\delta W_s$ is
\begin{align}
 D(\tau)&=g^{(0)}(\tau)+
       \int_0^\beta f(\overline{\tau-s})D(s)\,\mathrm ds,
       \label{c2:dlr:screened-equation}\\
 f(\tau)&=\frac{\mathcal V\mathcal P(\tau)}{N_c},\qquad
 g^{(0)}(\tau)=-\frac{\mathcal V\mathcal P(\tau)\mathcal V}
                         {N_c\sqrt{N_c}}.
       \label{c2:dlr:screened-source}
\end{align}
Momentum arguments are suppressed because the equation is solved
separately at each $\mathbf q$. A scalar onsite interaction commutes with the other matrix factors;
the formulation above retains the order required for general matrices.

For interpolation nodes $\tau_j$ and $T_{jl}=k_l(\tau_j)$, the expansion
coefficients are obtained by solving $T\mathbf p=\mathbf D$.
Similarly expand $f_{ab}(\tau)=\sum_m f_{ab,m}k_m(\tau)$ and introduce
the periodic convolution of two basis functions,
\begin{equation}
 C_{ml}(\tau)=\int_0^\beta
 k_m(\overline{\tau-s})k_l(s)\,\mathrm ds.
 \label{c2:dlr:basis-convolution}
\end{equation}
Collocating Eq.~\eqref{c2:dlr:screened-equation} gives, for each source
index $b$,
\begin{equation}
 \sum_{c,l}\left[\delta_{ac}T_{jl}
    -\sum_m f_{ac,m}C_{ml}(\tau_j)\right]p_{l,cb}
 =g^{(0)}_{ab}(\tau_j).
 \label{c2:dlr:collocation}
\end{equation}
There are $\nu r$ unknowns. Multiplying each time block by $T^{-1}$
expresses the same system as $(I-\mathsf C_f)\mathbf p=\mathbf g^{(0)}$
in coefficient space. The direct DLR implementation uses this coefficient-space form, with
linear solves and a bosonic convolution matrix.

The elementary convolution is itself analytic. Set
$a_l=(1+e^{-\beta\omega_l})^{-1}$ and define
\begin{equation}
 \mathcal F(z;L,R)=\int_L^R e^{zs}\,\mathrm ds
 =\begin{cases}
 e^{zL}\bigl(e^{z(R-L)}-1\bigr)/z,&z\ne0,\\
 R-L,&z=0.
 \end{cases}
 \label{c2:dlr:elementary-integral}
\end{equation}
Splitting the integral at $s=\tau$ then yields
\begin{equation}
 C_{ml}(\tau)=a_ma_l e^{-\omega_m\tau}
 \left\{\mathcal F(\omega_m-\omega_l;0,\tau)
 +e^{-\beta\omega_m}
  \mathcal F(\omega_m-\omega_l;\tau,\beta)\right\}.
 \label{c2:dlr:convolution-explicit}
\end{equation}
For coincident frequencies the two integrals become their interval
lengths. Factors proportional to $\tau e^{-\omega\tau}$ then appear;
they are obtained by taking $\omega_m\to\omega_l$, using the $z=0$
branch of Eq.~\eqref{c2:dlr:elementary-integral}. The elementary integral
is analytic, but projection of the convolution onto the finite DLR basis
introduces a representation error.

For the densities in Section~\ref{sec:author-density-results}, I solved for
$W$ on an imaginary-time grid before fitting its DLR representation
\cite{ShiLargeNf2025}. The
coefficient-space formulation above solves the same equation without
the intermediate grid. Either solution can be tested using the residual
\begin{equation}
 R(\tau)=D(\tau)-g^{(0)}(\tau)-(f*D)(\tau),
 \label{c2:dlr:residual}
\end{equation}
evaluated at additional times, including near the boundaries. The residual at the collocation nodes is small by construction. When the
screening equation is poorly conditioned,
$(I-\mathsf C_f)^{-1}$ can amplify a small representation error.

\subsection{Related compact representations}

Compact bases reduce the cost of storing and manipulating imaginary-time
propagators. The Legendre representation of Boehnke et al.
\cite{BoehnkeLegendre2011} uses orthogonal polynomials and can suppress
noisy high-order components. The intermediate representation (IR) instead
uses the singular functions of the spectral kernel
\cite{ShinaokaIR2017}. Its compression depends on temperature and the
spectral representation, as analysed by Chikano et al.
\cite{ChikanoIR2018}. Sparse sampling chooses time or frequency nodes
from which the basis coefficients can be recovered \cite{LiSparse2020};
the \texttt{sparse-ir} library implements these operations
\cite{WallerbergerSparseIR2023}. A recent preprint relates the IR kernel
to a finite Laplace transform and studies its compression through
special-function structure \cite{MisawaIR2026}.

The DLR uses selected exponential kernels \cite{KayeDLR2022}, with
numerical building blocks provided by \texttt{libdlr} and \texttt{cppdlr}
\cite{KayeLibDLR2022,KayeCppDLR2024}. Exponential separability allows products
to be integrated analytically over internal times. Kaye et al. used a separable DLR basis to decompose diagrams of the
Anderson-impurity strong-coupling expansion into products and
one-dimensional convolutions \cite{KayeSeparable2024}. The $L$ blocks that I derive in Section~\ref{c2:dlr:l-function} use the
same separability for the screened-interaction expansion. Each block
eliminates the internal variables of one vertex with the contact term
and momentum routing appropriate to that expansion.

Quantics tensor trains offer an alternative by resolving coordinates into
bits at different scales and approximating the resulting tensor as a
low-rank chain \cite{ShinaokaQTT2023}. Applications include compressed
two-time operations in nonequilibrium diagrammatic calculations
\cite{MurrayQTT2024} and local imaginary-time propagators, for which the
truncation norm and the choice of time or frequency representation affect
compression \cite{TakahashiQTT2024}. I use the DLR for the analytic
internal integrals derived below.

\section{Analytical integration of an internal vertex}
\label{c2:dlr:l-function}

The $L$ function is obtained by integrating out a vertex shared by two
fermion propagators and one screened interaction. For three retained vertices with absolute times
$t_a,t_b,t_c\in[0,\beta)$, define
\begin{equation}
 \mathcal L_{abc}=
 \sum_{d=1}^{\nu}\int_0^\beta\mathrm ds\,
 g_H^{ad}(t_a-s,\mathbf k_1)
 \mathcal W_{db}(s-t_b,\mathbf q)
 g_H^{dc}(s-t_c,\mathbf k_2),
 \qquad\mathbf q=\mathbf k_1-\mathbf k_2.
 \label{c2:dlr:l-definition}
\end{equation}
Momentum conservation at the integrated vertex fixes the transfer.
Changing the integration origin to $t_b$ reproduces the two-time form
with $\tau_1=t_a-t_b$ and $\tau_2=t_b-t_c$. The complete integrand is
periodic in the integration variable because its two fermionic signs
cancel under a shift by $\beta$. The three absolute times provide one
expression valid for every ordering of the external vertices.

\subsection{The contact contribution}

Equation~\eqref{c2:dlr:contact-split} separates
$\mathcal L=\mathcal L^{\mathrm{ct}}+\mathcal L^{\mathrm{dyn}}$.
The delta distribution can be integrated immediately:
\begin{equation}
 \mathcal L^{\mathrm{ct}}_{abc}
 =-\sum_d g_H^{ad}(t_a-t_b,\mathbf k_1)
       \mathcal V_{db}(\mathbf q)
       g_H^{dc}(t_b-t_c,\mathbf k_2).
 \label{c2:dlr:l-contact}
\end{equation}
For onsite $\mathcal V_{db}=2u\delta_{db}$ this becomes a product of
two propagators with a factor $-2u$. If vertex identification makes a
propagator exactly equal-time, Eq.~\eqref{c2:dlr:equal-time} applies.
Distinct integration variables coincide only on a set of zero measure,
whereas a contact identification holds algebraically and contributes
with finite weight.

\subsection{Band and exponential expansion of the dynamical contribution}

Insert Eqs.~\eqref{c2:dlr:bands} and \eqref{c2:dlr:expansion}, and let
\begin{equation}
 B^{\alpha\gamma d}_{ac}=
 Q_{a\alpha}(\mathbf k_1)Q_{d\alpha}(\mathbf k_1)^*
 Q_{d\gamma}(\mathbf k_2)Q_{c\gamma}(\mathbf k_2)^*.
 \label{c2:dlr:band-prefactor}
\end{equation}
The remaining scalar integral is
\begin{align}
 \mathcal L^{\mathrm{dyn}}_{abc}
 &=\sqrt{N_c}\sum_{d,\alpha,\gamma,l}
 B^{\alpha\gamma d}_{ac}p_{l,db}(\mathbf q)
 I(\varepsilon,\varepsilon',\omega_l),
 \label{c2:dlr:l-expanded}\\
 I(\varepsilon,\varepsilon',\omega)
 &=\int_0^\beta\mathrm ds\,
 s_F(t_a-s)s_F(s-t_c)
 F(\beta-\overline{t_a-s},\varepsilon)
 F(\beta-\overline{s-t_c},\varepsilon')
 k_\omega(\overline{s-t_b}),
 \label{c2:dlr:l-scalar}
\end{align}
where $\varepsilon=\varepsilon_{\alpha\mathbf k_1}$ and
$\varepsilon'=\varepsilon_{\gamma\mathbf k_2}$. This step is exact for the specified band data and finite DLR
representation; the approximation error already present in $\delta W_s$
is carried through unchanged.

Sorting $\{0,t_a,t_b,t_c,\beta\}$ and removing duplicates partitions the
integration range into intervals $[L,R]$ on which all wrapping integers
and fermion signs are constant. Let $\bar s=(L+R)/2$ and $h=(R-L)/2$. Increasing
$s$ by $\delta s$ multiplies the first fermion factor by
$e^{\varepsilon\delta s}$, the second by
$e^{-\varepsilon'\delta s}$ and the interaction kernel by
$e^{-\omega\delta s}$. Their product consequently has slope
\begin{equation}
 z=\varepsilon-\varepsilon'-\omega.
 \label{c2:dlr:resonance}
\end{equation}
If $A_{\bar s}$ denotes the full signed scalar product in
Eq.~\eqref{c2:dlr:l-scalar} evaluated at the midpoint, the interval
contribution is
\begin{equation}
 I_{[L,R]}=A_{\bar s}\,\mathcal J(z;L,R),\qquad
 \mathcal J=\int_L^R e^{z(s-\bar s)}\,\mathrm ds
 =\begin{cases}2\sinh(zh)/z,&z\ne0,\\2h,&z=0.\end{cases}
 \label{c2:dlr:midpoint-integral}
\end{equation}
Summing the interval contributions (at most four) completes the integral
for any ordering of the external times, including repeated vertices and
$z=0$. At fixed sublattice dimension, the cost of this evaluation is
linear in the DLR rank.

\subsection{Resonance limits and stable floating-point evaluation}
\label{c2:dlr:resonance-limits}

The apparent singularity of Eq.~\eqref{c2:dlr:midpoint-integral} at $z=0$
is removable. With $x=zh$,
\begin{equation}
 \mathcal J=2h\left(1+\frac{x^2}{6}+\frac{x^4}{120}
                          +O(x^6)\right).
 \label{c2:dlr:resonance-series}
\end{equation}
Near resonance, Eq.~\eqref{c2:dlr:resonance-series} evaluates the
removable limit with a stable Taylor expansion. At large $|x|$, the
individual exponential factors can exceed the floating-point range,
although their product with the Fermi factors is finite. Combining their
logarithms before exponentiation preserves these cancellations of scale.

Define the auxiliary function
$\phi(y)=\log(1+e^y)=\max(0,y)+\log(1+e^{-|y|})$.
The second form uses a nonpositive exponential argument and remains
finite at large $|y|$. The required logarithms are
\begin{align}
 \log F(t,\varepsilon)&=t\varepsilon-\phi(\beta\varepsilon),
 \label{c2:dlr:log-fermi}\\
 \log k_\omega(t)&=
 \begin{cases}
 -t\omega-\log(1+e^{-\beta\omega}),&\omega\geq0,\\
 (\beta-t)\omega-\log(1+e^{\beta\omega}),&\omega<0,
 \end{cases}
 \label{c2:dlr:log-kernel}\\
 \log\mathcal J&=|zh|+\log(1-e^{-2|zh|})-\log|z|,
 \quad z\ne0.
 \label{c2:dlr:log-integral}
\end{align}
For small $|zh|$, the last expression is replaced by
$\log(2h)+\log[1+x^2/6+x^4/120+O(x^6)]$.
The scalar sign is $s_F(t_a-\bar s)s_F(\bar s-t_c)$ because the remaining
scalar factors are nonnegative. Complex phases belong to the band
prefactor and DLR coefficient. These phases accompany the logarithmically
evaluated magnitudes in the sum over bands and spectral terms, where
cancellation determines the arithmetic precision of the final $L$ value.

\section{Graph reduction and representation errors}
\label{c2:dlr:graph-control}

An $L$-cover selects subgraphs of the form in
Eq.~\eqref{c2:dlr:l-definition}, with disjoint fermion and interaction
lines in distinct selected blocks. As in Section~\ref{c2:gr:l-cover}, each integrated vertex lies inside its
block with all three incident lines, leaving the remaining factors
independent of that vertex's time and sublattice.

Remaining screened lines are expanded into their contact and dynamical
parts. For the onsite problem, a contact line identifies both the time
and the sublattice of its two endpoints. If a graph has $n_L$ integrated
$L$ blocks and $n_c$ remaining contact contractions, the number of retained
time and sublattice labels is $2E-n_L-n_c$. Mapping each retained time
to a uniform variable in $[0,1)$ and each sublattice to a uniform discrete
choice therefore gives
\begin{equation}
 J_{\tau,a}=(\nu\beta)^{2E-n_L-n_c}.
 \label{c2:dlr:jacobian}
\end{equation}
Each integrated $L$ includes its internal time integral and sublattice
sum, so only the remaining labels contribute to the Jacobian. The contact
rule in Eq.~\eqref{c2:dlr:jacobian} applies to onsite interactions.
For a spatially nonlocal instantaneous interaction, the contact condition
identifies the times while leaving the spatial endpoints distinct.

Uniform selection of one of the $N_g$ graph configurations gives an
estimator of their sum after multiplication by $N_g$. The
complete normalization includes this selection factor, the momentum
measure in Eq.~\eqref{c2:gr:momentum-measure}, and the Jacobian in
Eq.~\eqref{c2:dlr:jacobian}.

The effect of a representation error can also be bounded before the
outer integration. Keep the Hartree data fixed and suppose that each
entry of the restored dynamical interaction has a uniform absolute error
at most $\epsilon_W$. Since the eigenvalues of the finite-temperature
propagator in Eq.~\eqref{c2:dlr:band-green} have magnitude at most one,
its matrix elements also have magnitude at most one. The defining
integral then implies the conservative bound
\begin{equation}
 |\Delta\mathcal L_{abc}^{\mathrm{dyn}}|
 \leq\sum_d\int_0^\beta\mathrm ds\,
 |g_H^{ad}|\,\epsilon_W\,|g_H^{dc}|
 \leq\nu\beta\epsilon_W.
 \label{c2:dlr:l-error-bound}
\end{equation}
Here $\epsilon_W$ bounds the error in the reconstructed physical kernel.
The bound is absolute and can be loose for rapidly decaying propagators;
near a zero of $\mathcal L$, the relative error can still be large.
Relating $\epsilon_W$ to a requested DLR tolerance requires a bound on
the reconstruction, and errors in the Hartree data add further terms.

At the graph level, an error in one factor is multiplied by all the
others. To first order, a product obeys
\begin{equation}
 \Delta\!\left(\prod_{j=1}^M f_j\right)
 =\sum_{j=1}^M\Delta f_j\prod_{i\ne j}f_i+O(\Delta f^2).
 \label{c2:dlr:product-error}
\end{equation}
Written in absolute errors, Eq.~\eqref{c2:dlr:product-error} also applies
when an individual factor vanishes. Cancellations
between graph contributions can further increase the relative sensitivity
of the final coefficient. Tolerances should be chosen by varying the
representation and checking the final observable, in addition to testing
local residuals.

For a fixed Hartree solution and screened kernel, the piecewise $L$
integral has no time-quadrature discretization. Its numerical error comes
from the kernel representation and arithmetic. The outer integral adds
Monte Carlo uncertainty, while summation over a finite set of flavour
orders introduces truncation error. Section~\ref{c2:mc:observables}
collects the corresponding convergence parameters.

\section{The remaining integral and the Monte Carlo estimator}
\label{c2:mc:integral}

\subsection{Graph configurations on a common integration domain}

Let $\mathcal I_{E,k}$ denote the contribution per physical site of one
homogeneous graph sector. After isomorphism reduction, $L$ compression
and the instantaneous/dynamical expansion, label its graph configurations by
$r=1,\ldots,N_g$. Each configuration contains $n_L(r)$ analytic blocks,
$n_G(r)$ explicit fermion propagators and the remaining screened lines.
For the onsite contraction rule, let $n_c(r)$ be the number of
instantaneous remaining lines whose endpoints have been identified.
The number of retained time and external sublattice labels is
\begin{equation}
 d_r=2E-n_L(r)-n_c(r).
 \label{c2:mc:retained-labels}
\end{equation}
As in Eq.~\eqref{c2:dlr:jacobian}, each $L$ block already includes its
internal sublattice sum.

To integrate the graph configurations together, embed them in a common
unit hypercube of dimension $D$, including the coordinate that selects
the configuration. Lower-dimensional terms are extended by auxiliary
coordinates with unit integral on $[0,1)$. Time labels are mapped by
$\tau_j=\beta x_j$, sublattice labels by $a_j=1+\lfloor\nu y_j\rfloor$,
and each component of a free momentum by
$k_{j\ell}=2\pi\lfloor N_\ell z_{j\ell}\rfloor/N_\ell$.
The additional coordinate $u\in[0,1)$ selects the configuration index
$r=1+\lfloor N_g u\rfloor$. Let $c_r=M_{\Gamma(r)}c_{p(r)}$ be its
physical weight from Eq.~\eqref{c2:gr:raw-weight}. The resulting integrand is
\begin{equation}
 \begin{aligned}
 f_{E,k}(x)&=\frac{N_g}{\nu}\,c_r
 (\nu\beta)^{d_r}
 \operatorname{Re}\!\left[
 \prod L\;\prod g_H\;\prod\Delta\mathcal W\;
 \prod\mathcal W^{\mathrm{inst}}\right],\\
 \mathcal I_{E,k}&=\int_{[0,1]^D}f_{E,k}(x)\,\mathrm dx.
 \end{aligned}
 \label{c2:mc:integrand}
\end{equation}
The last product refers to the evaluated instantaneous coefficients,
after their delta constraints have been applied. The discrete momentum
averages are implemented by the maps themselves, as in
Eq.~\eqref{c2:gr:momentum-measure}. The factor $N_g$ restores the sum
over uniformly sampled graph configurations, and $1/\nu$ converts the result from
per cell to per physical site.

On the square lattice $\nu=1$, whereas the two-sublattice onsite
implementation has the measure
$\tfrac12(2\beta)^{d_r}$, before the graph and flavour factors.
The contact identification is the onsite rule of
Eq.~\eqref{c2:dlr:jacobian}.

The propagator matrices can be complex even when the final
thermodynamic quantity is real. The real part in
Eq.~\eqref{c2:mc:integrand} is taken after multiplying all factors
of the graph, preserving interference between their phases. Hermiticity
and the momentum-reversal relations give consistency conditions for the
reality of the complete integral.

\subsection{Adaptive mapping of the integration variables}

A large integration domain may contain a contribution concentrated
near particular relative times or momenta. Uniform independent
sampling then spends most evaluations in regions of little weight.
For an importance density $q(x)$, the variance of an independent
estimator of $I=\int f(x)\,\mathrm dx$ is
\begin{equation}
 \operatorname{Var}\widehat I
 =\frac1M\left[\int\frac{f(x)^2}{q(x)}\,\mathrm dx-I^2\right].
 \label{c2:mc:importance-variance}
\end{equation}
At fixed support and normalization, the ideal choice is proportional
to $|f|$. A practical mapping approximates this ideal density using samples of the
high-dimensional integrand.

VEGAS represents the mapping by adaptive one-dimensional grids for
the coordinates \cite{LepageVEGAS1978,LepageVEGAS2021}. For
$x=x(y)$ with Jacobian $J(y)$,
\begin{equation}
 I=\int_{[0,1]^D}\widetilde f(y)\,\mathrm dy,
 \qquad\widetilde f(y)=f(x(y))J(y).
 \label{c2:mc:vegas-map}
\end{equation}
Grid training redistributes the intervals in response to measured
weight. Damping their changes helps prevent poorly resolved
intervals from collapsing. A product mapping can substantially
improve coordinate-wise localization, while strongly correlated
directions in configuration space can remain difficult. Analytic $L$
integration removes variables exactly for a fixed representation. The
adaptive map changes the sampling of the remaining variables.

For the densities in Section~\ref{sec:author-density-results}, I combined
adaptive mapping with the normalization chain described below
\cite{ShiLargeNf2025}. The ratio construction itself does
not require a map and can act directly on the unit-cube integrand. When the map is used, both the
transformed variables and its Jacobian enter the integrand in
Eq.~\eqref{c2:mc:vegas-map} before the normalization ratios are measured.

\subsection{Separating the sign from the absolute normalization}

For a real integrand on the common domain with
$0<\int|f(x)|\,\mathrm dx<\infty$, define
\begin{equation}
 A=\int|f(x)|\,\mathrm dx,\qquad
 \langle s\rangle_{|f|}
 =\frac{\int f(x)\,\mathrm dx}{A},\qquad
 I=A\langle s\rangle_{|f|}.
 \label{c2:mc:sign-factorization}
\end{equation}
The two factors expose different sources of statistical cost. A small
average sign demands many samples for relative accuracy, even though
the sign estimator is bounded. Independently, sharply localized weight
can make the absolute normalization $A$ expensive to integrate.
An effective reference calculation must resolve both the sign cancellation
and the concentration of the absolute weight.

Let $g_0=|f_{\theta_0}|$ be the target absolute integrand, where
$\theta$ includes temperature, lattice size and couplings.
Choose an auxiliary $g_1=|f_{\theta_1}|$ that is easier to
integrate and define $A_j=\int g_j$. Every reference normalization
satisfies $0<A_j<\infty$. Sampling the mixture
$\pi_{01}=(g_0+g_1)/(A_0+A_1)$ gives the bounded observables
below on its support:
\begin{equation}
 R_0=\frac{g_0}{g_0+g_1},\qquad
 R_1=\frac{g_1}{g_0+g_1},\qquad
 \frac{\langle R_0\rangle_{\pi_{01}}}
 {\langle R_1\rangle_{\pi_{01}}}
 =\frac{A_0}{A_1}.
 \label{c2:mc:ratio-identity}
\end{equation}
Integrating each numerator against this mixture yields
$\langle R_j\rangle_{\pi_{01}}=A_j/(A_0+A_1)$, proving the ratio
identity.

Repeating the construction produces
\begin{equation}
 I_{\theta_0}
 =\langle s\rangle_{g_0}
 \prod_{j=0}^{J-1}\frac{A_j}{A_{j+1}}\;A_J.
 \label{c2:mc:ratio-product}
\end{equation}
The endpoint is chosen at smaller volume, higher temperature or weaker
interaction so that its absolute integral can be estimated directly.
In my implementation the telescoping product runs successively through
system size, inverse temperature and interaction strength. Each ratio
uses two integrands embedded in the same normalized domain.

I refer to this procedure as renormalization-group Markov-chain Monte Carlo
(RG-MCMC) \cite{ShiLargeNf2025}, after the progressive change of physical
scales in the auxiliary problems. Its mathematical basis is the statistical
ratio identity in Eq.~\eqref{c2:mc:ratio-identity}. Its efficiency
depends on overlap between neighbouring absolute-weight distributions.
Poor overlap or separated modes can lead to long autocorrelation times,
so each ratio estimate needs checks of mixing and sensitivity to the
spacing between auxiliary problems.

\subsection{Updates, production measurements and uncertainty}

For a symmetric proposal $x\to x'$, the sign-chain Metropolis
acceptance probability is
\[
 \min[1,|f(x')|/|f(x)|].
\]
A ratio chain uses the sum of its two absolute weights instead.
The coordinate update used here selects one coordinate and gives it a
Gaussian displacement. Points outside the unit cube are rejected, and a
measurement is taken after every production step, accepted or rejected,
as required by the Metropolis measure.

After warm-up, $M$ production measurements are averaged. Independent
complete runs, with independently initialized random streams, give
independent estimates of the product in Eq.~\eqref{c2:mc:ratio-product}.
The uncertainty of their mean is the sample standard deviation of the
complete-run estimates divided by the square root of the number of runs.
Blocking within an equilibrated run can estimate the uncertainty of
averages of correlated measurements \cite{FlyvbjergPetersen1989}.
The Gelman--Rubin convergence diagnostic instead compares within- and
between-sequence variation across multiple independent sequences started
from an overdispersed distribution \cite{GelmanRubin1992}. When factors
share randomness, the product is formed within each run before the
independent complete products are averaged.

The two observables in Eq.~\eqref{c2:mc:ratio-identity} are
strongly correlated: for absolute weights, $R_0+R_1=1$
pointwise. If $r=\overline R_0/\overline R_1$, linearized
error propagation gives
\begin{equation}
 \operatorname{Var}r\simeq
 \frac{\operatorname{Var}\overline R_0}{\overline R_1^2}
 +\frac{\overline R_0^2}{\overline R_1^4}
  \operatorname{Var}\overline R_1
 -\frac{2\overline R_0}{\overline R_1^3}
  \operatorname{Cov}(\overline R_0,\overline R_1).
 \label{c2:mc:ratio-error}
\end{equation}
The uncertainty of each mean must itself include its Markov
correlations. Finite-sample ratio bias can be assessed by increasing run
lengths and comparing independent runs.

\section{Thermodynamic derivatives and numerical control}
\label{c2:mc:observables}

\subsection{Density as a derivative of the complete approximation}

The expansion is first organized for $\ln Z$. The total
density per physical site is
\begin{equation}
 n=\frac{1}{\beta N_s}\frac{\dd\ln Z}{\dd\mu}
 =n_{\mathrm{sp}}+\delta n_{\mathrm G}
 +\delta n_{1/N_f}+\cdots.
 \label{c2:mc:density-expansion}
\end{equation}
The saddle scales as $N_f$ at fixed flavour-scaled kernel,
the Gaussian/RPA term as $N_f^0$, and the first
non-Gaussian contribution as $N_f^{-1}$. These order labels refer to
the density summed over flavours, which is the quantity used below.

Stationarity removes implicit derivatives of the saddle
fields from the derivative of the saddle action itself.
It gives
\begin{equation}
 n_{\mathrm{sp}}=\frac{N_f}{\nu N_c}
 \sum_{\mathbf k,\alpha}n_F(\varepsilon_{\alpha\mathbf k}).
 \label{c2:mc:saddle-density}
\end{equation}
The Gaussian and higher terms must instead be differentiated
along the moving Hartree solution. Their dependences on the propagators,
screening, integration measure and explicit parameters all contribute to
the derivative.

For illustration, in a uniform scalar problem with
$\rho=F(\mu-v_0\rho)$ and
$\chi_0=F'(\mu-v_0\rho)$,
\begin{equation}
 \frac{\dd\rho}{\dd\mu}
 =\frac{\chi_0}{1+v_0\chi_0}.
 \label{c2:mc:hartree-response}
\end{equation}
For several density components the corresponding linear
system is $(I+\chi_0v)\,\dd\bm\rho/\dd\mu
=\chi_0\bm 1$, with the matrix order retained. The same implicit dependence enters the derivative of a Gaussian
determinant or a higher graph.

The numerical method can evaluate the derivative by
combining shifted integrands at the same point of the
unit cube. If $f(x;\mu)$ is already a contribution to
$\ln Z/N_s$, then
\begin{equation}
 \delta n_{E,k}\simeq\int\mathrm dx\,
 \frac{f(x;\mu+h_\mu)-f(x;\mu-h_\mu)}
 {2\beta h_\mu}.
 \label{c2:mc:density-difference}
\end{equation}
Both backgrounds are prepared self-consistently. The
difference is formed before Monte Carlo averaging, so
the covariance between its terms is retained. For positively correlated shifted integrands, this covariance reduces
the variance of the difference relative to independent estimates. The finite-difference error remains
and is tested by varying $h_\mu$.

The energy per site follows from the physical Hamiltonian,
which excludes $-\mu\widehat N$:
\begin{equation}
 e=-\partial_\beta\left(\frac{\ln Z}{N_s}\right)+\mu n.
 \label{c2:mc:energy}
\end{equation}
The $\beta$ derivative in Eq.~\eqref{c2:mc:energy}, taken at fixed
$\mu$ and Hamiltonian parameters, includes the time-variable rescaling
and its Jacobian. Analogous differences give a density
response or the observable conjugate to the physical
interaction. If the internal variable is
$u=N_fU_{\mathrm{phys}}/2$, a physical interaction
derivative includes
$\partial_{U_{\mathrm{phys}}}=(N_f/2)\partial_u$
at fixed $N_f$. The double-occupancy estimator uses the derivative
with respect to $U_{\mathrm{phys}}$ after this conversion.

\subsection{Representation, sampling and truncation errors}

The numerical error has contributions from the Hartree solution, the
screened-kernel representation, arithmetic, sampling and truncation of the
$1/N_f$ series. Residuals measure errors in the Hartree and screening
equations. To assess their effect on a full diagram, the spectral cutoff
and DLR tolerance must be varied and screening residuals evaluated away
from interpolation nodes. Derivative estimates also depend on the
difference steps $h_\mu$ and $h_\beta$. Stable arithmetic is needed to
avoid overflow and reduce cancellation error. Longer equilibrated runs
reduce the statistical uncertainty; representation accuracy and the
number of computed flavour orders determine the other errors.

For the density calculations, my numerical specification in
Ref.~\cite{ShiLargeNf2025} gives $2^{16}$ warm-up steps,
chemical-potential and inverse-temperature difference steps
$h_\mu/t=0.01$ and $t h_\beta=0.02$, and an adaptive grid of $100$
intervals per coordinate. Their role in the error budget is set by the
residual, difference-step and sampling checks described above.

The saddle and Gaussian terms are naturally obtained as total $\ln Z$,
whereas the higher-order Monte Carlo integral is evaluated per physical
site. These quantities are brought to the same normalization before
addition or differentiation. 

At fixed $k$, all sectors in
Eq.~\eqref{c2:gr:order-window} must be included before
the result is called complete to that flavour order.
For the first non-Gaussian density term, both the
quartic-vertex and two-cubic-vertex contributions
and their full parameter derivatives are required.
These contributions define the successive columns analysed in
Section~\ref{sec:author-density-results}.

\section{Density results for the six-flavour square-lattice model}
\label{sec:author-density-results}

\subsection{Density estimates and successive contributions}

Table~\ref{tab:table3-author-data} lists the densities that I calculated
for the $\mathrm{SU}(6)$ Hubbard model on a $70\times70$ square lattice at
$T/t=0.15$, first reported in Ref.~\cite{ShiLargeNf2025}. The density $n$
sums all six flavours per site, and the successive contributions resolve
the change from the Hartree prediction to the non-Gaussian result.
There are 31 state points: six at
$U/t=2.3$, eight at $U/t=4$, ten at $U/t=8$, and seven at $U/t=12$.

For the uniform saddle on this finite lattice, the self-consistency
equation is
\begin{equation}
 \rho=\frac{1}{70^2}\sum_{j_1,j_2=0}^{69}
 f\!\left[-2t\left(\cos\frac{2\pi j_1}{70}
                   +\cos\frac{2\pi j_2}{70}\right)
             -\mu+6U_{\mathrm{phys}}\rho\right],
 \qquad n_{\mathrm{sp}}=6\rho.
 \label{c2:results:saddle-recheck}
\end{equation}
Let $\Phi(\rho)$ denote the left-hand side minus the right-hand side of
Eq.~\eqref{c2:results:saddle-recheck}. For repulsive $U_{\mathrm{phys}}$,
$\Phi'(\rho)=1+6\beta U_{\mathrm{phys}}\langle f(1-f)\rangle>0$ and
$\Phi(0)<0<\Phi(1)$, ensuring a unique uniform scalar root.
The solution $n_{\mathrm{sp}}=6\rho$ is the saddle reference at each
of the 31 state points in Table~\ref{tab:table3-author-data}.

The table separates the saddle density and the additive RPA correction
from the cumulative density $n_{\leq1/N_f}$, which also includes the
first non-Gaussian contribution. Their relation is
\begin{align}
 n_{\mathrm{sp+RPA}}&=n_{\mathrm{sp}}+\delta n_{\mathrm{RPA}},
 \label{eq:author-density-gaussian}\\
 \delta n_{1/N_f}
 &=n_{\leq1/N_f}-n_{\mathrm{sp}}-\delta n_{\mathrm{RPA}}.
 \label{eq:author-density-reconstruction}
\end{align}
The increment $\delta n_{1/N_f}$ is evaluated at the physical flavour
number $N_f=6$ and includes its factor $1/N_f$. The saddle and RPA
contributions are determined numerically before the non-Gaussian term
is sampled, so their subtraction preserves the Monte Carlo uncertainty
of the cumulative result.

\subsection{Cumulative estimates and additive corrections}

\begin{figure}[p]
\centering
\begin{tikzpicture}
\begin{groupplot}[
group style={group size=2 by 2,horizontal sep=12mm,vertical sep=21mm},
width=0.45\textwidth,height=57mm,
xlabel={$\mu/t$},ylabel={$n$},
ymin=0.25,ymax=1.9,
grid=major,grid style={gray!15},axis line style={gray!60},
tick label style={font=\scriptsize},label style={font=\small},title style={font=\small},
legend style={font=\scriptsize,draw=none,fill=white,at={(0.03,0.97)},anchor=north west},
scaled ticks=false,
every axis plot/.append style={line width=0.7pt,mark size=1.6pt}
]
\nextgroupplot[title={$U/t=2.3$}]
\addplot+[gray!70!black,dashed,mark=square*] coordinates {
(-2.300000,0.39798)
(-1.725000,0.53662)
(-1.150000,0.67783)
(0.000000,0.96806)
(0.575000,1.11728)
(1.725000,1.42457)
};
\addlegendentry{$n_{\mathrm{sp}}$}
\addplot+[blue!70!black,mark=o] coordinates {
(-2.300000,0.48022)
(-1.725000,0.64040)
(-1.150000,0.80201)
(0.000000,1.13082)
(0.575000,1.29854)
(1.725000,1.64186)
};
\addlegendentry{$n_{\mathrm{sp}}+\delta n_{\mathrm{RPA}}$}
\addplot+[orange!85!black,mark=triangle*] coordinates {
(-2.300000,0.487)
(-1.725000,0.651)
(-1.150000,0.815)
(0.000000,1.148)
(0.575000,1.318)
(1.725000,1.665)
};
\addlegendentry{$n_{\leq 1/N_f}$}
\addplot+[orange!85!black,only marks,mark=none,forget plot,error bars/.cd,y dir=both,y explicit] coordinates {
(-2.300000,0.487) +- (0,0.001)
(-1.725000,0.651) +- (0,0.001)
(-1.150000,0.815) +- (0,0.001)
(0.000000,1.148) +- (0,0.001)
(0.575000,1.318) +- (0,0.001)
(1.725000,1.665) +- (0,0.001)
};
\nextgroupplot[title={$U/t=4.0$}]
\addplot+[gray!70!black,dashed,mark=square*] coordinates {
(-1.679207,0.38851)
(-0.886256,0.52397)
(-0.088440,0.66202)
(0.714382,0.80267)
(1.522465,0.94604)
(2.336111,1.09224)
(3.155679,1.24145)
(3.981595,1.39384)
};
\addplot+[blue!70!black,mark=o] coordinates {
(-1.679207,0.51314)
(-0.886256,0.67575)
(-0.088440,0.83818)
(0.714382,1.00121)
(1.522465,1.16542)
(2.336111,1.33123)
(3.155679,1.49904)
(3.981595,1.66931)
};
\addplot+[orange!85!black,mark=triangle*] coordinates {
(-1.679207,0.533)
(-0.886256,0.700)
(-0.088440,0.865)
(0.714382,1.030)
(1.522465,1.195)
(2.336111,1.362)
(3.155679,1.530)
(3.981595,1.701)
};
\addplot+[orange!85!black,only marks,mark=none,forget plot,error bars/.cd,y dir=both,y explicit] coordinates {
(-1.679207,0.533) +- (0,0.001)
(-0.886256,0.700) +- (0,0.001)
(-0.088440,0.865) +- (0,0.001)
(0.714382,1.030) +- (0,0.001)
(1.522465,1.195) +- (0,0.001)
(2.336111,1.362) +- (0,0.001)
(3.155679,1.530) +- (0,0.001)
(3.981595,1.701) +- (0,0.001)
};
\nextgroupplot[title={$U/t=8.0$}]
\addplot+[gray!70!black,dashed,mark=square*] coordinates {
(-0.218517,0.37908)
(1.087260,0.51151)
(1.746251,0.57872)
(2.409349,0.64660)
(3.076639,0.71515)
(3.748223,0.78440)
(5.104735,0.92502)
(6.479902,1.06863)
(7.874923,1.21540)
(9.291230,1.36554)
};
\addplot+[blue!70!black,mark=o] coordinates {
(-0.218517,0.56877)
(1.087260,0.73232)
(1.746251,0.81310)
(2.409349,0.89355)
(3.076639,0.97382)
(3.748223,1.05407)
(5.104735,1.21493)
(6.479902,1.37688)
(7.874923,1.54050)
(9.291230,1.70635)
};
\addplot+[orange!85!black,mark=triangle*] coordinates {
(-0.218517,0.6184)
(1.087260,0.781)
(1.746251,0.866)
(2.409349,0.941)
(3.076639,1.021)
(3.748223,1.098)
(5.104735,1.263)
(6.479902,1.417)
(7.874923,1.579)
(9.291230,1.745)
};
\addplot+[orange!85!black,only marks,mark=none,forget plot,error bars/.cd,y dir=both,y explicit] coordinates {
(-0.218517,0.6184) +- (0,0.0003)
(1.087260,0.781) +- (0,0.003)
(1.746251,0.866) +- (0,0.004)
(2.409349,0.941) +- (0,0.003)
(3.076639,1.021) +- (0,0.003)
(3.748223,1.098) +- (0,0.004)
(5.104735,1.263) +- (0,0.004)
(6.479902,1.417) +- (0,0.006)
(7.874923,1.579) +- (0,0.001)
(9.291230,1.745) +- (0,0.001)
};
\nextgroupplot[title={$U/t=12.0$}]
\addplot+[gray!70!black,dashed,mark=square*] coordinates {
(1.242173,0.37508)
(3.980462,0.57287)
(4.442917,0.60643)
(5.840947,0.70813)
(8.687005,0.91637)
(10.623692,1.05898)
(14.600865,1.35416)
};
\addplot+[blue!70!black,mark=o] coordinates {
(1.242173,0.60531)
(3.980462,0.84755)
(4.442917,0.88728)
(5.840947,1.00600)
(8.687005,1.24322)
(10.623692,1.40245)
(14.600865,1.72653)
};
\addplot+[orange!85!black,mark=triangle*] coordinates {
(1.242173,0.674)
(3.980462,0.92)
(4.442917,0.95)
(5.840947,1.06)
(8.687005,1.30)
(10.623692,1.44)
(14.600865,1.75)
};
\addplot+[orange!85!black,only marks,mark=none,forget plot,error bars/.cd,y dir=both,y explicit] coordinates {
(1.242173,0.674) +- (0,0.005)
(3.980462,0.92) +- (0,0.01)
(4.442917,0.95) +- (0,0.01)
(5.840947,1.06) +- (0,0.01)
(8.687005,1.30) +- (0,0.01)
(10.623692,1.44) +- (0,0.02)
(14.600865,1.75) +- (0,0.02)
};
\end{groupplot}
\end{tikzpicture}
\caption[Cumulative density estimates for the six-flavour model]{Cumulative density estimates in Table~\ref{tab:table3-author-data}, first reported in Ref.~\cite{ShiLargeNf2025}, for the $70\times70$ square-lattice $\mathrm{SU}(6)$ Hubbard model at $T/t=0.15$. The curves show the saddle result, the saddle plus the RPA correction, and the sum through $O(1/N_f)$. Error bars give the final-sum uncertainties listed in the table; the saddle and saddle-plus-RPA curves are deterministic numerical values. Lines are guides to the eye. Vertical separations give the density change produced by each correction.}
\label{fig:table3-author-cumulative}
\end{figure}
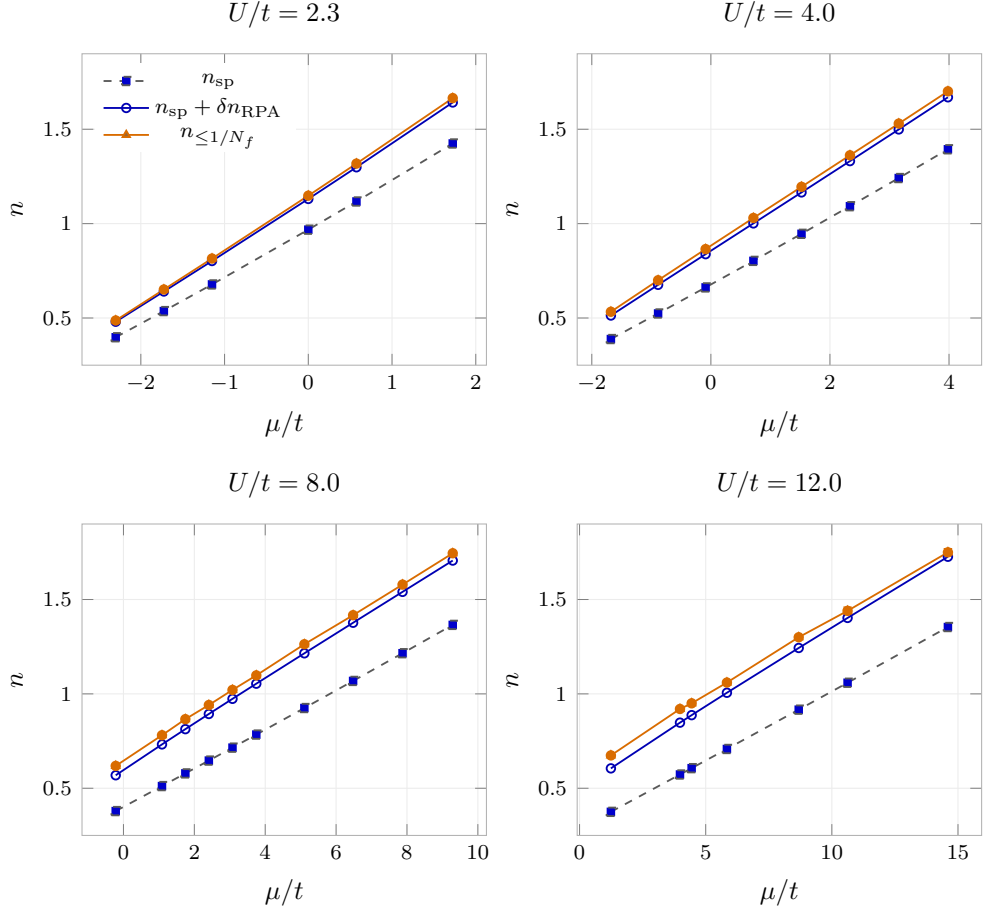

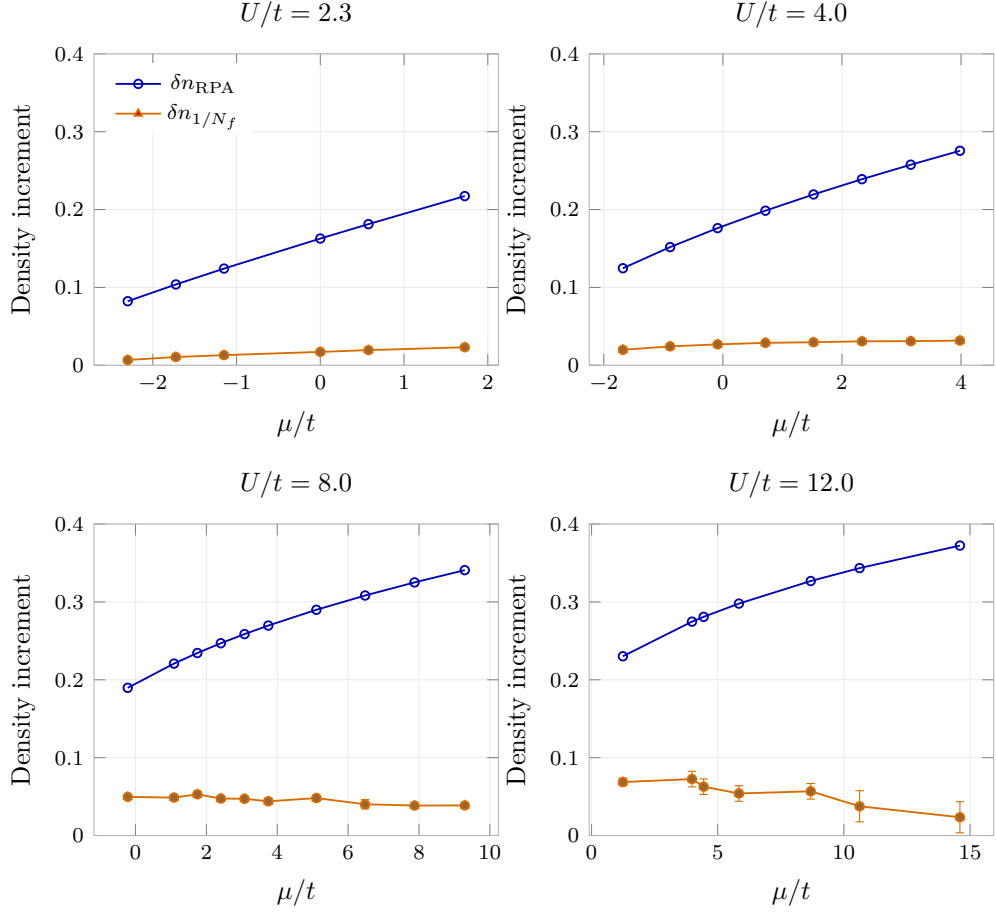
\begin{figure}[p]
\centering
\begin{tikzpicture}
\begin{groupplot}[
group style={group size=2 by 2,horizontal sep=12mm,vertical sep=21mm},
width=0.45\textwidth,height=57mm,
xlabel={$\mu/t$},ylabel={Density increment},
ymin=0,ymax=0.40,
grid=major,grid style={gray!15},axis line style={gray!60},
tick label style={font=\scriptsize},label style={font=\small},title style={font=\small},
legend style={font=\scriptsize,draw=none,fill=white,at={(0.03,0.97)},anchor=north west},
scaled ticks=false,
every axis plot/.append style={line width=0.7pt,mark size=1.6pt}
]
\nextgroupplot[title={$U/t=2.3$}]
\addplot+[blue!70!black,mark=o] coordinates {
(-2.300000,0.08224)
(-1.725000,0.10378)
(-1.150000,0.12418)
(0.000000,0.16276)
(0.575000,0.18126)
(1.725000,0.21729)
};
\addlegendentry{$\delta n_{\mathrm{RPA}}$}
\addplot+[orange!85!black,mark=triangle*] coordinates {
(-2.300000,0.00678)
(-1.725000,0.01060)
(-1.150000,0.01299)
(0.000000,0.01718)
(0.575000,0.01946)
(1.725000,0.02314)
};
\addplot+[orange!85!black,only marks,mark=none,forget plot,error bars/.cd,y dir=both,y explicit] coordinates {
(-2.300000,0.00678) +- (0,0.001)
(-1.725000,0.01060) +- (0,0.001)
(-1.150000,0.01299) +- (0,0.001)
(0.000000,0.01718) +- (0,0.001)
(0.575000,0.01946) +- (0,0.001)
(1.725000,0.02314) +- (0,0.001)
};
\addlegendentry{$\delta n_{1/N_f}$}
\nextgroupplot[title={$U/t=4.0$}]
\addplot+[blue!70!black,mark=o] coordinates {
(-1.679207,0.12463)
(-0.886256,0.15178)
(-0.088440,0.17616)
(0.714382,0.19854)
(1.522465,0.21938)
(2.336111,0.23899)
(3.155679,0.25759)
(3.981595,0.27547)
};
\addplot+[orange!85!black,mark=triangle*] coordinates {
(-1.679207,0.01986)
(-0.886256,0.02425)
(-0.088440,0.02682)
(0.714382,0.02879)
(1.522465,0.02958)
(2.336111,0.03077)
(3.155679,0.03096)
(3.981595,0.03169)
};
\addplot+[orange!85!black,only marks,mark=none,forget plot,error bars/.cd,y dir=both,y explicit] coordinates {
(-1.679207,0.01986) +- (0,0.001)
(-0.886256,0.02425) +- (0,0.001)
(-0.088440,0.02682) +- (0,0.001)
(0.714382,0.02879) +- (0,0.001)
(1.522465,0.02958) +- (0,0.001)
(2.336111,0.03077) +- (0,0.001)
(3.155679,0.03096) +- (0,0.001)
(3.981595,0.03169) +- (0,0.001)
};
\nextgroupplot[title={$U/t=8.0$}]
\addplot+[blue!70!black,mark=o] coordinates {
(-0.218517,0.18969)
(1.087260,0.22081)
(1.746251,0.23438)
(2.409349,0.24695)
(3.076639,0.25867)
(3.748223,0.26967)
(5.104735,0.28991)
(6.479902,0.30825)
(7.874923,0.32510)
(9.291230,0.34081)
};
\addplot+[orange!85!black,mark=triangle*] coordinates {
(-0.218517,0.04963)
(1.087260,0.04868)
(1.746251,0.05290)
(2.409349,0.04745)
(3.076639,0.04718)
(3.748223,0.04393)
(5.104735,0.04807)
(6.479902,0.04012)
(7.874923,0.03850)
(9.291230,0.03865)
};
\addplot+[orange!85!black,only marks,mark=none,forget plot,error bars/.cd,y dir=both,y explicit] coordinates {
(-0.218517,0.04963) +- (0,0.0003)
(1.087260,0.04868) +- (0,0.003)
(1.746251,0.05290) +- (0,0.004)
(2.409349,0.04745) +- (0,0.003)
(3.076639,0.04718) +- (0,0.003)
(3.748223,0.04393) +- (0,0.004)
(5.104735,0.04807) +- (0,0.004)
(6.479902,0.04012) +- (0,0.006)
(7.874923,0.03850) +- (0,0.001)
(9.291230,0.03865) +- (0,0.001)
};
\nextgroupplot[title={$U/t=12.0$}]
\addplot+[blue!70!black,mark=o] coordinates {
(1.242173,0.23023)
(3.980462,0.27468)
(4.442917,0.28085)
(5.840947,0.29787)
(8.687005,0.32685)
(10.623692,0.34347)
(14.600865,0.37237)
};
\addplot+[orange!85!black,mark=triangle*] coordinates {
(1.242173,0.06869)
(3.980462,0.07245)
(4.442917,0.06272)
(5.840947,0.05400)
(8.687005,0.05678)
(10.623692,0.03755)
(14.600865,0.02347)
};
\addplot+[orange!85!black,only marks,mark=none,forget plot,error bars/.cd,y dir=both,y explicit] coordinates {
(1.242173,0.06869) +- (0,0.005)
(3.980462,0.07245) +- (0,0.01)
(4.442917,0.06272) +- (0,0.01)
(5.840947,0.05400) +- (0,0.01)
(8.687005,0.05678) +- (0,0.01)
(10.623692,0.03755) +- (0,0.02)
(14.600865,0.02347) +- (0,0.02)
};
\end{groupplot}
\end{tikzpicture}
\caption[Additive RPA and non-Gaussian density contributions]{Additive density contributions at the state points of Fig.~\ref{fig:table3-author-cumulative}. The RPA increment is deterministic. The first non-Gaussian increment is $\delta n_{1/N_f}=n_{\leq1/N_f}-n_{\mathrm{sp}}-\delta n_{\mathrm{RPA}}$, with the same absolute Monte Carlo uncertainty as the cumulative density. All panels share the same vertical scale, and lines are guides to the eye. The non-Gaussian contribution varies with filling as well as interaction strength.}
\label{fig:table3-author-increments}
\end{figure}

The cumulative curves in Fig.~\ref{fig:table3-author-cumulative} show how
successive fluctuation orders change the equation of state.
Figure~\ref{fig:table3-author-increments} resolves those changes into the
Gaussian and first non-Gaussian contributions. The RPA correction and
the central non-Gaussian increment are positive at all 31 sampled state
points; the latter is $0.063$--$0.298$ of the RPA correction. The computed hierarchy therefore
moves the density upward from the Hartree reference at fixed chemical
potential. Locally, a small positive density correction on a monotone
branch corresponds to a lower chemical potential for the same filling.

The terms responsible for this change are specified by the collective
action. The Hartree saddle retains the direct density field, Gaussian
fluctuations include its exchange/contact contribution and screened
density bubbles, and higher trace vertices couple those fluctuations
beyond Gaussian order. In particular, the contact term in
Eq.~\eqref{c2:th:regularized-gaussian} restores the normal-ordered atomic
expansion to first order in the interaction. The density curves measure
the combined chemical-potential response of these contributions; their
sign at the sampled points follows from the calculated free-energy
derivatives.

At $U/t=8$ and $\mu/t=1.746251$, the first non-Gaussian contribution
raises the saddle-plus-RPA density from $0.81310$ to $0.866(4)$, giving
$\delta n_{1/N_f}=0.05290$. This is a visible correction to the Gaussian
equation of state, with a statistical uncertainty of $0.004$. The density
near this state point is also sensitive to lattice size
\cite{ShiLargeNf2025}, so the comparison concerns the stated
$70\times70$ system.

Filling dependence is particularly apparent at $U/t=12$: among the
sampled points the central non-Gaussian increment is largest near $n\simeq0.92$,
where it is about $0.072$, and falls to about $0.023$ near $n\simeq1.75$.
The correction thus changes the shape of the equation of state as well
as its level. A comparison of different interactions at a common density
would involve interpolation between the distinct chemical-potential grids.

\subsection{Statistical precision and numerical scope}
\label{c2:results:precision}

The saddle equation and the RPA integrals are evaluated deterministically.
For fixed numerical values of these terms, the only sampled contribution
to the cumulative density at this truncation is the first non-Gaussian
term. Subtracting the lower orders therefore gives
\begin{equation}
 \sigma_{\rm MC}(\delta n_{1/N_f})
 =\sigma_{\rm MC}(n_{\leq1/N_f}).
 \label{c2:results:increment-error}
\end{equation}
Both density figures use these statistical errors. In
Table~\ref{tab:table3-author-data}, the cumulative and non-Gaussian
columns have the same absolute uncertainty. The non-Gaussian central
values follow by subtracting the deterministic lower-order contributions
from the cumulative densities.

The deterministic RPA calculation uses imaginary-time and auxiliary-parameter
quadrature. In Ref.~\cite{ShiLargeNf2025}, I checked its time-discretization
errors to be much smaller than the reported statistical errors. This
numerical check is separate from Eq.~\eqref{c2:results:increment-error},
which concerns sampling variance at fixed lower-order values.
Representation, finite-difference and flavour-truncation errors also
remain distinct from the plotted Monte Carlo bars.

These results describe the uniform branch on the stated finite lattice.
Stability against ordered solutions, the omitted flavour orders and
finite-size dependence are separate questions of physical convergence;
the size sensitivity near $U/t=8$ noted above is one example.

\clearpage
\begingroup
\small
\setlength{\tabcolsep}{5pt}
\setlength{\LTcapwidth}{\textwidth}
\begin{longtable}{rrrrrr}
\caption[Six-flavour density data and successive contributions]{Saddle density, additive RPA correction, cumulative density and first non-Gaussian increment for the six-flavour Hubbard model, using Table~III of Ref.~\cite{ShiLargeNf2025}. The 31 state points have $T/t=0.15$ on a $70\times70$ square lattice. Parentheses give the cumulative Monte Carlo uncertainty in the last quoted digits; the same absolute uncertainty applies to $\delta n_{1/N_f}$. The eight cumulative uncertainties absent from the original table are one unit in the last digit. Saddle and RPA values are deterministic. Their numerical errors and the statistical-error relation are discussed in Section~\ref{c2:results:precision}.}\label{tab:table3-author-data}\\
\toprule
$\mu/t$ & $U/t$ & $n_{\mathrm{sp}}$ & $\delta n_{\mathrm{RPA}}$ & $n_{\leq1/N_f}$ & $\delta n_{1/N_f}$\\
\midrule
\endfirsthead
\multicolumn{6}{c}{\tablename~\thetable\ (continued)}\\
\toprule
$\mu/t$ & $U/t$ & $n_{\mathrm{sp}}$ & $\delta n_{\mathrm{RPA}}$ & $n_{\leq1/N_f}$ & $\delta n_{1/N_f}$\\
\midrule
\endhead
\midrule
\multicolumn{6}{r}{Continued on the next page}\\
\endfoot
\bottomrule
\endlastfoot
$-2.300000$ & 2.3 & 0.39798 & 0.08224 & 0.487(1) & 0.00678\\
$-1.725000$ & 2.3 & 0.53662 & 0.10378 & 0.651(1) & 0.01060\\
$-1.150000$ & 2.3 & 0.67783 & 0.12418 & 0.815(1) & 0.01299\\
0.000000 & 2.3 & 0.96806 & 0.16276 & 1.148(1) & 0.01718\\
0.575000 & 2.3 & 1.11728 & 0.18126 & 1.318(1) & 0.01946\\
1.725000 & 2.3 & 1.42457 & 0.21729 & 1.665(1) & 0.02314\\
\addlinespace[4pt]
$-1.679207$ & 4.0 & 0.38851 & 0.12463 & 0.533(1) & 0.01986\\
$-0.886256$ & 4.0 & 0.52397 & 0.15178 & 0.700(1) & 0.02425\\
$-0.088440$ & 4.0 & 0.66202 & 0.17616 & 0.865(1) & 0.02682\\
0.714382 & 4.0 & 0.80267 & 0.19854 & 1.030(1) & 0.02879\\
1.522465 & 4.0 & 0.94604 & 0.21938 & 1.195(1) & 0.02958\\
2.336111 & 4.0 & 1.09224 & 0.23899 & 1.362(1) & 0.03077\\
3.155679 & 4.0 & 1.24145 & 0.25759 & 1.530(1) & 0.03096\\
3.981595 & 4.0 & 1.39384 & 0.27547 & 1.701(1) & 0.03169\\
\addlinespace[4pt]
$-0.218517$ & 8.0 & 0.37908 & 0.18969 & 0.6184(3) & 0.04963\\
1.087260 & 8.0 & 0.51151 & 0.22081 & 0.781(3) & 0.04868\\
1.746251 & 8.0 & 0.57872 & 0.23438 & 0.866(4) & 0.05290\\
2.409349 & 8.0 & 0.64660 & 0.24695 & 0.941(3) & 0.04745\\
3.076639 & 8.0 & 0.71515 & 0.25867 & 1.021(3) & 0.04718\\
3.748223 & 8.0 & 0.78440 & 0.26967 & 1.098(4) & 0.04393\\
5.104735 & 8.0 & 0.92502 & 0.28991 & 1.263(4) & 0.04807\\
6.479902 & 8.0 & 1.06863 & 0.30825 & 1.417(6) & 0.04012\\
7.874923 & 8.0 & 1.21540 & 0.32510 & 1.579(1) & 0.03850\\
9.291230 & 8.0 & 1.36554 & 0.34081 & 1.745(1) & 0.03865\\
\addlinespace[4pt]
1.242173 & 12.0 & 0.37508 & 0.23023 & 0.674(5) & 0.06869\\
3.980462 & 12.0 & 0.57287 & 0.27468 & 0.92(1) & 0.07245\\
4.442917 & 12.0 & 0.60643 & 0.28085 & 0.95(1) & 0.06272\\
5.840947 & 12.0 & 0.70813 & 0.29787 & 1.06(1) & 0.05400\\
8.687005 & 12.0 & 0.91637 & 0.32685 & 1.30(1) & 0.05678\\
10.623692 & 12.0 & 1.05898 & 0.34347 & 1.44(2) & 0.03755\\
14.600865 & 12.0 & 1.35416 & 0.37237 & 1.75(2) & 0.02347\\
\end{longtable}
\endgroup

\clearpage
\section{Collective expansion and independent benchmarks}
\label{c2:closing}

The collective-field formulation connects a flavour order to a finite
family of diagram topologies and a specified integration measure.
Representative multiplicities fix their physical coefficients, while
momentum conservation and analytic $L$ blocks reduce the variables left
for Monte Carlo integration. The density calculation then resolves the
contribution of each computed fluctuation order to the equation of state.
Independent graph enumeration, direct quadrature of the $L$ blocks and
the atomic partition sum test the combinatorial weights, analytic
integrals and normal-ordering prescription along this construction.

For the six-flavour system, the results retain the first non-Gaussian
order. Their comparison with CoS \cite{KozikCoS2024} was reported in
Ref.~\cite{ShiLargeNf2025}; here the figures resolve the successive
$1/N_f$ contributions and their statistical precision. The three-flavour
partition logarithm used in Chapter~\ref{c3:chapter} is reported through
the second non-Gaussian order in Ref.~\cite{ShiCoS2025}. At that order,
the full expansion contains the five partitions in
Eq.~\eqref{c2:gr:k2-partitions}, spanning $E=3$ to $6$.

Figure~\ref{c3:results:sun}(d) compares this three-flavour result with
an interaction-order CoS++ series at the same Hamiltonian, state point,
lattice geometry and normalization $\psi=\ln Z/N_s$. The expansions
organize the interaction effects differently: the collective method
starts from a self-consistent saddle and screened lines, while CoS++
combines the connected contractions at each order of the shifted
interaction. Chapter~\ref{c3:chapter} develops the fixed-coordinate
summation algorithms that make this second route computationally accessible.

\chapter[CoS++: Generalized Combinatorial Summation]{CoS++: Generalized Combinatorial Summation and Monte Carlo Sampling}
\label{c3:chapter}

The expansion developed in the preceding chapter organizes a many-body
calculation by powers of the inverse flavour number. This chapter turns to a second problem: how to evaluate, at a fixed set of interaction
coordinates, the connected sum required by a diagrammatic expansion.
The number of labelled Wick contractions is factorial, but many contractions
share the same products of propagators or the same remaining computation.
An efficient summation algorithm identifies this common structure before
the spacetime integral is sampled.

At fixed positions and imaginary times, each allowed Wick contraction
contributes a product of propagators. An arithmetic circuit evaluates the
connected sum by sharing repeated additions and multiplications. Monte
Carlo then integrates this value over the interaction coordinates.

Combinatorial summation (CoS) expresses this operation as a directed acyclic arithmetic circuit \cite{KozikCoS2024}.
In CoS++, I extend this construction to contact interactions, multiple sublattices,
flavour-dependent propagators and counterterms, and the subset outputs required by many-configuration samplers \cite{ShiCoS2025}.
Sections~\ref{c3:sb:section}--\ref{c3:poly:section} give the central
constructions in detail: typed-head symmetry-broken CoS, exact circuit compression, all-subset terminal readout, and a determinant-polynomial sieve. A compressed single-output circuit, a multi-output circuit and
a determinant-polynomial calculation can represent the same final
coefficient, while exposing different intermediate quantities and having
different memory requirements.

The sieve in Section~\ref{c3:fast:section} computes all connected subset
values from principal-minor tables. Its fixed-configuration timing comparison reaches speedup factors
of about 24--25 at order nine for three flavours.
Section~\ref{c3:model:interfaces} defines the algebraic connected sum,
its physically normalized integrand and the statistical estimator used
to integrate it.

The numerical examples cover two-spin fermions on the honeycomb lattice
and three flavours on the square lattice. I use ``three-flavour'' for the
latter throughout, since unequal chemical potentials or reference shifts
can break SU(3) symmetry. Figures~\ref{c3:results:honey} and
\ref{c3:results:sun} show the results reported in Ref.~\cite{ShiCoS2025}.

Chapter~\ref{ch:benchmarks} reviews GPU acceleration of CoS,
normalizing-flow proposals and deterministic integration of compressed
diagrammatic integrands \cite{Sturt2024Parallel,Leoni2024Flow,Nunez2022TensorTrain}.
Here the focus is the connected evaluator and the subset outputs that these
integration strategies require.

\section{The shifted expansion and the quantity to be computed}
\label{c3:model:section}

\subsection{Physical parameters and auxiliary parameters}

Let $a=1,\ldots,N_f$ label flavours and let $N_s=\nu N_c$ be the number
of sites, with $N_c$ unit cells and $\nu$ sites per cell. For an onsite repulsive interaction, take the grand Hamiltonian
\begin{equation}
 K=\sum_{ij,a}t_{ij}c^\dagger_{ia}c_{ja}
 -\sum_{i,a}\mu_a n_{ia}
 +U\sum_i\sum_{a<b}n_{ia}n_{ib}.
 \label{c3:model:K}
\end{equation}
Set $k_{\mathrm B}=1$ and use the nearest-neighbour hopping as the
energy unit when numerical parameters are quoted. A site-dependent
one-body shift $\alpha_{ia}$ defines a family
\begin{align}
 K(\xi)&=K_0+\xi K_1, \nonumber\\
 K_0&=\sum_{ij,a}t_{ij}c^\dagger_{ia}c_{ja}
 -\sum_{i,a}(\mu_a-\alpha_{ia})n_{ia},\nonumber\\
 K_1&=U\sum_i\sum_{a<b}n_{ia}n_{ib}
 -\sum_{i,a}\alpha_{ia}n_{ia}.
 \label{c3:model:homotopy}
\end{align}
The physical model is recovered at $\xi=1$. Varying $\alpha_{ia}$ together with its compensating term alters the reference problem and the series, whereas varying $\mu_a$ alters the physical Hamiltonian.
At a finite truncation the answer can depend on the reference shift;
independence is recovered only by an exact evaluation or a controlled
continuation to the physical point. A staggered choice of $\alpha_{ia}$ selects the reference state; establishing spontaneous symmetry breaking requires a physical order parameter and the appropriate thermodynamic limit.

In the interaction representation,
\begin{equation}
 \frac{Z(\xi)}{Z_0}
 =\left\langle {\cal T}_\tau
 \exp\!\left[-\xi\int_0^\beta K_1(\tau)\,\mathrm d\tau\right]
 \right\rangle_0,\qquad
 \psi(\xi)=\frac{\ln Z(\xi)}{N_s}
 =\psi_0+\sum_{n\geq1}c_n\xi^n .
 \label{c3:model:series}
\end{equation}
By the linked-cluster theorem, the coefficients $c_n$ of $\ln Z$ contain connected correlators.
For a general interaction insertion $\mathcal V(X)$, with
$X=(i,\tau)$, the coefficient is
\begin{equation}
 c_n=\frac{(-1)^n}{N_s n!}
 \int\mathrm dX_1\cdots\mathrm dX_n\,
 \left\langle{\cal T}_\tau
 {\cal V}(X_1)\cdots{\cal V}(X_n)\right\rangle_{0,c},
 \quad
 \int\mathrm dX=\sum_i\int_0^\beta\mathrm d\tau .
 \label{c3:model:cumulant}
\end{equation}
Here $\mathcal V$ includes the couplings and counterterms. Sections~\ref{c3:sb:section}--\ref{c3:fast:section} remove these physical factors and compute a connected polynomial at fixed coordinates; Section~\ref{c3:num:prefactors} restores the coupling, factorial and integration factors.

\subsection{Propagators, equal times and the first-order anchor}

The contraction convention is
\begin{equation}
 g_a(X,Y)=\langle{\cal T}_\tau c_a(X)c^\dagger_a(Y)\rangle_0,
 \qquad g_a(0^-;i,i)=-\rho_{ia},
 \qquad \rho_{ia}=\langle n_{ia}\rangle_0 .
 \label{c3:model:g}
\end{equation}
For a free level of energy $\epsilon$ relative to the reference chemical
potential, $g(\tau)=e^{-\epsilon\tau}/(1+e^{-\beta\epsilon})$
on $0<\tau<\beta$. The negative-time value follows by antiperiodicity.
This convention differs by a minus sign from the commonly defined fermion Green function. Normal ordering selects the $\tau\to0^-$ equal-time limit.

The first coefficient follows directly from the Hamiltonian:
\begin{equation}
 c_1=-\frac{\beta U}{N_s}\sum_i\sum_{a<b}\rho_{ia}\rho_{ib}
 +\frac{\beta}{N_s}\sum_{i,a}\alpha_{ia}\rho_{ia}.
 \label{c3:model:first}
\end{equation}
Equation~\eqref{c3:model:first} checks the first-order normalization
of a shifted-matrix formulation. Combining the two-body interaction and
one-body counterterms can introduce an auxiliary local constant. Its
contribution changes only the linear term of $\ln Z$ and must be included
in the conversion to the physical coefficient. 

For two spins, define $\gamma_{i\uparrow}=\alpha_{i\downarrow}/U$
and $\gamma_{i\downarrow}=\alpha_{i\uparrow}/U$. Then
\begin{equation}
\begin{aligned}
 U n_{i\uparrow}n_{i\downarrow}
 -\alpha_{i\uparrow}n_{i\uparrow}
 -\alpha_{i\downarrow}n_{i\downarrow}
 &=
 U(n_{i\uparrow}-\gamma_{i\uparrow})
  (n_{i\downarrow}-\gamma_{i\downarrow})\\
 &\quad-\frac{\alpha_{i\uparrow}\alpha_{i\downarrow}}{U}.
\end{aligned}
 \label{c3:model:twospinshift}
\end{equation}
Wick's theorem therefore produces the shifted matrices
$M^\uparrow_{ij}=g_\uparrow(X_i,X_j)+
\gamma_{i\uparrow}\delta_{ij}$ and
$M^\downarrow_{ij}=g_\downarrow(X_i,X_j)+
\gamma_{i\downarrow}\delta_{ij}$.
The apparent reversal of the spin on the shift is required by
Eq.~\eqref{c3:model:twospinshift}. The constant contributes
$\beta\xi N_s^{-1}\sum_i\alpha_{i\uparrow}\alpha_{i\downarrow}/U$
to $\psi$ and cancels the auxiliary constant in the determinant
coefficient at first order.

\subsection{Algebraic sums and stochastic integration}
\label{c3:model:interfaces}

At fixed interaction coordinates, the propagator matrices and shift
weights determine an algebraic connected value $C(S)$ for each retained
set $S$. Both CoS circuits and determinant polynomials evaluate this
quantity. Restoring the coupling powers, perturbative factorial, volume
normalization and coordinate-map Jacobian gives the physical integrand
$f_n(x)$ on a unit cube, whose integral
$c_n=\int f_n(x)\,\mathrm dx$ is estimated by Monte Carlo sampling.
This separation allows the same sampling measure to be used with either
connected-sum construction, provided that its subset values carry the
same physical normalization. It also identifies the object compared in
fixed-coordinate timings: the algebraic sum, before the surrounding
spacetime integration.

\section{Determinants, cycle covers and connected constructions}
\label{c3:base:section}

\subsection{The combinatorial determinant}

For an $m\times m$ matrix,
\begin{equation}
 \det A=\sum_{\pi\in{\mathfrak S}_m}
 (-1)^{m-k(\pi)}\prod_{v=1}^{m}A_{v,\pi(v)},
 \label{c3:base:det}
\end{equation}
where $k(\pi)$ counts permutation cycles. A cycle of length $p$
has sign $(-1)^{p-1}$, so multiplying over cycles gives the displayed
sign. For example, $(1)(2)(3,4)$ has three cycles and contributes
$-A_{11}A_{22}A_{34}A_{43}$. The minus sign belongs to the two-cycle.
This sign is easy to lose when a closing edge
carries a minus sign and an initialization carries another.

Mahajan--Vinay evaluation replaces permutation enumeration by a
division-free dynamic program over closed walks, called clows
\cite{MahajanVinay1997}. The head of a clow is its smallest vertex and occurs only once in it, whereas other vertices may repeat; successive clow heads increase. A clow sequence has total length $m$. Sequences that are not cycle covers cancel through a sign-reversing split/merge pairing that preserves the matrix product and changes the number of clows by one.

Consider the first repeated vertex selected by the canonical walk
ordering. If its occurrences lie in one clow, cutting at them
separates a closed segment; if they lie in two clows, splicing at
that vertex joins the walks. With the prescribed head ordering the
split and merge operations are inverse on the non-cycle-cover set.
The distinguished-head restrictions are essential: an arbitrary
deletion of walks need not preserve this cancellation. The surviving
cycle covers have the sign in Eq.~\eqref{c3:base:det}.

\subsection{A recurrence with an explicit overall sign}

Let $V_\ell(h,e)$ contain paths with $\ell$ chosen edges, current head
$h$ and endpoint $e$. Initialize $V_0(h,h)=1$ for every $h$, with all other entries zero.
At each layer, sum all predecessors of the destination state:
\begin{align}
 V_{\ell+1}(h,w)&=\sum_{e=h}^{m}V_\ell(h,e)A_{ew},
 &&w>h,\nonumber\\
 V_{\ell+1}(w,w)&=-\sum_{h=1}^{w-1}\sum_{e=h}^{m}V_\ell(h,e)A_{eh},
 &&1\leq w\leq m .
 \label{c3:base:mvstep}
\end{align}
The off-diagonal entries continue the current clow, while the diagonal
entries collect closures followed by the choice of a new head. Empty
sums vanish, and entries with endpoint below the head remain zero. After $m-1$ transitions,
\begin{equation}
 \det A=(-1)^m\sum_{h,e}V_{m-1}(h,e)(-A_{eh}).
 \label{c3:base:mvfinal}
\end{equation}
Every clow now contributes one closing minus sign. Multiplication by
$(-1)^m$ converts this to $(-1)^{m-k}$.
Equivalently, the initial value can be set to $(-1)^{m+1}$ and
the final closing step taken with a positive sign.

For $m=2$, the surviving contributions are $A_{11}A_{22}$ and
$-A_{12}A_{21}$. At higher order, repeated-vertex walks enter intermediate sums and cancel in the final determinant.
There are $O(m^3)$ entries indexed by layer, head and endpoint, and at most $O(m^4)$ scalar updates in this direct implementation. Streaming two layers
reduces working storage to $O(m^2)$. The state spaces introduced below to impose connectedness and flavour structure grow exponentially.

\subsection{The connectedness criterion}

At interaction order $n$, introduce $2n$ endpoints and the matching
\begin{equation}
 {\cal M}=\{\{i,i+n\}:1\leq i\leq n\},\qquad
 P(i)=i+n,\quad P(i+n)=i .
 \label{c3:base:matching}
\end{equation}
A fermionic contraction is a permutation of the endpoints. A diagram
is connected when the undirected union of its permutation edges and
$\mathcal M$ is connected. An interaction line can connect two fermion cycles even when each cycle
visits only part of the diagram. The matching therefore enters the
connectedness test explicitly; a product of two determinants alone
contains disconnected diagrams as well.

The original CoS constructions use different state information to enforce
this condition \cite{KozikCoS2024}. CoS-1 tracks connected components
during a determinant-like construction. CoS-2 uses a canonical ordering
of cycles in which a new cycle opens at an interaction partner of the
already constructed component. The CoS-2 matching frontier is the starting point for the contact-interaction extensions in Sections~\ref{c3:unif:section} and \ref{c3:sb:section}.

\subsection{Open components in the two-spin recurrence}

For two spins, the reduced CoS-1 state is $(b,h,e,\mathcal C)$,
where $b$ identifies the spin sector. The component data can be written
$\mathcal C=(\mathcal C_{\rm inactive},c_{\rm active})$: an unordered
collection of inactive components and one distinguished active component.
Each component is specified by its open interaction labels. An interaction
is open when exactly one endpoint has been visited; reaching its partner
completes it and joins the two components if they were distinct.

The traversal visits the full first spin block before entering the second. Let $O$ be the union
of the open component sets and $F=[n]$ the full interaction set. The visited
sets are reconstructed as
\begin{equation}
 (R_\uparrow,R_\downarrow)=
 \begin{cases}
 (O,\varnothing),&b=0,\\
 (F,F\setminus O),&b=1.
 \end{cases}
 \label{c3:base:openrecover}
\end{equation}
The specified spin-block order determines these two visited sets from $O$.

If a closed component has no open interaction, no future fermion edge
can reopen its completed cycles and no unconsumed interaction endpoint
can join it to another component. It therefore contributes only when it is the sole component at termination.
Permutations of the inactive components describe the same partition and
are identified, with the active component retaining its distinguished role.

The open-component recurrence extends the active component, starts a
new one or rejects a component that can no longer connect to the rest.
Its state records which remaining interaction can join which components;
the component count alone loses this information. The reductions in
Section~\ref{c3:dag:section} use these component labels.

\section{Uniform flavours and the contact-interaction quotient}
\label{c3:unif:section}

\subsection{Endpoint data versus future-relevant data}

Let $\bar v=1+((v-1)\bmod n)$ denote the interaction residue of endpoint
$v$. In a flavour-uniform contact model the propagator matrix depends on residues, $A_{\bar v,\bar w}$, and every closed fermion cycle carries $N_f$. Compression identifies states whose remaining computations coincide; mathematically it is a quotient of the state space. Before this reduction, a CoS-2 state contains a head $h$, endpoint $e$ and a
visited endpoint set $R$.

Define $m_i=|R\cap\{i,i+n\}|$. The raw visited subset has $4^n$
possibilities, whereas $\mathbf m$ has $3^n$. The replacement is exact
when the state also keeps every head, endpoint or local label that affects
a later edge. Different histories reaching the same canonical state may
have different accumulated values. These values can be added because all
of those histories have the same remaining computation.

Entering an unvisited interaction has two equivalent endpoint choices;
entering a singly visited interaction has one. Thus a uniform continuation carries the factor
\begin{equation}
 (2-m_j)A_{\bar e,j},\qquad m_j<2.
 \label{c3:unif:mult}
\end{equation}
For $m_j=0$, the two endpoint histories are represented either by two
parallel transitions or by a single transition of multiplicity two.
The condition $m_j<2$ restricts continuation to interactions with an
unvisited endpoint.

\subsection{Canonical opening and the closing sign}

After closing a fermion cycle, choose the canonical first unvisited
endpoint whose matching partner is already visited. At the occupancy
level this is the first eligible $m_j=1$ interaction, with the head and endpoint residues retained. Close with $-N_f A_{\bar e,\bar h}$
and consume the selected partner. If no frontier exists before all
interactions have been consumed, the construction cannot be completed
as a connected diagram.

The recurrence proceeds as follows:
\begin{enumerate}
 \item Initialize the designated root endpoint and its head.
 \item Continue into an eligible unvisited endpoint, then canonicalize
 and accumulate the child state.
 \item As an alternative to continuation, close the current cycle and open the next one
 at the canonical matching frontier.
 \item At the final layer, close the remaining cycle and add its value
 to the output.
\end{enumerate}
A completed order-$n$ path contains $2n$ fermionic factors and one
closing minus sign per cycle. Since $2n$ is even,
$(-1)^{2n-k}=(-1)^k$, so these signs give the permutation sign.
The factor $(-U/2)^n$ is restored with the physical integration factors in Section~\ref{c3:num:prefactors}.

Two cycles on disjoint interaction sets have no matching frontier
between them and cannot form a connected terminal path. Two cycles
joined by a matching edge can: the visited endpoint identifies the
interaction through which the second cycle opens. The full existence
and uniqueness argument is developed for the typed-head extension in
Section~\ref{c3:sb:section}.

\subsection{Placement of the uniform shift}

A matrix with two copies of each residue has both true endpoint-diagonal
entries and entries joining different endpoints with the same residue.
A one-cycle counterterm belongs only to the former:
\begin{equation}
 {\cal A}_{vw}=A_{\bar v,\bar w}
 +\frac{\alpha}{N_f U}\delta_{vw}.
 \label{c3:unif:shift}
\end{equation}
The full-endpoint delta isolates a one-cycle, leaving contractions
between the two distinct endpoints of one interaction unchanged.
The factor $1/N_f$ compensates the subsequent flavour sum on a closed
one-cycle. In a flavour-resolved representation the shift is attached to
a specified partner flavour with dimensionless weight $s_a=\alpha_a/U$.
Flavour-dependent diagonal shifts require the typed heads of Section~\ref{c3:sb:section}, which retain the flavour needed at closure.

\subsection{Complexity and the domain of the quotient}

There are $O(n^2)$ retained head and endpoint choices and $3^n$ occupancy patterns.
The number of visited endpoints determines the layer, and each state has
at most $O(n)$ candidate continuations. This gives $O(n^3 3^n)$ arithmetic operations at fixed coordinates before
further pruning. Kozik's CoS-2 algorithm already reaches this scaling for
the $\mathrm{SU}(N_f)$ Hubbard contact interaction by ordering the two
ends of each interaction \cite{KozikCoS2024}. Here I write this algorithm in terms of interaction occupancies,
the form to which the additional local labels below are added.
Unequal blocks, sublattice labels, typed shifts and multiple outputs each
require extra state information, so the $3^n$ count cannot be carried over
unchanged. Sections~\ref{c3:bip:section}, \ref{c3:sb:section} and
\ref{c3:sub:section} specify the enlarged states.

\section{Spin blocks and sublattice-resolved contact vertices}
\label{c3:bip:section}

\subsection{Unequal spin blocks}

For two spins, normal propagation preserves spin. The all-diagram
weight is $\det M^\uparrow\det M^\downarrow$. When these matrices differ,
a block label is part of the state. A continuation remains in the
current block, while opening a new cycle can change blocks according
to the matching structure and ordering.

If $M^\uparrow=M^\downarrow$, block-exchange symmetry can identify
additional states, with the appropriate multiplicities. Different chemical potentials or reference fields generally give unequal matrices. Block-exchange reduction is available only when $M^\uparrow=M^\downarrow$ holds for every configuration covered by the constructed recurrence.

\subsection{Sublattice-resolved occupancy states}

On the honeycomb lattice a vertex is $X=(\mathbf R,b,\tau)$ with
$b\in\{A,B\}$. Both endpoints of an onsite interaction have the same
$b$. A deterministic sublattice sum therefore requires
\begin{equation}
 m_i\in\{0,(1,A),(1,B),2\}.
 \label{c3:bip:mask}
\end{equation}
The first visit to an interaction branches over its two sublattice
choices, and the partner endpoint carries the same choice. Once both
endpoints have been visited, that interaction's sublattice has entered
the propagator product and no longer affects subsequent transitions.
Only sublattice indices belonging to an open interaction, the current
head or the current endpoint remain in the state specification.

Each transition uses the sublattice-resolved propagator
$g_a^{bb'}(X_i,X_j)$. The sublattice attached to a singly visited
interaction ensures that its matching partner lies on the same site,
as specified by the onsite interaction.

\subsection{Deterministic and sampled sublattice sums}

At fixed time and unit-cell coordinates, sublattices can be summed
deterministically or sampled with the Monte Carlo coordinates:
\begin{equation}
 \sum_{b_1,\ldots,b_n}C(X_1,\ldots,X_n)
 =\mathbb E_{b_i\ {\rm uniform}}
 \left[\nu^n C(X_1,\ldots,X_n)\right].
 \label{c3:bip:sum}
\end{equation}
Equation~\eqref{c3:bip:sum} fixes the compensation factor $\nu^n$.
Deterministic summation requires more arithmetic and combines cancelling
terms before sampling; sampling the labels adds their fluctuations to the
Monte Carlo variance.
A fast-minor evaluator can perform the same deterministic sum through
a three-choice recursion, described in Section~\ref{c3:fpm:section}.

The honeycomb CDet and sublattice-resolved CoS evaluators compute the
same connected subset table, so either can be used by the
many-configuration sampler.

\section{Symmetry-broken combinatorial summation with typed heads}
\label{c3:sb:section}

The uniform CoS construction can replace the flavour sum over a closed fermion loop by a scalar factor $N_f$ because every flavour gives
the same product of propagators. This simplification ceases to apply
when the free propagators differ between flavours. A counterterm adds
a further constraint: it removes one density endpoint of an interaction
but retains the flavour of the density on which that one-body term
acts. In CoS++ I carry this information in the dynamic-programming state \cite{ShiCoS2025}; the resulting construction is symmetry-broken CoS (SB-CoS). It retains the connected-cycle construction of CoS \cite{KozikCoS2024} with the additional labels needed by flavour-dependent contractions.

The construction sums all connected Wick contractions at fixed interaction
coordinates, including the one-body counterterms. I first specify the local
insertion and its first-order convention, then prove that the recurrence
produces this sum.

\subsection{The local insertion and the first-order convention}
\label{c3:sb:target}

Fix $n$ labelled interaction positions $x_1,\ldots,x_n$. For each
flavour $a\in[N_f]$, let $A^{(a)}$ be the $n\times n$ matrix of normal
propagator factors at these positions. The entries may be complex.
The diagonal entries are $A^{(a)}_{ii}=g_a(0^-)=-\rho_{ia}$, where
$\rho_{ia}$ is the occupation of flavour $a$ in the reference state.
Propagators preserve flavour, with a separate matrix for each flavour. The dimensionless shift weight $s_a=\alpha_a/U$ is determined by the shifted Hamiltonian in Eq.~\eqref{c3:model:homotopy}.

The dimensionless local insertion defined by this Hamiltonian is
\begin{equation}
 \mathcal J_i
 =\sum_{a,b=1}^{N_f}:\widehat n_a(x_i)\widehat n_b(x_i):
   -2\sum_{a=1}^{N_f}s_a\widehat n_a(x_i)
 =2\sum_{a<b}\widehat n_a(x_i)\widehat n_b(x_i)
   -2\sum_a s_a\widehat n_a(x_i).
 \label{c3:sb:local-insertion}
\end{equation}
Normal ordering removes the product of two same-flavour densities at a single insertion. Equivalently,
the coherent-state density monomial squares to zero at one insertion.
The physical interaction density is $U\mathcal J_i/2$. Define
\begin{equation}
 S_n(x_1,\ldots,x_n)
 =\bigl\langle\mathcal T_\tau
       \mathcal J_1\cdots\mathcal J_n\bigr\rangle_{0,c},
 \qquad n\geq1.
 \label{c3:sb:connected-target}
\end{equation}
Here the subscript $c$ denotes the connected cumulant of the composite
insertions in the flavour-diagonal Gaussian reference. Powers of $U$,
the factorial $1/n!$, spacetime measures and the chosen
extensive normalization are external to this algebraic definition.

Equation~\eqref{c3:sb:local-insertion} contains a two-body insertion
and a one-body counterterm, but no field-independent term. At first
order, Wick contraction therefore gives
\begin{align}
 S_1&=\left(\sum_a A^{(a)}_{11}\right)^2
       -\sum_a\left(A^{(a)}_{11}\right)^2
       +2\sum_a s_a A^{(a)}_{11}\notag\\
 &=2\sum_{a<b}\rho_{1a}\rho_{1b}
       -2\sum_a s_a\rho_{1a}.
 \label{c3:sb:first-order}
\end{align}
The shifted Hamiltonian fixes the sign of the linear shift term. As in
Section~\ref{c3:model:section}, a constant added by an auxiliary polynomial
or circuit must be removed from the first-order value. Its joint cumulant with other insertions vanishes for $n\geq2$, so only the first-order value requires this correction.

\subsection{Typed contractions and the partner-attached shift}
\label{c3:sb:contractions}

Represent each interaction by two labelled endpoints. Define
\begin{equation}
 \begin{gathered}
 V=[2n],\qquad M=\{\{i,i+n\}:i\in[n]\},\\
 P(v)=\begin{cases}v+n,&v\leq n,\\v-n,&v>n,\end{cases}
 \qquad \bar v=1+((v-1)\bmod n).
 \end{gathered}
 \label{c3:sb:endpoints}
\end{equation}
The matching $M$ records which endpoints belong to the same contact
interaction. A permutation $\pi$ of $V$ specifies the directed
contraction edges. Each cycle is either normal or, if it is a
one-cycle, replaced by a shift. Let
$\mathcal D\subseteq\{v:\pi(v)=v\}$ be the set of shifted endpoints.
The physical target requires
\begin{equation}
 \{v,P(v)\}\not\subseteq\mathcal D
 \quad\hbox{for every }v.
 \label{c3:sb:no-double-shift}
\end{equation}
A normal cycle $C$ carries one flavour $\lambda(C)$. The normal cycle
containing the matching partner fixes a shifted endpoint's flavour,
giving the weight $s_{\lambda(C_\pi(P(v)))}$.

This rule follows directly from the local insertion. Choosing a
counterterm leaves one density of flavour $a$ with coefficient $-2s_a$.
There are two ways to assign the retained density to the two endpoint
slots. The other slot is represented by a shifted one-cycle of weight
$-s_a$. Summing the two endpoint choices therefore reproduces the
coefficient two with the flavour fixed by the retained density. Since
Eq.~\eqref{c3:sb:local-insertion} contains a one-body counterterm and
no scalar constant, every such insertion retains exactly one density
bilinear.

The augmented undirected graph is
\begin{equation}
 \Gamma(\pi)=
 \left(V,\{\{v,\pi(v)\}:v\in V\}\cup M\right).
 \label{c3:sb:augmented-graph}
\end{equation}
The matching edges enter the connectivity test with unit weight in
this normalized contraction sum. For a typed
contraction $\Omega=(\pi,\mathcal D,\lambda)$, define
\begin{equation}
 w(\Omega)=(-1)^{c(\pi)}
 \prod_{\substack{C\in\operatorname{Cyc}(\pi)\\C\cap\mathcal D=\varnothing}}
 \prod_{v\in C}A^{(\lambda(C))}_{\bar v,\overline{\pi(v)}}
 \prod_{d\in\mathcal D}s_{\lambda(C_\pi(P(d)))}.
 \label{c3:sb:weight}
\end{equation}
Here $c(\pi)$ counts all cycles, including shifted one-cycles.
Since the endpoint count is even,
$\operatorname{sgn}\pi=(-1)^{2n-c(\pi)}=(-1)^{c(\pi)}$.
Equivalently, every normal fermion loop contributes a minus sign and
every counterterm contributes its own minus sign. This gives the same
weight after a normal one-cycle is replaced by a shift.

The connected target is the sum of Eq.~\eqref{c3:sb:weight} over
admissible typed contractions with connected $\Gamma(\pi)$.
To see its equivalence to Eq.~\eqref{c3:sb:connected-target}, expand
each local insertion into its normal and counterterm alternatives,
then apply Wick's theorem separately to each flavour. The two slots
retain the ordered flavour-pair sum in the normal term. Same-flavour terms cancel through determinant antisymmetry provided normal edges between equal residues are retained in the sum.
Diagram weights factor over disconnected components of the composite
insertions. The joint cumulant removes exactly these disconnected
products. A shifted endpoint is a leaf attached to its retained
density through the matching, so the augmented graph gives the same
component partition as the physical counterterm insertion.

\subsection{The unquotiented state and its transitions}
\label{c3:sb:raw-dp}

First retain every endpoint label. A state in layer $\ell$ is
\begin{equation}
 Q=(h,e,t,a,R,\mathcal U),\qquad |R|=\ell+1,
 \qquad \ell=0,\ldots,2n-1.
 \label{c3:sb:raw-state}
\end{equation}
The head $h$ and current endpoint $e$ belong to the visited set $R$.
The current head type is $t\in\{\mathrm N,\mathrm S\}$ and its
flavour label is $a$. For a normal head this is the flavour of its
cycle; for a shift head it is the flavour required at its partner.
The frontier labels form the set
\begin{equation}
 \mathcal U=\{(p,t_p,a_p):p\in R,\ P(p)\notin R\}.
 \label{c3:sb:frontier-payload}
\end{equation}
The stored value $F_\ell(Q)$ is the sum of the weights of all
partial histories reaching this state. This accumulated value already contains the propagators and closing
signs of completed cycles, so only the active cycle and unresolved
frontier labels enter subsequent transitions.

Initialize a normal and a shift state at endpoint $1$ for each
$a\in[N_f]$. Both have $h=e=1$, $R=\{1\}$ and
$\mathcal U=\{(1,t,a)\}$, with value one. A shift at the initial
endpoint requires a provisional label because its normal partner has
not yet been processed. This retained label fixes the flavour of the normal partner when that
partner is reached.

Only a normal head can continue. For each $w\notin R$, append the
edge $e\to w$ with factor $A^{(a)}_{\bar e,\bar w}$.
Replace $e$ by $w$ and add $w$ to $R$. If $P(w)$ is unvisited,
insert $(w,\mathrm N,a)$ into $\mathcal U$; otherwise remove the
label for $P(w)$. A shift head represents a replaced one-cycle and closes immediately.

Closing the current cycle contributes
\begin{equation}
 b(Q)=
 \begin{cases}
 -A^{(a)}_{\bar e,\bar h},&t=\mathrm N,\\
 -s_a,&t=\mathrm S.
 \end{cases}
 \label{c3:sb:closure}
\end{equation}
If $R=V$, add $F_\ell(Q)b(Q)$ to the output. Otherwise a new cycle
must start at the matching frontier
\begin{equation}
 \partial_MR=\{w\notin R:P(w)\in R\}.
 \label{c3:sb:matching-frontier}
\end{equation}
Order the interactions once and for all by their residues. Choose
the smallest interaction with one visited endpoint and let $w$ be
its unvisited partner. This interaction-based ordering is invariant
under exchanging the two endpoint slots; it is the ordering needed
for the quotient below.

The frontier label $(P(w),t_p,a_p)$ determines the new head type and
flavour. Visiting $w$ completes this interaction, so its label is
removed from $\mathcal U$ and the new head and endpoint are both $w$.
A normal head may carry any label in
\begin{equation}
 \Lambda(t_p,a_p)=
 \begin{cases}
 [N_f],&t_p=\mathrm N,\\
 \{a_p\},&t_p=\mathrm S.
 \end{cases}
 \label{c3:sb:normal-colours}
\end{equation}
For $t_p=\mathrm N$, a shift head with label $a_p$ is also allowed.
For $t_p=\mathrm S$, Eq.~\eqref{c3:sb:no-double-shift} restricts the
successor to a normal head. Each allowed successor receives the closed
value $F_\ell(Q)b(Q)$.

If the frontier is empty before all endpoints have been visited,
the close-and-open branch is absent. A normal state retains its continuation branches: its open cycle can still reach an untouched interaction through a propagator.

\subsection{Correctness before state compression}
\label{c3:sb:raw-proof}

\noindent\textbf{Proposition.}
The unquotiented physical recurrence establishes a weight-preserving
bijection between complete traversal paths and the connected admissible
typed contractions defined in Section~\ref{c3:sb:contractions}.

For a connected typed contraction, start at endpoint $1$ and traverse
its permutation cycle. After any collection of complete cycles has been
visited, no permutation edge crosses from that visited set to its
complement. If endpoints remain, connectedness of $\Gamma(\pi)$ supplies
a matching edge across this cut. The smallest eligible interaction then
selects a unique next endpoint and an unprocessed cycle. Repetition gives
a canonical order of all cycles, terminating because each cycle enlarges
the visited set. Throughout this traversal the frontier in
Eq.~\eqref{c3:sb:frontier-payload} is exactly the set of singly visited
interactions: the first visit adds its label and the partner visit
completes it. The label records either a known normal-cycle flavour or
the flavour required by a future normal partner. An unmatched shift
label occurs only in the initial shift state, whose immediate closure
opens its normal partner with the specified flavour. Every later shift
head opens after its partner has already been visited.

Conversely, every complete recurrence path visits each endpoint once.
Continuation and closure assign one outgoing and one incoming edge to
each endpoint, so the completed cycles form a permutation. A shift head
closes as a one-cycle, and the opening rules enforce the partner flavour
and exclude double-shift pairs. Each new cycle attaches to the visited
set through its opening matching edge; hence the augmented graph is
connected. Given an admissible target contraction, its canonical order
constructs the inverse path: the target permutation determines all
continuations and closures, and its types and flavours select allowed
successors in Eq.~\eqref{c3:sb:normal-colours}. Two normal cycles have
independent flavour labels, while a shift and its normal partner share
the label fixed by the counterterm. The prescribed root and frontier
order leave one path per typed contraction, with no cycle-order or
root-choice multiplicity.

Finally, continuation supplies all normal propagators except each
cycle's closing edge. A normal closure supplies that propagator and one
minus sign; a shift closure supplies $s_a$ and one minus sign. Their
product is exactly Eq.~\eqref{c3:sb:weight}. Histories reaching a common
state have identical allowed subsequent transitions, so their accumulated
weights can be summed there. This proves that the recurrence evaluates
$S_n$ for arbitrary complex input matrices. The quotient below combines
these paths only after their multiplicities have been determined.

\subsection{Canonicalization by contact-interaction residues}
\label{c3:sb:quotient}

The two endpoints of interaction $i$ refer to the same physical
insertion. Normal edge weights are functions of these interaction residues. The
occupancies corresponding to $R$ are
\begin{equation}
 \begin{gathered}
 m_i=|\{i,i+n\}\cap R|\in\{0,1,2\},\qquad
 \mathfrak F(\mathbf m)=\{i:m_i=1\},\\
 \boldsymbol\upsilon:\mathfrak F(\mathbf m)\longrightarrow
 \{\mathrm N,\mathrm S\}\times[N_f],\qquad
 i\longmapsto(t_i,a_i).
 \end{gathered}
 \label{c3:sb:local-record}
\end{equation}
The frontier map is defined only on singly visited interactions. Its
domain changes with the occupancies, so completed and untouched
interactions carry no frontier label. The compressed state is
\begin{equation}
 \widehat Q=(\bar h,\bar e,t,a,\mathbf m,\boldsymbol\upsilon),
 \qquad \sum_i m_i=\ell+1.
 \label{c3:sb:compressed-state}
\end{equation}
The head type identifies a shifted one-cycle and must be retained during compression because equal head and endpoint residues can represent different original endpoints.

Let $T_{\mathrm N}(\widehat Q,j)$ be the state obtained by extending a
normal cycle into interaction $j$. The continuation contribution is
\begin{equation}
 \widehat F^{\mathrm{cont}}_{\ell+1}(\widehat Q')
 =\sum_{\substack{(\widehat Q,j):\,t=\mathrm N,\ m_j<2\\
                  T_{\mathrm N}(\widehat Q,j)=\widehat Q'}}
 (2-m_j)A^{(a)}_{\bar e,j}\widehat F_\ell(\widehat Q).
 \label{c3:sb:quotient-continuation}
\end{equation}
The transition increases $m_j$ and sets the endpoint residue to $j$.
On the first visit, the frontier map is extended by
$\upsilon_j=(\mathrm N,a)$; a partner visit restricts its domain by
removing $j$. Closing uses Eq.~\eqref{c3:sb:closure} with the retained
residues. For a new head, take $j=\min\mathfrak F(\mathbf m)$ and use
its label $(t_j,a_j)$ to apply the same type and flavour rules. The
completed interaction then has $m_j=2$ and leaves the frontier domain.
The frontier fixes one unvisited partner, giving unit opening multiplicity.
The full next-layer value is the sum of the continuation contribution
above and all allowed close-and-open contributions,
$\widehat F_{\ell+1}=\widehat F^{\mathrm{cont}}_{\ell+1}
+\widehat F^{\mathrm{close}}_{\ell+1}$.

To prove equivalence of the compressed and unquotiented recurrences, group endpoint histories according to
Eq.~\eqref{c3:sb:compressed-state}. If $m_j=0$, either labelled
endpoint can be visited next. Both transitions have the same
matrix factor and the same projected destination, so their
contributions add with multiplicity two. If $m_j=1$, only its
unvisited partner is available, giving multiplicity one. If
$m_j=2$, neither endpoint is available. The type and flavour update
is identical for every history in the group.

The ordered frontier depends only on which occupancies equal one.
The compressed state stores the frontier label and closing factor. Thus every
history in a compressed state has the same weighted outgoing
transitions to compressed states. Initialization also agrees, with
no factor of two because the first endpoint is fixed to $1$.
Induction over the layers proves that each compressed value equals
the sum of its unquotiented values. Final closure depends only on
the compressed state, so the output is unchanged. Endpoint multiplicities enter only through $2-m_j$ in
Eq.~\eqref{c3:sb:quotient-continuation}.

\subsection{Evaluation of the physical recurrence}
\label{c3:sb:pseudocode}

The following procedure returns the physical fixed-configuration
sum, including its first-order convention.
\begin{enumerate}
 \item Given $N_f$ matrices of size $n\times n$ and $N_f$ shift weights,
 define the state sums
 $\widehat F_0,\ldots,\widehat F_{2n-1}$, initially zero before
 the source contributions are inserted.
 \item For each flavour and each type, initialize
 $\bar h=\bar e=1$, $m_1=1$, all other occupancies zero and
 $\upsilon_1=(t,a)$, with value one.
 \item At layer $\ell=0,\ldots,2n-2$, evaluate
 $\widehat F^{\mathrm{cont}}_{\ell+1}$ by summing all allowed normal
 continuations in Eq.~\eqref{c3:sb:quotient-continuation}.
 Retain these terms even when the source state's matching frontier is empty.
 \item For each current state, compute the closing factor. If no occupancy equals one,
 omit the close-and-open updates. Otherwise choose the smallest
 such interaction. Its frontier label $(t_j,a_j)$ specifies the next
 head; completing that interaction sets $m_j=2$ and restricts the
 frontier map to the remaining singly visited interactions.
 \item Open a normal head with every flavour allowed by
 Eq.~\eqref{c3:sb:normal-colours}. Also open a shift head of
 the partner flavour only when $t_j=\mathrm N$.
 Add the closed value into every allowed destination state.
 \item In layer $2n-1$, close each remaining cycle and sum
 $\widehat F_{2n-1}(\widehat Q)b(\widehat Q)$.
\end{enumerate}
At the final layer every interaction has occupancy two and the frontier
is empty. Contributions arriving at a common state are summed before
closure. The recurrence defines a DAG whose vertices carry the residues,
head type, flavour and occupancy data above. Each edge is weighted by
an integer multiplicity and a symbolic factor $A^{(a)}_{ij}$ or $s_a$.
Its algebraic structure is therefore reusable across configurations at
fixed $n$ and $N_f$. Section~\ref{c3:dag:section} derives the equivalence
relations that combine vertices and share repeated multiplications.

\subsection{Uniform limit and an explicit two-interaction path}
\label{c3:sb:limits}

Suppose $A^{(a)}=A$ and $s_a=s$ for every flavour. In an admissible
typed contraction the flavour sum gives one factor $N_f$ for
each normal cycle, with every shifted cycle inheriting its partner's
flavour. An ordinary flavour-summed endpoint matrix gives the same
counting:
\begin{equation}
 M_{vw}=A_{\bar v,\bar w}+\frac{s}{N_f}\delta_{vw}.
 \label{c3:sb:uniform-matrix}
\end{equation}
Expanding the $\delta_{vw}$ contribution selects the one-cycles replaced by shifts.
For $d=|\mathcal D|$, its factor
$N_f^{c(\pi)}(s/N_f)^d=N_f^{c(\pi)-d}s^d$
is precisely the typed flavour sum. The full-endpoint delta confines
the counterterm to one-cycles, preserving the normal edge between the
two different endpoints of the same interaction.

The matching flavour factors give the same connected contributions at
$n\geq2$. At $n=1$, the endpoint-matrix expression contains the scalar
$s^2$; its subtraction yields Eq.~\eqref{c3:sb:first-order}, the physical
normalization already enforced by the typed recurrence.

As an explicit path at $n=2$, take matching pairs $\{1,3\}$ and
$\{2,4\}$ and the typed permutation
$\pi=(1\,2\,3)(4)$, with normal flavour $a$ on the three-cycle
and a shift at endpoint $4$. Initialize a normal head at $1$,
continue to $2$, and continue to $3$. The first interaction is
now complete, while the frontier label at interaction two records
$(\mathrm N,a)$. Closing the normal cycle selects endpoint
$4$ as the next head. Its shift label must be $a$.
The accumulated factor is
\begin{equation}
 A^{(a)}_{12}A^{(a)}_{21}
 \bigl[-A^{(a)}_{11}\bigr](-s_a)
 =s_a A^{(a)}_{12}A^{(a)}_{21}A^{(a)}_{11}.
 \label{c3:sb:two-interaction-example}
\end{equation}
The final shift factor uses the flavour already fixed by its normal partner.
Other continuations and endpoint choices generate other typed
contractions, whose cancellations are recovered by summation.

\subsection{Domain, complexity and algebraic checks}
\label{c3:sb:scope}

The construction covers flavour-dependent normal propagators and
flavour-dependent local one-body counterterms for the stated
contact interaction. The matrices may be unequal and complex; each normal line preserves flavour. Off-diagonal flavour mixing, anomalous
propagators and a non-contact endpoint identification require
different contraction objects or a modified state. Fixed lattice or sublattice labels enter through the input matrices. Summing sublattices within the circuit requires the state of Section~\ref{c3:bip:section}.

The calculations here use one shift per flavour and evaluate the full
interaction set. For position-dependent counterterms, the closure weight
also depends on the interaction coordinate. Extending the construction
to several subset outputs entails preserving their identities during
compression, using the rooting and readout conditions derived for the
uniform case in Section~\ref{c3:sub:section}. The unequal-spin-block
construction additionally carries the block label of
Section~\ref{c3:bip:section}.

A loose structural estimate assigns to each interaction an
empty state, a full state, or a singly visited state with one
of $2N_f$ labels. This gives at most $(2+2N_f)^n$ local-state
patterns, accompanied by at most $2N_fn^2$ head, endpoint and
current-type/flavour choices. This count includes unreachable combinations and is only an upper
bound. A normal continuation examines at most $n$ interactions, and
closure opens at most $N_f+1$ successors. Section~\ref{c3:dag:cost} distinguishes graph-construction costs from the cost of repeated evaluation.

At low order I compare the connected sum with direct permutation
enumeration, the residue quotient with the endpoint recurrence, and the
counterterms with Eq.~\eqref{c3:sb:first-order}. The uniform limit tests
the flavour factors. The configurations include zero shifts, unequal
and complex matrices, and exact cancellations. Near a vanishing sum I
use an absolute tolerance. Section~\ref{c3:results:interface-checks}
checks the physical prefactors in the atomic limit.

A second low-order reference calculation uses the local density
insertions directly. Expand the product of the local insertions into density
monomials. If flavour $a$ occurs at the positions in a list $I_a$,
its Gaussian density moment is
\begin{equation}
 \left\langle\mathcal T_\tau
   \prod_{i\in I_a}\widehat n_a(x_i)\right\rangle_0
 =(-1)^{|I_a|}\det A^{(a)}[I_a,I_a].
 \label{c3:sb:density-moment}
\end{equation}
The sign follows from the chosen negative equal-time propagator.
Multiply the determinants over flavours, then multiply by the coefficients of the selected local monomials to obtain a moment
$\mathcal M(B)=\langle\mathcal T_\tau\prod_{i\in B}\mathcal J_i\rangle_0$.
The ordinary moment--cumulant inversion then gives
\begin{equation}
 S_n=\sum_{\mathcal P\in\operatorname{Part}([n])}
 (-1)^{|\mathcal P|-1}(|\mathcal P|-1)!
 \prod_{B\in\mathcal P}\mathcal M(B).
 \label{c3:sb:independent-cumulant}
\end{equation}
This reference calculation uses the physical local polynomial directly. It is
therefore sensitive to a wrong counterterm sign or an auxiliary
singleton constant even when two endpoint implementations share
the same mistake. Its combinatorial cost restricts practical use to low-order verification.

I compared these cumulants with the endpoint and compressed recurrences
for complex, non-uniform inputs through order three, including the
singleton correction and the uniform limit.

\section{Exact compression of the computational DAG}
\label{c3:dag:section}

\subsection{Arithmetic circuits independent of numerical inputs}

The same CoS recurrence can be represented by several arithmetic circuits. Two implementations may sum exactly the same connected
polynomial while storing different intermediate states and performing
different numbers of operations. The purpose of DAG compression is to
exploit this freedom without changing the polynomial. In the present
construction, compression includes canonical state generation, removal of
unproductive states, identification of equivalent computations, and
sharing of repeated factor applications. I apply these reductions within CoS++ \cite{ShiCoS2025} to the connected recurrence based on CoS \cite{KozikCoS2024}.

For a specified order, flavour structure, block symmetry and set of
outputs, the recurrence defines a directed graph with integer edge
multiplicities and symbolic propagator or shift factors. Each sampled
configuration supplies the numerical values of those factors. The graph
can therefore be reused across configurations sharing the same model
symmetries, with its layer ordering specifying the sequence of operations.

Let the graph have layers $V_0,\ldots,V_L$. Associate each state $v$ with a value
$x_v$, and an edge carries an integer coefficient $c_e$ and a linear map
$\mathcal L_{\lambda_e}$. Summing the incoming edges defines the next layer:
\begin{equation}
 x_w=\sum_{e:\,v\longrightarrow w}c_e\mathcal L_{\lambda_e}[x_v],
 \qquad w\in V_{\ell+1}.
 \label{c3:dag:linear-edge}
\end{equation}
For scalar state values, a label can identify multiplication by a particular
matrix element or by a particular shift. For vector and matrix state values,
it also identifies the projections, contractions and component placement.
The operator label specifies the mathematical action of the edge.

Each value sums the histories reaching its state. The
observable is a sum of source-to-output paths, with the operators on each
path composed in their traversal order. Every reduction below must therefore preserve the linear map from the
initialized state values to every requested output. 

\subsection{Canonical states and connectivity constraints}

States are canonicalized as they are generated, before the pruning
pass. The state specification contains every label that affects an allowed
continuation or its operator. In the uniform contact case, an occupancy tuple represents
equivalent endpoint histories, together with their head and endpoint
residues. Entering an untouched interaction gives two equivalent endpoint
choices with the same destination and matrix factor. Their weights add,
so one edge of multiplicity two represents both choices.

For interaction $i$ with occupancy $m_i$, the continuation multiplicity
is \mbox{$2-m_i$}. It counts the labelled transitions represented by the
canonical edge, so state canonicalization and edge multiplicities must be
updated together.

The model determines these retained labels. Flavour
labels are needed when $A^{(a)}$ or $s_a$ depends on $a$; a singly
occupied interaction needs its sublattice when that fixes the partner
endpoint. Normal and shift heads have different closure operators and
therefore distinct types. Equal-spin-block or zero-diagonal specializations
can remove more labels, provided these assumptions hold at every
configuration evaluated with the constructed graph.

Testing local necessary conditions avoids generating states that cannot contribute.
For a component-tracking construction, a closed component with no remaining
endpoint through which it can connect to the other components is stranded.
It cannot contribute to a connected output. If $c(v)$ components remain
and only $r(v)$ unused links can join different components, then
\begin{equation}
 r(v)<c(v)-1\quad\Longrightarrow\quad
 v\text{ cannot complete a connected graph}.
 \label{c3:dag:component-test}
\end{equation}
This inexpensive necessary test can reject states before their continuations are generated; additional dead states may survive it.

Structural zeros give another pruning test. An entry excluded by a
declared block structure vanishes identically. In a proper-colouring
representation, each fermion loop has one flavour and each interaction
joins two different flavours. An interaction whose endpoints lie on the
same loop then has zero multiplicity. A representation that includes
same-flavour terms before their antisymmetric cancellation must keep
those terms until the cancellation is performed.

\subsection{Paths from initial states to the requested outputs}

After construction, a state must both receive initialized information and
be capable of passing it to an output. Let $I$ be the initialized states,
and let $R_\ell$ contain the layer-$\ell$ states with a direct readout edge
to at least one requested output. Forward reachability and backward
productivity obey
\begin{align}
 F_0&=I, &
 F_{\ell+1}&=\{w:\exists v\in F_\ell, v\to w\},\label{c3:dag:forward}\\
 B_L&=R_L, &
 B_\ell&=R_\ell\cup\{v:\exists w\in B_{\ell+1}, v\to w\},\label{c3:dag:backward}\\
 K_\ell&=F_\ell\cap B_\ell.&&\label{c3:dag:live}
\end{align}
Edges with identically zero operators have already been removed from these
relations. The retained graph consists of the states in $K_\ell$ and
their connecting edges.

A state outside the forward set has no
initialized ancestor and therefore carries no source-dependent contribution.
A state outside the backward set has no requested descendant, so none of
the histories passing through it enters an output. Deleting either kind
removes no contributing source-to-output path. The argument uses only the graph structure, so the pruning holds for any numerical state values.

The output set determines which paths survive this restriction. For a
full-order scalar value, readout edges usually occur only in the final
layer. An all-subset calculation also has outputs at intermediate layers,
so the backward sets $B_\ell$ include paths ending at each of those
readouts. A state can then contribute to a subset value even when it
has no path to the full-order output. 

\subsection{Exact edge coalescing and future equivalence}

Before comparing states, parallel edges with the same source, destination
and exact operator label are coalesced. If their coefficients are
$c_1,\ldots,c_q$, the replacement coefficient is their integer sum.
A zero sum removes that edge identically. Edges with different labels
remain distinct even when their evaluated factors happen to coincide.
For a typed scalar circuit, define the two families of labelled operators by
\begin{equation}
 \mathcal L_{\lambda^A_{aij}}[x]=A^{(a)}_{ij}x,
 \qquad
 \mathcal L_{\lambda^s_a}[x]=s_ax,
 \label{c3:dag:labels}
\end{equation}
where the labels distinguish propagator factors from shifts and retain
their flavour and position indices. A model symmetry can identify two
labels when their factors coincide for every admissible configuration.

Future equivalence is determined backwards. After the child classes have
been assigned, group a state's edges by operator label and child class
and sum their coefficients. Include its complete direct-readout map,
with output identities preserved. Two states are merged only if these
entire signatures coincide. 

The combination rule follows from linearity. If states $u$ and $v$ have the
same continuation map $T$ to all outputs, their contribution is
\begin{equation}
 T[x_u]+T[x_v]=T[x_u+x_v].
 \label{c3:dag:future-proof}
\end{equation}
The quotient state must accumulate both incoming state values, but it keeps
only one copy of the common outgoing signature. Equivalently, all incoming
edges are redirected to one representative and its common continuation
is evaluated once. The initialized contributions are added if the merged
class lies in the initial layer.

For example, the common factor $f$ is applied once to the sum
$x_u+x_v$, reproducing the two original contributions $fx_u+fx_v$.

\subsection{Past equivalence and shared intermediate values}

A complementary reduction proceeds forwards. Once parent computations
are identified, compare the full incoming operator signatures and the
initialization annotations. States with the same past computation carry
the same state value for every admissible input.  Their
different outgoing uses may be combined while the common incoming
computation is retained once.

If $x_u=x_v=x$ identically and their continuation maps are $T_u$ and
$T_v$, the valid reduction is
\begin{equation}
 T_u[x]+T_v[x]=(T_u+T_v)[x].
 \label{c3:dag:past-proof}
\end{equation}
The merged state retains the common value $x$ and one representative
incoming computation. Combining its outgoing uses adds their integer
multiplicities: if both original states send $fx$ to the same child,
the merged edge has coefficient two. Thus past equivalence shares the
computation of $x$ while preserving both uses of its result.

The two reductions place the sum at different stages of the same linear
calculation. Equivalent continuations allow incoming values to be added
before applying their common outgoing map. Equivalent preceding
computations instead supply one shared value to the sum of its outgoing
maps. These identities determine the edge multiplicities in each case.

Both rules apply to matrix-valued states when the signatures also
specify operator order, type and shape. Equality of past signatures gives
the same matrix expression; equality of future signatures gives the same
linear map from that expression to every output. Matrix multiplication
and row or column projections must keep their prescribed order.

\subsection{Termination of the reduction sequence}

The reductions interact. Redirecting incoming edges during future
merging can create parallel edges, and combining outgoing uses during
past merging can cancel contributions and leave states unproductive. The reductions can be organized as follows:
\begin{enumerate}
 \item Generate canonical states and exact labelled edges, retaining
 all initialization and requested-output annotations.
 \item Coalesce parallel equal-label edges and delete zero coefficients.
 \item Restrict the graph to forward-reachable and backward-productive states.
 \item Sweep backwards, merging equal future signatures with summed
 state values and one copy of the common continuation.
 \item Sweep forwards, merging equal past signatures with one common
 state value and the combined outgoing uses.
 \item Repeat the reductions until the canonical graph no longer changes,
 then evaluate the resulting graph in layer order.
\end{enumerate}
Each merge lowers the state count, while deletions lower the edge count;
the graph is finite. Redirecting and coalescing do not introduce new
algebraic histories. Repeated reduction therefore terminates after finitely many changes. The iteration stops when the canonical signatures no longer change.
For these monotone operations, unchanged state and edge counts provide
a practical stopping test.

The fixed point depends on the chosen recurrence, basis for the state values and signature identities, so further factorizations may still produce a smaller circuit for the same polynomial.

\subsection{Scalar, vector and matrix state values}

State count also depends on where information is stored. A literal scalar
state can include the current head and endpoint. For compatible block
symmetries, a tail-vector stores the endpoint in its state value, and a
matrix-valued version stores both indices there:
\begin{equation}
 x_v(e),\qquad X_v(h,e).
 \label{c3:dag:payloads}
\end{equation}
The remaining state labels specify occupancy, spin block and connected
components; the head and endpoint are indices of the vector or matrix value.

For a common propagator block $A$, several continuations may use
\begin{equation}
 Y_v=X_v A.
 \label{c3:dag:matrix-product}
\end{equation}
Different outgoing maps select columns, rows or selected components of
$Y_v$. A closing operation can contract the appropriate diagonal and
insert its result in the coordinate of the new head. These maps are linear in the state value for fixed propagators, so the
path-restriction argument and both merging identities apply.

The scalar, tail-vector and matrix versions of the open-component algorithm evaluate the same recurrence with different state values. Their costs include the arithmetic
and storage within each state value: a single matrix state can represent many
scalar entries. Evaluation time therefore depends on matrix dimensions,
sparsity and the efficiency of the available linear-algebra routines.

\subsection{Operation counts for shared factors}\label{c3:dag:cost}

Even after state reduction, multiple edges can share the same numerical
factor. Grouping the edges of one layer by their scalar factor $f$ gives
\begin{equation}
 y_d=\sum_f f\left(\sum_s c_{sd}^{(f)}x_s\right).
 \label{c3:dag:factor-sum}
\end{equation}
The common factor can be applied before the sum, once to each distinct
$x_s$, or after the sum, once to each subtotal
$\sum_s c_{sd}^{(f)}x_s$. Let $S_f$ and $D_f$ be the sets of participating
source and destination states. Choosing the less expensive factorization
gives
\begin{equation}
 M_f=\min\{|S_f|,|D_f|\}
 \label{c3:dag:min-side}
\end{equation}
applications of the common scalar factor. Final-closing edges are grouped
by their factor in the same way. The typed-head calculation uses this minimum-count factorization.

Equation~\eqref{c3:dag:min-side} counts multiplications by the shared
factor. Applying integer coefficients, accumulating partial sums and
moving data add further cost. For operator-valued states, the corresponding
operation shares $X_vA$ before several projections; its cost depends on
the dimensions and sparsity of the matrix product.

The total memory includes the numerical state values, the labelled edges
and the requested outputs. Streaming two layers reduces the
working state storage, but vector and matrix states carry their own
internal dimensions.

These identities preserve the polynomial in exact arithmetic. Their
floating-point evaluation can differ when a changed addition order
alters cancellation between paths.

\section{Terminal readouts for all interaction subsets}
\label{c3:sub:section}

\subsection{Subset values required by the sampler}

A many-configuration sampler \cite{MCMCMC2021} compares several retained
subsets of a fixed envelope of interaction coordinates. These connected
sums share propagator products and partial contractions, which CoS++
reuses through terminal readouts of a common circuit \cite{ShiCoS2025}.
The underlying propagation rules then serve all requested subsets, rather
than being evaluated independently for each candidate.

Write the envelope as $\mathcal V=(X_1,\ldots,X_n)$ and let
$\mathcal C_S(\mathcal V_S)$ be the connected integrand required for a
nonempty subset $S\subseteq[n]$. Define the flux and local normalization
\begin{align}
 \Phi_S(\mathcal V)&=\lambda_{|S|}\mathcal C_S(\mathcal V_S)
 Q(\mathcal V_{\bar S}\mid\mathcal V_S),\label{c3:sub:flux}\\
 W(\mathcal V)&=\sum_{\substack{\varnothing\ne S\subseteq[n]\\|S|\in\mathcal U}}
 \frac{|\Phi_S(\mathcal V)|}{\binom n{|S|}}.
 \label{c3:sub:heatbath}
\end{align}
Here $\mathcal U$ specifies the allowed orders and $Q$ is the conditional density for generating the complement. The sampler of Section~\ref{c3:mc:section} uses the connected values
computed at one envelope to choose the retained subset and regenerate
its complement.

Each subset value requires its own order-specific normalization.
The circuit returns an algebraic polynomial before coupling powers,
factorials and integration Jacobians are applied. The output for $S$
therefore receives the physical prefactors of its standalone order-$|S|$
calculation. In particular, its factorial is $|S|!$, so the relative
weights in Eq.~\eqref{c3:sub:heatbath} retain the order dependence of
the physical expansion.

\subsection{Prefixes, exact labels and the minimum-root convention}

An intermediate layer of an ordered circuit may contain the connected value for a prefix of its interaction labels. Arbitrary subsets require additional rooted branches. The subsets $\{1,2\}$ and $\{1,3\}$ have equal cardinality
and generally use different matrix entries. The subset $\{2,3\}$ does
not contain the first interaction at all. A circuit initialized only at
interaction one cannot represent that subset while retaining its initial
vertex. Each requested result must therefore be identified by its full subset label.

The construction keeps the original interaction labels and introduces a
root $r=\min S$. There is one initialized branch for each possible root.
Within that branch, only labels $j\geq r$ can be visited. This assigns
every nonempty subset to exactly one branch.  Relabelling the envelope and the corresponding
matrix indices may change the representative, but leaves the connected polynomial of the relabelled subset unchanged.

The admissible-label and canonical opening rules transform together
with the root. For example, choosing the largest label as root changes
the admissible set to $j\leq r$.

For the uniform contact kernel, all flavour blocks are the same matrix
$A$, and the occupancy-quotiented state is
\begin{equation}
 v=(r,h,e,\bm m),\qquad m_i\in\{0,1,2\}.
 \label{c3:sub:state}
\end{equation}
The occupancy tuple records unused, singly used and completed interactions. The head and endpoint are stored as interaction residues. At layer zero, branch $r$ has $h=e=r$, $m_r=1$, all other
occupancies zero, and state value one. With this indexing, the total occupancy
in a layer-$\ell$ state is $\ell+1$.

\subsection{Propagation and the terminal closing map}

The continuation rule retains the root and the head. For an admissible
interaction $j$ with $j\geq r$ and $m_j<2$, it reads
\begin{equation}
 (r,h,e,\bm m)\longrightarrow
 (r,h,j,\bm m+\bm e_j),
 \qquad\text{weight }(2-m_j)A_{ej}.
 \label{c3:sub:continue}
\end{equation}
The integer multiplicity represents the equivalent endpoint choices in
the uniform contact quotient. It is inherited from the single-output algorithm. Different histories reaching the same state add their state values
before further propagation.

Closing the current cycle and opening the next cycle uses the canonical
frontier of singly occupied interactions,
\begin{equation}
 j=\min\{i\geq r:m_i=1\}.
 \label{c3:sub:frontier}
\end{equation}
If this set is nonempty, set $m_j$ to two, make $j$ the new head and
endpoint, and multiply by $-N_fA_{eh}$. This operation continues the
internal recurrence. A terminal closing operation uses the same factor
but sends the value to an output without opening another cycle.

For a subset $S$, both endpoints of every included interaction must have
been consumed and no excluded interaction may have been visited. Therefore
the feeding states lie at
\begin{equation}
 \ell(S)=2|S|-1,\qquad r=\min S,\qquad
 \bm m=2\bm 1_S.
 \label{c3:sub:interface-condition}
\end{equation}
If $R_S$ is this set of states, the algebraic output is
\begin{equation}
 O_S=\sum_{(r,h,e,\bm m)\in R_S}
 x_{\ell(S)}(r,h,e,\bm m)(-N_fA_{eh}).
 \label{c3:sub:output}
\end{equation}
There is one output identity for each subset, but generally several
feeding states with different heads, endpoints and histories. 

The terminal readout closes a copy of the current contribution.
The original internal state remains open and can continue into an unused
interaction, eventually contributing to a larger subset. Thus a completed
occupancy tuple can feed both a terminal output and further propagation.
Only the internal state carries the head and endpoint data needed for
that propagation; the accumulated scalar $O_S$ is a finished result.

\subsection{Proof that every readout equals the standalone computation}

Fix a nonempty subset $S$ and compare its envelope readout with a
standalone circuit built on the induced matrix $A_{S,S}$. Enumerate the
members of $S$ in increasing order and map local labels to those original
labels. This map preserves relative order, endpoint partners and all
matrix elements used by the recurrence. The standalone root maps to
$\min S$, and its initialized state value maps to the initialized state of
that rooted branch.

Occupancies increase monotonically along a propagation path. Since the
readout condition in Eq.~\eqref{c3:sub:interface-condition} vanishes outside
$S$, every contributing path has used only labels in $S$. Restricting
the envelope calculation to those labels therefore preserves its sum.

Within the restricted paths, continuation agrees with the standalone transition in its local occupancy, factor and endpoint multiplicity. The canonical frontier also agrees, because its minimum is
taken among the same occupied labels in the same relative order. These
observations prove equality of corresponding state values by induction
on the layer. At the feeding layer, both calculations apply the same
final factor and sum the same terminal histories. Hence
\begin{equation}
 O_S(A)=C_{|S|}(A_{S,S}),
 \label{c3:sub:standalone-equality}
\end{equation}
where $C_{|S|}$ denotes the standalone algebraic connected sum with the
same convention. Restoring the common physical prefactor preserves this
identity for the corresponding integrands.

The proof requires a compressed state to identify the visited
interactions and each requested subset output. Keeping only the number
of visited interactions loses the restriction to a specified $S$.
All-subset compression therefore preserves the exact occupancy tuple and output
identity, even when a single-output calculation admits further merging.

\subsection{A three-interaction example}

For an envelope with three interactions, seven nonempty subsets are
requested. Table~\ref{c3:sub:three-table} gives their roots, occupancy patterns and
feeding layers. In particular, $\{1,3\}$ is a non-prefix subset read
from the first branch, whereas $\{2,3\}$ is read from the second.
The third branch yields only $\{3\}$, since labels 1 and 2 are
inadmissible in that branch.
Each subset output uses the matrix entries at its original envelope labels.

\begin{table}[htbp]
\centering
\begin{tabular}{@{}cccc@{}}
\toprule
Subset $S$ & Root $r$ & Occupancy $\bm m$ & Feeding layer $\ell$\\
\midrule
$\{1\}$ & $1$ & $(2,0,0)$ & $1$\\
$\{2\}$ & $2$ & $(0,2,0)$ & $1$\\
$\{3\}$ & $3$ & $(0,0,2)$ & $1$\\
$\{1,2\}$ & $1$ & $(2,2,0)$ & $3$\\
$\{1,3\}$ & $1$ & $(2,0,2)$ & $3$\\
$\{2,3\}$ & $2$ & $(0,2,2)$ & $3$\\
$\{1,2,3\}$ & $1$ & $(2,2,2)$ & $5$\\
\bottomrule
\end{tabular}
\caption[Connected outputs for every interaction subset]{Connected outputs for three interactions in the uniform contact construction. Each row specifies an output fed by internal states with the indicated occupancy tuple and potentially different heads and endpoints.}
\label{c3:sub:three-table}
\end{table}

The singleton output provides a short algebraic check. Starting with
$h=e=i$ and $m_i=1$, the two ways to complete its occupancy are a
continuation weighted by $A_{ii}$ and a close-and-open step weighted by
$-N_fA_{ii}$. Both reach the same residue state. Its accumulated state value
is $(1-N_f)A_{ii}$, and the terminal closing gives
\begin{equation}
 O_{\{i\}}=N_f(N_f-1)A_{ii}^2.
 \label{c3:sub:singleton}
\end{equation}
The factor $N_f(N_f-1)$ counts ordered pairs of different flavours. 

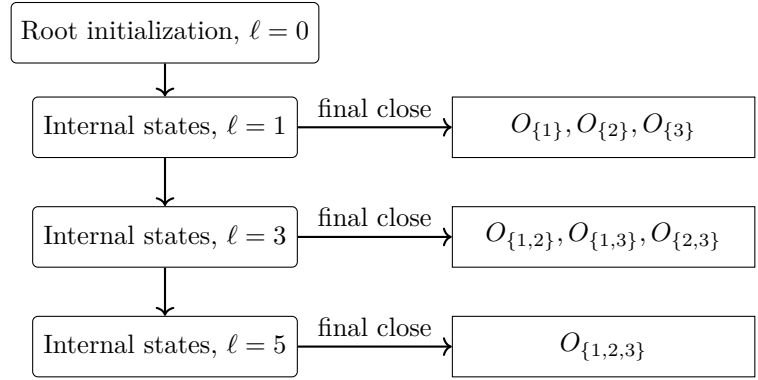
\begin{figure}[htbp]
\centering
\begin{tikzpicture}[font=\small,
 state/.style={draw,rounded corners=2pt,align=center,minimum width=35mm,minimum height=8mm},
 outputbox/.style={draw,align=center,minimum width=40mm,minimum height=8mm}]
\node[state] (init) at (0,0) {Root initialization, $\ell=0$};
\node[state] (one) at (0,-1.25) {Internal states, $\ell=1$};
\node[state] (two) at (0,-2.7) {Internal states, $\ell=3$};
\node[state] (three) at (0,-4.15) {Internal states, $\ell=5$};
\node[outputbox] (o1) at (5.8,-1.25) {$O_{\{1\}},O_{\{2\}},O_{\{3\}}$};
\node[outputbox] (o2) at (5.8,-2.7) {$O_{\{1,2\}},O_{\{1,3\}},O_{\{2,3\}}$};
\node[outputbox] (o3) at (5.8,-4.15) {$O_{\{1,2,3\}}$};
\draw[->,thick] (init)--(one);
\draw[->,thick] (one)--(two);
\draw[->,thick] (two)--(three);
\draw[->,thick] (one)--node[above]{final close}(o1);
\draw[->,thick] (two)--node[above]{final close}(o2);
\draw[->,thick] (three)--node[above]{final close}(o3);
\end{tikzpicture}
\caption[All-subset readout from one envelope]{All-subset readout for a three-interaction envelope. Vertical arrows
propagate internal states; layers $\ell=2,4$ are omitted. Horizontal
arrows close copies of those states into the named subset outputs.
The original open states continue to the next layer.}
\label{c3:sub:three-figure}
\end{figure}

For the subset $\{1,3\}$, the initial occupancy tuple is $(1,0,0)$ and every
contributing path leaves its second component zero. At layer three its
occupancy tuple is $(2,0,2)$, while the open head and endpoint vary over the states
in $R_{\{1,3\}}$. Closing and summing those states gives precisely the
two-interaction polynomial evaluated with rows and columns one and three.
 The full three-interaction output continues to layer five and is assembled from its own completed occupancy tuple. Figure~\ref{c3:sub:three-figure} shows the readout structure.

Figure~\ref{c3:sub:three-figure} separates propagation from terminal
closure. At layer three, for example, the occupancy tuple $(2,0,2)$ contributes to
$O_{\{1,3\}}$, while its open internal states can still enter interaction
2 and contribute to $O_{\{1,2,3\}}$. The sampler uses these independently
labelled values to compare retained subsets.

\subsection{Construction, pruning and memory for several outputs}

The implementation can accumulate outputs as soon as their feeding layer
is evaluated. The construction proceeds as follows:
\begin{enumerate}
 \item Initialize one branch for every root, keeping original interaction
 labels and the corresponding occupancy tuple.
 \item Generate the canonical continuation and close-and-open transitions
 to the next layer, accumulating the weights of histories that reach identical canonical states.
 \item Whenever a state has only zero and two occupancies, attach its
 final-closing contribution to the output named by that completed subset.
 \item Continue the internal recurrence along every valid transition.
 The completed output $O_S$ is a terminal sink; propagation uses the
 open state and its head and endpoint labels.
 \item Prune with backward productivity initialized from every requested
 readout, and preserve output identities in subsequent equivalence signatures.
 \item Evaluate the constructed layers and accumulate each output separately,
 then apply its order-specific physical normalization.
\end{enumerate}
Steps 1 and 2 guarantee the root and layer conditions in
Eq.~\eqref{c3:sub:interface-condition}. Step 3 then applies the
completed-occupancy test.

The requested outputs determine the retained graph through step 5.
Backward propagation from their union selects every state contributing
to at least one readout, so changing that output set changes the
states retained by pruning.

Layerwise evaluation stores only the current and next state value layers,
together with the graph topology and output accumulators. The
$2^n-1$ nonempty-subset outputs still require an exponential number of
entries when all are stored.

\subsection{Readout cost and implementation checks}\label{c3:sub:checks}

Readout operations must be counted in addition to propagation. If
$q_S=|R_S|$, a direct terminal evaluation uses
\begin{equation}
 M_{\mathrm{read}}=\sum_S q_S,\qquad
 A_{\mathrm{read}}=\sum_S\max(q_S-1,0)
 \label{c3:sub:readout-cost}
\end{equation}
closing-factor multiplications and within-output additions, before any
factor sharing. Grouping equal closing factors can lower the first count,
but cannot merge the named outputs. The overhead relative to a full-order-only circuit depends on which states are removed by pruning for that smaller output set.

I verified the uniform subset readouts against independent connected
endpoint enumeration, and the raw, merged and scheduled typed-head
circuits against flavour-resolved permutation sums.
For a typed-head all-subset calculation, the root and readout conditions
proved above must be combined with the flavour, head-type and sublattice
labels on which its operators depend.

The simulation cost depends on the terminal-readout work as well as
the variance and mixing of the sampler in Section~\ref{c3:mc:section}.

\section{A determinant-polynomial sieve for connected sums}
\label{c3:fast:section}
\label{c3:poly:section}

The connected sum at fixed interaction coordinates can be evaluated
from principal-minor polynomials, which already contain the fermionic
exchange signs within each flavour. Combining these polynomials reduces
the remaining task to selecting the two different flavours at each
interaction. For three flavours, evaluations at zero and at two opposite
points isolate this local quadratic contribution exactly. Small linear
transforms of the principal-minor tables perform this selection for all
interaction subsets together.

I first establish the physical normalization and the relation between
full moments and connected coefficients, then derive the coefficient
filters and their implementation. The construction accommodates unequal
flavour matrices and shifts. For more than three flavours, the symmetric
filter retains higher even powers, which I remove with a one-dimensional
Fourier projection. Including the connected recursion, the resulting
arithmetic bounds are $O(3^n)$ for three flavours and $O(n3^n)$ for any
fixed $N_f\geq4$.

\subsection{Local insertions and the physical normalization}

Fix $n$ labelled interaction coordinates $X_1,\ldots,X_n$ and let
$a=1,\ldots,N_f$, with $N_f\geq2$. Each coordinate labels one two-body
interaction insertion, with two fermionic endpoints. Define the unscaled
propagator matrices $A^{(a)}_{ij}=g_a(X_i,X_j)$ using the equal-time
prescription of Section~\ref{c3:sb:target}, and write
$\eta_{ai}=\psi_a(X_i)\bar\psi_a(X_i^+)$. In this ordering,
$\langle\eta_{ai}\rangle_0=A^{(a)}_{ii}$ and the physical density
bilinear is $-\eta_{ai}$. The local interaction and counterterm are
\begin{align}
 K_{1,i}&=U\sum_{a<b}\eta_{ai}\eta_{bi}
                  +\sum_a\alpha_a\eta_{ai}
            =\frac U2 B_i,
 \label{c3:poly:local-insertion}\\
 B_i&=\sum_{a\ne b}\eta_{ai}\eta_{bi}
                 +2\sum_{a=1}^{N_f}s_a\eta_{ai},
 \qquad s_a=\alpha_a/U.
 \label{c3:fast:target}
\end{align}
The factor two counts the two endpoint orderings of a distinct-flavour
pair. The linear term weights a single flavour bilinear, and the physical
insertion contains no scalar constant. Here $s_a$ uses the
partner-attached convention of Section~\ref{c3:sb:target}.

For an interaction subset $S$, let
$H(S)=\langle\prod_{i\in S}B_i\rangle_0$, with
$H(\varnothing)=1$, and denote its connected cumulant by $C(S)$.
The contribution to the logarithmic expansion is
\begin{equation}
 \Delta\ln Z(\xi)=
 \sum_{n\geq1}\frac{(-\xi U/2)^n}{n!}
 \int dX_1\cdots dX_n\,C([n]).
 \label{c3:poly:physical-series}
\end{equation}
The evaluator returns $C(S)$ at fixed coordinates. The coupling powers,
expansion factorial and spacetime measure are applied as in
Eq.~\eqref{c3:poly:physical-series}. The parameter $\xi$ varies at fixed
$U$ and shifts. Multiplying every $A^{(a)}$ and every $s_a$ by
$\lambda$ multiplies $H(S)$ and $C(S)$ by $\lambda^{2|S|}$.
The normalized $B_i$ assumes $U\ne0$; the unfactored $K_{1,i}$ also
covers $U=0$.

\subsection{Absorbing counterterms into shifted determinants}
\label{c3:fast:shift}

To absorb the linear counterterms into separate flavour determinants,
choose $c_a$ such that $\sum_{b\ne a}c_b=s_a$. Solving these equations
gives
\begin{equation}
 c_a=\frac{\sum_b s_b}{N_f-1}-s_a,\qquad
 \widehat A^{(a)}=A^{(a)}+c_a I,\qquad
 \Gamma=\left(\sum_a c_a\right)^2-\sum_a c_a^2 .
 \label{c3:fast:shift-map}
\end{equation}
Expanding the product proves the local identity
\begin{equation}
 \sum_{a\ne b}(\eta_{ai}+c_a)(\eta_{bi}+c_b)=B_i+\Gamma.
 \label{c3:fast:completion}
\end{equation}
The coefficient of $\eta_{ai}$ is $2\sum_{b\ne a}c_b=2s_a$;
the only extra term is the scalar $\Gamma$.
The identity holds for unequal matrices $A^{(a)}$.

In the flavour-uniform case, $A^{(a)}=A$ and $s_a=s$, so
\begin{equation}
 c_a=\frac{s}{N_f-1},\qquad
 \Gamma=\frac{N_f}{N_f-1}s^2 .
 \label{c3:fast:uniform}
\end{equation}
The uniform CoS construction adds $s/N_f$ to entries diagonal in the
full endpoint labels, whereas the present construction adds $s/(N_f-1)$
to each flavour matrix. With their respective scalar corrections, both
represent the same physical insertion; each increment belongs to its own
construction. For $N_f=2$,
$c_\uparrow=s_\downarrow$ and $c_\downarrow=s_\uparrow$, consistent
with the reversed spin on the shift in Eq.~\eqref{c3:model:twospinshift}.

\subsection{Principal-minor polynomials and shared variables}
\label{c3:fast:determinants}

For a single flavour, column multilinearity gives
\begin{equation}
 D_a(\mathbf z)=\det\!\left[I+A^{(a)}
                 \operatorname{diag}(z_1,\ldots,z_n)\right]
 =\sum_{S\subseteq[n]}\det A^{(a)}[S,S]\,z_S,
 \qquad z_S=\prod_{i\in S}z_i.
 \label{c3:poly:minor-polynomial}
\end{equation}
Choosing the identity column outside $S$ fixes the corresponding rows
and leaves the principal minor on $S$. The empty minor is one, and each
variable occurs at most once in each determinant. Gaussian flavour
independence and Wick's theorem identify these minors with
$\langle\prod_{i\in S}\eta_{ai}\rangle_0$, including all exchange
signs within the flavour. Principal-minor algorithms can prepare all
coefficients together \cite{FastPrincipalMinors2022}.

Introduce ordinary commuting variables $z_i$, shared by every flavour,
and $Z=\operatorname{diag}(z_1,\ldots,z_n)$. Define
\begin{equation}
 \begin{aligned}
 P_a(\mathbf z)&=\det[I-\widehat A^{(a)}Z]
       =\sum_{S\subseteq[n]}p_a(S)z_S,\\
 p_a(S)&=(-1)^{|S|}\det\widehat A^{(a)}[S,S],
 \qquad z_S=\prod_{i\in S}z_i .
 \end{aligned}
 \label{c3:fast:detpoly}
\end{equation}
This is the principal-minor identity of
Eq.~\eqref{c3:poly:minor-polynomial} with $A$ replaced by
$\widehat A^{(a)}$ and each $z_i$ replaced by $-z_i$. The extracted
monomials have even total degree, so choosing plus signs in every
determinant gives the same coefficients. Here the $z_i$ are ordinary
commuting variables: their quadratic terms represent the two bilinears
of an interaction.

Let $Q(\mathbf z)=\prod_{a=1}^{N_f}P_a(\mathbf z)$. The auxiliary
moment of an interaction subset $S$ is
\begin{equation}
 \widehat H(S)=2^{|S|}[z^{2\mathbf1_S}]Q(\mathbf z)
       =\left\langle\prod_{i\in S}(B_i+\Gamma)\right\rangle_0 .
 \label{c3:fast:moments}
\end{equation}
The notation $[z^{2\mathbf1_S}]$ selects the monomial with degree two
in each variable belonging to $S$ and degree zero in all other variables.
To verify the equality, write
$P_a=\langle\prod_i[1-z_i(\eta_{ai}+c_a)]\rangle_0$.
Because the Gaussian flavours are independent, the product of these
averages equals the average of the product. Multilinearity within each
flavour selects two different flavours at each occupied interaction.
The two endpoint orderings give the factor $2^{|S|}$, and
Eq.~\eqref{c3:fast:completion} gives the local insertion $B_i+\Gamma$.
The coefficients thus have the Wick signs and weights of the original
shifted expansion.

\subsection{Connected coefficients and scalar completion}

The moments $H(S)$ contain disconnected diagrams. Every such contribution
has a unique partition of its interaction labels into connected
components, so
\begin{equation}
 H(S)=\sum_{\pi\in\mathfrak P(S)}\prod_{T\in\pi}C(T),
 \label{c3:poly:set-partitions}
\end{equation}
where $\mathfrak P(S)$ is the set of unordered set partitions. Choosing
a root $r(S)\in S$ and identifying the component that contains it gives
\begin{equation}
 C(S)=H(S)-
 \sum_{\substack{\varnothing\ne T\subsetneq S\\r(S)\in T}}
       C(T)H(S\setminus T).
 \label{c3:poly:cdet}
\end{equation}
Each disconnected partition is removed once, according to its unique
root component. Processing subsets in increasing size makes every
quantity on the right available. This is the anchored connected-determinant
recursion \cite{RossiCDet2017}. Once the moments are given, it takes
$O(3^n)$ operations and $O(2^n)$ storage to obtain all connected subsets.

A generating-function formulation also explains the scalar correction.
Introduce interaction markers $x_i$ in the squarefree algebra
\begin{equation}
 \mathcal R_n=\mathbb C[x_1,\ldots,x_n]/(x_1^2,\ldots,x_n^2),
 \qquad H(\mathbf x)=\sum_{S\subseteq[n]}H(S)x_S,
 \qquad x_S=\prod_{i\in S}x_i.
 \label{c3:poly:ring}
\end{equation}
These markers record whether an interaction is present. They differ from
the ordinary determinant variables $z_i$, whose quadratic powers count
its two bilinears. Multiplication in $\mathcal R_n$ suppresses overlapping
interaction supports. Since $H(\mathbf0)=1$, its finite formal logarithm is
\begin{equation}
 C(\mathbf x)=\log_{\mathcal R_n}H(\mathbf x)
 =\sum_{j=1}^{n}\frac{(-1)^{j+1}}{j}
       [H(\mathbf x)-1]^j,\qquad H=\exp_{\mathcal R_n}C.
 \label{c3:poly:nilpotent-log}
\end{equation}
Expanding the exponential recovers Eq.~\eqref{c3:poly:set-partitions}:
the factorial cancels the orderings of the connected blocks. The
$1/n!$ in Eq.~\eqref{c3:poly:physical-series} instead accounts for
integration over labelled interaction coordinates.

The shifted determinants return the auxiliary moments $\widehat H$ of
Eq.~\eqref{c3:fast:moments}. Their relation to the physical moments is
\begin{equation}
 \widehat H(S)=\sum_{T\subseteq S}H(T)\Gamma^{|S\setminus T|},
 \qquad
 \widehat H(\mathbf x)=H(\mathbf x)\prod_i(1+\Gamma x_i)
 \quad\hbox{in }\mathcal R_n.
 \label{c3:poly:q-factor}
\end{equation}
Consequently,
\begin{equation}
 \log_{\mathcal R_n}\widehat H
 =\log_{\mathcal R_n}H+\Gamma\sum_i x_i.
 \label{c3:poly:q-log}
\end{equation}
If $\widehat C$ is obtained by applying the anchored recursion to the
complete auxiliary moment table, the desired connected values are
\begin{equation}
 C(S)=\widehat C(S)-\Gamma\,\mathbf1_{|S|=1}.
 \label{c3:fast:connected}
\end{equation}
A scalar insertion therefore changes only the singleton connected term.
The calculation uses the complete auxiliary moment table $\widehat H$
in the recursion and applies the subtraction $\Gamma$ to its singleton
outputs. This order retains the same moment convention in every
recursive product.

For one interaction, the physical value is
\begin{equation}
 C(\{i\})=2\sum_{a<b}A^{(a)}_{ii}A^{(b)}_{ii}
                 +2\sum_a s_a A^{(a)}_{ii}.
 \label{c3:poly:singleton}
\end{equation}
With $A^{(a)}_{ii}=-\rho_a$ in a homogeneous background, this gives the
first-order contribution per physical site
\begin{equation}
 \psi_1=-\beta U\sum_{a<b}\rho_a\rho_b
                  +\beta\sum_a\alpha_a\rho_a.
 \label{c3:poly:first-physical}
\end{equation}
Equation~\eqref{c3:poly:singleton} gives the common physical normalization
for the sieve and the endpoint recurrences.

\subsection{Three-point filtering and the three-flavour case}
\label{c3:fast:ternary}

A full coefficient array for $Q$ has $(N_f+1)^n$ entries, but only
monomials of degree zero or two in each $z_i$ are required. For a
polynomial $q$ define
the local filter
\begin{equation}
 \Delta_{i,\lambda}q=
 \frac{q|_{z_i=\lambda}+q|_{z_i=-\lambda}-2q|_{z_i=0}}{2}.
 \label{c3:fast:local-filter}
\end{equation}
It removes the constant and all odd powers, leaving only positive
even degrees. For a subset $S$, let
\begin{equation}
 Y_S(\lambda)=
 \left.
 \left(\prod_{i\in S}\Delta_{i,\lambda}\right)Q
 \right|_{z_i=0\ (i\notin S)} .
 \label{c3:fast:support}
\end{equation}
Setting $z_i=0$ outside $S$ fixes the support. The symmetric
differences keep positive even powers of the variables in $S$.

When $N_f=3$, the degree in each $z_i$ is at most three. The only
positive even degree is therefore two, and
\begin{equation}
 \widehat H(S)=2^{|S|}Y_S(1).
 \label{c3:fast:three}
\end{equation}
All subsets follow from the same grid $\{0,+1,-1\}^n$. For $N_f=2$
the same degree argument holds, although the usual product of two
principal-minor tables is a simpler specialized evaluator.

\subsection{Removing excess degree for more flavours}
\label{c3:fast:fourier-section}

For $N_f\geq4$, put $h=\lfloor N_f/2\rfloor$. The filter can retain
degrees $2,4,\ldots,2h$, so
\begin{equation}
 Y_S(\lambda)=\lambda^{2|S|}R_S(\lambda^2),\qquad
 \deg R_S\leq(h-1)|S|.
 \label{c3:fast:rank-bound}
\end{equation}
The desired local-degree coefficient is $R_S(0)$.
Choose $L=(h-1)n+1$ and
$\lambda_j=\exp(\pi\ii j/L)$, $0\leq j<L$. Roots-of-unity
orthogonality gives
\begin{equation}
 \widehat H(S)=\frac{2^{|S|}}{L}
       \sum_{j=0}^{L-1}\lambda_j^{-2|S|}Y_S(\lambda_j).
 \label{c3:fast:fourier}
\end{equation}
To verify Eq.~\eqref{c3:fast:fourier}, expand $R_S(w)$ in powers of $w$.
The average of $\exp(2\pi\ii jk/L)$ vanishes for $1\leq k<L$.
The degree bound excludes multiples of $L$ other than zero,
so no unwanted coefficient aliases into the answer.

After the phase correction, the required zero Fourier mode is the
direct average over $j$.
This projection removes the excess even degree left by the local
filters. The connected coefficients are subsequently obtained from the
moment table by Eq.~\eqref{c3:poly:cdet}. For $N_f=6$,
$L=2n+1$. Real input matrices and shifts give $R_S$ real coefficients,
allowing conjugate-pair reuse with $n+1$ grid evaluations. For complex matrices or shifts all $L$ samples are required unless
another symmetry relates them.

\subsection{Rectangular transforms and the evaluation procedure}
\label{c3:fast:implementation}

I compute the $3^n$ grid values by transforming the principal-minor
tables. For each Fourier
sample, first scale $p_a(S)$ by $\lambda_j^{|S|}$ and apply
the local evaluation transform
\begin{equation}
 E=\begin{pmatrix}1&0\\1&1\\1&-1\end{pmatrix},
 \qquad
 W=\begin{pmatrix}2&0&0\\-2&1&1\end{pmatrix}.
 \label{c3:fast:transforms}
\end{equation}
$E^{\otimes n}$ converts a binary coefficient table into the ternary
evaluation grid. The flavour grids are multiplied pointwise;
uniform inputs instead use one grid raised to the $N_f$th power.
$W^{\otimes n}$ converts the product grid back into a binary table.
The first row of $W$, used when $i\notin S$, is twice evaluation
at zero. The second row, used when $i\in S$, is twice the filter
in Eq.~\eqref{c3:fast:local-filter}.
If this output is $b_j(S)$, then $b_j(S)=2^nY_S(\lambda_j)$ and
\begin{equation}
 \widehat H(S)=\frac{2^{|S|-n}}{L}
         \sum_j\lambda_j^{-2|S|}b_j(S).
 \label{c3:fast:normalization}
\end{equation}
The factor $2^{-n}$ removes the factor of two in each row of $W$,
while $2^{|S|}$ counts the two endpoint orderings at each occupied
interaction.

Successive one-coordinate maps evaluate the tensor-product transform
using the small matrices $E$ and $W$. The Fourier sum
then gives a binary moment table, from which the anchored recursion
computes $C(S)$ for every subset. This includes every proper subset even
when the desired final observable uses only $C([n])$.

\subsection{Correctness and arithmetic complexity}
\label{c3:fast:complexity}

By Eq.~\eqref{c3:fast:moments}, the extracted coefficients are the
auxiliary Wick moments. Equations~\eqref{c3:fast:three} and
\eqref{c3:fast:fourier} extract every required coefficient. The anchored
recursion then removes disconnected diagrams by identifying the unique
component containing the root, and Eq.~\eqref{c3:fast:connected}
removes the scalar completion. The result is the physical connected
sum for each nonempty subset.

Computing every principal minor separately costs at most
 $O(N_f n^3 2^n)$. The bound below assumes this separate-minor
calculation; faster principal-minor algorithms are available
\cite{FastPrincipalMinors2022}. One rectangular transform processes
\begin{equation}
 \sum_{j=0}^{n-1}3^j2^{n-1-j}=3^n-2^n
 \label{c3:fast:fibres}
\end{equation}
small fibres. Their arithmetic cost is constant, so a complete
transform costs $O(3^n)$ rather than $O(n3^n)$.
Each Fourier sample requires $N_f$ forward transforms, one inverse
transform and $N_f-1$ pointwise products. Scaling the binary tables and accumulating the moments costs
$O(N_f L2^n)$, which is included in the bounds below.
The anchored connected recurrence has
\begin{equation}
 \sum_{k=1}^n\binom nk(2^{k-1}-1)
       =\frac{3^n-2^{n+1}+1}{2}
 \label{c3:fast:connected-ops}
\end{equation}
multiply--subtract terms.

Adding these costs gives the following bounds on the number of
arithmetic operations $T_{\rm nonuniform}$ and the working memory
${\cal M}_{\rm work}$, measured in stored numbers:
\begin{equation}
 T_{\rm nonuniform}
 =O\!\left(N_f n^3 2^n+N_f L3^n+3^n\right),
 \qquad
 {\cal M}_{\rm work}=O(N_f2^n+3^n).
 \label{c3:fast:cost}
\end{equation}
For equal flavour matrices and shifts, one minor table and one forward transform
suffice. Binary powering gives
\begin{equation}
 T_{\rm uniform}
 =O\!\left(n^3 2^n+L(1+\log N_f)3^n+3^n\right).
 \label{c3:fast:uniform-cost}
\end{equation}
For $N_f=3$, a single sample suffices ($L=1$), and both bounds
reduce to $O(3^n)$.
For any fixed $N_f\geq4$, they give $O(n3^n)$.
This operation count requires a grid of $3^n$ stored values.

Compared with the published $O(n^3 3^n)$ bound for CoS with ordered
contact endpoints \cite{KozikCoS2024}, the determinant-polynomial sieve
removes the factor $n^3$ for $N_f=3$ and reduces it to $n$ for fixed
$N_f\geq4$. These bounds include the anchored recursion used here;
a cheaper connected-coefficient extraction would reduce that contribution.
The $O(n^2 2^n)$ all-vertices-ordered CoS variant evaluates a
nonsymmetrized integrand, so it has a different fixed-configuration target.
The present counts apply to dense matrices with nonzero diagonals and
use no numerical pruning threshold.

\subsection{Exact examples with unequal shifts}
\label{c3:fast:examples}

For a concrete three-flavour check, take
\begin{equation}
 \begin{gathered}
 A^{(1)}=\begin{pmatrix}1&1&2\\1&2&1\\2&1&3\end{pmatrix},\quad
 A^{(2)}=\begin{pmatrix}2&1&1\\2&3&2\\1&1&2\end{pmatrix},\quad
 A^{(3)}=\begin{pmatrix}3&2&1\\1&1&2\\2&1&3\end{pmatrix},\\
 (s_1,s_2,s_3)=(1,2,3),\qquad(c_1,c_2,c_3)=(2,1,0),\qquad\Gamma=4.
 \end{gathered}
 \label{c3:fast:example-input}
\end{equation}
I choose integer matrices so that the algebra can be checked by hand.
At order two use their upper-left submatrices. The three
determinant polynomials become
\begin{align*}
 P_1&=1-3z_1-4z_2+11z_1z_2,\\
 P_2&=1-3z_1-4z_2+10z_1z_2,\\
 P_3&=1-3z_1-z_2+z_1z_2.
\end{align*}
Their product on the grid is
\begin{equation}
 \begin{array}{c|rrr}
 z_1\backslash z_2&0&1&-1\\ \hline
 0&1&0&50\\
 1&-8&-40&-144\\
 -1&64&220&2052
 \end{array}.
 \label{c3:fast:example-grid}
\end{equation}
The four corner entries sum to $2088$, the four entries with exactly
one zero coordinate sum to $106$, and the centre entry is one. Applying the filters gives
$\widehat H(\{1,2\})=2088-2(106)+4=1880$.
The singleton auxiliary moments are $54$ and $48$, so
$C(\{1,2\})=1880-54\cdot48=-712$.
Their physical singleton values are $50$ and $44$ after subtracting
$\Gamma$.

\begin{table}[tbp]
\centering
\setlength{\belowcaptionskip}{6pt}
\caption[Exact values of the three-flavour sieve]{Auxiliary moments
$\widehat H(S)$ and physical connected sums $C(S)$ for
Eq.~\eqref{c3:fast:example-input}. The moments include the scalar
$\Gamma=4$ at each interaction; this scalar has been removed from
the connected singleton values.}
\label{c3:fast:example-table}
\begin{tabular}{crr}
\toprule
$S$&$\widehat H(S)$&$C(S)$\\
\midrule
$\{1\}$&54&50\\
$\{2\}$&48&44\\
$\{3\}$&78&74\\
$\{1,2\}$&1880&$-712$\\
$\{1,3\}$&3116&$-1096$\\
$\{2,3\}$&2824&$-920$\\
$\{1,2,3\}$&97024&52672\\
\bottomrule
\end{tabular}
\end{table}

At order three the grid has $27$ points. Grouping their values
by the number of zero coordinates yields
\begin{align}
 \widehat H(\{1,2,3\})&=113392-2(8552)+4(186)-8=97024,\notag\\
 C(\{1,2,3\})&=97024-54(2824)-48(3116)-78(1880)
                         +2(54)(48)(78)=52672.
 \label{c3:fast:example-three}
\end{align}
Table~\ref{c3:fast:example-table} lists both quantities for every nonempty
subset. I checked them independently by enumerating all $4!$ endpoint
permutations at order two and all $6!$ at order three. I include the
interaction links when testing connectivity and replace allowed
singleton cycles by partner-attached counterterms. The enumeration
includes same-flavour contact terms, which cancel by antisymmetry.
Scaling the shifts by $t$ resolves the result into counterterm
sectors:
\begin{align}
 C_{\{1,2\}}(t)&=-264-340t-108t^2,\notag\\
 C_{\{1,2,3\}}(t)&=14704+23312t+12424t^2+2232t^3 .
 \label{c3:fast:counterterm-sectors}
\end{align}
Evaluating the sieve in exact rational arithmetic gives the same result
as endpoint enumeration for the full set and every proper subset.
Agreement in each power of $t$ also checks the linear counterterms and
their mixed contributions.

\subsection{Double-precision comparison with direct CoS}
\label{c3:fast:benchmark}

I benchmarked the sieve and two direct CoS evaluators in C++ at orders
one to nine, for $N_f=3$ and $N_f=6$. In the flavour-uniform case I
compare with the faster of the original shifted uniform CoS DAG and an
optimized connected recurrence. The original DAG is faster at
every tested order. For unequal flavour matrices and shifts I use the
constructed recurrence, which carries flavour labels and endpoint
multiplicities and shares products among continuation and closure edges.
These baselines implement the CoS recurrences discussed in
Sections~\ref{c3:unif:section}--\ref{c3:dag:section}.

For each order and each flavour-uniform or non-uniform case, I generated
sixteen dense real matrices with entries of both signs. Entries are drawn
uniformly from $[-0.6,0.6)$, with $0.3$ added to each diagonal entry.
In the uniform case every flavour uses the same matrix and the same
shift, drawn uniformly from $[0.08,0.20)$. Non-uniform cases have
independently generated matrices and shifts
$s_a=0.03+0.11(a-1)+0.04u_a$, where $u_a$ is uniform on $[0,1)$.

I measured the times on one core of an Intel Xeon Platinum 8358 on
Imperial College's HX1 cluster, using an optimized GCC 12.3 build
with standard floating-point semantics.
All timed arithmetic is double precision. Each timed evaluation includes
initialization, shift preparation, every principal minor and transform,
the anchored recursion and the singleton correction, ending with
$C([n])$. The input-independent graph or transform tables are prepared
before timing, and each evaluation recomputes all matrix-dependent
results. Seven rounds each evaluate the sixteen test cases in a randomized
method order. Allocation of reusable numerical arrays precedes these timed evaluations.

The sieve first computes principal minors with the pivot-corrected
Schur recursion of Section~\ref{c3:fpm:section}. If the absolute value of a corrected pivot
is at or below the numerical threshold, the complete minor table is
recomputed by direct LU factorization.
For $N_f=3$ I use the real three-point filter. For $N_f=6$ I use the
unit-circle projection and evaluate each conjugate pair once.

\begin{figure}[tbp]
\centering
\begingroup
\begin{tikzpicture}
\begin{groupplot}[
  group style={group size=2 by 1,horizontal sep=12mm},
  width=0.45\textwidth,height=52mm,
  xlabel={Expansion order $n$},
  ylabel={$t_{\mathrm{CoS}}/t_{\mathrm{sieve}}$},
  xmin=0.6,xmax=9.4,xtick={1,...,9},
  ymode=log,log basis y=2,
  grid=major,grid style={gray!15},axis line style={gray!60},
  tick label style={font=\scriptsize},
  label style={font=\small},title style={font=\small},
  legend style={font=\scriptsize,draw=none,fill=white,
                at={(0.03,0.97)},anchor=north west},
  scaled ticks=false,
  every axis plot/.append style={line width=0.7pt,mark size=1.6pt}
]
\nextgroupplot[
  title={(a) $N_f=3$},ymin=0.20,ymax=36,
  ytick={0.25,0.5,1,2,4,8,16,32},
  yticklabels={$0.25$,$0.5$,$1$,$2$,$4$,$8$,$16$,$32$}
]
\addplot[gray!65,dashed,mark=none,line width=0.6pt,forget plot]
  coordinates {(0.6,1) (9.4,1)};
\addplot+[blue!70!black,mark=o,mark options={solid,fill=white},
  error bars/.cd,y dir=both,y explicit,
  error bar style={line width=0.55pt},error mark options={rotate=90,mark size=1.5pt}]
  table[x=order,y=speedup,y error minus=error_minus,y error plus=error_plus]
  {figures/chapter3/figure32_pgfplots/data/nf3_uniform.dat};
\addlegendentry{Uniform}
\addplot+[orange!85!black,mark=square*,mark options={solid,fill=orange!85!black},
  error bars/.cd,y dir=both,y explicit,
  error bar style={line width=0.55pt},error mark options={rotate=90,mark size=1.5pt}]
  table[x=order,y=speedup,y error minus=error_minus,y error plus=error_plus]
  {figures/chapter3/figure32_pgfplots/data/nf3_nonuniform.dat};
\addlegendentry{Non-uniform}

\nextgroupplot[
  title={(b) $N_f=6$},ylabel={},ymin=0.12,ymax=2.25,
  ytick={0.125,0.25,0.5,1,2},
  yticklabels={$0.125$,$0.25$,$0.5$,$1$,$2$}
]
\addplot[gray!65,dashed,mark=none,line width=0.6pt,forget plot]
  coordinates {(0.6,1) (9.4,1)};
\addplot+[blue!70!black,mark=o,mark options={solid,fill=white},
  error bars/.cd,y dir=both,y explicit,
  error bar style={line width=0.55pt},error mark options={rotate=90,mark size=1.5pt}]
  table[x=order,y=speedup,y error minus=error_minus,y error plus=error_plus]
  {figures/chapter3/figure32_pgfplots/data/nf6_uniform.dat};
\addlegendentry{Uniform}
\addplot+[orange!85!black,mark=square*,mark options={solid,fill=orange!85!black},
  error bars/.cd,y dir=both,y explicit,
  error bar style={line width=0.55pt},error mark options={rotate=90,mark size=1.5pt}]
  table[x=order,y=speedup,y error minus=error_minus,y error plus=error_plus]
  {figures/chapter3/figure32_pgfplots/data/nf6_nonuniform.dat};
\addlegendentry{Non-uniform}
\end{groupplot}
\end{tikzpicture}
\endgroup
\caption[Speedup of the determinant-polynomial sieve]{Time ratio
$t_{\rm CoS}/t_{\rm sieve}$ for complete double-precision evaluations
at fixed interaction coordinates: (a) $N_f=3$ and (b) $N_f=6$.
Each point divides the median CoS time over seven rounds by the median
sieve time for the same test cases. Vertical ranges are formed from the
round extrema and are mostly smaller than the symbols. Values above one
favour the sieve. The times exclude Green-function table look-up and
Monte Carlo sampling.}
\label{c3:fast:speedup}
\end{figure}
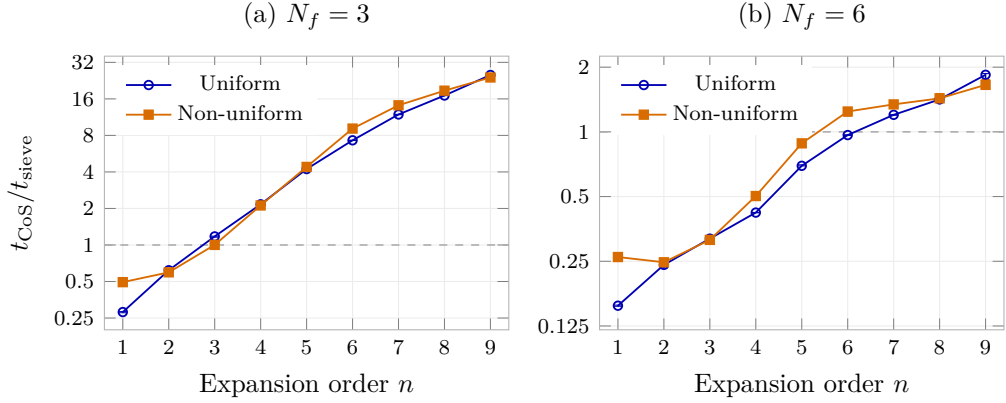

\begin{table}[tbp]
\centering
\setlength{\belowcaptionskip}{6pt}
\caption[Order-nine CoS and sieve timings]{Median time in
microseconds for one complete evaluation of $C([n])$ at order nine.
The speedup is $t_{\rm CoS}/t_{\rm sieve}$.}
\label{c3:fast:timings}
\begin{tabular}{clrrr}
\toprule
 $N_f$&Flavour case&CoS&Sieve&Speedup\\
\midrule
3&Uniform&2428.05&96.41&$25.18\times$\\
3&Non-uniform&3930.81&163.61&$24.03\times$\\
6&Uniform&2468.52&1339.40&$1.84\times$\\
6&Non-uniform&6227.20&3763.72&$1.65\times$\\
\bottomrule
\end{tabular}
\end{table}

For $N_f=3$, the sieve is faster in both flavour cases from order four
onwards, reaching factors of $25.18$ and $24.03$ at order nine
(Fig.~\ref{c3:fast:speedup}). With separate LU factorizations in place of
fast principal minors, the corresponding order-nine factors are
$14.32$ and $10.60$. For $N_f=6$, the additional Fourier projection
moves the crossover to order seven in the uniform case and order six
in the non-uniform case. Table~\ref{c3:fast:timings} lists the absolute
times. The modest six-flavour advantage is also sensitive to the
transform implementation.

I compared the timed results with extended-precision connected references
through order nine and with endpoint enumeration at low order. The
unit-circle result $x$ satisfies
$|x-x_{\rm ref}|\leq10^{-10}+10^{-8}|x_{\rm ref}|$
against its extended-precision reference $x_{\rm ref}$ throughout this
test set. Integer interpolation nodes, however,
produce failures in the six-flavour calculation from order five onwards,
illustrating why the mathematically equivalent projections can differ
in floating-point stability.

The benchmark measures complete evaluations of the same connected
integrand at fixed coordinates. The resulting subset table supplies the
conditional updates and estimators of Section~\ref{c3:mc:section};
the cost of an integrated result additionally includes their sampling
and correlation costs.

\subsection{Physical derivatives and normalization checks}
\label{c3:fast:derivatives}

The same polynomial identities determine derivatives, provided that
the physical parameters held fixed are specified. Let $C_n=C([n])$ and
$b_n=(-U/2)^n C_n(A,\boldsymbol\alpha/U)$ at fixed interaction positions.
Keeping the physical shifts and unscaled matrices fixed gives
\begin{equation}
 \left.\partial_U b_n\right|_{\boldsymbol\alpha,A}
 =\frac{(-U/2)^n}{U}
 \left[n C_n-\sum_a s_a\partial_{s_a}C_n\right].
 \label{c3:poly:interaction-derivative}
\end{equation}
Both the coupling factor and the dimensionless counterterms contribute.
At first order their derivatives cancel for a pure one-body counterterm,
as expected when its physical coefficient is held fixed. After the
remaining integrations and addition of the reference contribution,
the derivative of $\ln Z$ gives the interaction-conjugate occupancy,
summed over all distinct-flavour pairs. Any parameter dependence of the
background matrices contributes through the chain rule as well.
Holding $s_a$ fixed and holding $\alpha_a$ fixed therefore define
different response directions. A common chemical potential probes total
density, while a flavour-resolved chemical potential probes that
flavour's occupation.

The singleton formula, degree-$2n$ scaling and independent-block limit
provide checks at fixed coordinates: block-diagonal propagators for
independent groups give zero connected coefficients spanning those
groups. In the uniform three-flavour limit, the physical values agree
with the CoS endpoint shift $s/3$ after each construction's own scalar
subtraction. The exact examples and comparison with CoS above test these
normalizations before integration. Continuation to the physical coupling
is treated in Section~\ref{c3:num:pade-section}.

\section{Determinant and principal-minor reference evaluators}
\label{c3:fpm:section}

\subsection{Removing disconnected components with a fixed root}

Direct determinant evaluation, which also underlies the sieve of
Section~\ref{c3:fast:section}, provides an independent reference for
the circuit constructions \cite{RossiCDet2017,FastPrincipalMinors2022}.
For two spin blocks and a retained set $S$, define
\begin{equation}
 F(S)=\det M^\uparrow_S\det M^\downarrow_S,\qquad F(\varnothing)=1 .
 \label{c3:fpm:full}
\end{equation}
Every full diagram has a unique connected component containing a
designated root $r\in S$. If that component occupies $T$, the rest is an
arbitrary full diagram on $S\setminus T$. Therefore
\begin{equation}
 C(S)=F(S)-
 \sum_{\substack{T\subsetneq S\\r\in T}}C(T)F(S\setminus T),
 \qquad r=\min S .
 \label{c3:fpm:connected}
\end{equation}
Computing subsets in increasing cardinality makes the recursion
triangular. Choosing the component containing the root counts each decomposition exactly once. With two labels the identity reads
$C(\{1,2\})=F(\{1,2\})-F(\{1\})F(\{2\})$.
At three labels, the root-containing component can contain one, two or
three labels; these mutually exclusive cases exhaust the full sum.

The all-subset cost of a direct recursion is $O(3^n)$: each vertex is
absent, in the root component or in its complement. This count excludes preparation of the determinant table. The local auxiliary
constant introduced by shifted vertices must be handled according to
Section~\ref{c3:model:section}; the corrected singleton gives the physical first-order term.

\subsection{Fast principal minors by Schur complements}

Write an active matrix as
\begin{equation}
 M=\begin{pmatrix}p&r\\c&B\end{pmatrix},\qquad
 S=B-\frac{cr}{p}.
 \label{c3:fpm:schur}
\end{equation}
The two branches of the recursion separate minors according to the
first index: those omitting it are minors of $B$, and those including it
are $p$ times the corresponding minors of $S$. Each Schur complement
therefore serves the whole family of remaining subsets. At level $\ell$,
there are at most $2^\ell$ active branches with dense updates of size
$(n-\ell)^2$. Summing their costs gives an $O(2^n)$ arithmetic count
for this tree organization. This scaling matches the $2^n$ entries of the output table.

For deterministic sublattice summation there are three choices per
interaction: omit it, retain its $A$ entry, or retain its $B$ entry.
The selected entries come from a $2n\times2n$ block matrix, but no valid minor retains both alternatives for one interaction. A three-branch
Schur traversal generates this restricted family. The two spin
determinants must use the \emph{same} sublattice assignment:
\begin{equation}
 F(S)=\sum_{\mathbf b\in\{A,B\}^{|S|}}
 \det M^\uparrow_{S,\mathbf b}\,
 \det M^\downarrow_{S,\mathbf b}.
 \label{c3:fpm:sublattice}
\end{equation}
For each term, both spin determinants share the same onsite coordinates.
After their products are summed over the common sublattice labels,
Eq.~\eqref{c3:fpm:connected} applies because the sums on disjoint
components factorize.

\subsection{Small pivots and exact corrections}

Although the determinant is a polynomial that remains defined at $p=0$,
the Schur representation involves division by $p$. Consequently, a small
pivot can make this representation ill-conditioned even for a regular
diagrammatic weight.

For one active diagonal index $i$, replace $M$ by
$\widetilde M=M+\delta e_ie_i^{\mathsf T}$. Multilinearity gives
\begin{equation}
 \det\widetilde M_T=
 \begin{cases}
 \det M_T,&i\notin T,\\
 \det M_T+\delta\det M_{T\setminus\{i\}},&i\in T .
 \end{cases}
 \label{c3:fpm:pivotcorrection}
\end{equation}
The unshifted minor is recovered by subtracting the second term.
If shifts are introduced in nested Schur branches, corrections must be
propagated with the branch's stored prefactors and in an order that
restores descendant results before their ancestors use them. The identity applies both to a common pivot shift and to shifts chosen
for individual branches. The $O(2^n)$ count above concerns the Schur tree;
back-propagating corrections at every node can add $O(n2^n)$ arithmetic
\cite{FastPrincipalMinors2022}.

Even after the correction identity has restored the unshifted minors, subtracting nearly equal contributions can still lose relative precision. The loss of relative precision in one connected subtraction is measured by
\begin{equation}
 \kappa_C(S)=
 \frac{|F(S)|+
 \displaystyle\sum_{\substack{T\subsetneq S\\r\in T}}
 |C(T)F(S\setminus T)|}{|C(S)|}.
 \label{c3:fpm:condition}
\end{equation}
Large $\kappa_C$ signals amplification of intermediate roundoff; at a
true zero, absolute rather than relative error is appropriate.
The same principle applies to cancellations inside a CoS circuit:
division-free evaluation avoids pivot division but does not eliminate
all floating-point cancellation.

For the honeycomb calculation, principal minors can be obtained by
direct matrix factorization or Schur recursion at the chosen arithmetic
precision. The propagator tables are constructed at high precision,
then rounded to the working precision of the subset calculation,
including double precision when that is selected.

\section{Sampling connected integrals}
\label{c3:mc:section}

\subsection{The coefficient and the average sign}

Let $x$ denote normalized integration coordinates and $f_n(x)$ the real
integrand at order $n$. For a finite, positive absolute integral, define
\begin{equation}
 I_n=\int f_n(x)\,\mathrm dx,\qquad
 A_n=\int|f_n(x)|\,\mathrm dx,\qquad
 \sigma_n=\frac{I_n}{A_n}.
 \label{c3:mc:sign}
\end{equation}
At fixed coordinates, the connected sum and physical prefactors specify
$f_n$. Sampling then determines the mean sign $\sigma_n$ and the
normalization $A_n$. The ratios below are defined on the support of the
sampled measure, with positive finite absolute references or nonzero
signed references. Resolving a signed reference statistically is also
necessary for a usable normalization estimate.

Sampling $|f_n|/A_n$ measures $\sigma_n$. For a proposal density
$q(x'|x)$ the Metropolis--Hastings acceptance is
\begin{equation}
 {\cal A}(x\to x')=
 \min\!\left(1,
 \frac{|f_n(x')|q(x|x')}{|f_n(x)|q(x'|x)}\right).
 \label{c3:mc:accept}
\end{equation}
Wrapped Gaussian displacements and uniform resets are symmetric under the usual periodic convention, whereas learned histogram proposals require the full Hastings ratio. After a rejected move, the old state is measured again to preserve the stationary sampling measure.

VEGAS-type mappings \cite{LepageVEGAS1978,LepageVEGAS2021} first transform $x=x(y)$, giving the sampled integrand 
\[\widetilde f_n(y)=f_n(x(y))|\det(\partial x/\partial y)|.\]

The target weight and estimator both include this Jacobian. In the
three-flavour fixed-order calculations, this adaptive coordinate map is
prepared before sampling.

\subsection{Reference chains based on branch occupations}

Suppose $g_0$ and $g_1$ are two nonnegative functions on a common
normalized domain. Introduce a discrete branch $b\in\{0,1\}$ and sample
the extended weight $g_b(x)$. Then
\begin{equation}
 \frac{\Pr(b=0)}{\Pr(b=1)}
 =\frac{\int g_0(x)\,\mathrm dx}{\int g_1(x)\,\mathrm dx}.
 \label{c3:mc:branchratio}
\end{equation}
I represent the branch label by two equal-measure halves of an
auxiliary unit coordinate and estimate Eq.~\eqref{c3:mc:branchratio}
from their occupations. A different division of that coordinate requires
the corresponding measure compensation.

The three-flavour reference chain varies the lattice size and then the inverse temperature, ending at a domain where a direct absolute-value integral
can be obtained. With $A_j$ denoting consecutive absolute integrals,
\begin{equation}
 I_n=\sigma_n
 \left(\prod_{j=0}^{J-1}\frac{A_j}{A_{j+1}}\right)A_J .
 \label{c3:mc:referencechain}
\end{equation}
The ratios give an exact normalization identity when each branch uses its own propagators, maps and volume factors. The intermediate lattice sizes and temperatures determine the overlap
between neighbouring branches and the mixing of the chain. Uncertainty in the endpoint Monte Carlo integral is propagated with that of the ratios.

\subsection{An envelope and its retained subsets}

Many-configuration MCMC evaluates several candidate configurations in each step \cite{MCMCMC2021}. Let an envelope contain $N$ vertices,
$V=(X_1,\ldots,X_N)$, with each coordinate measured relative to a
normalized base measure. For a nonempty retained set $S$, let
$Q_{\bar S|S}(V_{\bar S}|V_S)$ be a normalized conditional density
that regenerates the complement. Let $\lambda_k>0$ adjust the relative
frequency of order $k$. Define
\begin{align}
 w_S(V)&=\frac{\lambda_{|S|}}{\binom N{|S|}}\,
 |f_{|S|}(V_S)|Q_{\bar S|S}(V_{\bar S}|V_S),\nonumber\\
 W(V)&=\sum_{\substack{S\subseteq[N]\\|S|\in{\cal O}}}w_S(V),
 \qquad {\cal N}=\int W(V)\,\mathrm dV
 =\sum_{k\in{\cal O}}\lambda_k A_k .
 \label{c3:mc:envelope}
\end{align}
The allowed orders form $\mathcal O$. Each physical integrand $f_k$
contains its perturbative factor $1/k!$, and the binomial normalization
of the joint measure compensates for the equivalent embeddings of a
$k$-vertex configuration in an $N$-vertex envelope.

The joint density $w_S(V)/{\cal N}$ has two simple conditionals.
At fixed $V$, choose $S$ with probability $w_S/W$.
At fixed $S,V_S$, redraw $V_{\bar S}$ with density $Q_{\bar S|S}$.
These are conditional updates of the same joint distribution and
therefore preserve it. Marginalizing the subset gives the stationary
envelope density $W/{\cal N}$. 

Ergodicity requires adequate proposal support and a choice of retained
sets that eventually permits every coordinate to change. Assume that
$f_k$ is symmetric under vertex permutations and that $Q$ transforms
equivariantly under relabelling. The retained vertices may then be stored
first and the generated vertices last. Uniform random relabelling after
regeneration gives the labelled density $W/{\cal N}$. The all-subset
estimators are invariant under these permutations, so their expectations
also follow from this labelled measure when the vertex tuple is kept
in the retained-first ordering.

\subsection{Conditional regeneration and the sum over histories}

The honeycomb regeneration proposal, called the seed density below, selects a parent uniformly from the retained and already generated vertices, then draws a child from a conditional histogram density $q_0(X|Y)$. For a nonempty current set $T$,
\begin{equation}
 q(X|V_T)=\frac{1}{|T|}\sum_{j\in T}q_0(X|X_j),
 \qquad \int q_0(X|Y)\,\mathrm dX=1.
 \label{c3:mc:seed}
\end{equation}
The seed is a product of normalized conditional distributions for time,
unit cell and sublattice. Mixing it with a uniform distribution supplies
the proposal support. This remains a valid proposal for a correlated
model, with its efficiency set by how closely the conditional
histograms describe the sampled correlations.

The complement density used in the labelled product measure is the
symmetrized average over every possible generation order. If $B$ is
the set to be generated, then
\begin{equation}
 Q(B)=\frac{1}{|B|}
 \sum_{i\in B}
 \left[\frac{1}{N-|B|}\sum_{j\notin B}q_0(X_i|X_j)\right]
 Q(B\setminus\{i\}),\qquad Q(\varnothing)=1 .
 \label{c3:mc:Qrec}
\end{equation}
The recursion is used only for $|B|<N$, because the retained set is
nonempty. The bracket averages the first child's density over parents in the original retained set; the recursive factor allows that child
to act as a parent thereafter. Expanding the recursion produces the
average over all generation permutations and their normalized
conditional probabilities. With a uniform seed every $Q(B)$ equals
one. The heat-bath weight uses this sum over generation histories,
including the factor $1/|B|$ at each recursive step. Together they give
the normalized density in the labelled product measure, with no
additional permutation factorial.

\subsection{Measuring every order at each envelope}

After the connected subset table is available, define
\begin{align}
 B_k(V)&=\frac{1}{\lambda_k W(V)}
 \sum_{|S|=k}w_S(V),\nonumber\\
 D_k(V)&=\frac{1}{\lambda_k W(V)}
 \sum_{|S|=k}w_S(V)\operatorname{sgn}f_k(V_S).
 \label{c3:mc:RB}
\end{align}
Integration against $W/{\cal N}$ gives
\begin{equation}
 \langle B_k\rangle=\frac{A_k}{\cal N},\qquad
 \langle D_k\rangle=\frac{I_k}{\cal N}.
 \label{c3:mc:RBmean}
\end{equation}
Averaging conditionally over all subsets reduces the variance associated with selecting one subset. Measurements within an envelope remain correlated.

The honeycomb two-stage workflow first uses the nonnegative signals
to obtain
\begin{equation}
 A_k=A_{\rm ref}
 \frac{\langle B_k\rangle}{\langle B_{\rm ref}\rangle},
 \qquad I_k=A_k\sigma_k .
 \label{c3:mc:twostage}
\end{equation}
A separate fixed-order chain measures each $\sigma_k$.
Here $B_{\rm ref}=B_{k_0}$ for a reference order $k_0$ whose absolute
integral $A_{\rm ref}=A_{k_0}$ is known. The factor $1/\lambda_k$
in the definitions of $B_k$ and $D_k$ gives the cross-order normalization
in Eq.~\eqref{c3:mc:RBmean}.
The low-order signed reference can exploit analytic time integration,
whereas the absolute-value reference integrates $|f|$ and is obtained by numerical quadrature.
Its quadrature uncertainty enters in addition to the Monte Carlo uncertainty.

I first prepare the propagator tables and reference integrals, then
tune the conditional histogram seed and the order weights $\lambda_k$.
Independent envelope chains measure all $B_k$ with those quantities held
fixed. Fixed-order absolute-weight chains, with their proposal widths
also tuned and frozen, measure the signs. Each normalization estimate
is then combined with its sign estimate, accounting for correlations
introduced by the shared reference.

Tuned quantities are frozen before production because the
stationary-chain error analysis assumes a fixed measurement kernel.

\subsection{Single-configuration variable-order sampling}

The three-flavour calculations also admit single-configuration
variable-order sampling, whose ideal target is the union of order sectors with weight $|f_n(x)|$.
With no additional order weights, its normalization is
$A_{\rm tot}=\sum_n A_n$.
The measured order-resolved signed occupation is
\begin{equation}
 M_n=\left\langle
 \mathbf1_{\mathrm{order}=n}\operatorname{sgn}f_n
 \right\rangle=\frac{I_n}{A_{\rm tot}} .
 \label{c3:mc:varorder}
\end{equation}
It is averaged over all production steps. By contrast, the conditional
mean sign divides the same signed total by the number of visits to that
order and equals $I_n/A_n$.

One known nonzero signed reference gives
$I_n=I_{n_0}M_n/M_{n_0}$. An independently calculated absolute
normalization instead gives $I_n=A_{\rm tot}M_n$.

For insertion of a vertex $z$ at a uniformly selected position, with
density $r(z|x)$, the forward and reverse probabilities contain the
insertion/deletion move probabilities and the probabilities of choosing
the corresponding positions. In the symmetric position convention
they cancel, giving
\begin{equation}
 {\cal A}_{n\to n+1}
 =\min\!\left(1,
 \frac{|f_{n+1}(x,z)|p_-(n+1)}
 {|f_n(x)|p_+(n)r(z|x)}\right).
 \label{c3:mc:birth}
\end{equation}
The reverse move uses the reciprocal ratio, with $p_\pm$ normalized over
the moves available at each order boundary. The physical definition of
$f_n$ already contains its perturbative factorial and measure factors.

For a general positive sampling weight $\widetilde w_n$, the signed
observable is $f_n/\widetilde w_n$. The sign measurement in
Eq.~\eqref{c3:mc:varorder} follows when $\widetilde w_n=|f_n|$.
The numerical sampling weight has a positive lower threshold; with the
sign measurement used here, Eq.~\eqref{c3:mc:varorder} describes the
zero-threshold limit. Finite-threshold effects enter the numerical
uncertainty discussed below.

\subsection{Correlations and finite-sample ratios}

The reference-ratio and sign-product relations hold for expectations;
their finite-sample ratios can be biased. For $R=\bar X/\bar Y$, the
linearized variance follows from Eq.~\eqref{c2:mc:ratio-error} with the
numerator and denominator replaced by $\bar X$ and $\bar Y$.
The numerator and denominator from one envelope chain are correlated,
and a reference shared by several orders correlates their final
coefficients. Blocking, independent complete chains, or a block
jackknife/bootstrap must preserve those dependencies
\cite{FlyvbjergPetersen1989}.
When normalization and sign estimates come from independent runs,
their leading product variance is
$\sigma_k^2\operatorname{Var}A_k+
A_k^2\operatorname{Var}\sigma_k$.
The reference-integral quadrature uncertainty also contributes to the error budget.

The uncertainty of a sampled coefficient includes both normalization
and sign measurements. Comparing samplers at a specified coefficient
precision therefore involves their correlation times and these joint
errors, in addition to the cost of the connected sum.

\section{Propagator tables and thermodynamic observables}
\label{c3:num:section}

\subsection{Band propagators and the free contribution}

For a translationally invariant reference, diagonalize its flavour-resolved
Bloch matrix, $h_a(\mathbf k)=\sum_\lambda\epsilon_{a\lambda}(\mathbf k)
P_{a\lambda}(\mathbf k)$. The energies include the shifted chemical
potentials. On the positive-time branch,
\begin{equation}
 g_a^{bb'}(\mathbf R,\tau)=\frac{1}{N_c}
 \sum_{\mathbf k,\lambda}e^{-\mathrm i\mathbf k\cdot\mathbf R}
 [P_{a\lambda}(\mathbf k)]_{bb'}
 \frac{e^{-\tau\epsilon_{a\lambda}(\mathbf k)}}
 {1+e^{-\beta\epsilon_{a\lambda}(\mathbf k)}} .
 \label{c3:num:green}
\end{equation}
Here $\mathbf R=\mathbf R_i-\mathbf R_j$ defines the Fourier convention used for the honeycomb tables.
The square lattice has one band with dispersion
$-2t(\cos k_x+\cos k_y)-\mu_a+\alpha_a$.
The honeycomb reference has Bloch matrix
\begin{equation}
 h_a(\mathbf k)=
 \begin{pmatrix}
 -\mu_a+\alpha_a+\Delta_a&-t(1+e^{\mathrm ik_1}+e^{\mathrm ik_2})\\
 -t(1+e^{-\mathrm ik_1}+e^{-\mathrm ik_2})&-\mu_a+\alpha_a-\Delta_a
 \end{pmatrix}.
 \label{c3:num:honey}
\end{equation}
The site-dependent counterterm shift is $\alpha_a+\Delta_a$ on $A$
and $\alpha_a-\Delta_a$ on $B$. For two spins, $\Delta_\uparrow=-\Delta_\downarrow$ produces a staggered spin reference, while $\Delta_\uparrow=\Delta_\downarrow$ produces a staggered charge reference. The compensating counterterm restores the same physical Hamiltonian at $\xi=1$.

The free logarithmic partition density is
\begin{equation}
 \psi_0=\frac{1}{\nu N_c}\sum_{\mathbf k,a,\lambda}
 \log\!\left(1+e^{-\beta\epsilon_{a\lambda}(\mathbf k)}\right).
 \label{c3:num:free}
\end{equation}
The sum over flavour-resolved energies $\epsilon_{a\lambda}(\mathbf k)$ also covers unequal chemical potentials.

For numerical stability, logarithms and propagator exponentials are
evaluated with scaled formulas; the exponent
$-\tau\epsilon-\log(1+e^{-\beta\epsilon})$ remains nonpositive on the
positive-time interval.

\subsection{Chebyshev representation and runtime evaluation}

The free propagators are tabulated using Chebyshev representations and spatial Fourier transforms. With $z=2\tau/\beta-1$,
\begin{equation}
 g_a^{bb'}(\mathbf R,\tau)
 \simeq\sum_{\ell=0}^{p}d_{a,\mathbf R,bb',\ell}T_\ell(z).
 \label{c3:num:cheb}
\end{equation}
Sampling at Chebyshev nodes permits efficient coefficient construction.
At runtime the Clenshaw recurrence
\begin{equation}
 b_{\ell}=2zb_{\ell+1}-b_{\ell+2}+d_\ell,\quad
 \ell=p,\ldots,1,\qquad
 g\simeq d_0+zb_1-b_2,\quad b_{p+1}=b_{p+2}=0
 \label{c3:num:clenshaw}
\end{equation}
evaluates the truncated expansion through successive auxiliary coefficients.
The branch convention accounts for the equal-time discontinuity,
so each Chebyshev fit stays within a continuous branch. For the same site and flavour,
$g(0^+)-g(0^-)=1$; crossing the thermal boundary introduces the
fermionic sign.

The three-flavour calculation uses a distinct interpolation for each flavour. Interpolation is checked by
increasing $p$ and comparing with direct band sums, including near the
time-interval boundaries.

\subsection{From coordinates to a physical coefficient}
\label{c3:num:prefactors}

An order-$n$ square-lattice vertex set has one time and two spatial
coordinates per interaction. A unit-cube map gives time measure
$\beta^n$ and spatial sums $N_s^n$. Dividing by $N_s$ converts the
extensive vacuum contribution to $\psi$. Fixing one vertex by spatial and imaginary-time translation invariance restores a factor $N_s\beta$ for that vertex.

In the three-flavour normalization, each propagator and each shift
replacing a one-cycle propagator is multiplied by $\sqrt{U N_s\beta/2}$.
Every complete typed contraction contains $2n$ such factors.
Multiplication by $(-1)^n/(N_s n!)$ then includes the coupling and
normalized-coordinate measure. The degree-$2n$ homogeneity accounts
for the complete Jacobian in this convention.

The honeycomb two-spin calculation uses the local vertex
$(n_\uparrow-\gamma_\uparrow)(n_\downarrow-\gamma_\downarrow)$, with
$\gamma_{i\uparrow}=\alpha_{i\downarrow}/U$ and
$\gamma_{i\downarrow}=\alpha_{i\uparrow}/U$ as in
Eq.~\eqref{c3:model:twospinshift}. It scales each spin matrix by
$\sqrt{U\beta N_c}$ and multiplies an order-$k$ connected determinant
by $(-2)^k/(k!\,2N_c)$. Since $N_s=2N_c$, the product is
$(-U\beta N_s)^k/(k!N_s)$, as required for uniformly sampled
sublattices. The singleton uses the corrected first-order expression.
The typed-head convention instead uses the ordered-pair insertion
$\mathcal J_i/2$ of Eq.~\eqref{c3:sb:local-insertion}; each convention
has its corresponding matrix scale and external factor.

\subsection{Periodic difference coordinates}
\label{c3:num:time-cutoff}

The SU(3) integrand fixes one vertex as an origin and constructs the
others from successive spatial and temporal increments. On a periodic
lattice, uniformly sampled spatial increments give a uniform set of
relative positions: the triangular cumulative-sum map is a bijection
on the finite translation group. In imaginary time, the complete density-insertion polynomial is periodic
with period $\beta$ in each insertion time. Translating an insertion
through one thermal period multiplies each incident fermion operator by
minus one, and the even number of operators in every density bilinear
makes their product periodic.

Consequently, for any such periodic function $F$,
\begin{equation}
 \frac{1}{2\beta}\int_{-\beta}^{\beta}F(\delta\tau)\,d\delta\tau
 =\frac{1}{\beta}\int_0^\beta F(\delta\tau)\,d\delta\tau.
 \label{c3:num:two-cover}
\end{equation}
Normalized increments on $(-\beta,\beta)$ give the uniform
relative-time measure through this two-to-one covering. Including the
translated origin restores the physical time volume $\beta^n$, with
the matrix scaling stated above. Individual propagators retain their
antiperiodic signs.

This normalized periodic measure defines the target integral. The finite
time cutoff used in numerical evaluation changes the measure at nonzero
cutoff; the identity above describes the zero-cutoff target.

\subsection{Differentiating the complete approximation}
\label{c3:num:derivatives}

At the physical point, let $n_a=\langle N_a\rangle/N_s$,
$\kappa=\langle K\rangle/N_s$ and $e=\langle H\rangle/N_s$.
Then
\begin{align}
 n_a&=\frac{1}{\beta}\frac{\partial\psi}{\partial\mu_a},
 &\kappa&=-\frac{\partial\psi}{\partial\beta},
 &e&=\kappa+\sum_a\mu_a n_a,\nonumber\\
 D_{\rm pair}&=-\frac{1}{\beta}\frac{\partial\psi}{\partial U},
 &D_{\rm pair}&=\frac1{N_s}\sum_i\sum_{a<b}\langle n_{ia}n_{ib}\rangle .
 \label{c3:num:thermo}
\end{align}
Other physical parameters are fixed in each derivative.
Total density follows from a common shift of all $\mu_a$.
Unequal increments instead measure a different flavour-weighted
response.

At intermediate $\xi$, the interaction derivative at fixed auxiliary
shifts gives $\xi D_{\rm pair}(\xi)$; ordinary double occupancy is
recovered at $\xi=1$. When comparing energy results, the reported quantity must be identified as grand energy $\kappa$ or internal energy $e$.
The derivative includes the parameter dependence of the reference propagators, counterterms and free contribution. Evaluating nearby-parameter integrands
on common configurations can reduce finite-difference noise, but the
step size still requires a deterministic error check.

For a common chemical potential,
\begin{equation}
 \frac{\partial\kappa}{\partial\mu}
 =-n-\beta\frac{\partial n}{\partial\beta}.
 \label{c3:num:mixed}
\end{equation}
When density is nearly temperature independent, the slope of grand
energy with chemical potential is approximately $-n$.
This sign follows from $\kappa=-\partial_\beta\psi$ and explains the slope in Fig.~\ref{c3:results:honey}(d).

\subsection{Rational continuation and coefficient covariance}
\label{c3:num:pade-section}

A Pad\'e approximant $P_L(\xi)/Q_M(\xi)$, with $Q_M(0)=1$,
matches the available series through order $L+M$.
Writing $Q_M=1+\sum_{j=1}^M q_j\xi^j$ gives
\begin{equation}
 \sum_{j=1}^M q_j c_{k-j}=-c_k,\qquad k=L+1,\ldots,L+M .
 \label{c3:num:pade}
\end{equation}
Coefficients with negative indices vanish. The numerator consists of
the low-order coefficients of $Q_M\psi$. An ill-conditioned coefficient system amplifies input errors, and nearby pole-zero pairs may be numerical artefacts; a pole near the evaluation point makes continuation sensitive
to coefficient changes.

For a result $R(\mathbf c)$ and coefficient covariance $\Sigma$,
\begin{equation}
 \operatorname{Var}R\simeq
 \sum_{ij}\frac{\partial R}{\partial c_i}
 \Sigma_{ij}\frac{\partial R}{\partial c_j}.
 \label{c3:num:padeerror}
\end{equation}
Replacing $\Sigma$ by its diagonal is exact only for independent
coefficient errors. Shared normalization and all-subset measurements
correlate different orders. When the rational map is nonlinear over the
uncertainty range, resampling complete coefficient vectors captures
effects beyond the linearized expression above.

The bands in Figs.~\ref{c3:results:honey} and \ref{c3:results:sun}
combine coefficient uncertainties with the spread of selected
approximants. They quantify sensitivity to that selection. Their
interpretation as a continuation uncertainty also depends on stability
under changes of order, shift and approximant; the bands have not been
calibrated as confidence intervals or rigorous omitted-order bounds.
Finite-threshold and time-cutoff biases are excluded from the plotted
errors, and the full cross-order covariance is undetermined for these
continuation comparisons.

\section{Numerical results and independent benchmarks}
\label{c3:results:section}

\subsection{Figures and normalization conventions}

Figure~\ref{c3:results:honey} shows the two-spin honeycomb results,
and Fig.~\ref{c3:results:sun} shows the three-flavour square-lattice
results. An axis labelled $\ln Z$
denotes the intensive quantity $\psi=\ln Z/N_s$. For example,
Eq.~\eqref{c3:num:free} for the $32\times32$ square lattice at
$\beta t=3$, $\boldsymbol\mu/t=(0,0.5,1)$ and
$\boldsymbol\alpha/t=(1.2,1.5,1.8)$ gives $\psi_0=4.0334$, the zeroth-order point of the lower curve in Fig.~\ref{c3:results:sun}(b).
The extensive logarithm is $N_s\psi$, and the grand-potential density
is $\omega=-\psi/\beta$.

\subsection{Two-spin honeycomb illustration}

\begin{figure}[p]
 \centering
 \includegraphics[width=\textwidth]{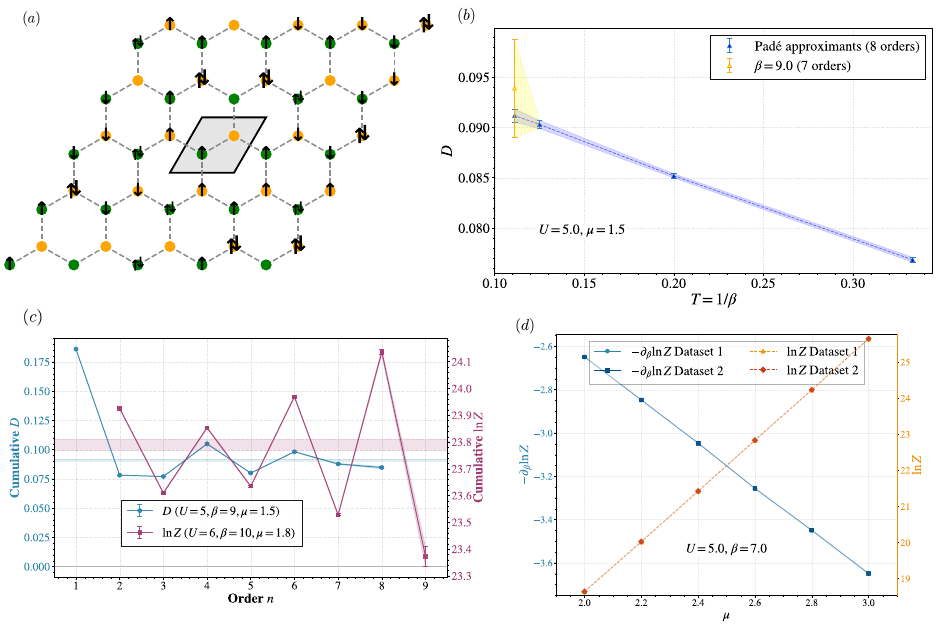}
 \caption[CoS++ honeycomb benchmark]{Thermodynamic results for the honeycomb lattice.
 (a) Two-sublattice geometry.
 (b) Double occupancy versus temperature at $U/t=5$, $\mu/t=1.5$,
 including the comparison of seventh- and eighth-order continuations at $\beta t=9$.
 (c) Cumulative double occupancy at $U/t=5$, $\beta t=9$, $\mu/t=1.5$,
 and cumulative partition logarithm $\psi$ at $U/t=6$,
 $\beta t=10$, $\mu/t=1.8$.
 (d) Logarithmic partition and grand-energy quantities for two shift
 choices at $U/t=5$, $\beta t=7$. Axes labelled $\ln Z$ denote $\psi=\ln Z/N_s$. The bands combine coefficient uncertainties with the spread of selected Pad\'e approximants. }
 \label{c3:results:honey}
\end{figure}

The honeycomb calculation uses the principal-minor and CDet recursions
of Section~\ref{c3:fpm:section} on $32\times32$ unit cells, or $2048$
sites. Its double occupancy is the probability of finding both spin
components on one site, so its temperature dependence measures the local
effect of repulsion. Panel (b) of Fig.~\ref{c3:results:honey} gives this
quantity at $U/t=5$, $\mu/t=1.5$ and $\beta t=3,5,8,9$. At
$\beta t=9$, adding the eighth order narrows the selected Pad\'e envelope
relative to the seventh-order result.

Panel (c) displays cumulative order dependence for double occupancy
and the partition logarithm $\psi$ at different state points.
The oscillating partial sums motivate rational continuation and the sensitivity analysis of Section~\ref{c3:num:pade-section}. The double-occupancy series starts at first order, consistent with the factor $\xi$ in Section~\ref{c3:num:derivatives}.

Panel (d) varies chemical potential at $U/t=5$, $\beta t=7$.
The two datasets use different auxiliary shifts. The resummed results agree within the displayed uncertainties, providing a check on the choice of reference. The decreasing grand-energy curve is
consistent with Eq.~\eqref{c3:num:mixed}: near a temperature-insensitive
density plateau its slope is negative.

\FloatBarrier

\subsection{Three flavours and non-uniform references}

\begin{figure}[p]
 \centering
 \includegraphics[width=\textwidth]{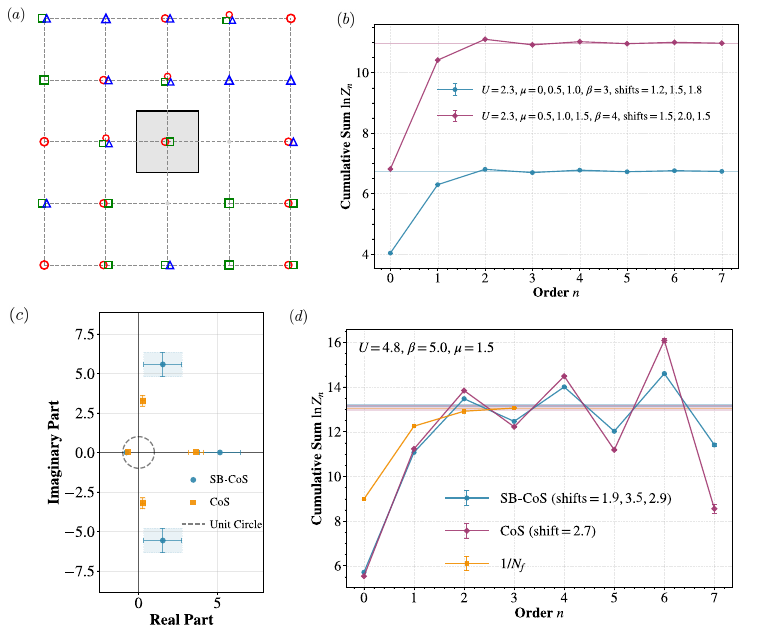}
 \caption[CoS++ three-flavour benchmark]{Three-flavour square-lattice results.
 (a) Flavour and lattice schematic.
 (b) Cumulative partition logarithms $\psi$ with non-uniform
 chemical potentials and shifts; both parameter sets are specified
 in the text.
 (c) Complex poles of selected Pad\'e approximants and the unit circle.
 (d) Uniform CoS and SB-CoS series at $U/t=4.8$, $\beta t=5$,
 $\mu/t=1.5$, together with the large-flavour comparison.
 The large-flavour markers are my cumulative saddle, Gaussian, first
 and second non-Gaussian approximations. Their horizontal order labels
 count large-flavour corrections; those of the CoS curves count
 interaction orders.
 For SB-CoS, $\boldsymbol\alpha/t=(1.9,3.5,2.9)$. Late partial sums change little in (b) and oscillate in (d).}
 \label{c3:results:sun}
\end{figure}

Figure~\ref{c3:results:sun} concerns three flavours on a
$32\times32$ square lattice. Panel (b) contains two physically
non-uniform examples. The lower curve has
$U/t=2.3$, $\boldsymbol\mu/t=(0,0.5,1)$,
$\beta t=3$ and $\boldsymbol\alpha/t=(1.2,1.5,1.8)$.
The upper curve has the same coupling,
$\boldsymbol\mu/t=(0.5,1,1.5)$, $\beta t=4$ and
$\boldsymbol\alpha/t=(1.5,2,1.5)$.

With flavour-resolved shifts, the late partial sums in panel (b)
change little on the scale of the figure. Panel (c) shows poles of the
selected rational approximants used in panel (d). Negative-real-axis
poles inside the unit circle indicate possible failure of direct Taylor
summation at $\xi=1$. Establishing an exact singularity or a physical
phase boundary requires information beyond a finite-order Pad\'e pole.

At $U/t=4.8$, $\beta t=5$ and $\mu/t=1.5$, panel (d) compares
uniform CoS and SB-CoS (typed-head) series with the large-flavour
expansion.
The uniform reference uses $\alpha/t=2.7$; the non-uniform shifts are
$\boldsymbol\alpha/t=(1.9,3.5,2.9)$.
The markers show partial sums of the computed series, and the bands show the selected rational continuations.

\FloatBarrier

\subsection{Connection to the preceding chapter}

The large-flavour curve uses the method of Chapter~\ref{c2:chapter}
\cite{ShiLargeNf2025}. The four markers report the cumulative saddle-point,
RPA/Gaussian, first non-Gaussian and second non-Gaussian approximations.
In Ref.~\cite{ShiCoS2025}, the last marker is the truncation through
$O(1/N_f^2)$;
its final increment is $0.1434(93)$ in $\psi$, giving $13.0571(95)$
for the cumulative result.
The final truncation is close to the CoS continuation, with a different
sequence of omitted orders.

Figure~\ref{c3:results:sun}(d) compares the same observable $\psi$
at one three-flavour state point. The large-flavour expansion organizes
terms around a collective saddle, while CoS sums coefficients of the
shifted interaction expansion. This is a direct comparison of the two
methods, unlike the six-flavour density application in
Chapter~\ref{c2:chapter}.

\subsection{Atomic-limit and sampling-measure checks}
\label{c3:results:interface-checks}

The low-order contraction checks test the diagram polynomial, but a
physical coefficient also requires its sign, coupling, factorial and
time measure. These factors can be checked together in the atomic
limit. For three flavours on one site, the reference and interaction
Hamiltonians commute and the trace contains only eight occupation
states. With reference probabilities $p_0(\mathbf n)$, define
\begin{equation}
 k_1(\mathbf n)=U\sum_{a<b}n_an_b-\sum_a\alpha_a n_a,
 \qquad
 \log\frac{Z(\xi)}{Z_0}
 =\log\sum_{\mathbf n}p_0(\mathbf n)
                       e^{-\beta\xi k_1(\mathbf n)}.
 \label{c3:results:atomic-cumulants}
\end{equation}
The eight-state trace gives the coefficient of $\xi^n$ directly as
$(-\beta)^n\kappa_n(k_1)/n!$, where $\kappa_n$ is the ordinary
cumulant of its probability distribution.
Using atomic propagators and the factor $(-\beta U/2)^n/n!$, I recover
this trace from the typed-head calculation through order four with the
physical singleton normalization of Eq.~\eqref{c3:sb:first-order}. This comparison tests the physical normalization
as well as the contraction polynomial. Endpoint enumeration also
reproduces the sieve's full-set and subset examples in
Section~\ref{c3:fast:examples}; a finite-state calculation checks the
envelope transition measure.

\subsection{Summary and numerical limitations}

The typed-head recurrence extends connected-cycle summation to
flavour-dependent propagators and counterterms. Its occupancy quotient
and DAG reductions share repeated arithmetic, and the subset readouts
make the resulting values usable in a many-configuration sampler.
The determinant-polynomial sieve in Section~\ref{c3:fast:section}
evaluates the same local insertion through shared coefficient extraction.
At order nine for three flavours it is about 24--25 times faster than
the direct CoS baselines in the fixed-configuration benchmark.

In the honeycomb example, eighth-order information narrows the selected
continuation range relative to order seven. The square-lattice example
compares interaction and large-flavour expansions at matched parameters.
Chapter~\ref{c4:chapter} uses connected-determinant evaluation in a replica
geometry, where the spatial cut changes the Gaussian boundary condition
and the propagator depends on two absolute imaginary times.

\chapter[Fermionic R\'enyi Entropy with Diagrammatic Monte Carlo]{Fermionic R\'enyi Entropy with Diagrammatic Monte Carlo}
\label{c4:chapter}

I use the diagrammatic methods developed in Chapters 2 and 3 to calculate
the R\'enyi entropy of a fermionic subsystem, an observable that is
nonlocal in space. The principal
change is already present in the Gaussian reference theory. Tracing out a
complement and multiplying reduced density matrices sew different copies
of the system along a chosen spatial region. The resulting free propagator
is a matrix on the replicated lattice and depends, in general, on two
absolute imaginary times.

This chapter gives a full account of my work in
Ref.~\cite{ShiRenyi2026}, from the fermionic trace to the numerical
estimator. I derive the replica path integral with its parity signs and
boundary conditions, then the stable free Green function used in the
calculation. Interpolation, connected-determinant summation and Monte
Carlo sampling yield the coefficients whose continuation gives the
entropy.

The numerical analysis begins with a \mbox{$3\times3$} cluster, where
independent deterministic integration checks the first three coefficients
and exact diagonalization gives the entropy at the physical coupling.
Monte Carlo extends this series through tenth order, providing a finite
system in which to examine continuation before applying the construction
to lattices of \mbox{$40\times40$} and \mbox{$50\times50$} sites.
The comparisons in Sections~\ref{c4:continue:branches} and
\ref{c4:bench:section} connect the coefficient precision to the final
entropy estimates; the larger lattices are analysed through their order
and reference-shift dependence.

\section{Observable, physical setting and expansion conventions}
\label{c4:target:section}

Tracing out the complement of a spatial region produces a reduced
density matrix whose mixedness reflects correlations across the cut.
Even a pure global state gives a mixed reduced state when the two regions
are entangled. The purity, equal to one for a pure state and smaller than
one for a mixed state, quantifies this effect; its negative logarithm is
the second R\'enyi entropy. At finite temperature the same observable
also includes thermal mixedness. To calculate it, I express the quadratic
function of the reduced density matrix as a trace over two replicas,
which admits a diagrammatic expansion.

\subsection{A reduced density matrix at finite temperature}
\label{c4:target:thermal}

Let $H_{\rm phys}$ exclude the chemical-potential term and define
\begin{equation}
 K=H_{\rm phys}-\mu\widehat N,\qquad
 Z_1=\operatorname{Tr}e^{-\beta K},\qquad
 \rho=\frac{e^{-\beta K}}{Z_1}.
 \label{c4:target:rho}
\end{equation}
The ordinary trace is taken on fermionic Fock space with a fixed mode
ordering. Choose lattice sites $A$ and their complement $B$.
The subsystem contains both spin modes at each selected site.
With the tensor identification associated with this ordering, define
$\rho_A=\operatorname{Tr}_B\rho$ and
\begin{equation}
 {\cal R}_n(A)=\operatorname{Tr}_A\rho_A^n,\qquad
 S_n(A)=\frac{1}{1-n}\ln{\cal R}_n(A),\qquad n>0,\quad n\ne1.
 \label{c4:target:renyi}
\end{equation}
The replica construction uses integer $n\ge2$; the Gaussian
correlation-matrix formula also allows general positive real $n$.
The interacting calculation uses $n=2$.

At zero temperature with a nondegenerate ground state, the total state
is pure and the subsystem entropy characterizes entanglement across the
bipartition. At finite temperature, it contains thermal mixedness as well
as quantum correlations. I calculate $\mathcal R_2(A)$ itself. The
combination $S_2(A)+S_2(B)-S_2(A\cup B)$ can remove an extensive thermal
contribution in suitable geometries \cite{Melko2010Mutual}, but is generally
not a mixed-state entanglement measure \cite{LuHsiehGrover2020}.
For von Neumann mutual information, Wolf et~al. proved a thermal area law
for finite-range Hamiltonians \cite{WolfMutual2008}.

For spinful fermions on $|A|$ sites, $d_A=4^{|A|}$.
Positivity and unit trace imply
\begin{equation}
 d_A^{-1}\le {\cal R}_2(A)\le1,\qquad
 0\le S_2(A)\le |A|\ln4 .
 \label{c4:target:bounds}
\end{equation}
For eigenvalues $p_j\ge0$ with $\sum_jp_j=1$, the lower purity bound
follows from $(\sum_jp_j)^2\le d_A\sum_jp_j^2$; the upper bound follows
from $p_j^2\le p_j$. These inequalities provide inexpensive endpoint checks.

\subsection{The unnormalized replica numerator}

Let $R_A=\operatorname{Tr}_B e^{-\beta K}$. Then
\begin{equation}
 \widetilde Z_n[A]=\operatorname{Tr}_A R_A^n,\qquad
 {\cal R}_n(A)=\frac{\widetilde Z_n[A]}{Z_1^n}.
 \label{c4:target:ratio}
\end{equation}
A fermionic sewing construction must reproduce this numerator exactly. For the second moment,
\begin{equation}
 S_2(A)=2\ln Z_1-\ln\widetilde Z_2[A],\qquad
 f_A=\frac{\ln{\cal R}_2(A)}{N_s}=-\frac{S_2(A)}{N_s}.
 \label{c4:target:density}
\end{equation}
Here $N_s$ is the number of physical sites. The numerical results in
Sections~\ref{c4:bench:section} and \ref{c4:large:section} are reported for
$f_A$. Multiplication by $N_s$ converts its uncertainty to that of the
total entropy.

If $A=\varnothing$, $R_A=Z_1$ is a scalar,
$\mathcal R_n=1$ and the entropy vanishes. For the full system,
\begin{equation}
 {\cal R}_n(A_{\rm full})
 =\frac{Z_1(n\beta)}{Z_1(\beta)^n}.
 \label{c4:target:full}
\end{equation}
At finite temperature this is generally less than one. A mixed state has
positive full-system entropy, which tends to zero in the nondegenerate
ground-state limit. At $\beta=0$, maximal mixing gives
$S_2(A)=|A|\ln4$.

\subsection{The finite Hubbard model and its shifted family}

The square-lattice Hubbard grand Hamiltonian is
\begin{equation}
 K=-t\sum_{\langle ij\rangle,\sigma}
 (c^\dagger_{i\sigma}c_{j\sigma}+c^\dagger_{j\sigma}c_{i\sigma})
 +U\sum_i n_{i\uparrow}n_{i\downarrow}
 -\mu\sum_{i,\sigma}n_{i\sigma}.
 \label{c4:target:hubbard}
\end{equation}
Each undirected nearest-neighbour bond is counted once.
Numerical tables use $t=1$ and $k_{\rm B}=1$.
There are $N_s=N_1N_2$ sites with periodic boundaries.
The homotopy family is
\begin{align}
 K(\xi)&=K_0+\xi V,\nonumber\\
 K_0&=K_{\rm kin}
 -\sum_{i,\sigma}(\mu-\alpha_\sigma)n_{i\sigma},\nonumber\\
 V&=U\sum_i n_{i\uparrow}n_{i\downarrow}
 -\sum_{i,\sigma}\alpha_\sigma n_{i\sigma}.
 \label{c4:target:shift}
\end{align}
At $\xi=1$ the shifts cancel. At $\xi=0$ the Gaussian reference has
effective chemical potential $\mu-\alpha_\sigma$.
All benchmark calculations use equal spin shifts.

Write the intensive expansion as
\begin{equation}
 f_A(\xi)=a_0+\sum_{k\ge1}a_k\xi^k,\qquad
 a_k=\left.\frac{1}{k!}\frac{\mathrm d^k f_A}{\mathrm d\xi^k}\right|_{\xi=0}.
 \label{c4:target:series}
\end{equation}
These Taylor coefficients depend on $U$ and the shifts; differentiation
in $\xi$ changes the interaction and its counterterms together. The
extensive coefficients $c_k$ of Ref.~\cite{ShiRenyi2026} satisfy
$a_k=c_k/N_s$. Their shift dependence can be strong even though the exact
endpoint is fixed, as examined in Sections~\ref{c4:bench:section} and
\ref{c4:large:section}.

\subsection{The replica boundary in diagrammatic expansions}

An entangling region inserts a spatial projector at the temporal seam.
Because this projector generally fails to commute with hopping, the
propagator depends on two absolute times and on the chosen region.
The local interaction retains the connected-determinant structure of
Chapter~3, with this modified Gaussian contraction.

A direct expansion of the replica numerator complements formulations
that reconstruct entropy from ordinary connected correlators
\cite{MoitraSensarma2023} or sample a replicated auxiliary-field ensemble
\cite{Grover2013,BroeckerTrebst2014}. Wang and Troyer sampled the
interaction correction to the free-fermion R\'enyi entropy using a
continuous-time interaction expansion, with updates between sewn and
ordinary ensembles \cite{WangTroyer2014}. Their updates between ensembles differ from the continuation of a finite
series in $\xi$ used here.

Sections~\ref{c4:swap:section}--\ref{c4:cdet:section} derive the trace,
Gaussian kernel and connected integrands.
Sections~\ref{c4:mc:section}--\ref{c4:continue:section} develop sampling,
coefficient covariance and continuation, followed by numerical tests in
Sections~\ref{c4:bench:section} and \ref{c4:large:section}.

\subsection{Notation}

Table~\ref{c4:target:notation} collects the notation used below.

\begin{table}[htbp]
\centering
\begin{tabular}{ll}
\toprule
Symbol&Meaning in this chapter\\
\midrule
$N_s=N_1N_2$&number of physical sites\\
$A,B$&subsystem and complement\\
$n$&replica number\\
$r,s$&replica labels\\
$\sigma$&physical spin\\
$k$&interaction-expansion order\\
$\xi$&homotopy parameter; physical point $\xi=1$\\
$\alpha_\sigma$&auxiliary energy shift\\
$p$&largest retained order\\
$\mathcal R_n$&normalized reduced-density-matrix moment\\
$f_A$&$\ln\mathcal R_2/N_s=-S_2/N_s$\\
\bottomrule
\end{tabular}
\caption[Replica, order and size conventions]{Notation used in this chapter.}
\label{c4:target:notation}
\end{table}

The numerical results are from my calculations in
Ref.~\cite{ShiRenyi2026}.

\subsection{Entropy observables and reference limits}
\label{c4:lit:context}

\paragraph{Boundary terms and subleading contributions.}

Subsystem entropy probes spatial correlations beyond the equation of
state, with a leading contribution that often scales with the region's
boundary. Srednicki's free-field vacuum calculation illustrates this
behaviour \cite{Srednicki1993Area}; the conditions supporting boundary
laws across different states and Hamiltonians are reviewed by Eisert,
Cramer and Plenio \cite{Eisert2010Area}. Smaller terms can distinguish
correlated phases \cite{Laflorencie2016Entanglement}. Extracting them
from the total entropy sets a quantitative accuracy target: its absolute
error must lie below the correction of interest.

Calabrese and Cardy used partition functions on sewn geometries to
derive conformal scaling forms for critical one-dimensional systems
\cite{CalabreseCardy2004,CalabreseCardy2009}. The lattice construction below
uses the same trace topology to compute the nonuniversal entropy of a
finite interacting two-dimensional system directly.

The geometry of the region determines which subleading contributions
can be resolved. For free fermions, a Fermi surface enhances the boundary
term by a logarithm: Wolf established violations of the strict area law,
and Gioev and Klich related the leading asymptotic coefficient to the
real-space boundary and Fermi surface
\cite{Wolf2006FermionArea,GioevKlich2006}. These results give Gaussian
references for a large-region area-plus-logarithm fit. Corners
\cite{CasiniHuerta2007Corners} and Goldstone modes in states with broken
continuous symmetry \cite{MetlitskiGrover2011} produce further logarithmic
terms. Topological contributions can be isolated through combinations of
region entropies that cancel local boundary terms, as in the
Kitaev--Preskill and Levin--Wen constructions
\cite{KitaevPreskill2006,LevinWen2006}. A set of absolute entropies at
several sizes and shapes thus provides the inputs for distinguishing
these geometric and phase-dependent contributions.

\paragraph{Gaussian limits and fluctuation observables.}

For Gaussian fermions, one-particle correlations determine the reduced
density matrix and its effective single-particle entanglement spectrum,
from which entropy formulas follow \cite{PeschelEisler2009}.
Grassmann representations extend this Gaussian algebra to states,
operators and maps \cite{Bravyi2005Gaussian}.

Purity selects one moment of the reduced density matrix, whereas the
full entanglement spectrum can distinguish additional features, as in
Li and Haldane's fractional-quantum-Hall analysis \cite{LiHaldane2008}.
For a free-fermion quantum point contact, the full counting statistics
of transmitted charge encodes the generated entanglement, establishing
a route from experimentally accessible fluctuations to entropy
\cite{KlichLevitov2009}. This connection was extended to bipartite
fluctuation scaling in one dimension by Song, Rachel and Le Hur
\cite{Song2010Fluctuations}, followed by a broader treatment of
particle-number cumulants, entropies and entanglement spectra for systems
that map to noninteracting fermions \cite{Song2012Bipartite}.
In an interacting system, joint measurements of number fluctuations and
purity can examine information that the number variance alone leaves
unresolved. R\'enyi mutual information combines region entropies in
a common thermal ensemble to remove a leading thermal volume term in
suitable settings (Section~\ref{c4:target:thermal}); Monte Carlo studies
have applied it to interacting spin systems \cite{Melko2010Mutual}.

\paragraph{Mixed-state entanglement.}

Vidal and Werner's negativity uses a partial transpose to detect quantum
correlations in mixed states \cite{VidalWerner2002}. For fermions, Eisler
and Zimbor\'as expressed the partial transpose of a Gaussian state as a
combination of Gaussian operators \cite{EislerZimboras2015}.
Shapourian, Shiozaki and Ryu introduced a partial time-reversal formulation
\cite{Shapourian2017TimeReversal}; the choice between twisted and
untwisted fermionic partial transposes affects their spectra
\cite{Shapourian2019Spectrum}.

A weighted Gaussian decomposition and incremental sampling also make
interacting-fermion R\'enyi negativity accessible in DQMC, as shown by
Wang and Xu \cite{WangXu2025Negativity}. A diagrammatic formulation of
this observable would use its partial-transpose or time-reversal
insertion to define the replica boundary kernel.

\subsection{Replica sampling and experimental measurements}
\label{c4:lit:replica-methods}

Swap estimators measure subsystem moments through replicated traces
\cite{Hastings2010,Melko2010Mutual}. Humeniuk and Roscilde developed a
quantum Monte Carlo scheme for spin and rotor models and studied the
crossover from low-temperature entanglement to thermal entropy
\cite{HumeniukRoscilde2012}. In auxiliary-field determinant methods, the replicated trace is sampled
through field configurations weighted by fermion determinants.

For interacting fermions, the auxiliary-field decomposition of the
reduced density matrix gives determinant estimators. Assaad, Lang and
Parisen Toldin extended them to higher R\'enyi orders and entanglement
spectra and introduced a stabilization procedure
\cite{AssaadLangToldin2014}. Assaad subsequently established the formal
equivalence of two determinant-QMC formulations with different sampling
strategies \cite{Assaad2015Stable}. The two formulations nevertheless differ in conditioning, ensemble
overlap and estimator distributions, which determine the practical cost.

Even in a sign-free model, a poorly sampled entropy estimator can have
large variance. Two approaches address this at the level of the
estimator or its sampling measure: relating entropy to participation
entropies \cite{Luitz2014Improving}, and incorporating entanglement
information into a hybrid-Monte-Carlo measure, as demonstrated for the
half-filled one-dimensional Hubbard model
\cite{DrutPorter2015Hybrid}.

Bauer compared factorization and inversion strategies for equal-time
and time-displaced DQMC Green functions and found an efficient stabilized
procedure based on pivoted QR decompositions \cite{BauerStable2020}.
The Gaussian replica kernel used here has a different matrix structure,
but faces a related problem: large intermediate spectral scales can
destroy accuracy before the final inverse is formed. The bounded
exponentials in Section~\ref{c4:stable:section} address this difficulty.

\paragraph{Incremental estimators.}

Intermediate ensembles reduce a difficult replica ratio to smaller
changes in the trace geometry. In the nonequilibrium-work approach,
D'Emidio expressed entropy as the free-energy difference associated with
a change of trace topology \cite{DEmidio2020Work}. Zhao et~al.\ combined
this sampling strategy with incremental changes to obtain universal
terms in two-dimensional spin and topological models
\cite{Zhao2022Increment}. The improvement comes from measuring ratios
between better-overlapping ensembles, with the accuracy of the work
distribution controlling the remaining statistical cost.

Srdin\v{s}ek, Casula and Vuilleumier optimized thermodynamic-integration
paths between uncoupled and sewn ensembles
\cite{Srdinsek2022Integration}. Optimizing that path changes the variance and quadrature cost together.
Incremental SWAP sampling has also resolved boundary, Goldstone and
corner contributions in antiferromagnetic Heisenberg models
\cite{Zhou2024Swap}.

The intermediate ensemble can include the entangling region itself.
D'Emidio, Or\'us, Laflorencie and de Juan sampled that region in
auxiliary-field QMC, resolving subleading ground-state contributions in
the Dirac, critical and antiferromagnetic regimes of the half-filled
honeycomb Hubbard model \cite{DEmidio2024Honeycomb}. Related approaches
address the exponential form of the measured observable directly:
Zhang et~al.\ developed an integral algorithm for free energies and
R\'enyi entropies within DQMC \cite{Zhang2024Integral}, and Liao's
power-incremental method reduced the coefficient of variation in direct
entropy estimation, with an application to the universal term of the
$\pi$-flux Hubbard model \cite{Liao2025PiFlux}.

Parameter paths can also connect different physical systems. The bipartite reweight-annealing method of Wang et~al.\ obtains entropy curves
and derivatives \cite{Wang2025Reweight}. Jiang et~al.\ applied
a fermionic parameter-scanning algorithm to Hubbard models, including
Gross--Neveu criticality \cite{Jiang2026Scanning}. Given one reference normalization, these methods obtain further points
from differences between neighbouring ensembles, at a cost set by the
overlap along the path and by temperature or projection convergence. Shen, Yin and Li
instead extracted universal corner information from early imaginary-time
relaxation near a quantum critical point \cite{Shen2026Imaginary}.

Sequences of overlapping ensembles and expansions about a Gaussian
reference provide complementary routes to the replica ratio. The former
control the measurement through a path in ensemble space; the calculation
here organizes it into connected coefficients of the replica-ratio
logarithm and continues their series to the physical coupling. Temperature
and coupling affect ensemble overlap, coefficient variance and the
required expansion order differently, motivating the cost discussion in
Section~\ref{c4:large:section}.

\paragraph{Measurements of subsystem purity.}

Two-copy interference gives an operational interpretation of the replica
moment. Daley et~al.\ proposed extracting R\'enyi entropies
from tunnelling between copies and occupation-parity measurements in
optical lattices \cite{Daley2012Entanglement}. Abanin and Demler proposed
a quantum switch controlling the connections between replicated
subsystems \cite{AbaninDemler2012}. 

The local thermalization and entanglement growth observed by Kaufman
et~al.\ illustrate the distinction between global and reduced states
\cite{Kaufman2016Thermalization}: subsystems approached thermal reduced
states even as the isolated full system remained nearly pure. An
equilibrium reference for the measured region should therefore be
compared at its local thermodynamic parameters and degree of
equilibration, with the global purity recorded as a distinct property
of the preparation.

Brydges et~al.\ measured subsystem R\'enyi entropies through
randomized measurements in a trapped-ion simulator
\cite{Brydges2019Randomized}. Elben et~al.\ developed a correlation-based framework for randomized measurements
and analysed its statistical errors
\cite{Elben2019Correlations}. Classical shadows permit reuse of measurement
records for many observables \cite{Huang2020Shadows}; their relation to the
wider randomized-measurement toolbox is reviewed in
Ref.~\cite{Elben2023Toolbox}. The sample cost of estimating a subsystem
purity depends on its size, the measurement ensemble and the noise model.

Numerical calculations can test normalization and finite-temperature
corrections where their parameters overlap with those of an experiment.
Such a comparison combines experimental uncertainties from finite
sampling, imperfect operations and state preparation with the numerical
sampling, representation and continuation errors assessed in the
small-cluster and large-lattice calculations below.

\section{Fermionic replica trace and sewing operator}
\label{c4:swap:section}

The replica method converts a nonlinear function of a reduced density
matrix into a linear trace on several copies of the system. This is the
starting point of swap estimators and of replica formulations of
interacting-fermion entropies
\cite{Hastings2010,Grover2013,BroeckerTrebst2014,MoitraSensarma2023}.
For fermions, permutations of ordinary tensor factors and of fermionic
modes can have different matrix elements. I relate both to the sewing
insertion inside the trace before deriving the boundary condition for
the perturbative expansion.

Consider a finite set of fermionic modes and a parity-even grand
Hamiltonian $K$. The thermal state is
$\rho=e^{-\beta K}/Z_1$, where $Z_1=\operatorname{Tr}e^{-\beta K}$.
The Hubbard Hamiltonian considered here also conserves particle number.
Parity conservation is sufficient for the trace identities in this
section; the number-conserving quadratic form will be introduced when
constructing the free propagator. The subsystem $A$ is a specified set of
modes, and $B$ contains the remaining modes. Spin is included among the mode labels, so a spatial subsystem contains
both spin components of each selected site.

\subsection{A fixed occupation basis and the ordinary replica identity}

I use a fixed mode ordering in which all modes in $A$ precede those in
$B$, with a fixed internal order in each region. Applying creation
operators in this sequence defines the occupation states $|a,b\rangle$,
where $a$ and $b$ specify the two sets of occupations. The associated
Hilbert-space identification
$\mathcal H\simeq\mathcal H_A\otimes\mathcal H_B$ determines the
partial trace:
\begin{equation}
 (\rho_A)_{aa'}=\sum_b\rho_{ab,a'b},\qquad
 \rho_{ab,a'b}=\langle a,b|\rho|a',b\rangle.
 \label{c4:swap:partial-trace}
\end{equation}
Fermionic operators on these modes still obey the canonical
anticommutation relations.

Let $p(a)$ and $p(b)$ denote the parities of the corresponding occupation
states. Since $\rho$ commutes with total parity, its nonzero matrix
elements satisfy
\begin{equation}
 \rho_{ab,a'b'}\ne0
 \quad\Longrightarrow\quad
 p(a)+p(b)=p(a')+p(b')\pmod 2.
 \label{c4:swap:even-selection}
\end{equation}
Setting $b'=b$ in this condition proves that $\rho_A$ is parity even.
Off-diagonal matrix elements within a parity sector remain allowed in
both the derivation below and the resulting trace identity.

Introduce $n$ replicas and let $r=1,\ldots,n$ denote the sheet label.
For a right cyclic permutation, define $a_0=a_n$. Direct multiplication
of reduced density matrices gives
\begin{equation}
 \mathcal R_n(A)=\operatorname{Tr}\rho_A^n
 =\sum_{\{a_r,b_r\}}\prod_{r=1}^n
       \rho_{a_r b_r,a_{r-1}b_r}.
 \label{c4:swap:component-cycle}
\end{equation}
Reversing the cycle gives the same trace. The ordinary tensor permutation
$(a_1,\ldots,a_n)\mapsto(a_n,a_1,\ldots,a_{n-1})$ reproduces this formula
with every $b_r$ fixed and with no sign. Its fermionic implementation
requires additional parity factors, derived next.

\subsection{Fermionic regrouping and the signs it introduces}

The replicated occupation basis initially has the interleaved order
\begin{equation}
 A_1,B_1,A_2,B_2,\ldots,A_n,B_n.
 \label{c4:swap:interleaved-order}
\end{equation}
It is convenient to group the $A$ modes first and the $B$ modes second.
There are two different maps implementing that change of labels.
The ordinary tensor regrouping $U_0$ introduces no sign. The fermionic
regrouping $U_F$, obtained by reordering creation operators, introduces a sign:
\begin{align}
 U_0|a_1,b_1;\ldots;a_n,b_n\rangle
 &=|a_1,\ldots,a_n;b_1,\ldots,b_n\rangle,\notag\\
 U_F|a_1,b_1;\ldots;a_n,b_n\rangle
 &=(-1)^{\chi(\mathbf a,\mathbf b)}
   |a_1,\ldots,a_n;b_1,\ldots,b_n\rangle,\notag\\
 \chi(\mathbf a,\mathbf b)
 &=\sum_{r<s}p(b_r)p(a_s)\pmod2.
 \label{c4:swap:regrouping}
\end{align}
The exponent counts crossings of occupied modes in $B_r$ with occupied
modes in later blocks $A_s$, each contributing a minus sign. The internal
mode order of each block is preserved.

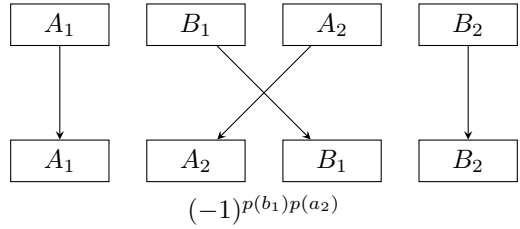
\begin{figure}[htbp]
\centering
\begin{tikzpicture}[font=\small,>=stealth]
 \node[draw,minimum width=13mm] (a1) at (0,0) {$A_1$};
 \node[draw,minimum width=13mm] (b1) at (1.8,0) {$B_1$};
 \node[draw,minimum width=13mm] (a2) at (3.6,0) {$A_2$};
 \node[draw,minimum width=13mm] (b2) at (5.4,0) {$B_2$};
 \node[draw,minimum width=13mm] (aa1) at (0,-1.8) {$A_1$};
 \node[draw,minimum width=13mm] (aa2) at (1.8,-1.8) {$A_2$};
 \node[draw,minimum width=13mm] (bb1) at (3.6,-1.8) {$B_1$};
 \node[draw,minimum width=13mm] (bb2) at (5.4,-1.8) {$B_2$};
 \draw[->] (a1)--(aa1);
 \draw[->] (b1)--(bb1);
 \draw[->] (a2)--(aa2);
 \draw[->] (b2)--(bb2);
 \node at (2.7,-2.45) {$(-1)^{p(b_1)p(a_2)}$};
\end{tikzpicture}
\caption[Fermionic regrouping of two replicas]{Regrouping two replicas in an occupation basis. Exchanging blocks $B_1$
and $A_2$ contributes $(-1)^{p(b_1)p(a_2)}$, which is negative when both
blocks have odd parity. The ordinary tensor relabelling $U_0$ carries
unit weight.}
\label{c4:swap:regroup-figure}
\end{figure}

The regrouping in Figure~\ref{c4:swap:regroup-figure} relates the
occupation-basis partial trace to the canonical fermionic mode order.
Its parity factor must accompany the mode transformation when the
replica trace is written with fermionic operators.

Independent replicas can be embedded in one fermionic Fock space with
the interleaved mode order. An odd operator in replica $r$ then contains
the parity string of the preceding replicas. An even operator contains
an even number of these strings, so they cancel. Consequently the
replicated Hamiltonian $K^{[n]}=\sum_rK_r$ represents independent even
copies, and its thermal operator is the ordinary tensor product of
their thermal operators. This factorization permits occupation-basis tensor products for the
thermal operator, with subsequent mode transformations carried out
using canonical fermionic operators.

\subsection{The graded cycle and parity compensation}

In the grouped basis, let $F_A$ be the second-quantized right cyclic
permutation of the $A$ mode blocks. Its action on homogeneous states is
\begin{equation}
 F_A|\mathbf a;\mathbf b\rangle
 =(-1)^{p(a_n)\sum_{r=1}^{n-1}p(a_r)}
   |a_n,a_1,\ldots,a_{n-1};\mathbf b\rangle.
 \label{c4:swap:bare-cycle}
\end{equation}
This is the Koszul sign obtained by moving the last occupied block
past all the preceding occupied blocks. It is required by fermionic
mode permutation, whereas the ordinary tensor permutation in Eq.~\eqref{c4:swap:component-cycle} is unsigned.

For one mode with $A$ equal to the full system and
$\rho=|1\rangle\langle1|$, the desired second moment is one.
The bare fermionic swap instead sends $|1,1\rangle$ to $-|1,1\rangle$
and gives a trace of minus one. A parity compensation is therefore
required in the ordinary Fock-space trace.

Let $P_{A,n}=(-1)^{\widehat N_{A,n}}$ be the parity operator of $A$ on
the last replica, and introduce the compensated cycle
\begin{equation}
 \mathcal T_A^{\rm raw}=F_A P_{A,n}^{\,n-1}.
 \label{c4:swap:compensated-cycle}
\end{equation}
The parity factor acts before the permutation. The compensated cycle acts as
\begin{equation}
 \mathcal T_A^{\rm raw}|\mathbf a;\mathbf b\rangle
 =(-1)^{\vartheta(\mathbf a)}
   |a_n,a_1,\ldots,a_{n-1};\mathbf b\rangle,\qquad
 \vartheta(\mathbf a)=
 p(a_n)\left[n-1+\sum_{r=1}^{n-1}p(a_r)\right]\pmod2.
 \label{c4:swap:compensated-sign}
\end{equation}
The extra parity factor is the identity for odd $n$ and multiplies the
odd-parity sector by $-1$ for even $n$. In both cases the complete replica contour on $A$ is antiperiodic,
as derived in Section~\ref{c4:path:sectors}.

\paragraph{Trace identity including the regrouping signs.}
Evaluate the trace of the compensated cycle after fermionic regrouping.
Inserting the occupation basis gives the products in
Eq.~\eqref{c4:swap:component-cycle}, multiplied by
\begin{equation}
 (-1)^{
 \vartheta(\mathbf a)+
 \chi(\mathbf a,\mathbf b)+
 \chi(\mathcal C\mathbf a,\mathbf b)},
 \qquad
 \mathcal C\mathbf a=(a_n,a_1,\ldots,a_{n-1}).
 \label{c4:swap:total-sign}
\end{equation}
Each nonzero factor in that product has the same $b_r$ on its two
sides. Equation~\eqref{c4:swap:even-selection} then forces
$p(a_r)=p(a_{r-1})$ for every $r$. All $A$ states in a surviving cycle
have a common parity $p$. Therefore
$\vartheta=p(n-1)(1+p)=0\pmod2$. Also, the list of $A$ parities is
unchanged by $\mathcal C$, so the two regrouping exponents are equal
and cancel modulo two. The total sign is one, term by term.

This proves
\begin{equation}
 \mathcal R_n(A)=
 \operatorname{Tr}\!\left[
 U_F^\dagger\mathcal T_A^{\rm raw}U_F\,\rho^{\otimes n}\right].
 \label{c4:swap:raw-trace-identity}
\end{equation}
The cancellation uses the $B$ labels and total parity conservation.
It applies to the thermal state considered here, including mixtures of
even and odd particle-number sectors.

\subsection{From the compensated cycle to a convenient Gaussian twist}

For a number-conserving one-particle transformation $T$, denote its
Fock-space lift by $\Gamma(T)$. It fixes the vacuum and acts on creation
operators according to
\begin{equation}
 \Gamma(T)c_j^\dagger\Gamma(T)^{-1}
 =\sum_i c_i^\dagger T_{ij},\qquad
 \Gamma(T_1)\Gamma(T_2)=\Gamma(T_1T_2).
 \label{c4:swap:second-quantization}
\end{equation}
For unitary $T$, this is a unitary mode rotation. On a fixed-particle
sector its matrix elements are minors of $T$. The minors already include the permutation signs of the occupied states.

For any one mode in $A$, the compensated right cycle has the replica
matrix $R$ with entries
\begin{equation}
 R_{r+1,r}=1\quad(1\leq r<n),\qquad
 R_{1,n}=(-1)^{n-1},\qquad
 R^n=(-1)^{n-1}I_n.
 \label{c4:swap:raw-one-particle}
\end{equation}
It is the identity on modes in $B$. For $n=1$ the cycle is simply the
identity. Introduce the replica phase matrix
$D=\operatorname{diag}(1,-1,\ldots,(-1)^{n-1})$, acting on every mode
of a replica, including its $B$ modes. Its Fock-space lift is a product
of whole-replica parity operators and commutes with $\rho^{\otimes n}$.
Conjugating the insertion by this transformation therefore leaves the
trace unchanged. The whole-replica action of $D$ is essential here: a phase transformation
restricted to $A$ generally fails to commute with hopping between $A$ and $B$.

Define the signed left shift $K_n$ by
\begin{equation}
 (K_nx)_r=x_{r+1}\quad(r<n),\qquad
 (K_nx)_n=-x_1,\qquad K_n^n=-I_n.
 \label{c4:swap:K}
\end{equation}
Direct multiplication gives $DRD^{-1}=-K_n^{-1}$. The phase-transformed one-particle sewing insertion and its inverse are
\begin{align}
 T&=P_B\otimes I_n-P_A\otimes K_n^{-1},\notag\\
 \Omega&=T^{-1}=P_B\otimes I_n-P_A\otimes K_n.
 \label{c4:swap:T-Omega}
\end{align}
Here $P_A$ and $P_B$ act on the physical one-particle space, and the
tensor order is physical mode followed by replica. Reversing the tensor-factor order only permutes matrix indices. The relation that defines the
unnormalized replica partition function is now
\begin{equation}
 \widetilde Z_n[A]
 =\operatorname{Tr}\!\left[\Gamma(T)e^{-\beta K^{[n]}}\right],
 \qquad
 \mathcal R_n(A)=\frac{\widetilde Z_n[A]}{Z_1^n}.
 \label{c4:swap:sewn-trace}
\end{equation}
The trace contains $\Gamma(T)=\Gamma(\Omega^{-1})$, and the column-field
boundary condition contains $-\Omega=-T^{-1}$. This inverse fixes the
phase orientation in Section~\ref{c4:path:sectors}.

\subsection{Elementary checks and the meaning of a spatial cut}

For a single physical mode with occupation probability $p$, the reduced
state is $\operatorname{diag}(1-p,p)$. The compensated trace gives
\begin{equation}
 \mathcal R_n=(1-p)^n+p^n.
 \label{c4:swap:single-mode}
\end{equation}
Only the all-empty and all-occupied replica occupations contribute to
the cyclic trace. Both have positive sewing weight. Vacuum and filled
pure states consequently have moment one, while a maximally mixed mode
has $\mathcal R_n=2^{1-n}$. 

If $A=\varnothing$, the insertion is the identity and
$\widetilde Z_n=Z_1^n$, so $\mathcal R_n=1$. If $A$ contains the complete
system, cyclic contraction gives
\begin{equation}
 \mathcal R_n(\text{whole system})=
 \frac{\operatorname{Tr}e^{-n\beta K}}
      {(\operatorname{Tr}e^{-\beta K})^n}.
 \label{c4:swap:whole-system}
\end{equation}
At finite temperature this is the R\'enyi moment of the full thermal
state, Eq.~\eqref{c4:target:full}. It tends to one for a nondegenerate
pure ground state. With degenerate ground states, the limiting moment depends on their statistical weights.

The spatial cut determines the sewing at the temporal boundary while
leaving the Hamiltonian, including hopping across the $A$--$B$ boundary,
unchanged on each sheet. Particles can therefore propagate between
regions with different temporal boundary conditions. This coupling is
the origin of the two-time dependence derived below.

\subsection{Parity selection and few-mode examples}

A two-mode example illustrates the parity assumption. Let one mode belong to $A$ and one to $B$,
and consider
\begin{equation}
 |\Phi\rangle=\frac{|1,0\rangle+|0,1\rangle}{\sqrt2},
 \qquad \rho=|\Phi\rangle\langle\Phi|,
 \qquad \rho_A=\frac12 I_A.
 \label{c4:swap:odd-state-example}
\end{equation}
The state vector has odd total parity. Its density operator nevertheless
commutes with total parity, because the same parity acts on its bra
and ket. The preceding proof therefore applies and gives
$\mathcal R_2(A)=1/2$. The condition is on the density operator: it must commute with total
parity. It admits odd-parity pure states and thermal mixtures of both
parity sectors, while excluding coherent superpositions across total
parity sectors.

The off-diagonal elements in Eq.~\eqref{c4:swap:odd-state-example}
change the occupations of $A$ and $B$ together, as hopping across the cut
can do. Within each factor of the cyclic trace, equal $B$ indices and
total parity conservation force the two $A$ parities to agree. Repeating
this argument around the cycle gives the common parity $p$ used above.

For two replicas, the regrouping exponents before and after the cycle
are $p(b_1)p(a_2)$ and $p(b_1)p(a_1)$. Their sum,
$p(b_1)[p(a_1)+p(a_2)]$, vanishes on every nonzero cyclic term by the
selection rule. Thus regrouping signs cancel inside this trace even
though they occur in the individual transformation matrices.

The compensated fermionic cycle and the ordinary tensor permutation
have equal traces against $\rho^{\otimes n}$ under the stated parity
assumption. This identity gives the replica partition function. Correlation functions
with additional odd insertions require their corresponding crossing signs.

\section{Coherent-state replica path integral}
\label{c4:path:section}

\subsection{Coherent states, measure and the fermionic trace}

For an ordered set of modes, introduce independent Grassmann variables
$\eta_j,\bar\eta_j$ and coherent states
\begin{equation}
 |\eta\rangle=e^{-\sum_j\eta_jc_j^\dagger}|0\rangle,\qquad
 \langle\bar\eta|=\langle0|e^{-\sum_jc_j\bar\eta_j},\qquad
 \langle\bar\eta|\eta'\rangle=e^{\bar\eta\eta'}.
 \label{c4:path:coherent-states}
\end{equation}
The coherent states satisfy $c_j|\eta\rangle=\eta_j|\eta\rangle$, with
$\eta_j$ and $\bar\eta_j$ integrated independently. The ordering of the
Grassmann differentials is fixed by
\begin{equation}
 I=\int d(\bar\eta,\eta)\,e^{-\bar\eta\eta}
       |\eta\rangle\langle\bar\eta|,\qquad
 \int d(\bar\eta,\eta)\,e^{-\bar\eta M\eta}=\det M.
 \label{c4:path:measure}
\end{equation}
These identities fix the integration convention and its overall sign.
Cycling states inside the fermionic trace gives the coherent-state
representation
\begin{equation}
 \operatorname{Tr}O
 =\int d(\bar\eta,\eta)\,e^{-\bar\eta\eta}
       \langle-\bar\eta|O|\eta\rangle.
 \label{c4:path:trace}
\end{equation}
The minus sign in $\langle-\bar\eta|$ accounts for Grassmann
anticommutation and distinguishes the ordinary trace from the
fermion-parity-weighted supertrace. For one mode, the overlap
$1+\bar\eta\eta'$ separates into the vacuum kernel one and the occupied
kernel $\bar\eta\eta'$. The latter becomes $-\bar\eta\eta$ in the trace,
so each occupation projector has trace one and their sum has trace two.
With the measure in Eq.~\eqref{c4:path:measure},
\begin{equation}
 \int d(\bar\eta,\eta)e^{-\bar\eta\eta}=1,\qquad
 \int d(\bar\eta,\eta)e^{-\bar\eta\eta}
                         (-\bar\eta\eta)=1.
 \label{c4:path:projector-traces}
\end{equation}
This elementary calculation fixes the overall sign independently of
replica conventions. For many modes, expand an occupation projector in
the prescribed mode order. Reordering its Grassmann variables gives
the determinant signs encoded by the same Gaussian integration rule.

The Gaussian sewing operator has the coherent-state kernel
\begin{equation}
 \langle\bar\eta|\Gamma(T)|\eta'\rangle
 =e^{\bar\eta T\eta'},\qquad
 \langle\bar\eta|\Gamma(T)=\langle\bar\eta T|.
 \label{c4:path:sewing-kernel}
\end{equation}
Combining this kernel with Eq.~\eqref{c4:path:trace} gives
\begin{equation}
 \widetilde Z_n[A]
 =\int d(\bar\eta,\eta)\,e^{-\bar\eta\eta}
   \langle-\bar\eta T|e^{-\beta K^{[n]}}|\eta\rangle.
 \label{c4:path:trace-with-twist}
\end{equation}
The matrix multiplying the final coherent bra fixes the inverse in the
column-field boundary condition, Eq.~\eqref{c4:path:column-boundary}.

\subsection{Finite time slicing before the continuum limit}

Divide the imaginary-time interval into $M$ slices of width
$\epsilon=\beta/M$. Insert the coherent-state identity between
successive factors of $e^{-\epsilon K^{[n]}}$. With slice variables
indexed from zero to $M-1$, the exact kernel expression is
\begin{align}
 \widetilde Z_n[A]
 =\int\prod_{\ell=0}^{M-1}d(\bar\eta_\ell,\eta_\ell)\,
 e^{-\sum_\ell\bar\eta_\ell\eta_\ell}
 &\langle-\bar\eta_0T|
              e^{-\epsilon K^{[n]}}|\eta_{M-1}\rangle\notag\\
 {}\times&
 \prod_{\ell=1}^{M-1}
 \langle\bar\eta_\ell|
              e^{-\epsilon K^{[n]}}|\eta_{\ell-1}\rangle.
 \label{c4:path:finite-slicing}
\end{align}
Equation~\eqref{c4:path:finite-slicing} uses exact short-time kernels.
Replacing them by normal-ordered symbols introduces the time-slice
approximation, whose limit is taken with the seam matrix $T$ retained
in the final link.

For a normal-ordered operator, replace creation operators by the
left coherent variable and annihilation operators by the right
one. At an interior link,
\begin{equation}
 \langle\bar\eta_\ell|
 e^{-\epsilon K^{[n]}}|\eta_{\ell-1}\rangle
 =e^{\bar\eta_\ell\eta_{\ell-1}}
 \left[1-\epsilon
 K^{[n]}(\bar\eta_\ell,\eta_{\ell-1})
 +O(\epsilon^2)\right].
 \label{c4:path:short-time}
\end{equation}
For a fixed finite-dimensional Hilbert space, this error vanishes as $M\to\infty$. The two Grassmann arguments lie on different sides
of the time link. That ordering specifies the contact prescription for density vertices:
each equal-time contraction takes the one-sided limit fixed by its time link.

The overlap factors and the factors from the resolution of the identity
combine into the discrete kinetic term. In the interior this is
$\bar\eta_\ell(\eta_\ell-\eta_{\ell-1})$. At the seam the final
overlap is $e^{-\bar\eta_0T\eta_{M-1}}$, so the column field is
closed by the identification
\begin{equation}
 \psi(0^+)=-T\psi(\beta^-),\qquad
 \psi(\beta^-)=-T^{-1}\psi(0^+)=-\Omega\psi(0^+).
 \label{c4:path:column-boundary}
\end{equation}
The row field closes with the inverse relation,
\begin{equation}
 \bar\psi(\beta^-)=-\bar\psi(0^+)T
                  =-\bar\psi(0^+)\Omega^{-1}.
 \label{c4:path:row-boundary}
\end{equation}
Together these closures cancel the boundary term obtained by integrating
$\bar\psi\partial_\tau\psi$ by parts. The column field carries $\Omega$
on the left, whereas the row field carries $\Omega^{-1}$ on the right.

The inverse in Eq.~\eqref{c4:path:column-boundary} can also be
derived directly from the finite-slice equations. In the
exact quadratic discretization, an interior column field evolves by the one-slice propagator
$B=e^{-\epsilon h}$ (Section~\ref{c4:path:normalization}). The seam row contains
$\eta_0+TB\eta_{M-1}$. A homogeneous vector in the null space of
the discrete kinetic operator therefore satisfies
$\eta_0=-TB^M\eta_0$. Equivalently, its final evolved value is
$B^M\eta_0=-T^{-1}\eta_0$. This yields the continuum column boundary for the ordered product
$TB^M$, including when $T$ and the evolution do not commute.

Taking the time-slice limit gives
\begin{equation}
 \widetilde Z_n[A]=
 \int_{\rm sewn}\mathcal D(\bar\psi,\psi)\,
 \exp\!\left\{-\sum_{r=1}^n\int_0^\beta d\tau\,
 \left[\sum_j\bar\psi_{jr}\partial_\tau\psi_{jr}
       +K_r(\bar\psi_r,\psi_r)\right]\right\}.
 \label{c4:path:continuum}
\end{equation}
The measure is the limit of the finite-slice measure, with the sewn
boundary conditions in Eqs.~\eqref{c4:path:column-boundary} and
\eqref{c4:path:row-boundary}. In a physical Hubbard copy, $K_r$ contains
hopping, chemical potential and onsite interaction. In the homotopy family it contains the chosen quadratic reference,
compensating one-body terms and interaction, with the same seam.

The coherent-state symbol of an onsite Hubbard term
on an interior link is
\begin{equation}
 U\,\bar\eta_{\ell,i\uparrow}\eta_{\ell-1,i\uparrow}
    \bar\eta_{\ell,i\downarrow}\eta_{\ell-1,i\downarrow}.
 \label{c4:path:Hubbard-symbol}
\end{equation}
The Grassmann-even density bilinears commute, and their two reorderings
to annihilator-first form give a positive overall prefactor multiplying
the spin-resolved Wick determinants.

At the seam, $T$ transforms the coherent variables in the short-time
symbol as well as in the overlap. In the continuum this transformation enters the boundary condition, so
the interaction vertices carry their sheet labels and the sewing is
contained in the free propagators.

\subsection{Absolute normalization from Gaussian time slices}
\label{c4:path:normalization}

A propagator equation leaves the multiplicative normalization of the
partition function undetermined. The finite-slice Gaussian integral
determines that normalization before the continuum limit. For one quadratic copy,
$K_0=c^\dagger h c$, let $B=e^{-\epsilon h}$; in replica space
$B$ acts identically on each sheet. The exact kernel is
\begin{equation}
 \langle\bar\eta|e^{-\epsilon K_0^{[n]}}|\eta'\rangle
 =e^{\bar\eta B\eta'}.
 \label{c4:path:exact-slice}
\end{equation}
The finite-slice action is then
\begin{equation}
 S_M=\sum_{\ell=0}^{M-1}\bar\eta_\ell\eta_\ell
 -\sum_{\ell=1}^{M-1}\bar\eta_\ell B\eta_{\ell-1}
 +\bar\eta_0TB\eta_{M-1}.
 \label{c4:path:discrete-Gaussian}
\end{equation}
For one slice, the first and last terms combine into
$\bar\eta_0(I+TB)\eta_0$. For several slices, eliminating the
interior variables or multiplying the exact coherent kernels gives
\begin{equation}
 \int\prod_\ell d(\bar\eta_\ell,\eta_\ell)e^{-S_M}
 =\det(I+TB^M)
 =\det(I+T e^{-\beta h^{[n]}}).
 \label{c4:path:finite-determinant}
\end{equation}
Here $h^{[n]}$ denotes the replicated one-particle Hamiltonian.
The seam kernel gives the ordered product $TB^M$. In the ordinary case $T=I$, the expression reduces
to the product of the familiar free-mode factors
$1+e^{-\beta\varepsilon}$.

In Fock space, $\operatorname{Tr}\Gamma(M)$ is the sum of the principal
minors of $M$ over all occupied subsets, equal to $\det(I+M)$
\cite{Klich2002}. Taking $M=T e^{-\beta h^{[n]}}$ gives
Eq.~\eqref{c4:path:finite-determinant} and the absolute Gaussian
normalization.

\subsection{Sources and the sign of the Gaussian contraction}

The relation between the finite-slice quadratic form and the Green
function can be established before invoking an equation of motion.
Let $\mathbb A_M$ denote the complete finite-slice kernel in
Eq.~\eqref{c4:path:discrete-Gaussian}, including its seam block.
Introduce independent Grassmann sources $J,\bar J$ in the same
ordered space. Shifting the integration variables gives
\begin{equation}
 \mathcal Z_M[\bar J,J]
 =\int d(\bar\eta,\eta)\,
 e^{-\bar\eta\mathbb A_M\eta+\bar J\eta+\bar\eta J}
 =\det\mathbb A_M\,
 e^{\bar J\mathbb A_M^{-1}J}.
 \label{c4:path:Gaussian-sources}
\end{equation}
The change of variables is a Grassmann translation and has unit
Jacobian. Equivalently, the identity follows by expanding the
finite polynomial in the sources. The sign follows by taking derivatives in a specified Grassmann order,
or by matching the coefficient of $\eta_i\bar\eta_j$. With that ordering,
\begin{equation}
 \langle\eta_i\bar\eta_j\rangle_M
 =(\mathbb A_M^{-1})_{ij},\qquad
 \langle\bar\eta_j\eta_i\rangle_M
 =-(\mathbb A_M^{-1})_{ij}.
 \label{c4:path:Gaussian-contraction}
\end{equation}
The normalized expectation divides by
$\det\mathbb A_M$. This discretization gives the free trace in
Eq.~\eqref{c4:path:finite-determinant} and the contraction sign in
Eq.~\eqref{c4:path:Gaussian-contraction}.

The annihilator-first convention gives a positive delta-function source
in the continuum Green-function equation and fixes the signs of the
contraction matrices used in the determinant vertices. In particular,
the lower equal-time ordinary contraction equals minus the occupation,
as expressed by the second relation in
Eq.~\eqref{c4:path:Gaussian-contraction}.

This lower-side prescription follows from the time-slice regularization.
The coherent density symbol places its creation variable at
the later side of a link and its annihilation variable at
the earlier side. After exchanging their order to match
Eq.~\eqref{c4:path:Gaussian-contraction}, the annihilator-first
kernel therefore has its second time infinitesimally later
than its first. This argument fixes the prescription at
ordinary and sewn vertices alike. 

With sources introduced before the continuum limit, the final overlap
closes the $M$-slice integral by relating the fields at zero and $\beta$
through the seam matrix. Vertices in the interior of the interval and
their one-sided contact limits then follow from the same discrete
construction.

\subsection{The sewn contour and its replica eigenmodes}
\label{c4:path:sectors}

\begin{figure}[htbp]
\centering
\begin{tikzpicture}[font=\small,>=stealth]
 \node at (-1.2,0) {$r=1$};
 \node at (-1.2,-1.6) {$r=2$};
 \node at (0,0.5) {$0$};
 \node at (5,0.5) {$\beta$};
 \draw[->,thick] (0,0)--(5,0);
 \draw[->,thick] (0,-1.6)--(5,-1.6);
 \fill (0,0) circle (1.6pt);
 \fill (5,0) circle (1.6pt);
 \fill (0,-1.6) circle (1.6pt);
 \fill (5,-1.6) circle (1.6pt);
 \draw[->,blue] (5,0) to[out=10,in=170] (0,-1.6);
 \draw[->,red] (5,-1.6) to[out=-20,in=200] (0,0);
 \node[blue,fill=white] at (2.5,-0.48) {$+1$};
 \node[red,fill=white] at (2.5,-1.13) {$-1$};
 \node at (2.5,-2.35) {$A:\ K_2^2=-I_2$};
\end{tikzpicture}
\caption[Two-sheet replica sewing]{Two-sheet sewing on $A$. Arrows along a sheet denote increasing
imaginary time. The seam joins the end of one sheet to the beginning of
the next, forming a cycle of length $2\beta$ with a minus sign on its
final return. On $B$, which is not drawn, each sheet closes
antiperiodically on itself.}
\label{c4:path:seam-figure}
\end{figure}
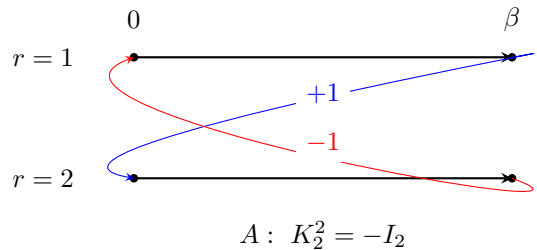

On $B$, Eq.~\eqref{c4:path:column-boundary} gives ordinary
antiperiodicity separately on every sheet. On $A$ it gives
\begin{align}
 \psi_{A,r}(\beta^-)&=\psi_{A,r+1}(0^+)
                                  &&(r<n),\notag\\
 \psi_{A,n}(\beta^-)&=-\psi_{A,1}(0^+),&
 \psi_{B,r}(\beta^-)&=-\psi_{B,r}(0^+).
 \label{c4:path:mixed-boundaries}
\end{align}
The $A$ contour closes antiperiodically after length $n\beta$
(Figure~\ref{c4:path:seam-figure}). Replica phase transformations can move
the minus sign to another junction while preserving the product of phases
around the cycle. 

Tracing over $B$ closes its indices within each copy. Multiplying the
reduced density matrices then joins the open $A$ indices between copies,
as in Figure~\ref{c4:path:seam-figure}. The free boundary condition
therefore depends on the selected region. Diagonalizing the replica
cycle replaces the connections between sheets by phases, reducing the
free calculation to matrices on the physical lattice.

The eigenvalues of $K_n$ are roots of minus one:
\begin{equation}
 \lambda_m=e^{iq_m},\qquad
 q_m=\frac{(2m+1)\pi}{n},\qquad m=0,\ldots,n-1.
 \label{c4:path:replica-phases}
\end{equation}
One convenient unitary eigenvector matrix is
\begin{equation}
 F_{rm}=\frac1{\sqrt n}e^{iq_m(r-1)},\qquad
 F^\dagger F=I_n,\qquad K_nF=F\operatorname{diag}(e^{iq_m}).
 \label{c4:path:Fourier}
\end{equation}
The sector fields are $\psi_m=\sum_rF_{rm}^*\psi_r$ and
$\bar\psi_m=\sum_r\bar\psi_rF_{rm}$.
The row transformation is the inverse of the column
transformation, as required for the kinetic bilinear.
Each quadratic sector obeys
\begin{equation}
 \psi_m(\beta^-)=-\Omega_m\psi_m(0^+),\qquad
 \Omega_m=P_B-e^{iq_m}P_A,\qquad
 T_m=\Omega_m^{-1}=P_B-e^{-iq_m}P_A.
 \label{c4:path:sector-boundary}
\end{equation}
The common one-particle Hamiltonian makes the quadratic action diagonal
in $m$.

The free trace therefore factorizes as
\begin{equation}
 \widetilde Z_{0,n}[A]
 =\prod_{\sigma}\prod_{m=0}^{n-1}
   \det\!\left(I+\Omega_m^{-1}e^{-\beta h_\sigma}\right).
 \label{c4:path:sector-partition}
\end{equation}
Equation~\eqref{c4:path:sector-partition} holds for a spin-diagonal
quadratic reference. A local interaction on a fixed sheet generally
couples the Fourier sectors. Individual sector determinants can be complex, while the
normalized moment is real and positive. Combining their logarithms
requires consistent branches.

This product fixes the free path-integral normalization.
Section~\ref{c4:free:section} relates it to the restricted correlation
matrix and uses it to calculate the zeroth-order term.

\subsection{A one-mode cancellation visible before perturbation theory}

For one mode wholly in $A$, let $z=e^{-\beta\varepsilon}$.
At $n=2$, the insertion $T$ of Eq.~\eqref{c4:swap:T-Omega} has the matrix
\begin{equation}
 T=K_2=\begin{pmatrix}0&1\\-1&0\end{pmatrix},\qquad
 \det(I+zT)=1+z^2.
 \label{c4:path:one-mode-determinant}
\end{equation}
In the Fock trace, the empty state contributes one and the
doubly occupied replica state contributes $z^2$. The two
singly occupied states are exchanged and have zero diagonal
matrix elements. In the Grassmann determinant the same result
comes from the negative sign of an off-diagonal determinant
permutation combined with the negative seam matrix element.
Their product is positive.

The ordinary denominator is $(1+z)^2$, giving the result in
Eq.~\eqref{c4:swap:single-mode} with $p=z/(1+z)$.
Replacing the compensated Gaussian insertion by the unsigned
one-particle swap would instead give $1-z^2$. At $z=1$ that
expression vanishes, although the correct normalized moment
is $1/2$. The discrepancy arises because the Fock-space lift of the unsigned
one-particle swap differs from the ordinary tensor swap by the
occupied-state sign in Eq.~\eqref{c4:swap:bare-cycle}.

\section{Exact two-time replica Green function}
\label{c4:green:section}

\subsection{The inverse boundary-value problem}

Consider a spin-diagonal quadratic reference with finite
Hermitian one-particle matrix $h_\sigma$. I omit $\sigma$ in this derivation. The convention throughout is
\begin{equation}
 \bar G_{rr'}(i,\tau;j,\tau')
 =\langle\mathcal T_\tau
       c_{ir}(\tau)c_{jr'}^\dagger(\tau')\rangle_{0,\rm sewn}.
 \label{c4:green:definition}
\end{equation}
This annihilator-first convention matches the determinant contractions above. The sewn expectation is normalized by its own free sewn partition function.

In a sector, the inverse quadratic operator satisfies
\begin{equation}
 (\partial_\tau+h)\bar G_m(\tau,\tau')
 =\delta(\tau-\tau')I,\qquad
 \bar G_m(\beta^-,\tau')=-\Omega_m\bar G_m(0^+,\tau').
 \label{c4:green:boundary-problem}
\end{equation}
The source is a delta function in the physical mode space.
Across the insertion time it gives
\begin{equation}
 \bar G_m(\tau'^+,\tau')-\bar G_m(\tau'^-,\tau')=I.
 \label{c4:green:jump}
\end{equation}
The jump is local in time and independent of $A$. The chosen region
enters through the homogeneous solution and the boundary matching.

For fixed $\tau'$, away from the source write
$\bar G_m(\tau,\tau')=e^{-\tau h}C_\pm(\tau')$ on the two
time-ordering domains. The boundary and jump conditions are
\begin{equation}
 e^{-\beta h}C_+=-\Omega_m C_-,\qquad
 C_+-C_-=e^{\tau'h}.
 \label{c4:green:two-constants}
\end{equation}
Define $E=e^{-\beta h}$ and
\begin{equation}
 B_m=I+\Omega_m^{-1}E,\qquad A_m=B_m^{-1}.
 \label{c4:green:A}
\end{equation}
Multiplying the first relation in
Eq.~\eqref{c4:green:two-constants} by $\Omega_m^{-1}$
on the left gives $C_-=-\Omega_m^{-1}EC_+$.
Substitution into the second relation then determines both
constants. The result is
\begin{equation}
 \bar G_m(\tau,\tau')=
 \begin{cases}
 e^{-\tau h}A_m e^{\tau'h},&\tau>\tau',\\
 e^{-\tau h}(A_m-I)e^{\tau'h},&\tau<\tau'.
 \end{cases}
 \label{c4:green:exact}
\end{equation}
For a cut with $[h,P_A]\ne0$, the middle factor generally commutes with
neither exponential, so both absolute times are needed. The two branches
and their unit jump are shown in Figure~\ref{c4:green:time-square}.

\subsection{The second argument and the contact prescription}

The boundary condition of the creation-field argument is
\begin{equation}
 \bar G_m(\tau,\beta^-)
 =-\bar G_m(\tau,0^+)\Omega_m^{-1}.
 \label{c4:green:right-boundary}
\end{equation}
It follows directly from the row-field closure, and it
can be checked algebraically. The identity
$A_m(I+\Omega_m^{-1}E)=I$ gives
$A_m-I=-A_m\Omega_m^{-1}E$. Multiplying on the right
by $E^{-1}$ reproduces
Eq.~\eqref{c4:green:right-boundary} from the lower and
upper branches in Eq.~\eqref{c4:green:exact}.
The inverse $\Omega_m^{-1}$ acts from the right.
For the present unitary phases,
$\Omega_m^{-1}=\Omega_m^\dagger$.

Normal-ordered density insertions require the second
argument to be infinitesimally later than the first:
\begin{equation}
 \bar G_m(\tau,\tau^+)
 =e^{-\tau h}(A_m-I)e^{\tau h},\qquad
 \bar G_m(\tau^+,\tau)
 =e^{-\tau h}A_m e^{\tau h}.
 \label{c4:green:contact}
\end{equation}
The upper value minus the lower value is $I$. In replica space,
the negative of the diagonal lower-limit contraction
is the sewn density, which generally depends on imaginary time. The
diagonal of a single complex sector is an algebraic component of this
trace; an occupation interpretation requires the reconstructed replica
quantity.

The interpolation tables in Section~\ref{c4:tables:section} keep the
one-sided branch label at coincident times. Density vertices use the lower value. The counterterm shifts are added
to the diagonals of the determinant matrices in
Eq.~\eqref{c4:free:shiftmatrix}.

\subsection{Replica reconstruction and sector conventions}
\label{c4:green:conventions}

With the Fourier matrix in Eq.~\eqref{c4:path:Fourier},
the replica-space propagator is
\begin{equation}
 \bar G_{rr'}(\tau,\tau')
 =\frac1n\sum_{m=0}^{n-1}
 e^{iq_m(r-r')}\bar G_m(\tau,\tau').
 \label{c4:green:replica-reconstruction}
\end{equation}
For two replicas, I label the conjugate phases by
$q_-=-\pi/2$ and $q_+=\pi/2$. They are the same roots of minus one as
in Eq.~\eqref{c4:path:replica-phases}. With columns ordered as $(-,+)$
and replica rows labelled $r=1,2$, Eq.~\eqref{c4:path:Fourier} becomes
\begin{equation}
 F_2=\frac1{\sqrt2}\begin{pmatrix}1&1\\-i&i\end{pmatrix},
 \qquad \Omega_\pm^{-1}=P_B\pm iP_A.
 \label{c4:green:two-sector-basis}
\end{equation}
Here $\bar G_\pm$ denotes the solution of
Eq.~\eqref{c4:green:exact} at $q=q_\pm$; time ordering continues to be
denoted by the superscripts $>$ and $<$. This phase labelling is used
throughout the two-replica calculation, giving
\begin{align}
 \bar G_{11}=\bar G_{22}
 &=\tfrac12(\bar G_-+\bar G_+),\notag\\
 \bar G_{12}&=\tfrac i2(\bar G_--\bar G_+),&
 \bar G_{21}&=-\tfrac i2(\bar G_--\bar G_+).
 \label{c4:green:n2}
\end{align}
For real symmetric $h$, $\bar G_+=\bar G_-^*$ at the same times and
on the same branch. Writing $\bar G_-=X+iY$ therefore gives equal
real diagonal blocks $X$ and opposite off-diagonal blocks $-Y$ and $Y$.
For complex Hermitian hopping, Eq.~\eqref{c4:green:n2} still reconstructs
the kernel from both sectors, which need not be entrywise conjugates.

The numerical calculations use real symmetric hopping. The scaled
sector solutions in Section~\ref{c4:stable:section} are combined using
this same $F_2$, and any residual imaginary part in the reconstructed
blocks measures numerical inconsistency between the conjugate sectors.

\subsection{Existence for the physical Hermitian reference}

The inverse in Eq.~\eqref{c4:green:A} exists
throughout the stated finite-temperature physical
domain. 
For finite Hermitian $h$ and real $\beta>0$,
$E$ is positive definite. If $B_mv=0$, then
\begin{equation}
 (E+\Omega_m)v=0,\qquad
 v^\dagger Ev+\|P_Bv\|^2
             -e^{iq_m}\|P_Av\|^2=0.
 \label{c4:green:null-vector}
\end{equation}
When $\sin q_m\ne0$, the imaginary part forces
$P_Av=0$. The real part then contradicts
positive definiteness unless $v=0$.
The remaining allowed case is $q_m=\pi$,
which occurs for odd $n$. In that case
$\Omega_m=I$, and $E+I$ is positive definite.
There is no allowed sector with $q_m=0$ modulo
$2\pi$. Therefore every physical sector is
nonsingular.

The kernel exists at finite physical parameters, but its numerical
accuracy depends on the condition number. The scaled solves in
Section~\ref{c4:stable:section} avoid large intermediate exponentials;
I estimate the reciprocal condition numbers of these scaled systems.

Outside this physical domain, other twists or complex parameter
continuations can make $B_m$ singular. This occurs precisely when the
homogeneous sewn boundary problem has a nonzero solution
$e^{-\tau h}v$.

\subsection{Uniqueness, thermal limits and the generic two-time geometry}

\subsubsection{Uniqueness and the equation in the second argument}

Suppose two matrix kernels satisfy the same left equation,
boundary condition and unit jump. Their difference has no
delta-function source and no jump. For fixed $\tau'$, this difference is $e^{-\tau h}C$, with one constant
matrix throughout the interval. The boundary condition
implies $(E+\Omega_m)C=0$. The nonsingularity proved above
forces $C=0$. Thus Eq.~\eqref{c4:green:exact} is the unique
kernel for the specified domain and boundary conditions.

The same kernel satisfies the equation in its second
argument,
\begin{equation}
 -\partial_{\tau'}\bar G_m(\tau,\tau')
       +\bar G_m(\tau,\tau')h
 =\delta(\tau-\tau')I.
 \label{c4:green:right-equation}
\end{equation}
Away from coincidence, differentiation of the right
exponential cancels the matrix product with $h$.
As the second argument increases through the first,
the kernel changes from the upper to the lower
branch and jumps by $-I$. The minus sign on the derivative gives the required positive source,
confirming the time ordering and the right action of $h$.

\subsubsection{Relative-time limits and the unfolded full-system contour}

For an empty region, $\Omega_m=I$ in every sector.
The sector kernels are identical, and Fourier
orthogonality gives
\begin{equation}
 \bar G_{rr'}(\tau,\tau')
 =\delta_{rr'}\,\bar G_\beta^{\rm th}(\tau-\tau').
 \label{c4:green:empty}
\end{equation}
There is no propagation between different replicas.
For any positive contour length $L$, the ordinary
kernel in the present sign convention is
\begin{equation}
 \bar G_L^{\rm th}(\delta)=
 \begin{cases}
 e^{-\delta h}(I+e^{-Lh})^{-1},&0<\delta<L,\\
 -e^{-(\delta+L)h}(I+e^{-Lh})^{-1},&-L<\delta<0.
 \end{cases}
 \label{c4:green:thermal-kernel}
\end{equation}
The equal-time upper value is one minus the
Fermi occupation matrix, while the lower value
is minus that occupation. These relations
fix the ordinary limit of both the replica
solver and any interpolation table built
from it.

When all physical modes belong to $A$, the
sheets can instead be unfolded into a single
interval of length $n\beta$. Set
$s=\tau+(r-1)\beta$ and
$s'=\tau'+(r'-1)\beta$. The sewing conditions
make the field continuous at the intermediate
junctions and antiperiodic at the final return.
Consequently,
\begin{equation}
 \bar G_{rr'}(\tau,\tau')
 =\bar G_{n\beta}^{\rm th}(s-s').
 \label{c4:green:full-unfolded}
\end{equation}
This can be proved without performing a
trigonometric sum: both sides obey the same
equation, source and sewn boundary conditions,
so uniqueness establishes equality.
For different replicas the equal-local-time
kernel has no unit jump, because the source
is proportional to $\delta_{rr'}$.

For a proper subregion coupled to its complement, hopping connects
$A$ modes on a contour of length $n\beta$ to $B$ modes on separate
$\beta$ circles. These connections join equal local times on each sheet, so the full
Hamiltonian cannot be placed on one unfolded contour.

\subsubsection{The generic cut}

Within either time-ordering domain, simultaneous
translation of the two arguments gives
\begin{equation}
 (\partial_\tau+\partial_{\tau'})
 \bar G_m(\tau,\tau')
 =e^{-\tau h}[A_m,h]e^{\tau'h}.
 \label{c4:green:time-translation}
\end{equation}
The identity part of the lower branch commutes
with $h$, so the same expression applies on
both domains. For a generic cut crossed by
hopping, $[h,P_A]\ne0$, and the middle matrix
does not commute with $h$. The kernel therefore changes under a common shift of both observation times.

If $A$ is empty or contains the whole system,
the twist is scalar within a sector and
commutes with $h$. A Hamiltonian that does
not couple $A$ and $B$ also commutes with
their projectors. Relative-time formulas hold in these limits and provide analytical
checks. For a generic cut, the kernel can lose invariance under both common
time shifts and spatial translations, even when the hopping matrix
is translation invariant.

\begin{figure}[htbp]
\centering
\begin{tikzpicture}[font=\small,>=stealth,scale=0.85]
 \draw[->] (0,0)--(5.4,0) node[right] {$\tau'$};
 \draw[->] (0,0)--(0,5.4) node[above] {$\tau$};
 \draw (0,5)--(5,5)--(5,0);
 \draw[dashed] (0,0)--(5,5);
 \node at (1.5,3.7) {$\tau>\tau'$};
 \node at (3.7,1.5) {$\tau<\tau'$};
 \node[left] at (0,5) {$\beta$};
 \node[below] at (5,0) {$\beta$};
 \node[below left] at (0,0) {$0$};
 \draw[->,thick] (2.7,2.3)--(2.3,2.7);
 \node[fill=white] at (3.7,3.0) {jump $I$};
\end{tikzpicture}
\caption[The two time-ordering domains]{The two time-ordering domains of the exact replica kernel. The diagonal
carries a unit jump, and each smooth branch is interpolated on its own
triangle. When the twist and $h$ do not commute, the kernel depends on
$\tau$ and $\tau'$ separately within each domain.}
\label{c4:green:time-square}
\end{figure}
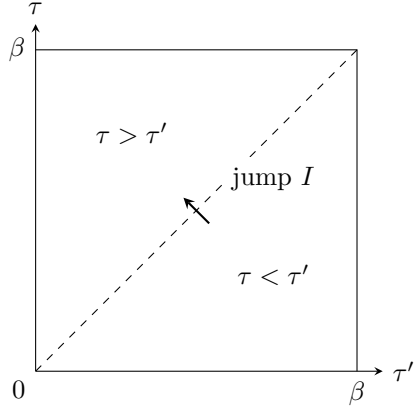

Moving parallel to the diagonal in Figure~\ref{c4:green:time-square}
keeps the time difference fixed while moving both observations relative
to the seam. I therefore tabulate both time coordinates within each
branch and store the two coincident-time limits separately.

\subsection{A two-mode cut and checks of the exact solution}

Consider one spin species on two modes with
$A=\{1\}$, $B=\{2\}$, zero chemical potential,
and hopping matrix
\begin{equation}
 h=\begin{pmatrix}0&-t\\-t&0\end{pmatrix},\qquad
 E=\begin{pmatrix}\cosh x&\sinh x\\
                   \sinh x&\cosh x\end{pmatrix},
 \qquad x=\beta t.
 \label{c4:green:two-mode}
\end{equation}
For nonzero hopping, the spatial projector and $h$ do not commute.
The $q_+$ sector has $\Omega_+^{-1}=\operatorname{diag}(i,1)$.
Writing $c=\cosh x$ and $s=\sinh x$ gives
\begin{equation}
 B_+=\begin{pmatrix}1+ic&is\\s&1+c\end{pmatrix},
 \qquad
 \det B_+=(1+c)(1+i),\qquad B_-=B_+^*.
 \label{c4:green:two-mode-sector}
\end{equation}
The normalized free replica moment is
\begin{equation}
 \mathcal R_{2,0}(A)=
 \frac{\det B_+\det B_-}{\det(I+E)^2}
 =\frac{2(1+c)^2}{4(1+c)^2}=\frac12.
 \label{c4:green:two-mode-ratio}
\end{equation}
The restricted correlation matrix, with $C_{11}=1/2$, gives the same
result. The subsystem entropy of this one-spin example is $\ln2$ at
every temperature, from a hot mixed state to the low-temperature
delocalized one-particle state. This temperature-independent value
checks the sewing normalization.

For this two-mode example, substitution of $A_\pm=B_\pm^{-1}$ into
Eq.~\eqref{c4:green:exact} verifies both boundary conditions and the
unit jump by matrix multiplication. Since $[A_\pm,h]\ne0$, equal time
differences can give different kernel matrices at different absolute
times.

Direct few-mode calculations test these trace and kernel identities for
complex coherences within parity sectors and for complex Hermitian
hopping. Together with the empty- and full-region limits, they check
normalization and both boundary conditions before stochastic integration.

\section{Stable evaluation of the replica kernel}
\label{c4:stable:section}

I now reorganize the exact replica kernel into bounded exponential
factors and linear systems. Treating the two time branches separately
allows their large spectral scales to cancel algebraically before
floating-point evaluation.

\subsection{The numerical problem created by a sewn boundary}

The direct expression in Eq.~\eqref{c4:green:exact} contains large
factors that cancel: occupied one-particle modes generate large
eigenvalues of $e^{-\beta h}$, while the sector inverse compensates for
them. Computing these factors separately can overflow or lose significant
digits. Combining their scales before numerical evaluation avoids these large
intermediate factors.

As above, the spin index is omitted. The matrix $h$ is the grand
one-body Hamiltonian of the shifted reference, including the chemical
potential and the energy-valued counterterm. It is real symmetric for the square-lattice calculations. There are $N_s$ physical sites;
replica labels do not enlarge this matrix after sector diagonalization.
The discussion assumes $\beta>0$ and $0\leq\tau,\tau'\leq\beta$.
The two open triangles of this time square are treated separately,
and their limits define the two values at coincident times.

I use the Green-function convention of Eq.~\eqref{c4:green:boundary-problem},
\begin{equation}
 \begin{gathered}
 (\partial_\tau+h)\bar G_m(\tau,\tau')=\delta(\tau-\tau')I,\\
 \bar G_m(\beta^-,\tau')=-\Omega_m\bar G_m(0^+,\tau'),\\
 \Omega_m=P_B-e^{iq_m}P_A,\qquad q_m=\frac{(2m+1)\pi}{n}.
 \end{gathered}
 \label{c4:stable:boundary}
\end{equation}

Away from the source time, left multiplication by $e^{\tau h}$ removes
the differential operator. Write the upper solution as
$e^{-\tau h}A_m e^{\tau'h}$. Integrating the equation across the source
fixes the lower solution, and the boundary condition fixes its constant:
\begin{align}
 \bar G_m^>(\tau,\tau')&=e^{-\tau h}A_m e^{\tau'h},&
 \bar G_m^<(\tau,\tau')&=e^{-\tau h}(A_m-I)e^{\tau'h},
 \nonumber\\
 e^{-\beta h}A_m&=-\Omega_m(A_m-I),&
 A_m&=(I+\Omega_m^{-1}e^{-\beta h})^{-1}.
 \label{c4:stable:direct}
\end{align}
The superscripts specify the branch and its one-sided boundary limits. Its two-time dependence follows from Eq.~\eqref{c4:green:time-translation}.

\subsection{Algebra in the energy basis}

The eigendecomposition of the real symmetric reference transforms the twist to the energy basis:
\begin{equation}
 h=V\Lambda V^T,\qquad V^TV=I,\qquad
 \Lambda_{ab}=\epsilon_a\delta_{ab},\qquad
 S_m=V^T\Omega_m^{-1}V .
 \label{c4:stable:eigenbasis}
\end{equation}
The matrix $S_m$ is unitary because $\Omega_m$ is unitary and $V$ is
orthogonal. The subregion projector generally mixes energy eigenmodes,
so $S_m$ is dense and scalar Fermi functions cannot replace the matrix
inverse. Complex Hermitian hopping admits the same algebra with unitary
eigenvectors and conjugate transposes.

Writing $\eta_m=-e^{-iq_m}$ gives $S_m=I+(\eta_m-1)Q_A$, with
$Q_A=V^TP_AV$ a Hermitian projector. Its eigenvalues are zero and one,
so those of $S_m$ are one and $\eta_m$. The projector is the Gram matrix
of the selected rows of $V$; its idempotence and the unitarity of $S_m$ are numerical consistency tests.

The factorization uses the formal diagonal matrices and dimensionless times
\begin{equation}
 D=e^{\beta\Lambda},\qquad H=D^{-1},\qquad
 u=\tau/\beta,\qquad v=\tau'/\beta,\qquad
 \mathcal A_m=V^TA_mV=(I+S_mD^{-1})^{-1}.
 \label{c4:stable:formalD}
\end{equation}
The matrices $D$ and $H$ are formal notation for the factorization.
The numerical expressions below use bounded combinations of their
entries. Factoring on the right in the stated matrix order gives
\begin{equation}
 I+S_mD^{-1}=(D+S_m)D^{-1},
 \qquad
 \mathcal A_m=D(D+S_m)^{-1}.
 \label{c4:stable:factorA}
\end{equation}
Because $S_m$ and $D$ generally do not commute, their order matters;
taking the inverse reverses that order.
Multiplication by the external propagators then yields the upper
energy-basis kernel,
\begin{equation}
 \bar G_m^>=VK_m^>V^T,\qquad
 K_m^>=D^{1-u}(D+S_m)^{-1}D^v,\qquad u>v.
 \label{c4:stable:upperraw}
\end{equation}
One large factor from the inverse has already been combined
algebraically with the left imaginary-time evolution. The diagonal scales are then distributed so that every numerically
evaluated exponential is bounded.

The energy-basis factorization preserves the full twist matrix as it
combines exponential scales. This is essential because small site-basis
entries can contain cancellations between large eigenmode contributions;
the factorization retains these cancellations through the matrix solve.

The lower branch contains $\mathcal A_m-I$. I carry out this subtraction
symbolically to preserve relative accuracy in modes where
$\mathcal A_m$ is close to one:
\begin{align}
 \mathcal A_m-I
 &=D(D+S_m)^{-1}-(D+S_m)(D+S_m)^{-1}
 =-S_m(D+S_m)^{-1},\nonumber\\
 I+DS_m^{-1}&=(D+S_m)S_m^{-1}
             =D(H+S_m^{-1}),\nonumber\\
 \mathcal A_m-I&=-(H+S_m^{-1})^{-1}H .
 \label{c4:stable:loweridentity}
\end{align}
Consequently,
\begin{equation}
 \bar G_m^<=VK_m^<V^T,\qquad
 K_m^<=-H^u(H+S_m^\dagger)^{-1}H^{1-v},\qquad u<v .
 \label{c4:stable:lowerraw}
\end{equation}
Unitarity gives $S_m^{-1}=S_m^\dagger$. The lower branch thus uses
the adjoint twist and the overall minus sign in
Eq.~\eqref{c4:stable:lowerraw}.

\subsection{Upper-branch diagonal scaling}

Let $x_a=\beta\epsilon_a$ and separate its positive part:
\begin{equation}
 c_a^+=\max(0,x_a),\qquad
 C_+=\operatorname{diag}(e^{c_a^+}),\qquad
 \theta_+\in[v,u].
 \label{c4:stable:upperchoice}
\end{equation}
The matrix $C_+$, like $D$, is only a formal object. Define
\begin{equation}
 M_m^+=C_+^{-\theta_+}(D+S_m)C_+^{-(1-\theta_+)}.
 \label{c4:stable:uppermiddle}
\end{equation}
The two scales are different unless the splitting parameter is one half.
Their placement matters because the central matrix is not diagonal.
Solving this equation for the unscaled inverse gives
\begin{align}
 (D+S_m)^{-1}
 &=C_+^{-(1-\theta_+)}(M_m^+)^{-1}C_+^{-\theta_+},
 \nonumber\\
 K_m^+&=L_+(M_m^+)^{-1}R_+,\nonumber\\
 L_+&=D^{1-u}C_+^{-(1-\theta_+)},&
 R_+&=C_+^{-\theta_+}D^v .
 \label{c4:stable:upperfactored}
\end{align}
Here and below the plus and minus labels on the scaled kernels denote
the upper and lower branches. I evaluate each exponential from its combined exponent:
\begin{align}
 \ell_a^+&=(1-u)x_a-(1-\theta_+)c_a^+,&
 r_a^+&=vx_a-\theta_+c_a^+,\nonumber\\
 (L_+)_{aa}&=e^{\ell_a^+},&
 (R_+)_{aa}&=e^{r_a^+},\nonumber\\
 (M_m^+)_{ab}
 &=e^{-\theta_+c_a^+}(S_m)_{ab}e^{-(1-\theta_+)c_b^+}
   +\delta_{ab}e^{x_a-c_a^+}.
 \label{c4:stable:upperelements}
\end{align}
The left and right exponents are nonpositive. If $x_a\geq0$, then
$\ell_a^+=(\theta_+-u)x_a\leq0$ because $\theta_+\leq u$.
If $x_a<0$, then $\ell_a^+=(1-u)x_a\leq0$ because $u\leq1$.
For a nonnegative $x_a$, the right exponent is
$r_a^+=(v-\theta_+)x_a\leq0$ because $\theta_+\geq v$.
For a negative $x_a$, it is $r_a^+=vx_a\leq0$ because $v\geq0$.
Thus both diagonal prefactors are bounded by one.

The three exponents entering the middle matrix are also nonpositive:
$-\theta_+c_a^+\leq0$,
$-(1-\theta_+)c_b^+\leq0$, and $x_a-c_a^+\leq0$.
Unitarity implies $|(S_m)_{ab}|\leq1$, so
\begin{equation}
 0<(L_+)_{aa}\leq1,\qquad 0<(R_+)_{aa}\leq1,\qquad
 |(M_m^+)_{ab}|\leq1+\delta_{ab}.
 \label{c4:stable:upperbound}
\end{equation}
These bounds hold in exact arithmetic. Very small entries can underflow
in floating point; conditioning of the remaining linear solve is treated
in Section~\ref{c4:stable:conditioning}.

\subsection{Lower-branch scaling and the diagonal limits}

For the lower branch use the negative energy variable and its positive
part:
\begin{equation}
 y_a=-x_a,\qquad c_a^-=\max(0,y_a),\qquad
 C_-=\operatorname{diag}(e^{c_a^-}),\qquad
 \theta_-\in[1-v,1-u].
 \label{c4:stable:lowerchoice}
\end{equation}
Define
\begin{align}
 M_m^-&=C_-^{-\theta_-}(H+S_m^\dagger)C_-^{-(1-\theta_-)},
 \nonumber\\
 K_m^-&=-L_-(M_m^-)^{-1}R_-,\nonumber\\
 L_-&=H^uC_-^{-(1-\theta_-)},&
 R_-&=C_-^{-\theta_-}H^{1-v}.
 \label{c4:stable:lowerfactored}
\end{align}
The expressions directly evaluated are
\begin{align}
 \ell_a^-&=uy_a-(1-\theta_-)c_a^-,&
 r_a^-&=(1-v)y_a-\theta_-c_a^-,\nonumber\\
 (L_-)_{aa}&=e^{\ell_a^-},&
 (R_-)_{aa}&=e^{r_a^-},\nonumber\\
 (M_m^-)_{ab}
 &=e^{-\theta_-c_a^-}(S_m^\dagger)_{ab}e^{-(1-\theta_-)c_b^-}
   +\delta_{ab}e^{y_a-c_a^-}.
 \label{c4:stable:lowerelements}
\end{align}
For $y_a\geq0$, the left exponent is
$(u+\theta_--1)y_a\leq0$ because $\theta_-\leq1-u$.
For $y_a<0$, it is $uy_a\leq0$.
For $y_a\geq0$, the right exponent is
$(1-v-\theta_-)y_a\leq0$ because $\theta_-\geq1-v$.
For $y_a<0$, it is $(1-v)y_a\leq0$.
The middle-matrix exponents are nonpositive by the same positive-part
construction. Since the adjoint is unitary, the lower branch has the
same entrywise bounds as Eq.~\eqref{c4:stable:upperbound}.

I use the interval midpoints,
\begin{equation}
 \theta_+=\frac{u+v}{2},\qquad
 \theta_-=1-\frac{u+v}{2}.
 \label{c4:stable:midpoints}
\end{equation}
Any value in the allowed interval gives the same kernel in exact
arithmetic. Varying it about the midpoint tests numerical sensitivity to
the distribution of scaling between the two sides.

At coincident times $u=v$, the intervals collapse to
$\theta_+=u$ and $\theta_-=1-u$. Both formulas still have well-defined
one-sided limits. Their difference must satisfy
\begin{equation}
 K_m^+(u,u)-K_m^-(u,u)=I,\qquad
 \bar G_m^+(\tau,\tau)-\bar G_m^-(\tau,\tau)=I .
 \label{c4:stable:jump}
\end{equation}
Before scaling, this identity follows by subtracting
$\mathcal A_m-I$ from $\mathcal A_m$; it is independent of the twist.
At equal times a branch must be specified. I construct an interpolant
for each branch in Section~\ref{c4:tables:section}.

\subsection{Scalar checks and extreme subregions}

For $S_m=I$, every energy mode decouples. The upper middle entry becomes
$e^{-c_a^+}+e^{x_a-c_a^+}$, independent of the splitting parameter.
Multiplying its left and right prefactors gives
\begin{align}
 K_a^+(u,v)
 &=\frac{e^{(1-u+v)x_a-c_a^+}}
 {e^{-c_a^+}+e^{x_a-c_a^+}}
 =\frac{e^{(1-u+v)x_a}}{1+e^{x_a}},
 \nonumber\\
 K_a^-(u,v)
 &=-\frac{e^{(v-u)x_a}}{1+e^{x_a}}.
 \label{c4:stable:scalar}
\end{align}
I evaluate the first expression in its scaled form: the numerator is
at most one and the denominator lies between one and two.
For $x_a\gg1$, the upper branch behaves as $e^{-(u-v)x_a}$.
For $x_a\ll-1$, it behaves as $e^{(1-u+v)x_a}$.
The corresponding lower expressions are also small or order one on
their valid triangle. At $x_a=0$, the upper and lower values are
one half and minus one half, respectively.

The matrix scaling generalizes the scalar shift used to evaluate Fermi
factors without overflow.
For $0\leq s\leq1$, scaling by $c=\max(0,x)$ gives
\begin{equation}
 \frac{e^{sx}}{1+e^x}
 =\frac{e^{sx-c}}{e^{-c}+e^{x-c}} .
 \label{c4:stable:logistic}
\end{equation}
The common scaling makes every exponential argument nonpositive.

If the entanglement region is empty, $\Omega_m=I$ in every sector.
All sector kernels coincide. The inverse replica transform therefore
gives the ordinary thermal kernel on each diagonal replica block
and zero between different replicas. This checks the normalization
of the transform as well as the scalar time dependence. With an empty region and the same shifted Hamiltonian, this procedure
also generates the ordinary kernel used for the partition-function table
in Section~\ref{c4:tables:section}.

If the region is the whole system, the twist is a scalar phase in
site space and commutes with $h$. In the two-replica basis of
Eq.~\eqref{c4:green:two-sector-basis}, $\Omega_-^{-1}=-iI$ and
$\Omega_+^{-1}=+iI$. Let $z_a=e^{-x_a}$. The upper-branch denominator
combinations entering $\bar G_{11}$ and $\bar G_{12}$ are
\begin{equation}
 \frac12\left(\frac1{1-iz_a}+\frac1{1+iz_a}\right)
 =\frac1{1+z_a^2},\qquad
 \frac{i}{2}\left(\frac1{1-iz_a}-\frac1{1+iz_a}\right)
 =-\frac{z_a}{1+z_a^2}.
 \label{c4:stable:fullA}
\end{equation}
The diagonal replica block has the thermal denominator at inverse
temperature $2\beta$; the off-diagonal block carries the propagation
between the two sheets with the graded sign. The lower diagonal
branch has coefficient $-z_a^2/(1+z_a^2)$. These analytic expressions are evaluated with the bounded scalar rearrangement when $z_a$ is large.

\subsection{Sector conventions and real replica matrices}
\label{c4:stable:sectors}

For the real symmetric reference, the scaled sector solutions obey
$\bar G_+=\bar G_-^*$. With the replica and phase labels defined in
Eq.~\eqref{c4:green:two-sector-basis}, the full matrix of physical-site
blocks is
\begin{equation}
 \begin{pmatrix}
 \bar G_{11}&\bar G_{12}\\
 \bar G_{21}&\bar G_{22}
 \end{pmatrix}
 =\begin{pmatrix}
 \operatorname{Re}\bar G_-&-\operatorname{Im}\bar G_-\\
 \operatorname{Im}\bar G_-&\operatorname{Re}\bar G_-
 \end{pmatrix}.
 \label{c4:stable:replicablocks}
\end{equation}
This is Eq.~\eqref{c4:green:n2} applied to the scaled kernels on either
time branch. The same transformation gives the jump $\delta_{rr'}I$,
including a zero jump between different replicas. These identities and
the residual imaginary parts provide numerical checks of the sector
reconstruction at the chosen working precision.

\subsection{Solves, conditioning and error diagnostics}
\label{c4:stable:conditioning}

Each branch is evaluated through a linear system with a diagonal
right-hand side, followed by the prescribed row scaling. 

At finite temperature the physical two-replica matrix is nonsingular
in exact arithmetic. In the notation above,
\begin{equation}
 S_m=V^T(P_B\pm iP_A)V,\qquad
 \operatorname{Re}(D+S_m)=D+V^TP_BV>0 .
 \label{c4:stable:nonsingular}
\end{equation}
If a nonzero vector were in its nullspace, its quadratic form would
have a strictly positive real part and also vanish, a contradiction.
The corresponding statement holds for $H+S_m^\dagger$. Invertible diagonal scaling preserves this property, although
conditioning still depends on temperature, system size and scaling.

For a computed solution $\widehat Y$ of $MY=R$, the normalized
residual and a first-order conditioning estimate are
\begin{equation}
 \eta_{\rm solve}=
 \frac{\|M\widehat Y-R\|}
 {\|M\|\,\|\widehat Y\|+\|R\|},\qquad
 \kappa(M)=\|M\|\,\|M^{-1}\| .
 \label{c4:stable:residual}
\end{equation}
A small residual controls backward error. Forward error also depends
on conditioning and errors in the input matrices. Further cancellation
can occur in the basis transforms, so near-zero kernel entries are best
compared using absolute errors.

The solver estimates condition numbers using the computed solution.
With nonsingular diagonal $R$, the computed solution contains
the columns of $M^{-1}$ scaled by the respective right-hand sides:
\begin{equation}
 \|M^{-1}\|_1
 =\max_j\frac{\sum_i|Y_{ij}|}{|R_{jj}|},\qquad
 {\rm rcond}_1=\frac1{\|M\|_1\|M^{-1}\|_1}.
 \label{c4:stable:rcond}
\end{equation}
The estimate is used as a diagnostic at the chosen working precision
and scaling. A small reciprocal condition number flags possible loss of
accuracy and motivates a comparison at higher precision.

\subsection{Algorithm and independent checks}
\label{c4:stable:algorithm}

The pointwise procedure starts by diagonalizing $h$ and forming $S_m$
for each spin. At every requested time pair and branch, the solver
assembles the bounded middle matrix and diagonal factors of
Eq.~\eqref{c4:stable:upperelements} or
\eqref{c4:stable:lowerelements}, solves $M_m^\pm Y=R_\pm$, scales the
solution and transforms it back to real replica blocks. The masks,
eigensystem and twist matrices are reused across time nodes, while the
scaled matrix generally changes with the splitting parameter.
Table~\ref{c4:tables:sources} summarizes the stages.

Independent small-matrix calculations compare the stable and direct
formulas wherever both are numerically usable, testing the jump, twisted
boundary, splitting-parameter independence and conjugate sectors.
Scalar spectra with large positive and negative energies provide a
further comparison with stable analytic Fermi expressions.

Bounded exponents prevent overflow when their inputs are representable.
Underflow can still remove small entries, and roundoff enters the
eigensystem, solves and basis transforms. Double-double arithmetic gives
about 106 significand bits, compared with 53 for double, but essentially
the same exponent range. It reduces roundoff without extending the
low-temperature range permitted by those exponential factors.

\section{Two-time interpolation and numerical precision}
\label{c4:tables:section}

The interpolation formulas below specify the propagator representation.

\subsection{Branchwise interpolation}

Repeating a dense solve for every Monte Carlo propagator evaluation would dominate
the integration cost. I therefore evaluate the stable kernel on a time
grid before sampling and represent it by reusable interpolation tables.
Grid refinement and comparisons with pointwise solves assess the
deterministic approximation shared by all samples.

The discontinuity at $\tau=\tau'$ is known analytically. I map each
smooth branch to its own reference square and interpolate within that
branch. Use $\zeta,\eta\in[-1,1]$, distinct from
the normalized physical times $u,v$ used above:
\begin{align}
 \tau&=\frac{\beta}{2}(1+\zeta),&
 \tau'&=\frac{\tau}{2}(1+\eta)
 &&(\tau\geq\tau'),\nonumber\\
 \tau'&=\frac{\beta}{2}(1+\zeta),&
 \tau&=\frac{\tau'}{2}(1+\eta)
 &&(\tau\leq\tau').
 \label{c4:tables:maps}
\end{align}
The edges $\eta=1$ and $\eta=-1$ correspond to equal times and zero
earlier time, respectively. At $\zeta=-1$ the map collapses to the
origin, where each branch has a definite limit.

The inverse map used in lookup is
\begin{equation}
 \zeta=\frac{2\tau_{\max}}{\beta}-1,\qquad
 \eta=\frac{2\tau_{\min}}{\tau_{\max}}-1,\qquad
 \tau_{\max}>0 .
 \label{c4:tables:inversemap}
\end{equation}
Under the same change of variables, the integration measure over a
triangle is
\begin{equation}
 \mathrm d\tau\,\mathrm d\tau'
 =\frac{\beta^2}{8}(1+\zeta)\,\mathrm d\zeta\,\mathrm d\eta .
 \label{c4:tables:jacobian}
\end{equation}
This Jacobian enters quadrature over a triangle; interpolation
coefficients represent the kernel without an integration weight.

\subsection{Chebyshev nodes and coefficient normalization}

For each spin, branch, replica pair and ordered pair of physical
sites, define a scalar function $\mathcal G^\pm(\zeta,\eta)$ by
composing the corresponding kernel entry with
Eq.~\eqref{c4:tables:maps}. Its truncated representation is
\begin{equation}
 \mathcal G^\pm(\zeta,\eta)\approx
 \sum_{a=0}^{N_u-1}\sum_{b=0}^{N_v-1}
 c_{ab}^\pm T_a(\zeta)T_b(\eta).
 \label{c4:tables:expansion}
\end{equation}
The integers $N_u,N_v$ count nodes and coefficients; the corresponding maximal polynomial degrees are $N_u-1$ and $N_v-1$. The two branches have independent coefficient sets, preserving their
distinct one-sided values at the same point on the time diagonal.

The nodes are roots of the first-kind Chebyshev polynomials:
\begin{equation}
 \theta_i=\frac{(2i+1)\pi}{2N_u},\quad
 \phi_j=\frac{(2j+1)\pi}{2N_v},\quad
 \zeta_i=\cos\theta_i,\quad \eta_j=\cos\phi_j,
 \quad
 0\leq i<N_u,\ 0\leq j<N_v .
 \label{c4:tables:nodes}
\end{equation}
They lie strictly inside the reference square, so table generation
avoids the collapsed edge and coincident times. At these nodes the discrete
cosine orthogonality gives
\begin{align}
 \sum_{i=0}^{N_u-1}\cos(a\theta_i)\cos(a'\theta_i)
 &=\frac{N_u}{\gamma_a}\delta_{aa'},&
 \gamma_a&=\begin{cases}1,&a=0,\\2,&a>0,\end{cases}
 \nonumber\\
 c_{ab}^\pm
 &=\frac{\gamma_a\gamma_b}{N_uN_v}
 \sum_{i=0}^{N_u-1}\sum_{j=0}^{N_v-1}
 \mathcal G^\pm(\zeta_i,\eta_j)
 \cos(a\theta_i)\cos(b\phi_j).
 \label{c4:tables:dct}
\end{align}
The orthogonality statement assumes both indices lie within the
retained degree range. The zero-frequency factors are essential:
$\gamma_0=1$ gives the correct weight of every term with a zero index. In the convention here the coefficients already
contain these factors, so no additional half-weight is attached
to the constant mode during evaluation.

The coefficients are obtained by two separable cosine transforms,
first along the second node index and then along the first. Precomputed
cosine sums yield the same tensor-product polynomial as the four-index
formula at lower arithmetic cost.

Smoothness on each branch permits rapid polynomial convergence. The
required degree depends on temperature, spectrum, region geometry and
desired accuracy, particularly near edges with sharp continuous
variation. All seven runs use $25\times25$ coefficients per time branch.

\subsection{Nested Clenshaw evaluation}

I evaluate the tensor-product polynomial by nested Clenshaw recurrence.
For a one-dimensional series $p(z)=\sum_{j=0}^{d}a_jT_j(z)$, use the recurrence
\begin{equation}
 b_{d+1}=b_{d+2}=0,\qquad
 b_j=2zb_{j+1}-b_{j+2}+a_j\quad(j=d,\ldots,1),
 \qquad
 p(z)=a_0+zb_1-b_2 .
 \label{c4:tables:clenshaw}
\end{equation}
This form uses coefficients whose constant term already has its full
weight, as in Eq.~\eqref{c4:tables:dct}. It follows by using
$T_{j+1}=2zT_j-T_{j-1}$ to eliminate successive higher-degree
contributions. At degree zero the result is $a_0$.

Apply the recurrence to each row at $\eta$ to obtain
$d_a(\eta)=\sum_b c_{ab}^\pm T_b(\eta)$, then at $\zeta$ to evaluate
$\sum_a d_a(\eta)T_a(\zeta)$. This costs $O(N_uN_v)$ arithmetic and
requires only a vector of row results. 

The selected time branch, including its one-sided diagonal limit,
maps into $[-1,1]^2$.

\subsection{Corners, equal times and interpolation checks}

At the common origin, the ratio in Eq.~\eqref{c4:tables:inversemap} is
indeterminate. Upper and lower corner values are computed directly with
the stable solver. Lookup uses the corresponding value for
$\tau_{\max}<\tau_c$ and the polynomial elsewhere. Here $\tau_c$ is the
corner threshold; the runs below use $\tau_c=10^{-13}$ in units with
$t=1$.

The corner substitution is exact at the origin and is an
approximation in a finite neighbourhood. On a smooth branch, a
local derivative bound would give, schematically,
\begin{equation}
 \|\bar G^\pm(\tau,\tau')-\bar G^\pm(0,0)\|
 \leq
 \tau\sup\|\partial_\tau\bar G^\pm\|
 +\tau'\sup\|\partial_{\tau'}\bar G^\pm\| .
 \label{c4:tables:cornerbound}
\end{equation}
The suprema are taken within the same branch. A convergence study of
the finite-neighbourhood approximation requires varying the corner
threshold together with the interpolation resolution; the table
construction does not evaluate the derivative bounds themselves.

Elsewhere on the diagonal, the branch polynomial is evaluated at
$\eta=1$. Comparing each one-sided value and their difference with the pointwise
solver tests both the contact limits and the jump. The empty region gives another check:
$G^<(0,0)=-[I+e^{\beta h}]^{-1}$ at $A=\varnothing$.

Interpolation error can be measured against the pointwise solver at
times outside the construction grid, including points near the diagonal,
thermal edges and origin. Changing the resolution tests the polynomial approximation, and varying
the construction or storage precision tests its floating-point representation. The deterministic coefficient comparison
in Section~\ref{c4:bench:section} tests their combined effect on low-order
integrals.

Since $|T_a(z)|\leq1$ on $[-1,1]$, the absolute sum of discarded
coefficients bounds the difference from a more highly resolved
polynomial. Aliasing can still move unresolved high-degree contributions
into retained coefficients, so this absolute truncation estimate must be checked against finer
grids or direct kernel evaluations.

\subsection{Spatial indexing and the size of the table}

The selected region is defined in real space and generally breaks
translation invariance. For each spin, the Chebyshev coefficients depend
on both absolute sites and the two replica labels:
\begin{equation}
 c^{\pm;rr'}_{ab}(R_1,R_2),\qquad
 r,r'\in\{1,2\},\qquad
 0\leq a<N_u,\quad 0\leq b<N_v .
 \label{c4:tables:indices}
\end{equation}
There are $4N_s^2$ replica-site pairs, each represented by $N_uN_v$
coefficients arranged for the nested recurrence.

The tables store real replica entries after combining the complex
sectors. Ordinary and twisted tables use the same shifted Hamiltonians.
The ordinary contribution in Eq.~\eqref{c4:cdet:difference} depends
only on the single-copy thermal kernel.

For one spin and one twisted table, both time branches contain
\begin{equation}
 N_{\rm coeff}=8N_s^2N_uN_v
 \label{c4:tables:memory}
\end{equation}
real coefficients, in addition to corner arrays. Total memory depends
on the number of simultaneous spin and ordinary tables, the polynomial
resolution, the scalar size and whether equal-spin tables share storage.
Equal shifts give identical spin tables, which need be constructed only
once.

Storage and I/O scale as $O(N_s^2N_uN_v)$. Dense eigendecomposition and
sector-matrix assembly are performed during setup. Solves and basis
transforms are then repeated at the time nodes, contributing
$O(N_s^3N_uN_v)$ in generic dense arithmetic, followed by the coefficient
transforms.

\subsection{Distributed table construction}
\label{c4:tables:distributed}

Dense linear algebra at the time nodes and cosine transforms for the
site pairs form two stages of table construction. Both admit a division
into independent blocks using common spectral data and a common
polynomial representation. This organization allows table preparation
to be distributed, while the large-lattice Monte Carlo calculation uses
the completed coefficients. Each propagator is obtained from the selected
site pair and time branch, using the corner value or the polynomial at
the inverse triangular coordinates. Propagators whose two vertices
remain in the next envelope can be reused.

\subsection{Arithmetic precision at successive stages}

The precision used to construct the table, store its coefficients and
evaluate determinants controls three different sources of roundoff.
At a fixed grid, wider construction arithmetic improves the sampled
kernel values and cosine transforms only to the extent retained by the
stored coefficients. Wider determinant arithmetic then reduces roundoff
in the connected subtraction of a fixed set of propagators. Grid
refinement reduces interpolation error until the accuracy of the
pointwise solves or coefficient representation becomes limiting.

The \mbox{$50\times50$} calculation with equal shifts
$\alpha_\uparrow=\alpha_\downarrow=\alpha=1.4$ drops ten mantissa bits
from table coefficients and corner values. Its comparison with
$\alpha=1.3$ therefore includes a change of stored precision
(Table~\ref{c4:large:controls}).

\subsection{Propagator checks and numerical approximations}

\begin{table}[tbp]
\centering\small
\begin{tabularx}{\textwidth}{@{}>{\raggedright\arraybackslash}p{0.31\linewidth}X@{}}
\toprule
Stage & Construction and evaluation\\
\midrule
Spectral preparation & Diagonalize $h$ and form $V^T\Omega_m^{-1}V$.\\
Pointwise kernel & Solve the scaled upper and lower systems and reconstruct real replica blocks.\\
Interpolation & Map each triangle to $[-1,1]^2$, apply cosine transforms and retain corner values.\\
Kernel evaluation & Select the branch and site pair, then evaluate the tensor-product polynomial.\\
Precision & Choose construction, stored-coefficient and integrand arithmetic separately.\\
Conditioning & Estimate reciprocal condition numbers and check conjugate-sector reality.\\
\bottomrule
\end{tabularx}
\caption[Construction and evaluation of Gaussian replica tables]{Construction and evaluation of the Gaussian replica tables.}
\label{c4:tables:sources}
\end{table}

The comparison in Table~\ref{c4:bench:lowtable} tests the low-order
integrals formed from replica propagators, shifted vertices,
normalization subtraction and connected determinants against ED
derivatives. It complements the pointwise algebraic checks in
Section~\ref{c4:green:section}. The interpolation-error assessment is
limited to these tests and does not give a bound uniform in lattice
size and temperature.

The branch identities and scaling are algebraically exact. The
numerical representation introduces eigensystem, solve, interpolation
and stored-precision errors before the Monte Carlo and continuation
uncertainties of Sections~\ref{c4:stats:section} and
\ref{c4:continue:section}.

\section{The free entropy, shifted vertices and deterministic first order}
\label{c4:free:section}

\subsection{The absolute free contribution}

The connected expansion gives only corrections around a shifted
Gaussian reference, so its absolute zeroth-order contribution must be
computed from the reference moment:
\begin{equation}
 a_0=\frac{1}{N_s}
 [\ln\widetilde Z_{0,2}[A]-2\ln Z_{0,1}].
 \label{c4:free:a0}
\end{equation}
The restricted-correlation formula \cite{Peschel2003} evaluates this
normalization through the eigenvalues of the reduced correlation matrix
and gives a positive free purity directly.

For one spin species let $C_{ij}=\langle c_j^\dagger c_i\rangle_0$ and
restrict both indices to $A$. Diagonalize
\begin{equation}
 C_A=W\,\operatorname{diag}(\nu_1,\ldots,\nu_{|A|})W^\dagger,
 \qquad 0\le\nu_j\le1.
 \label{c4:free:CA}
\end{equation}
The reduced state of a number-conserving Gaussian state is Gaussian.
To establish this, choose independent subsystem modes with occupations
$\nu_j$. Their Wick contractions reproduce $C_A$, so every subsystem
creation/annihilation monomial has the same expectation as in the
partial trace of the original state. These monomials span the
subsystem operator algebra; the density operators therefore agree.

In the transformed mode basis,
\begin{equation}
 \rho_{A,0}=\bigotimes_j
 [(1-\nu_j)|0_j\rangle\langle0_j|+\nu_j|1_j\rangle\langle1_j|].
 \label{c4:free:factor}
\end{equation}
At $\nu_j=0$ the mode is empty and at $\nu_j=1$ it is filled; both states are pure. For an
interior probability the equivalent entanglement energy is
$\epsilon_j^{\rm ent}=\ln[(1-\nu_j)/\nu_j]$.
This entanglement spectrum describes the reduced state and differs from
the one-particle dispersion of the original lattice.

\subsection{The moment and its limits}

Taking the power and trace gives
\begin{align}
 {\cal R}_{n,0}(A)&=\prod_j[(1-\nu_j)^n+\nu_j^n],\nonumber\\
 \ln{\cal R}_{n,0}(A)&=\ln\det[(I-C_A)^n+C_A^n].
 \label{c4:free:moment}
\end{align}
Spin contributions multiply in the moment and add in its logarithm:
\begin{equation}
 a_0=\frac1{N_s}\sum_{\sigma,j}
 \ln[\nu_{\sigma j}^2+(1-\nu_{\sigma j})^2].
 \label{c4:free:second}
\end{equation}
Each bracket lies between $1/2$ and $1$. A half-occupied reduced mode
contributes $\ln2$ to $S_2$; a pure empty or filled mode contributes zero.
The result is real even though a sector representation uses complex
twists.

Differentiating the moment at $n=1$ yields
\begin{equation}
 S_{1,0}(A)=
 -\sum_{\sigma,j}
 [\nu_{\sigma j}\ln\nu_{\sigma j}
 +(1-\nu_{\sigma j})\ln(1-\nu_{\sigma j})].
 \label{c4:free:vonNeumann}
\end{equation}
This $n\to1$ limit applies to the Gaussian reference. For the interacting
system I calculate $n=2$ and continue in the homotopy parameter $\xi$.

For an empty region the determinant is one. For the full lattice,
the eigenvalues are Fermi occupations and each mode satisfies
\begin{equation}
 (1-\nu)^n+\nu^n
 =\frac{1+e^{-n\beta\epsilon}}{(1+e^{-\beta\epsilon})^n}.
 \label{c4:free:fullmode}
\end{equation}
Multiplication recovers Eq.~\eqref{c4:target:full}, fixing the
normalization power and thermal boundary convention simultaneously.
The endpoint probabilities are treated by continuity, without
forming divergent entanglement energies.

\subsection{Equivalence to sewn Gaussian determinants}

For one spin write $E=e^{-\beta h}$, so $C=E(I+E)^{-1}$.
In Fourier sector $m$ the one-particle insertion is
$T_m=P_B-e^{-iq_m}P_A$. The Gaussian Fock-space identity
$\operatorname{Tr}\Gamma(T_m)\Gamma(E)=\det(I+T_mE)$
does not require commuting matrices \cite{Klich2002}.
Factoring on the right gives
\begin{align}
 I+T_mE
 &=\left[I+(T_m-I)E(I+E)^{-1}\right](I+E),\nonumber\\
 \frac{\det(I+T_mE)}{\det(I+E)}
 &=\det[I-(1+e^{-iq_m})P_AC]\nonumber\\
 &=\det_A[I_A-(1+e^{-iq_m})C_A].
 \label{c4:free:detlemma}
\end{align}
Order the one-particle basis as $A$ followed by $B$. Only the $A$ rows
of the middle determinant differ from the identity; block triangularity
therefore proves the final equality. The reduction holds under the determinant. Within a Green kernel,
the non-commuting factors $P_A$ and $C$ keep their stated order.

The phases $\lambda_m=e^{-iq_m}$ are all roots of $\lambda_m^n=-1$.
Consequently
\begin{equation}
 \prod_{m=0}^{n-1}(x-\lambda_m y)=x^n+y^n .
 \label{c4:free:rootpoly}
\end{equation}
For each eigenvalue of $C_A$, take $x=1-\nu_j$ and $y=\nu_j$.
Multiplying Eq.~\eqref{c4:free:detlemma} over sectors then reproduces
Eq.~\eqref{c4:free:moment}, including the denominator $Z_{0,1}^n$.
For $n=2$ the factors are conjugates. Taking their combined real
logarithm through the eigenvalues avoids a complex-log branch choice.

The correlation-matrix entropy and sewn-sector determinants give
independent evaluations of the Gaussian normalization for the same
$h$, temperature and region.

\subsection{Construction on a periodic lattice}

In the periodic reference,
\begin{equation}
 C_\sigma(\mathbf R_i,\mathbf R_j)=
 \frac1{N_s}\sum_{\mathbf k}
 \frac{e^{\mathrm i\mathbf k\cdot(\mathbf R_i-\mathbf R_j)}}
 {1+\exp\{\beta[\epsilon_{\mathbf k}-\mu+\alpha_\sigma]\}},
 \quad
 \epsilon_{\mathbf k}=-2(\cos k_x+\cos k_y).
 \label{c4:free:correlation}
\end{equation}
Only the block of $C_\sigma$ on the sites of $A$ is needed. The
calculation fills this Hermitian matrix, diagonalizes it and sums the
mode entropies. With more general boundaries, the matrix can be
constructed from occupied one-particle projectors of the hopping
eigensystem.

Before the mode entropies are evaluated, small endpoint excursions of
the eigenvalues are clamped to $[0,1]$ to absorb roundoff. The reference
uses the same $\mu-\alpha_\sigma$ as the interacting coefficients.

For the regions of all seven runs in Table~\ref{c4:large:runs},
Eq.~\eqref{c4:free:correlation} reproduces the free reference values used
in the simulations to within $3\times10^{-12}$.

\subsection{Shifted vertices on each replica}

The local insertion obeys
\begin{equation}
\begin{aligned}
 U n_\uparrow n_\downarrow-\alpha_\uparrow n_\uparrow
 -\alpha_\downarrow n_\downarrow
 &=
 U\left(n_\uparrow-\frac{\alpha_\downarrow}{U}\right)
  \left(n_\downarrow-\frac{\alpha_\uparrow}{U}\right)\\
 &\quad-\frac{\alpha_\uparrow\alpha_\downarrow}{U}.
\end{aligned}
 \label{c4:free:shiftidentity}
\end{equation}
With $\bar G=\langle{\cal T}_\tau cc^\dagger\rangle$, the determinant
matrices contain
\begin{align}
 M^\uparrow_{pq}&=\bar G_\uparrow(X_p,X_q)
 +\frac{\alpha_\downarrow}{U}\delta_{pq},\nonumber\\
 M^\downarrow_{pq}&=\bar G_\downarrow(X_p,X_q)
 +\frac{\alpha_\uparrow}{U}\delta_{pq}.
 \label{c4:free:shiftmatrix}
\end{align}
The product in Eq.~\eqref{c4:free:shiftidentity} places the opposite-spin
shift in each determinant, while every propagator remains spin diagonal.
The relation holds on each replica because the interaction is local to
one sheet.

All calculations in this chapter use $U>0$. At fixed nonzero shifts,
$\alpha_\sigma/U$ is singular as $U\to0$; that limit must be taken in the
unfactored insertion on the left of Eq.~\eqref{c4:free:shiftidentity}.

\subsection{A first-order expression with a fixed equal-time prescription}
\label{c4:free:first-order}

For $0<\tau<\beta$ define the lower-side contraction and shifted
diagonal by
\begin{align}
 g_{\sigma r}^-(i,\tau)
 &=\lim_{\varepsilon\downarrow0}
 \bar G_\sigma^{rr}(i,\tau-\varepsilon;i,\tau),\nonumber\\
 B_{\uparrow r}(i,\tau)&=g_{\uparrow r}^-(i,\tau)
 +\frac{\alpha_\downarrow}{U},\qquad
 B_{\downarrow r}(i,\tau)=g_{\downarrow r}^-(i,\tau)
 +\frac{\alpha_\uparrow}{U}.
 \label{c4:free:lower}
\end{align}
Time endpoints are defined by the corresponding one-sided branch
limits. The annihilation time is earlier than the creation time;
reversing the time ordering adds the equal-time jump.
The ordinary thermal diagonal satisfies $g_\sigma^-=-\rho_\sigma$,
whereas the sewn diagonal is a twisted Gaussian expectation and
need not have the interpretation of an ordinary occupation probability.

Let $\widetilde a_1$ and $a_1^{\rm ord}$ denote the first-order coefficients per physical site in the replica and ordinary logarithms. Then
\begin{align}
 \widetilde a_1
 &=-\frac{U}{N_s}\sum_{r=1}^{2}\sum_i
 \int_0^\beta
 \left[B_{\uparrow r}B_{\downarrow r}
 -\frac{\alpha_\uparrow\alpha_\downarrow}{U^2}\right]\mathrm d\tau,\nonumber\\
 a_1^{\rm ord}
 &=-\frac{U}{N_s}\sum_i
 \int_0^\beta
 \left[B_\uparrow^{\rm ord}B_\downarrow^{\rm ord}
 -\frac{\alpha_\uparrow\alpha_\downarrow}{U^2}\right]\mathrm d\tau,\nonumber\\
 a_1&=\widetilde a_1-2a_1^{\rm ord}.
 \label{c4:free:first}
\end{align}
The product subtraction removes the artificial constant from the
factorized vertex. Its replica integral is twice its ordinary
integral, so it cancels in the normalized difference. It must
nevertheless be included consistently when either absolute
logarithm is considered separately.
At higher orders a constant cannot connect different insertions
and drops out of the connected cumulant.

The sewn diagonal is generally time dependent, so its first-order
contribution requires a time integral. The ordinary diagonal is
stationary and its integral reduces to a Hartree-density product times
$\beta$.

\subsection{Deterministic integration of the diagonal table}

On the lower triangular Chebyshev branch, the diagonal edge sets
the second interpolation coordinate to one. Since $T_b(1)=1$,
\begin{equation}
 d_a=\sum_b c_{ab}^-,\qquad
 g^-(\tau,\tau)=\sum_a d_aT_a(2\tau/\beta-1).
 \label{c4:free:diagline}
\end{equation}
For the uniform time measure,
\begin{equation}
 J_\ell=\int_{-1}^{1}T_\ell(z)\,\mathrm dz
 =\begin{cases}
 2,&\ell=0,\\
 0,&\ell\ {\rm odd},\\
 2/(1-\ell^2),&\ell\ge2\ {\rm even}.
 \end{cases}
 \label{c4:free:chebweight}
\end{equation}
Substituting $z=\cos\theta$ proves the formula by integrating
$\cos(\ell\theta)\sin\theta$. The integral has uniform weight because it represents the physical time
measure; Chebyshev orthogonality uses a different weight.

If $g=\sum_a d_aT_a$ and $h=\sum_b e_bT_b$, then
\begin{equation}
 \int_0^\beta g(\tau)h(\tau)\,\mathrm d\tau
 =\frac{\beta}{4}\sum_{a,b}d_ae_b
 [J_{a+b}+J_{|a-b|}].
 \label{c4:free:diagproduct}
\end{equation}
The product identity $T_aT_b=(T_{a+b}+T_{|a-b|})/2$ gives this formula.
A constant shift changes only $d_0$ and $e_0$; the remaining scalar in
Eq.~\eqref{c4:free:first} is integrated directly. The diagonal polynomial
coefficients thus determine the integral without time sampling.

The first-order integral inherits table and arithmetic errors, with no
Monte Carlo noise. Increasing the Chebyshev resolution and construction
precision tests these deterministic errors.

\subsection{Normalization identities}

Three calculations constrain the normalization: the free moment from
the restricted correlation matrix, the first order from the lower-side
insertion in Eq.~\eqref{c4:free:first}, and higher orders from the
connected integrand with its factorial and replica measure.

The empty-region identity requires $a_0$ and every normalized
coefficient to vanish after the replica sum
(Section~\ref{c4:cdet:empty-check}). With $A$ equal to the full lattice,
the expansion must give the ordinary thermal identity at $2\beta$.
For any region, a change of auxiliary shift changes $a_0$ and the
individual $a_k$ while preserving their exact sum at $\xi=1$.

\section{Connected entropy integrands and the replica measure}
\label{c4:cdet:section}

The coefficient integrand uses absolute lattice positions, imaginary
times and replica labels. The entangling region breaks the translation invariance that
simplifies an ordinary homogeneous calculation. Since the
observable is $\ln\widetilde Z_2[A]-2\ln Z_1$, I construct the difference
of the two connected expansions on a common integration domain.

As in my calculation in Ref.~\cite{ShiRenyi2026}, I evaluate the
connected weights with CDet. Its connected-subset recursion is the same
as for ordinary partition functions
\cite{RossiCDet2017,FastPrincipalMinors2022}, now with the replica
propagator. The many-configuration sampler of Sections~\ref{c3:sub:section} and
\ref{c3:mc:section} uses these connected weights for all subsets.

\subsection{Fixed vertices and shifted determinant weights}

Let \(N_s=N_1N_2\) be the number of physical lattice sites. A physical interaction vertex is \(X_i=(x_i,\tau_i,r_i)\), with replica label \(r_i\in\{1,2\}\) in the twisted two-replica problem. The spin label \(\sigma\) is carried by the propagators; each interaction vertex enters both spin determinants. For a labelled set \(S\) of vertices, define the unscaled shifted matrices
\begin{equation}
 \mathsf M^{(n,A)}_{\sigma,ij}
 =\bar G^{(n,A)}_{\sigma;r_i r_j}(x_i,\tau_i;x_j,\tau_j)
 +\delta_{ij}\frac{\alpha_{\bar\sigma}}{U},
 \qquad i,j\in S.
 \label{c4:cdet:matrix}
\end{equation}
Here \(\bar\sigma\) denotes the opposite spin. The shifts and homotopy
parameter were introduced in Eq.~\eqref{c4:target:shift}; the first-order
constant subtraction is given in Eq.~\eqref{c4:free:first}. As in
Eq.~\eqref{c4:free:shiftmatrix}, the up-spin determinant receives
\(\alpha_\downarrow/U\), and the down-spin determinant receives
\(\alpha_\uparrow/U\). The ordinary matrix uses the same shifts and
physical parameters with the ordinary thermal Green function.

For either choice of propagator, introduce the full fixed-set weight
\begin{equation}
 \mathcal A(S)=\det\mathsf M_\uparrow[S]\det\mathsf M_\downarrow[S],
 \qquad \mathcal A(\varnothing)=1.
 \label{c4:cdet:moment}
\end{equation}
The determinants sum all Wick contractions, including disconnected ones. To isolate the logarithm, choose a distinguished vertex \(a(S)=\min S\). Every contraction has a unique connected component containing this vertex. Separating that component from all remaining components gives
\begin{equation}
 \mathcal A(S)=\sum_{\substack{T\subseteq S\\a(S)\in T}}
 D(T)\mathcal A(S\setminus T).
 \label{c4:cdet:decomposition}
\end{equation}
Consequently, processing subsets in increasing cardinality yields
\begin{equation}
 D(S)=\mathcal A(S)-
 \sum_{\substack{T\subsetneq S\\a(S)\in T}}
 D(T)\mathcal A(S\setminus T).
 \label{c4:cdet:connected}
\end{equation}
Fixing the anchor \(a(S)\) ensures that each decomposition is counted
once. Simultaneous relabelling of vertices and matrix indices leaves
the connected weight unchanged.

The ordinary and twisted propagators give \(D^{\mathrm{ord}}(S)\)
and \(D^{\mathrm{tw}}(S)\). The scalar constant in the shifted interaction
contributes only to the first cumulant, so its correction is applied to
the singleton outputs.

At two vertices the recursion can also be checked without any subset algorithm. Write the up-spin matrix as \(\bigl(\begin{smallmatrix}a&b\\c&d\end{smallmatrix}\bigr)\) and the down-spin matrix as \(\bigl(\begin{smallmatrix}e&f\\g&h\end{smallmatrix}\bigr)\). Subtracting the product of the two singleton weights gives
\begin{equation}
 D_2=(ad-bc)(eh-fg)-(ae)(dh)
 =-adfg-ehbc+bcfg.
 \label{c4:cdet:twoconnected}
\end{equation}
The first surviving term joins the vertices through the down-spin contractions, the second through the up-spin contractions, and the third through both. The term with no intervertex contraction cancels. Connectedness therefore concerns the full two-spin graph: requiring each spin determinant to be connected on its own would
discard the first two surviving terms. In a replica calculation the entries also carry their sheet dependence, but the algebra is unchanged.

\subsection{Principal minors and all-subset evaluation}

For an envelope containing \(N\) vertices, I construct the up- and down-spin matrices at the same coordinates for both the twisted and ordinary propagators. A principal minor indexed by \(S\) is the determinant in Eq.~\eqref{c4:cdet:moment}. Fast principal-minor evaluation obtains all full weights from these four matrices \cite{FastPrincipalMinors2022}. Its Schur-complement construction and pivot corrections were discussed in Section~\ref{c3:fpm:section}; here the input contains the replica Green function and absolute coordinates.

The recursion produces all connected outputs with direct subset cost
$O(3^N)$ and storage $O(2^N)$. 

Multiplying matrix entries by \(\sqrt{U\beta N_s}\) gives a factor
\((U\beta N_s)^k\) in the product of order-\(k\) determinants and its
connected part. The formulas below keep \(D\) unscaled and include this factor in the
integration prefactor of Eq.~\eqref{c4:cdet:difference}. Division by \(N_s\) then gives
the coefficient per physical site.

\subsection{A common unit-cube measure for both geometries}

The sampler stores four unit coordinates per vertex,
\begin{equation}
 \begin{gathered}
 z_i=(s_i,u_i,v_i,a_i)\in[0,1)^4,
 \quad
 \tau_i=\beta s_i,\\
 x_i=(\lfloor N_1u_i\rfloor,\lfloor N_2v_i\rfloor),
 \quad r_i=1+\lfloor2a_i\rfloor.
 \end{gathered}
 \label{c4:cdet:unitcoordinates}
\end{equation}
Write \(\mathrm d\mu(z)=\mathrm ds\,\mathrm du\,\mathrm dv\,\mathrm da\). This is a normalized measure. For a function that depends on the physical vertex but not on the position inside a discrete coordinate bin,
\begin{equation}
 \int \mathrm d\mu(z)\,F(X(z))
 =\frac{1}{2\beta N_s}\sum_{r=1}^{2}\sum_x
 \int_0^\beta\mathrm d\tau\,F(x,\tau,r).
 \label{c4:cdet:measure}
\end{equation}
Thus the conversion from a unit-cube integral to a two-replica vertex integral contains \(2\beta N_s\) for every vertex. The factor two counts the replicas.

The ordinary contribution uses the same four-coordinate domain, with
the single-copy thermal kernel independent of every \(a_i\).
Integration over these auxiliary replica coordinates contributes a factor
of one per vertex. Both terms can be evaluated at the same sampled
coordinates to exploit their covariance.

For \(k\ge2\), the sampled coefficient integrand is
\begin{align}
 f_k(z_1,\ldots,z_k)
 &=\frac{(-U\beta N_s)^k}{k!N_s}
 \Bigl[2^kD^{\mathrm{tw}}(X_1,\ldots,X_k)\nonumber\\
 &\qquad{}-2D^{\mathrm{ord}}(x_1,\tau_1;\ldots;x_k,\tau_k)\Bigr],
 \label{c4:cdet:difference}\\
 I_k&=\int f_k(z_1,\ldots,z_k)\prod_{i=1}^k\mathrm d\mu(z_i).
 \label{c4:cdet:integral}
\end{align}
The factor \(2^k\) converts the replica sampling measure, the factor two
multiplying the ordinary term implements the normalized logarithm, and
\(k!\) comes from the expansion in labelled interaction vertices.
Equivalently, the twisted prefactor is
\((-U)^k(2\beta N_s)^k/(k!N_s)\).

In the chapter-wide notation \(f_A(\xi)=a_0+\sum_{k\ge1}a_k\xi^k\), the intensive coefficient is \(I_k=a_k\). The symbol \(f_k\) in Eq.~\eqref{c4:cdet:difference} denotes its fixed-coordinate integrand, whereas \(f_A\) denotes the integrated normalized logarithm as a function of the auxiliary expansion parameter.

The singleton evaluator uses the corrected shifted product from Section~\ref{c4:free:section}. At fixed unit coordinates, define
\begin{equation}
 P^{g}(z)=B^{g}_\uparrow(z)B^{g}_\downarrow(z)
 -\frac{\alpha_\uparrow\alpha_\downarrow}{U^2},
 \qquad g\in\{\mathrm{tw},\mathrm{ord}\},
 \label{c4:cdet:singletonproduct}
\end{equation}
where each \(B\) is the corresponding shifted equal-time diagonal entry. Then
\begin{equation}
 f_1(z)=-2U\beta\left[P^{\mathrm{tw}}(z)-P^{\mathrm{ord}}(z)\right].
 \label{c4:cdet:singleton}
\end{equation}
As in Section~\ref{c4:free:first-order}, the constants cancel in the
normalized difference. The free logarithm is computed separately, and
the stochastic integrands give its interacting correction.

\subsection{The empty-region replica check}
\label{c4:cdet:empty-check}

For an empty region, every normalized coefficient vanishes after the replica sum. With \(A=\varnothing\), the two replicas are uncoupled and their propagator is block diagonal in replica space. At two vertices, write the ordinary connected determinant weight as \(D_2\). For fixed spatial and temporal coordinates,
\begin{equation}
 D^{\mathrm{tw}}_2(r_1,r_2)=\delta_{r_1r_2}D_2,
 \qquad D^{\mathrm{ord}}_2=D_2.
 \label{c4:cdet:emptytwo}
\end{equation}
If the labels differ, there is no propagator joining the vertices, and the connected subtraction removes the product of singleton components. If the labels agree, the remaining matrix block is the ordinary matrix. Defining \(p_2=(-U\beta N_s)^2/(2!N_s)\), the common-coordinate difference becomes
\begin{equation}
 f_2=p_2D_2(4\delta_{r_1r_2}-2)
 =\begin{cases}
 2p_2D_2,&(r_1,r_2)=(1,1),(2,2),\\
 -2p_2D_2,&(r_1,r_2)=(1,2),(2,1).
 \end{cases}
 \label{c4:cdet:emptytwovalues}
\end{equation}
Averaging the four equally probable replica assignments gives zero,
even though individual assignments generally have nonzero integrands.

The argument extends to any connected order. With uncoupled replicas, a connected graph must place every vertex on one replica. Among \(2^k\) assignments, exactly two do so. Hence
\begin{equation}
 2^{-k}\sum_{r_1,\ldots,r_k}
 2^kD^{\mathrm{tw}}_k(r_1,\ldots,r_k)=2D^{\mathrm{ord}}_k,
 \qquad A=\varnothing.
 \label{c4:cdet:emptygeneral}
\end{equation}
This identity checks the measure and replica factors. For an empty
region, the first-order coefficient vanishes, so this case cannot give
the nonzero reference required in Section~\ref{c4:mc:reference-section}.
For a nonempty region, common coordinates correlate the two terms;
the amount of variance reduction depends on the chosen parameters.

\subsection{Physical subtraction and numerical cancellation}

The connected recursion in Eq.~\eqref{c4:cdet:connected} removes
disconnected Wick contractions within each geometry. The subsequent
ordinary--twisted difference in Eq.~\eqref{c4:cdet:difference} then forms
the normalized R\'enyi logarithm from these connected contributions.

Near cancellation of large disconnected contributions can reduce the
significant digits in the connected remainder. The numerical calculation
filters these remainders using the local roundoff scale
\begin{equation}
 \Delta_{\mathrm{conn}}(S)=\epsilon_0\eta_{
 \mathrm{safe}}\left[
 |\mathcal A(S)|+
 \sum_{\substack{T\subsetneq S\\a(S)\in T}}
 |D(T)\mathcal A(S\setminus T)|\right].
 \label{c4:cdet:roundoffproxy}
\end{equation}
Here \(\epsilon_0\) is the roundoff scale of the working arithmetic:
$2\times10^{-16}$ for double, $10^{-31}$ for double-double and the
machine epsilon for \texttt{long double}. The subtraction safety factor
\(\eta_{\mathrm{safe}}\) is listed in Table~\ref{c4:large:controls}. Remainders below the threshold
are set to zero. Errors propagated from tables, minors and earlier recursion steps lie
beyond this local estimate. Threshold and precision sensitivity therefore
remain systematic error sources; the settings in
Table~\ref{c4:large:controls} have not been varied individually to isolate
their effects.

In the all-subset calculation, each geometry's connected coefficients
are converted to double before forming the ordinary--twisted difference. Cancellation in this final subtraction
can therefore limit accuracy even when intermediate arithmetic uses a
wider significand.

\section{Many-configuration sampling and normalization}
\label{c4:mc:section}

Many-configuration Markov-chain Monte Carlo (MCMCMC) samples this
family of signed integrands \cite{MCMCMC2021}. Each evaluated envelope
contributes to many retained subsets and several orders. I sample the
normalized-logarithm difference directly and normalize all orders by
one deterministic first-order reference
(Section~\ref{c4:mc:reference-section}).

\subsection{Envelopes and normalized conditional generators}

Let \(V=(z_1,\ldots,z_N)\) be an envelope, with measured orders
\(k\in\mathcal K\). Taking \(1\le\min\mathcal K\) and \(N>\max\mathcal K\)
ensures that every retained subset is nonempty and each update generates
at least one new vertex. For a subset \(S\), a conditional density \(Q_{\bar S|S}\) generates its complement from its retained vertices. It is normalized with respect to the same unit measure used by the integrand:
\begin{equation}
 \int Q_{\bar S|S}(V_{\bar S}\mid V_S)
 \prod_{i\notin S}\mathrm d\mu(z_i)=1.
 \label{c4:mc:Qnormalized}
\end{equation}
For proposals expressed in other coordinates, the density includes the
Jacobian to this unit measure, keeping the conditional generator and
coefficient estimator in the same normalization.

The generator starts from a normalized pair seed \(q(z'\mid z)\). It chooses a parent uniformly from all vertices currently present, generates a child and adds it to that set. Subsequent children may use earlier children as parents. For disjoint generated and retained sets \(B\) and \(R\), its symmetrized density obeys
\begin{align}
 P(\varnothing\mid R)&=1,\nonumber\\
 P(B\mid R)&=\frac{1}{|B||R|}
 \sum_{i\in B}\sum_{j\in R}
 q(z_i\mid z_j)
 P(B\setminus\{i\}\mid R\cup\{i\}).
 \label{c4:mc:generation}
\end{align}
The factor $1/|B|$ averages over which generated label becomes the first
child, and $1/|R|$ averages over its parent. Expanding the recursion sums
all sequential growth histories with their probabilities, including
their ordering multiplicity.

Normalization follows by induction on \(|B|\). With the first child
and parent fixed, the remaining conditional density integrates to unity
by the inductive hypothesis. Integrating the normalized first-child
seed and averaging the \(|B||R|\) choices preserves normalization.
This construction requires a nonempty retained set from which to choose
the first parent.

For a fixed envelope, write \(Q_V(B)=P(V_B\mid V_{\bar B})\). Equation~\eqref{c4:mc:generation} becomes the subset dynamic program
\begin{equation}
 Q_V(B)=\frac{1}{|B|(N-|B|)}
 \sum_{i\in B}\sum_{j\notin B}
 q(z_i\mid z_j)Q_V(B\setminus\{i\}),
 \qquad Q_V(\varnothing)=1.
 \label{c4:mc:Qrecursion}
\end{equation}
The recursion involves proper subsets and sums over candidate parents,
so all growth histories can be combined using precomputed pair densities.
Uniform seeds give \(Q_V(B)=1\), providing a direct normalization check.

For example, retain one vertex \(z_0\) and generate two vertices \(z_1,z_2\). With \(q_{ij}=q(z_i\mid z_j)\), the conditional density is
\begin{equation}
 P(\{z_1,z_2\}\mid\{z_0\})
 =\frac14\left[q_{10}(q_{20}+q_{21})
 +q_{20}(q_{10}+q_{12})\right].
 \label{c4:mc:twogenerated}
\end{equation}
The two terms correspond to the two choices of which child is generated
first. The second child can then use either the retained vertex or the
first child as parent. Each of the four products integrates to unity,
and a uniform seed gives $Q=1$. Asymmetric conditional densities without
translation invariance are allowed, as needed for the replica geometry.

Summing over all histories that reach the regenerated set gives the
symmetrized density entering the detailed-balance flux in
Section~\ref{c4:mc:heatbath}. This sum supplies the matching forward and
reverse factors in Eq.~\eqref{c4:mc:balance}.

\subsection{Learning a proposal in the replica geometry}

The learned seed factorizes into conditional distributions for time, position and replica labels. Let \(b(s)\) be one of \(B_\tau\) time bins and let \(j(x)\) enumerate the \(N_s\) absolute lattice sites. Three row-normalized probability tables, denoted \(p_\tau\), \(p_x\) and \(p_r\), specify
\begin{align}
 q(z'\mid z)
 &=B_\tau\,p_\tau\bigl(b(s')\mid b(s)\bigr)
 \;N_s\,p_x\bigl(j(x')\mid j(x)\bigr)
 \;2\,p_r(r'\mid r).
 \label{c4:mc:histdensity}
\end{align}
The child is uniform inside its selected coordinate bins. The density factors are therefore \(B_\tau\), \(N_s\) and two, respectively. Since \(s=\tau/\beta\), the time factor is \(B_\tau\), not \(B_\tau/\beta\), relative to \(\mathrm d\mu\). Integrating each factor over its unit coordinates sums a normalized row, proving normalization of the product.

The conditional tables use absolute parent and child coordinates,
since the entangling region generally breaks spatial and temporal
translation symmetries. Factorizing the proposal neglects correlations
among coordinate groups and can cost efficiency. Equation~\eqref{c4:mc:Qrecursion}
nevertheless evaluates the exact density of that factorized proposal,
so this factorization introduces no estimator bias.

Proposal learning starts from uniformly sampled envelopes. Each subset
of order at least two contributes the absolute normalized-logarithm
integrand divided by its envelope binomial multiplicity. This weight
is shared equally among the \(k(k-1)\) ordered vertex pairs to fill
the three conditional histograms. The resulting positive training
weights determine the proposal, while the measured coefficient retains
its sign.

Each row is normalized without a pseudocount. A row with no entries
falls back to a uniform distribution.
The uniform fraction is $\varepsilon=0.1$ for the \mbox{$3\times3$} runs and
$0.2$ for the four large-lattice runs. It is mixed into each normalized
row as follows:
\begin{equation}
 p_\ell(j\mid i)=(1-\varepsilon)p^{\mathrm{learned}}_\ell(j\mid i)
 +\frac{\varepsilon}{d_\ell},
 \quad
 (d_\tau,d_x,d_r)=(B_\tau,N_s,2).
 \label{c4:mc:uniformsupport}
\end{equation}
For $\varepsilon>0$, every coordinate bin has positive support. The
factorwise mixtures give a product of mixed marginal densities. The
trained histogram is fixed before measurement.

\subsection{The joint measure and the heat-bath step}
\label{c4:mc:heatbath}

Choose positive order weights \(\lambda_k\), fixed during production. For every eligible subset of the current envelope, define
\begin{align}
 \Phi_S(V)&=\lambda_{|S|}f_{|S|}(V_S)
 Q_{\bar S|S}(V_{\bar S}\mid V_S),\nonumber\\
 w_S(V)&=\frac{|\Phi_S(V)|}{\binom N{|S|}},
 \qquad W(V)=\sum_{\substack{S\subseteq[N]\\|S|\in\mathcal K}}w_S(V).
 \label{c4:mc:flux}
\end{align}
The binomial factor averages the subsets of a given order within an
envelope. The perturbative \(k!\) counts labelled interaction vertices
in the expansion, so both factors enter the normalization.

For a symmetric labelled representation, the joint density is
\begin{equation}
 \pi(V,S)=\frac{w_S(V)}{Z_W},
 \qquad Z_W=\int W(V)\prod_{i=1}^N\mathrm d\mu(z_i).
 \label{c4:mc:joint}
\end{equation}
Conditional on \(V\), select \(S\) with probability \(w_S/W\). Conditional on its retained coordinates, regenerate the complement with \(Q_{\bar S|S}\). Detailed balance follows because the flux between two envelopes sharing those coordinates contains the symmetric product
\begin{equation}
 \frac{\lambda_k|f_k(V_S)|}{Z_W\binom Nk}
 Q_{\bar S|S}(V_{\bar S}\mid V_S)
 Q_{\bar S|S}(V'_{\bar S}\mid V_S).
 \label{c4:mc:balance}
\end{equation}
Interchanging the old and new complements leaves this expression
unchanged, giving an always-accepted heat-bath update. The proof assumes
finite normalization, correct conditional densities and accessible
support. Mixing time still depends on the proposal and target weights.

With retained vertices placed first in the ordered tuple, exchangeability
under $W/Z_W$ is recovered by assigning the generated children uniformly
to complement labels and relabelling the envelope. The integrands,
subset sums and generator are permutation equivariant, so this relabelling
preserves measured expectations and defines the same process on unordered
vertex sets.

Normalizing each conditional generator and averaging over the equal-order subsets gives
\begin{equation}
 Z_W=\sum_{k\in\mathcal K}\lambda_k
 \int |f_k(z_1,\ldots,z_k)|\prod_{i=1}^k\mathrm d\mu(z_i).
 \label{c4:mc:absolutenormalizer}
\end{equation}
The proposal determines how the total weight is distributed among
envelopes, whereas $Z_W$ depends only on the absolute integrals and order
weights. A small signed coefficient can coexist with a large absolute
integral, limiting signal-to-noise. In the empty-region example, the
absolute two-vertex integral is finite while its signed coefficient
vanishes.

\subsection{Signed multi-output measurements}

A single evaluated envelope supplies every eligible retained subset.
Averaging their contributions gives
\begin{equation}
 A_k(V)=\frac{1}{\lambda_k\binom Nk}
 \sum_{|S|=k}\frac{\Phi_S(V)}{W(V)}.
 \label{c4:mc:signal}
\end{equation}
This is a signed Rao--Blackwell measurement. A measurement based only on the selected subset would be \(\mathbf1_{|S|=k}\operatorname{sgn}(f_k(V_S))/\lambda_k\). Its conditional expectation given the envelope is exactly Eq.~\eqref{c4:mc:signal}. Averaging over that conditional selection reduces its conditional measurement variance, while autocorrelation and evaluation cost determine the gain per unit computing time.

The expectation identity follows directly:
\begin{align}
 \mathbb E_\pi A_k
 &=\frac{1}{Z_W\binom Nk}\sum_{|S|=k}
 \int f_k(V_S)Q_{\bar S|S}(V_{\bar S}\mid V_S)
 \prod_{i=1}^{N}\mathrm d\mu(z_i)\nonumber\\
 &=\frac{I_k}{Z_W}.
 \label{c4:mc:signalmean}
\end{align}
The complement density integrates to unity. Summing the
\(\binom Nk\) equal retained-subset integrals gives the coefficient,
with its sign carried by the numerator throughout.

\subsection{Deterministic reference and finite-sample ratios}
\label{c4:mc:reference-section}

The first-order coefficient can be integrated deterministically from the lower-branch Chebyshev tables, as described in Section~\ref{c4:free:section}. Provided it is nonzero and is included among the measured orders, it sets the common normalization:
\begin{equation}
 \widehat I_k=I_1^{\mathrm{det}}
 \frac{\overline A_k}{\overline A_1},
 \qquad \widehat I_1=I_1^{\mathrm{det}}.
 \label{c4:mc:reference}
\end{equation}
The bar denotes the average after warm-up. Under the chain convergence
assumptions, Eq.~\eqref{c4:mc:signalmean} gives the limiting ratio.
At finite sample size the ratio is generally biased, especially when
the denominator is small relative to its uncertainty. A zero reference
coefficient requires another normalization order.

All order estimators \(A_k\) are measured on the same chain, and the
normalized coefficients share \(\overline A_1\). I therefore retain the
paired vector of order means for each chain and recompute the common
normalization when omitting a chain. This preserves numerator--denominator
and cross-order covariance in the analysis of
Section~\ref{c4:stats:section}.

The first-order line integral is exact for the finite Chebyshev
polynomial. Table-construction and interpolation errors remain systematic errors
of this reference. The tests in Section~\ref{c4:tables:section} and the
small-cluster comparison in Table~\ref{c4:bench:lowtable} assess these
errors outside the sampling uncertainty.

\subsection{Warm-up and sampling}

During warm-up, \(\lambda_k\) is adjusted multiplicatively from the
chosen-order histogram to balance the frequency of retained orders.
A positive count offset regularizes empty bins. The weights are fixed
before measurement, and each order estimator is divided by the
\(\lambda_k\) used in its sampling flux.

Each independent chain has its own pseudorandom stream, envelope and
warm-up weights. I fix the order weights and numerical filters before
measurement and use the paired chain means for uncertainty estimates.
Acceptance is automatic for the heat-bath step, so the acceptance rate
contains no information about mixing. Order frequencies, retained
fractions and comparisons between initializations instead probe how the
chain explores the configuration space.

If sampling weights are changed, an estimator can be reweighted provided
its support is adequate. An integrand cutoff changes the quantity being
integrated and must be included in the assessment of numerical bias.

Each chain stores the count, mean and second central moment of every
order estimator. The paired means determine between-chain covariance.
Table~\ref{c4:mc:codemap} summarizes the operations in
Sections~\ref{c4:free:section}--\ref{c4:mc:section}.

\begin{table}[tbp]
\centering\small
\caption[Construction and sampling of connected coefficients]{Construction and sampling of the connected entropy coefficients.}
\label{c4:mc:codemap}
\begin{tabularx}{\textwidth}{@{}>{\raggedright\arraybackslash}p{0.34\linewidth}X@{}}
\toprule
Stage & Mathematical construction\\
\midrule
Replica measure & Map four unit coordinates to a vertex; include the $2^k$ replica factor in the twisted term.\\
Connected coefficients & Apply the anchored subset recursion separately to the ordinary and twisted determinant weights.\\
Entropy difference & Combine the ordinary and twisted connected contributions before forming absolute sampling weights.\\
First-order reference & Integrate the lower-branch diagonal polynomial and subtract the ordinary contribution.\\
Conditional proposal & Normalize time, site and replica histograms to define sampling and density evaluation.\\
Envelope update & Select a subset by $w_S/W$, regenerate its complement and measure all signed orders after dividing by $\lambda_k$.\\
\bottomrule
\end{tabularx}
\end{table}

The cost of the subset algebra grows exponentially with envelope size.
I therefore use moderate expansion orders and measure each order from
all subsets of the envelope.

\section{Correlated perturbative coefficients}
\label{c4:stats:section}

\subsection{Paired chain means}

Each envelope contributes to all measured orders at once. Its subsets
share vertices and propagators, and the normalized coefficients share
the same stochastic denominator. The resulting correlations require
the complete vector of order means from each independent chain in the
error analysis.

The seven numerical runs each contain $R=128$ separately seeded chains
($896$ in total), with equal production length within each run. Because
production lengths vary between runs, each run is analysed as a separate
ensemble. Let $s_{rk}$ be the production mean for order $k$ in chain $r$,
after removal of the artificial order weight. Define
\begin{equation}
 \bar s_k=\frac1R\sum_{r=1}^{R}s_{rk},\qquad
 \widehat a_0=a_0^{\rm det},\quad
 \widehat a_1=a_1^{\rm det},\quad
 \widehat a_k=a_1^{\rm det}\frac{\bar s_k}{\bar s_1}\quad(k\ge2).
 \label{c4:stats:ratio}
\end{equation}
The deterministic inputs are obtained from the free calculation and
the integrated diagonal observable. They have deterministic representation errors and no Monte Carlo variance.

\subsection{The ratio estimator}

The sampling identity is a ratio of expectations. I estimate it by
first averaging each order over the equal-length chains and then taking
the ratio in Eq.~\eqref{c4:stats:ratio}. Averaging per-chain ratios gives
a different finite-sample bias and variance.

To display the role of correlation, write
$g(x,y)=a_1^{\rm det}x/y$ and expand around the population means
$(\mu_x,\mu_y)$, assuming $\mu_y\ne0$. The leading bias is
\begin{equation}
 \mathbb E\,g(\bar x,\bar y)-g(\mu_x,\mu_y)
 \simeq a_1^{\rm det}
 \left[
 \frac{\mu_x\,\operatorname{Var}\bar y}{\mu_y^3}
 -\frac{\operatorname{Cov}(\bar x,\bar y)}{\mu_y^2}
 \right].
 \label{c4:stats:bias}
\end{equation}
Both denominator variance and numerator--denominator covariance enter
the bias. When the reference mean is too small, the local expansion
can fail and the ratio distribution can develop long tails.

Let $\Sigma$ denote the covariance matrix of the vector of order means. Linearization of all coefficient ratios uses the Jacobian
\begin{equation}
 J_{ki}=a_1^{\rm det}
 \left(\frac{\delta_{ik}}{\bar s_1}
 -\frac{\bar s_k}{\bar s_1^2}\delta_{i1}\right),
 \qquad k\ge2,
 \qquad C\simeq J\Sigma J^{\mathsf T}.
 \label{c4:stats:Jacobian}
\end{equation}
The shared denominator couples all normalized coefficients, adding
correlations to those already present between the sampled order means.
Both contributions are retained by the full covariance propagation.

\subsection{Delete-one-chain reconstruction}

I propagate the nonlinear normalization through a delete-one-chain
jackknife \cite{EfronStein1981}, recomputing every ratio in each replicate. For every omitted
chain $r$, form
\begin{align}
 \bar s_k^{(-r)}
 &=\frac1{R-1}\sum_{j\ne r}s_{jk},\nonumber\\
 a_k^{(-r)}
 &=a_1^{\rm det}\frac{\bar s_k^{(-r)}}{\bar s_1^{(-r)}}
 \quad(k\ge2),\nonumber\\
 \bar{\mathbf a}^{(-)}
 &=\frac1R\sum_{r=1}^{R}\mathbf a^{(-r)}.
 \label{c4:stats:leaveout}
\end{align}
Orders zero and one retain their deterministic input values in every
replicate. The estimated coefficient covariance is
\begin{equation}
 C=\frac{R-1}{R}\sum_{r=1}^{R}
 [\mathbf a^{(-r)}-\bar{\mathbf a}^{(-)}]
 [\mathbf a^{(-r)}-\bar{\mathbf a}^{(-)}]^{\mathsf T}.
 \label{c4:stats:jackknife}
\end{equation}
The factor $(R-1)/R$ accounts for the strong overlap among
leave-one-out datasets.
Each deletion removes a complete paired order vector, including its normalization contribution.

For a smooth statistic, pseudovalues
\begin{equation}
 \mathbf p_r=R\widehat{\mathbf a}-(R-1)\mathbf a^{(-r)}
 \label{c4:stats:pseudo}
\end{equation}
give the equivalent covariance
$[R(R-1)]^{-1}\sum_r(\mathbf p_r-\bar{\mathbf p})
(\mathbf p_r-\bar{\mathbf p})^{\mathsf T}$.
Their mean also yields the usual first-order jackknife bias correction,
$R\widehat{\mathbf a}-(R-1)\bar{\mathbf a}^{(-)}$.
The reported central values use the uncorrected ratio estimator.
The jackknife correction removes only the leading finite-chain bias.

The statistical covariance has zero rows and columns for the fixed
inputs $a_0$ and $a_1$, and their roundoff residues are set to zero.
The deterministic checks of these inputs are the free-reference
comparison in Section~\ref{c4:free:section} and the low-order quadrature
comparison in Table~\ref{c4:bench:lowtable}.

\subsection{Correlation matrices and subsequent observables}

For orders with nonzero variance, define
\begin{equation}
 {\rm Corr}_{ij}=\frac{C_{ij}}{\sqrt{C_{ii}C_{jj}}}.
 \label{c4:stats:corr}
\end{equation}
The deterministic rows are excluded when visualizing correlations.
The matrix should be symmetric and positive semidefinite up to
roundoff, since Eq.~\eqref{c4:stats:jackknife} is a sum of outer
products. A negative eigenvalue exceeding roundoff indicates an error in construction or data alignment.

The raw partial sum through order $p$ and its variance are
\begin{equation}
 F_p^{\rm raw}=\sum_{k=0}^{p}a_k,\qquad
 \operatorname{Var}F_p^{\rm raw}
 =\sum_{i,j=0}^{p}C_{ij}.
 \label{c4:stats:partial}
\end{equation}
Cross terms can increase or decrease the variance of either a partial
sum or a continued estimate. Comparisons between observables derived
from the same coefficient vector likewise require their covariance.

This covariance is conditional on the simulation setup. Errors from
Green-function tables, finite-precision arithmetic, connected-value
filters and reference quadrature are systematic. The settings and selected convergence comparisons are given in
Table~\ref{c4:large:controls}, Table~\ref{c4:bench:lowtable} and
Section~\ref{c4:continue:quadrature}. These comparisons cover only part
of the numerical parameter space. An error in the deterministic
normalization is common to all coefficients and therefore correlates them.

\begin{figure}[htbp]
\centering
\includegraphics[width=\textwidth]{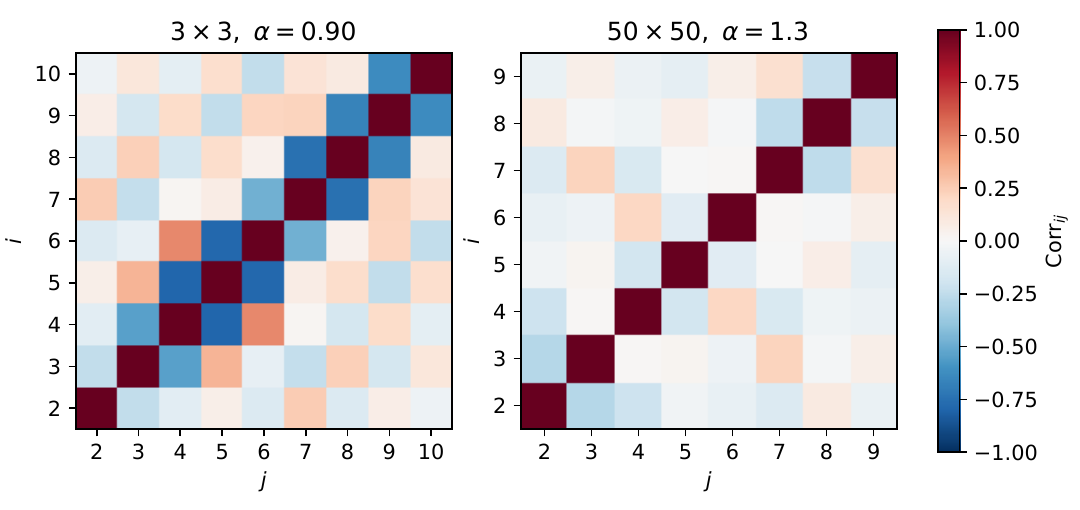}
\caption[Cross-order coefficient correlations]{Cross-order coefficient correlations from the jackknife covariance
for the \mbox{$3\times3$}, $\alpha=0.90$ run and the \mbox{$50\times50$}, $\alpha=1.3$
run. Axes start at order two; the fixed zeroth and first orders are
omitted. Sections~\ref{c4:bench:section} and \ref{c4:large:section}
describe these calculations. }
\label{c4:stats:correlation-figure}
\end{figure}

In Figure~\ref{c4:stats:correlation-figure}, several adjacent orders of
the $3\times3$ series are strongly anticorrelated. The off-diagonal
correlations are weaker in the $50\times50$ run. Negative covariance
reduces part of the noise in a raw partial sum, whose weights are all
positive. Continuation weights can have either sign, so I propagate the
full covariance through the chosen continuation.

\subsection{Chain assumptions and diagnostic limits}
\label{c4:stats:chainchecks}

I form the ratios and delete-one-chain covariance from paired means of
equal-length chains, weighting each chain equally. This treatment
assumes independent, equilibrated chains. It measures between-chain
variation; the analysis contains no time-bin series from which to
estimate integrated autocorrelation times or within-chain drift.
Agreement between initializations therefore provides an indirect
mixing diagnostic. Each continuation below uses the resulting
coefficient vector and its full covariance.

\section{Conformal Borel--Leroy continuation and its uncertainty}
\label{c4:continue:section}

\subsection{The finite series and the endpoint}

The physical answer is at $\xi=1$, while the measured coefficients
describe an expansion around $\xi=0$. Even when the finite-system
partition function is regular along the physical path, complex zeros
and the chosen reference can restrict the convergence of its logarithm.
A partial sum can therefore oscillate strongly. 

For a finite input vector $(a_0,\ldots,a_p)$ define the Borel--Leroy
transform with parameter $b>-1$,
\begin{equation}
 {\cal B}_{b,p}(t)=\sum_{k=0}^{p}
 \frac{a_k t^k}{\Gamma(k+b+1)}.
 \label{c4:continue:borel}
\end{equation}
Formally, a suitable analytic continuation of the complete transform
would reconstruct
\begin{equation}
 f_A(\xi)=\int_0^\infty e^{-t}t^b{\cal B}_b(\xi t)\,\mathrm dt.
 \label{c4:continue:inverse}
\end{equation}
Term-by-term use of the Gamma-function integral recovers each Taylor
coefficient. Interchanging the complete sum and integral additionally
requires convergence or a summability prescription. Integrating the
finite polynomial in Eq.~\eqref{c4:continue:borel} returns the raw partial
sum. Improvement therefore requires re-expansion in a different variable
before truncation, as achieved by the conformal variable below
\cite{RossiThesis2017,LeGuillouZinnJustin1990}.

\subsection{The conformal map and its assumptions}

I use the conformal variable
\begin{equation}
 w(t)=\frac{\sqrt{1+t/t_0}-1}{\sqrt{1+t/t_0}+1},
 \qquad
 t(w)=\frac{4t_0w}{(1-w)^2},\qquad t_0>0 .
 \label{c4:continue:map}
\end{equation}
With the principal square root, the positive integration axis maps
to $0\le w<1$. The cut beginning at $-t_0$ maps to the unit circle.
The branch point built into this map is a representation choice here: the exact finite-system Borel--Leroy transform is entire, as shown next.

At finite lattice size, both partition functions are analytic in
the coupling and nonzero near the reference point, so their chosen
logarithms define an analytic germ. For any circle inside its
analyticity domain, Cauchy's estimate gives
\begin{equation}
 |a_k|\leq\frac{M_R}{R^k},\qquad
 M_R=\max_{|\xi|=R}|f_A(\xi)|,\qquad R>0.
 \label{c4:continue:entire}
\end{equation}
Dividing this bound by $\Gamma(k+b+1)$ makes the Borel--Leroy
series converge for every finite complex argument. The exact
finite-system transform is therefore entire, although it can
have exponential growth. The Laplace integral at the physical endpoint additionally requires a bound on this growth.

In this finite-lattice setting, $t_0$ is therefore an effective parameter
for reorganizing finitely many coefficients.
Changing the reference shift can change the analytic structure of
the series while leaving its physical endpoint unchanged. Scanning
both $b$ and $t_0$ tests sensitivity to the representation.
When the available coefficients leave competing continuations, I report
each branch separately in Section~\ref{c4:bench:section}.

\subsection{The re-expansion matrix}

The inverse map has the series
\begin{equation}
 t(w)=4t_0\sum_{\ell\ge1}\ell w^\ell,\qquad
 t(w)^k=(4t_0)^k
 \sum_{j\ge k}\binom{k+j-1}{j-k}w^j,\quad k\ge1 .
 \label{c4:continue:powers}
\end{equation}
The second identity is the binomial expansion of
$w^k(1-w)^{-2k}$. Retaining terms through degree $p$ gives
\begin{equation}
 {\cal B}_{b,p}^{\rm conf}(t)
 =\sum_{j=0}^{p}d_j w(t)^j,
 \quad
 d_0=\frac{a_0}{\Gamma(b+1)},\quad
 d_j=\sum_{k=1}^{j}
 \frac{a_k(4t_0)^k}{\Gamma(k+b+1)}
 \binom{k+j-1}{j-k}.
 \label{c4:continue:coeffmap}
\end{equation}
All sums in this expression are finite. Each monomial of the input
series contributes to every allowed higher power of $w$.
Truncation in $w$ defines a new approximant after mapping back to $t$;
integrating the original degree-$p$ polynomial would return the raw
partial sum.

For instance, through degree two,
\begin{equation}
 d_1=\frac{4t_0a_1}{\Gamma(b+2)},\qquad
 d_2=\frac{8t_0a_1}{\Gamma(b+2)}
 +\frac{16t_0^2a_2}{\Gamma(b+3)}.
 \label{c4:continue:example}
\end{equation}
The transformed second-order coefficient contains both $a_1$ and $a_2$.
In general its uncertainty includes their covariance. Here $a_1$ is a
deterministic input, so only the variance of $a_2$ contributes to its
Monte Carlo uncertainty.

\subsection{Quadrature and weights on the original coefficients}
\label{c4:continue:quadrature}

Define
\begin{equation}
 J_j(b,t_0)=\int_0^\infty e^{-t}t^b w(t)^j\,\mathrm dt .
 \label{c4:continue:J}
\end{equation}
Then the continued endpoint is a linear functional,
\begin{equation}
 F_p(b,t_0)=\sum_{k=0}^{p}\kappa_k^{(p)}(b,t_0)a_k,
 \label{c4:continue:linear}
\end{equation}
with
\begin{align}
 \kappa_0^{(p)}&=1,\nonumber\\
 \kappa_k^{(p)}
 &=\frac{(4t_0)^k}{\Gamma(k+b+1)}
 \sum_{j=k}^{p}\binom{k+j-1}{j-k}J_j(b,t_0),
 \qquad k\ge1.
 \label{c4:continue:kappa}
\end{align}
The identity $J_0=\Gamma(b+1)$ reproduces the constant term and tests
the quadrature and the free anchor.

I evaluate $J_j$ by generalized Gauss--Laguerre quadrature with weight
$e^{-t}t^b$ and $140$ nodes, checking quadrature convergence with $100$
and $180$ nodes. Truncated polynomial convolution implements the
triangular transformation in Eq.~\eqref{c4:continue:coeffmap}.

The vector $\boldsymbol\kappa^{(p)}(b,t_0)$ depends only on $p,b,t_0$;
the physical coupling and shifts are already contained in the input
coefficients. It is computed once per choice and applied to coefficient
means, their covariance and every jackknife replicate.

\subsection{Statistical error for one map and a fixed window}

For one fixed map,
\begin{equation}
 \sigma_{\rm MC}^2(b,t_0)
 =\boldsymbol\kappa^{\mathsf T}C\boldsymbol\kappa .
 \label{c4:continue:variance}
\end{equation}
For a given $C$, Eq.~\eqref{c4:continue:variance} follows exactly from
the linearity of the functional.
The covariance estimator itself still relies on the chain assumptions
described in Section~\ref{c4:stats:section}.

Let a fixed window contain $M>1$ parameter pairs $\theta_\ell=(b_\ell,t_{0\ell})$.
For equally weighted window averaging,
\begin{equation}
 \bar{\boldsymbol\kappa}=\frac1M\sum_{\ell=1}^{M}
 \boldsymbol\kappa(\theta_\ell),\qquad
 \bar F=\bar{\boldsymbol\kappa}^{\mathsf T}\mathbf a,\qquad
 \sigma_{\rm MC,window}^2=
 \bar{\boldsymbol\kappa}^{\mathsf T}C\bar{\boldsymbol\kappa}.
 \label{c4:continue:window}
\end{equation}
Different maps share the same random coefficients:
$\operatorname{Cov}(F_\ell,F_m)=
\boldsymbol\kappa_\ell^{\mathsf T}C\boldsymbol\kappa_m$.
The variance of their mean is the quadratic form in
Eq.~\eqref{c4:continue:window}, including these cross-covariances.

The separately reported scan dispersion is
\begin{equation}
 \sigma_{\rm scan}=
 \left[\frac1{M-1}\sum_{\ell=1}^{M}(F_\ell-\bar F)^2\right]^{1/2}.
 \label{c4:continue:scan}
\end{equation}
The scan dispersion is the standard deviation across the chosen maps,
reported without division by $\sqrt M$. It measures variation within the chosen window. Its relation to the
unknown continuation bias has no statistical calibration.

\subsection{Truncation drift and nonunique branches}
\label{c4:continue:branches}

A flat region shared by adjacent truncation orders can indicate local
stability, yet two disjoint regions may predict different endpoints with
small internal scan dispersions. This occurs in the small-cluster results
in Section~\ref{c4:bench:section}.

I assess each small-system branch and large-lattice window through its
internal flatness and adjacent-order stability, then propagate the
coefficient covariance at fixed window. The resulting Monte Carlo error
is conditional on that choice. ED comparisons, shift dependence, scan
dispersion and order drift assess the remaining continuation sensitivity;
selection uncertainty and continuation bias have no calibrated error
estimate here. The ED comparison is retrospective.

\subsection{Numerical integration versus continuation systematics}

More samples reduce stochastic uncertainty, and finer quadrature
improves the numerical evaluation of a fixed approximant. Higher-order
coefficients are needed to change the information available for
continuation. I assess the remaining endpoint sensitivity through order
dependence, parameter windows and the ED comparison.

\section{Independent small-system benchmarks and continuation ambiguity}
\label{c4:bench:section}

\subsection{A benchmark with a known numerator and denominator}

The small system has $N_1=N_2=3$, $U=2$, $\mu=0$, $\beta=8$,
and $A$ equal to the first row of three sites. Three reference
expansions use $\alpha_\uparrow=\alpha_\downarrow=0.80,0.85,0.90$.
The three expansions have distinct Taylor coefficients at $\xi=0$
and a common physical Hamiltonian at $\xi=1$. The subsystem and full
spinful Hilbert spaces have dimensions $4^3=64$ and $4^9=262144$,
respectively. Conservation of particle number in each spin sector and
periodic translations reduce the cost of diagonalizing the full problem.

Exact diagonalization gives the reduced state directly. Deterministic
quadrature gives the first three connected coefficients. The Monte Carlo series reaches order ten, allowing both
the sampled low orders and the continued endpoint to be compared with
these references.

\subsection{Block diagonalization and the partial trace}

In a sector with fixed $(N_\uparrow,N_\downarrow)$, the occupation
basis has dimension
\begin{equation}
 d_{N_\uparrow,N_\downarrow}
 =\binom{N_s}{N_\uparrow}\binom{N_s}{N_\downarrow}.
 \label{c4:bench:blocksize}
\end{equation}
The largest particle-number sectors for nine sites, with four or five particles of each spin, have dimension $126^2=15876$.
Momentum blocks reduce the dense eigenproblems further.
I construct translation orbits with their fermionic translation signs
and project them into momentum sectors.

Let $\{E_\lambda(\xi),|\lambda;\xi\rangle\}$ denote the complete
eigensystem, including all contributing number and momentum blocks.
Then
\begin{align}
 Z_1(\xi)&=\sum_\lambda e^{-\beta E_\lambda(\xi)},\nonumber\\
 [R_A(\xi)]_{aa'}
 &=\sum_{\lambda,b}
 e^{-\beta E_\lambda(\xi)}
 \langle a,b|\lambda;\xi\rangle
 \langle\lambda;\xi|a',b\rangle.
 \label{c4:bench:EDtrace}
\end{align}
The same complement configuration $b$ occurs in the bra and ket.
The ED calculation groups basis entries by the complement configuration before
accumulating the reduced matrix. Contributions from different
sectors are added to the unnormalized $R_A$ before its square
is traced:
\begin{equation}
 \widetilde Z_2[A;\xi]=\operatorname{Tr}[R_A(\xi)^2],
 \qquad
 f_A^{\rm ED}(\xi)=\frac{
 \ln\operatorname{Tr}[R_A(\xi)^2]-2\ln Z_1(\xi)}{N_s}.
 \label{c4:bench:EDmoment}
\end{equation}
Summing before squaring includes the cross terms between sectors that
occupy the same subsystem states.

An occupation basis also fixes the subsystem tensor convention.
The ED basis groups all up-spin modes and then all down-spin modes,
with site index $yN_1+x$. For the first-row cut, $A$ consists of the contiguous sites $0,1,2$. The mode order is consequently
$A_\uparrow,B_\uparrow,A_\downarrow,B_\downarrow$.
Moving the $A_\downarrow$ modes before $B_\uparrow$ produces the
regrouping phase
\begin{equation}
 (-1)^{N_{B\uparrow}N_{A\downarrow}}.
 \label{c4:bench:regroup-phase}
\end{equation}
Within a fixed spin-number sector and a fixed complement
configuration, both numbers in this exponent are fixed. The
phase is therefore identical for the two coefficients in a
partial-trace matrix element and cancels. The partial trace for the first-row cut can therefore omit this
regrouping sign, provided the subsystem basis order is kept consistent.
The ED implementation encodes each subsystem site by its local up/down
occupation. Relating this encoding to a subsystem ordering with all up-spin modes first introduces an additional basis transformation confined
to $A$ (and independently to $B$). The $B$ transformation cancels
in the partial trace, while the $A$ transformation conjugates
$R_A$ and leaves $\operatorname{Tr}R_A^2$ invariant. It must be
applied consistently when matrix elements themselves, rather
than only the purity, are compared.

Particle-number conservation alone would not establish this
cancellation for an arbitrary noncontiguous region. For example,
take four spinless modes ordered $0,1,2,3$, with $A=\{0,2\}$ and
$B=\{1,3\}$. An equal superposition of the occupation states
$\{0,1\}$, $\{0,3\}$, $\{1,2\}$ and $\{2,3\}$ has fixed total
particle number two and one particle in each region. An unsigned
split gives purity one, whereas the correctly signed
$A$-before-$B$ regrouping gives purity one half. The first-row
benchmark does not have this obstruction; applying the same ED partial trace to other regions requires either
including the regrouping sign or proving that it cancels for the new
ordering.

Evaluating the ED partial trace directly from many-body eigenvectors
provides an independent representation of the reduced state. Agreement
with the diagrammatic coefficients therefore tests the sewing, shifted
insertion and normalization together.

\subsection{Finite-difference derivatives of the shifted family}

The coefficient to compare is
$a_k^{\rm ED}=(k!)^{-1}\partial_\xi^k f_A^{\rm ED}(0)$.
The derivative is taken along the shifted family: $K_0$ stays fixed,
and $\xi$ changes both the two-body interaction and its compensating
one-body term.

Let the integer stencil offsets be $j=-r,\ldots,r$ and step size
$\delta$. Weights $w_j^{(k)}$ are determined by
\begin{equation}
 \sum_{j=-r}^{r}w_j^{(k)}j^m
 =k!\,\delta_{mk},\qquad 0\le m\le2r .
 \label{c4:bench:FDweights}
\end{equation}
The moment equations follow by substituting a Taylor expansion:
they annihilate all represented powers except the desired one.
The coefficient estimate is
\begin{equation}
 a_k^{\rm ED}(\delta)=
 \frac{1}{k!\,\delta^k}
 \sum_{j=-r}^{r}w_j^{(k)}f_A^{\rm ED}(j\delta).
 \label{c4:bench:FD}
\end{equation}
If the unnormalized logarithmic difference is used in place of
$f_A^{\rm ED}$, one must additionally divide by $N_s$.

The benchmark uses a stencil radius $r=4$ and step sizes
$\delta=0.01,0.007,0.003,0.001$. The reported comparison uses
$\delta=0.007$.
Stencil symmetry implies $w_{-j}^{(k)}=(-1)^k w_j^{(k)}$.
For this nine-point stencil the first unconstrained, nonvanishing
Taylor moment gives errors of order $\delta^8$ for first and
second derivatives and $\delta^6$ for the third derivative,
provided the function is sufficiently smooth and the weights
are evaluated accurately.

Roundoff has the opposite tendency. If each function value has
an absolute numerical error of scale $\varepsilon_f$, a
worst-case derivative error is proportional to
$\varepsilon_f\sum_j|w_j^{(k)}|/\delta^k$.
A stable interval of step sizes balances truncation and roundoff; the
smallest tested step is already roundoff limited.
The ED uncertainty reflects finite-difference truncation and roundoff.

\subsection{Deterministic summation of the discrete labels}

An interaction vertex on the replicated \mbox{$3\times3$} lattice has
$18$ possible discrete site/sheet labels. At order $k$, the
benchmark sums over all these labels.
For the first-row subsystem, translation along the periodic row
and simultaneous replica translation leave the full integrand
unchanged. I check this symmetry numerically before using the sixfold reduction.

Fixing the first vertex's horizontal coordinate and replica leaves
$3\times18^{k-1}$ representative label assignments. Restoring
their orbit multiplicity and the unit-coordinate cell weights
gives the same discrete average as summing all $18^k$ assignments.
The reduction uses translations along the periodic row. The entangling
row breaks perpendicular translation invariance, and the temporal seam
prevents fixing a global imaginary time.

A midpoint unit coordinate selects the site and sheet label of each
discrete cell. Those sums are finite and exact; the remaining
approximations are propagator evaluation, arithmetic and continuous-time
quadrature.

\subsection{Ordered time simplices and their Jacobian}
\label{c4:bench:simplex-section}

Simultaneous permutations of complete interaction vertices leave
the integrand unchanged. The time cube can therefore be divided
into $k!$ equivalent ordered simplices. For normalized ordered
times $0<s_1<\cdots<s_k<1$, use
\begin{equation}
 s_j=\prod_{\ell=j}^{k}u_\ell,\qquad
 0<u_\ell<1,\qquad
 \left|\frac{\partial(s_1,\ldots,s_k)}
 {\partial(u_1,\ldots,u_k)}\right|
 =\prod_{\ell=2}^{k}u_\ell^{\ell-1}.
 \label{c4:bench:simplex}
\end{equation}
For two vertices this is $s_1=u_1u_2$, $s_2=u_2$, with
Jacobian $u_2$. For three vertices the Jacobian is $u_2u_3^2$.
The general expression follows because the Jacobian matrix is triangular when the coordinates are ordered in reverse.

The substitution $\tau_j=\beta s_j$ contributes $\beta^k$ to a
physical-time integral. Here that factor is already part of the
coefficient integrand on normalized coordinates. Quadrature therefore uses the Jacobian in Eq.~\eqref{c4:bench:simplex}
and the $k!$ simplex count to restore the full time cube. The
perturbative $1/k!$ is already included in the coefficient integrand.

Tensor-product Gauss--Legendre nodes are applied on the transformed
unit cube. Separating the time-ordering domains keeps the quadrature
within smooth branches. The tables use $18$ nodes per coordinate and
quote the absolute difference from $16$ nodes as an indicator of
quadrature convergence.

\begin{table}[htbp]
\centering\small
\begin{tabular}{ccrrr}
\toprule
$\alpha$&$k$&Deterministic $a_k$&$|a_k^{18}-a_k^{16}|$&$a_k^{18}-a_k^{\rm ED}$\\
\midrule
$0.80$&1&$-0.015959696296$&$8.35\times10^{-11}$&$2.25\times10^{-12}$\\
$0.80$&2&$0.014333628280$&$7.63\times10^{-8}$&$-5.18\times10^{-9}$\\
$0.80$&3&$0.013050002039$&$5.86\times10^{-8}$&$-2.99\times10^{-9}$\\
$0.85$&1&$-0.007194809377$&$7.16\times10^{-11}$&$1.24\times10^{-12}$\\
$0.85$&2&$0.014576117985$&$9.01\times10^{-8}$&$-6.55\times10^{-9}$\\
$0.85$&3&$-0.003545370122$&$7.42\times10^{-8}$&$2.21\times10^{-9}$\\
$0.90$&1&$-0.002312973195$&$5.65\times10^{-11}$&$5.07\times10^{-13}$\\
$0.90$&2&$0.021770951384$&$1.06\times10^{-7}$&$-7.78\times10^{-9}$\\
$0.90$&3&$-0.044639011175$&$1.09\times10^{-7}$&$8.54\times10^{-9}$\\
\bottomrule
\end{tabular}
\caption[Deterministic coefficients and ED comparisons]{Deterministic low-order coefficients and ED comparisons. Superscripts
18 and 16 denote the Gauss--Legendre node count in each transformed time
coordinate; the ED derivative step is $0.007$. Extra digits are shown
to permit comparisons between columns.}
\label{c4:bench:lowtable}
\end{table}

Across Table~\ref{c4:bench:lowtable}, the largest absolute difference
from ED is $8.54\times10^{-9}$ and the largest $16$-to-$18$-node change
is $1.09\times10^{-7}$. Each ED difference is smaller than its
corresponding quadrature indicator.

\subsection{The complete series and its independent endpoint}

\begin{figure}[p]
\centering
\includegraphics[width=\textwidth]{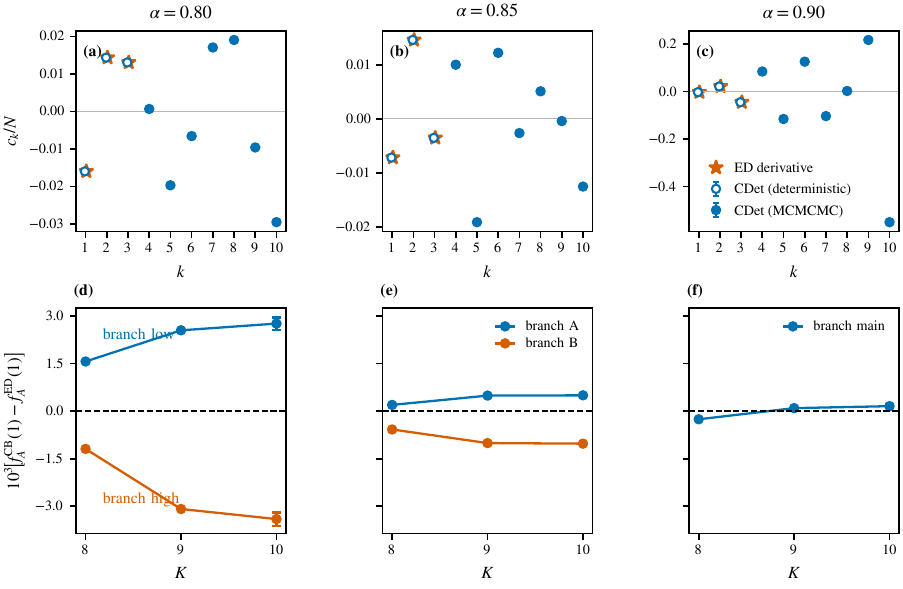}
\caption[Small-system entropy benchmark]{Periodic \mbox{$3\times3$} Hubbard lattice at $U=2$, $\mu=0$, $\beta=8$, with
the first row as subsystem. Reproduced from my calculation in
Ref.~\cite{ShiRenyi2026}. Panels (a)--(c): open circles give deterministic
CDet coefficients at orders 1--3, stars give ED derivatives, and filled
circles give MCMCMC estimates at orders 4--10. Panels (d)--(f): endpoint
deviations from ED, multiplied by $10^3$, using deterministic $a_0,a_1$
and MCMCMC orders 2--10 with full covariance. The bar at order ten on
each branch shows $\sigma_{\rm scan}$. In (d), blue labels the ``low''
branch and orange the ``high'' branch. The axis labels $N$ and $K$
correspond to $N_s$ and $p$. Zero in panels (d)--(f) is the ED endpoint. At $\alpha=0.80$ and $0.85$,
locally flat branches lie on opposite sides of that value.}
\label{c4:bench:original}
\end{figure}

The upper panels of Figure~\ref{c4:bench:original} display deterministic
coefficients at orders one to three and MCMCMC estimates at orders four
to ten. Changing the shift reorganizes the series. In Figures~\ref{c4:bench:original}
and \ref{c4:large:original}, reproduced from my work in
Ref.~\cite{ShiRenyi2026}, the axis labels $c_k/N$, $N$ and $K$ denote
$a_k$, $N_s$ and the retained order $p$.

The continuations in the lower panels and
Table~\ref{c4:bench:branches} use deterministic $a_0,a_1$ and \mbox{MCMCMC}
coefficients at orders two to ten with their full chain covariance.
The deterministic second- and third-order values in the upper panels
serve as independent checks.

Table~\ref{c4:bench:mc-low} compares the six second- and third-order
Monte Carlo estimates with deterministic quadrature. At $\alpha=0.80$,
the third-order estimate lies $2.755$ Monte Carlo standard errors below
the deterministic result. The quadrature indicator is
$5.86\times10^{-8}$ and the difference from the ED derivative is
$2.99\times10^{-9}$, both much smaller than the Monte Carlo standard
error of $1.37\times10^{-6}$.

\begin{table}[htbp]
\centering\small
\begin{tabular}{ccrrr}
\toprule
$\alpha$&$k$&MCMCMC $\widehat a_k$&$\sigma_{{\rm MC},k}$&
$(\widehat a_k-a_k^{18})/\sigma_{{\rm MC},k}$\\
\midrule
$0.80$&2&$0.014333858016$&$7.522\times10^{-6}$&$0.031$\\
$0.80$&3&$0.013046226166$&$1.371\times10^{-6}$&$-2.755$\\
$0.85$&2&$0.014573215172$&$4.928\times10^{-6}$&$-0.589$\\
$0.85$&3&$-0.003545380573$&$4.415\times10^{-7}$&$-0.024$\\
$0.90$&2&$0.021777223321$&$7.015\times10^{-6}$&$0.894$\\
$0.90$&3&$-0.044640924721$&$2.554\times10^{-6}$&$-0.749$\\
\bottomrule
\end{tabular}
\caption[MC coefficients used in small-system continuation]{Low-order Monte Carlo estimates used in the small-system continuation. The deterministic comparison values
$a_k^{18}$ are given in Table~\ref{c4:bench:lowtable}. The last column gives the difference in units of the marginal Monte Carlo standard error.}
\label{c4:bench:mc-low}
\end{table}

The six comparisons use marginal errors of correlated coefficient
estimates. The $2.755\sigma$ deviation at $\alpha=0.80$, $k=3$ alone
cannot distinguish statistical fluctuation from sampling bias. The
quadrature indicator is a deterministic convergence measure; a joint
assessment needs the full comparison covariance and an error model for
the reference. The limits of the chain diagnostics are discussed in
Section~\ref{c4:stats:chainchecks}.

The deterministic coefficients in Figure~\ref{c4:bench:original} agree
with the ED derivatives near the free reference. At the interacting endpoint, locally stable continuations
can predict different entropies. The branches are shown individually so
that their displacement from ED can be compared with their Monte Carlo
errors and internal scan dispersion.

The direct endpoint is
\begin{equation}
 f_A^{\rm ED}(1)=-0.275249813784967.
 \label{c4:bench:endpoint}
\end{equation}
Table~\ref{c4:bench:branches} lists the endpoints at the fixed parameter
windows. The error bars are conditional on these windows, as discussed in
Section~\ref{c4:continue:branches}.

\begin{table}[htbp]
\centering\small
\begin{tabular}{clrrrr}
\toprule
$\alpha$&Branch&$f_A^{\rm CB}(1)$&$\sigma_{\rm MC}$&$\sigma_{\rm scan}$&$f_A^{\rm CB}-f_A^{\rm ED}$\\
\midrule
$0.80$&low&$-0.2724919348$&$6.27\times10^{-6}$&$1.90\times10^{-4}$&$2.76\times10^{-3}$\\
$0.80$&high&$-0.2786528680$&$6.41\times10^{-6}$&$2.09\times10^{-4}$&$-3.40\times10^{-3}$\\
$0.85$&A&$-0.2747537476$&$4.26\times10^{-6}$&$5.07\times10^{-5}$&$4.96\times10^{-4}$\\
$0.85$&B&$-0.2762757346$&$4.36\times10^{-6}$&$6.56\times10^{-5}$&$-1.03\times10^{-3}$\\
$0.90$&main&$-0.2750946036$&$6.47\times10^{-6}$&$5.71\times10^{-5}$&$1.55\times10^{-4}$\\
\bottomrule
\end{tabular}
\caption[Small-system continuation branches]{Tenth-order conformal Borel--Leroy results at fixed parameter windows.
$\sigma_{\rm MC}$ is propagated from the coefficient covariance and
$\sigma_{\rm scan}$ is the dispersion over the window. The ED reference
is Eq.~\eqref{c4:bench:endpoint}.}
\label{c4:bench:branches}
\end{table}

At $\alpha=0.80$, two locally stable branches lie on opposite sides of
the ED value, separated by much more than their Monte Carlo errors or
internal scan dispersions. The two branches at $\alpha=0.85$ are also
incompatible. The ten available orders do not distinguish these alternatives.

The $\alpha=0.90$ series has one selected continuous branch. Its
tenth-order value differs from ED by $1.55\times10^{-4}$, about $0.056\%$
of the endpoint magnitude. This discrepancy is about fifteen times the
order-nine to order-ten drift, $1.05\times10^{-5}$, and about three times
the scan dispersion, despite agreement of the low-order coefficients.

\section{Large-lattice results and computational cost}
\label{c4:large:section}

\subsection{Geometry and run parameters}

The larger systems share $U=3$, $\mu=1$ and $\beta=8$, away from the
particle-hole-symmetric chemical potential $U/2$. For \mbox{$40\times40$}, $A$ is the $16\times16$ square centred on the
origin, with $r_A=8$ and $|A|=256$ sites. For \mbox{$50\times50$},
$r_A=13$ gives a $26\times26$ square with $|A|=676$. The side length is
$2r_A$ in the periodic-coordinate convention.

Because both the lattice and the region grow between these geometries,
the comparison concerns qualitative size trends. Within each geometry,
shift and order dependence provide the direct tests of endpoint
stability; a fixed-region finite-size extrapolation requires additional
sizes.

\begin{table}[htbp]
\centering\small
\begin{tabular}{ccccrr}
\toprule
Lattice&$\alpha$&$|A|$&$p$&Envelope&Production/chain\\
\midrule
3$\times$3&$0.80$&3&10&11&300\\
3$\times$3&$0.85$&3&10&11&500\\
3$\times$3&$0.90$&3&10&11&300\\
40$\times$40&$1.30$&256&9&10&300\\
40$\times$40&$1.40$&256&7&8&300\\
50$\times$50&$1.30$&676&9&10&300\\
50$\times$50&$1.40$&676&7&8&400\\
\bottomrule
\end{tabular}
\caption[Parameters of the seven R\'enyi-entropy Monte Carlo runs]{Parameters of the seven R\'enyi-entropy Monte Carlo runs. Each run has $128$ chains of equal
measurement length, preceded by ten million warm-up steps per chain.
The final column gives measurement lengths in millions of steps.
Small systems use $U=2$, $\mu=0$, $\beta=8$; large systems use $U=3$,
$\mu=1$, $\beta=8$.}
\label{c4:large:runs}
\end{table}

\begin{table}[htbp]
\centering\small
\begin{tabular}{ccrrr}
\toprule
Lattice&$\alpha$&Principal-minor tolerance&Subtraction safety&Dropped bits\\
\midrule
3$\times$3&$0.80$&$1\times10^{-7}$&$1\times10^{4}$&0\\
3$\times$3&$0.85$&$1\times10^{-5}$&$1\times10^{1}$&0\\
3$\times$3&$0.90$&$1\times10^{-7}$&$1\times10^{4}$&0\\
40$\times$40&$1.30$&$1\times10^{-6}$&$1\times10^{4}$&0\\
40$\times$40&$1.40$&$1\times10^{-6}$&$1\times10^{4}$&0\\
50$\times$50&$1.30$&$1\times10^{-6}$&$1\times10^{1}$&0\\
50$\times$50&$1.40$&$1\times10^{-6}$&$1\times10^{1}$&10\\
\bottomrule
\end{tabular}
\caption[Numerical tolerances and stored-table precision]{Numerical controls. The principal-minor tolerance applies to the pivot
procedure of Section~\ref{c3:fpm:section}; the subtraction safety factor
is $\eta_{\rm safe}$ in Eq.~\eqref{c4:cdet:roundoffproxy}. The last column
gives the number of mantissa bits dropped from the stored tables.}
\label{c4:large:controls}
\end{table}

All runs use \mbox{$25\times25$} Chebyshev coefficients per time branch, and
the envelope contains one more vertex than the highest measured order.
Pilot histogram learning and order-weight adaptation precede measurement.

Some numerical settings vary between shifts
(Table~\ref{c4:large:controls}). The \mbox{$3\times3$}, $\alpha=0.85$ run uses
a different principal-minor tolerance and subtraction safety factor;
the \mbox{$50\times50$}, $\alpha=1.4$ run drops ten table mantissa bits.
Those comparisons combine a change of shift with a change of numerical
settings, so their differences cannot be attributed to the shift alone.

\subsection{Endpoint estimates and common-order comparisons}

\begin{figure}[p]
\centering
\includegraphics[width=\textwidth]{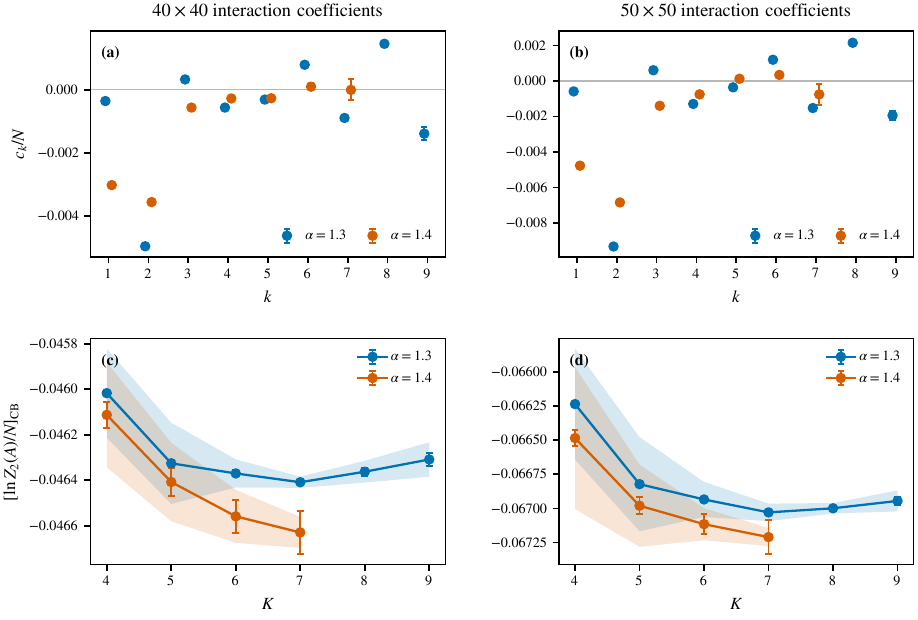}
\caption[Large-lattice entropy benchmark]{Large-lattice entropy results at $U=3$, $\mu=1$, $\beta=8$, reproduced
from my calculation in Ref.~\cite{ShiRenyi2026}. The \mbox{$40\times40$} region
contains $256$ sites and the \mbox{$50\times50$} region $676$ sites. Panels
(a),(b) show shifted coefficient series; (c),(d) show conformal
Borel--Leroy estimates versus retained order. Error bars are obtained by
propagating the complete equal-chain jackknife covariance. Shaded bands
show $\sigma_{\rm scan}$ over the fixed parameter window.
The axis labels $c_k/N$ and $K$ denote $a_k$ and $p$. }
\label{c4:large:original}
\end{figure}

Figure~\ref{c4:large:original} shows the large-lattice results.
For each size, shift $1.3$ reaches order nine and shift $1.4$
reaches order seven. The common parameter window consists of
$b=0.5,1,2,3$ and $t_0=5,10,20$, hence twelve parameter pairs.
It is held fixed across the displayed truncations.

For these lattices, I assess endpoint stability by comparing the two
shifts at common order seven and then following the $\alpha=1.3$
sequence through ninth order in Figure~\ref{c4:large:original}.
The common-order comparison isolates the effect of changing the
coefficient series, subject to the numerical-setting differences noted
above; the longer sequence probes the information added by higher
orders. No ED endpoint enters this comparison.

\begin{table}[htbp]
\centering\small
\begin{tabular}{cccrrr}
\toprule
Lattice&$\alpha$&$p$&$f_A^{\rm CB}(1)$&$\sigma_{\rm MC}$&$\sigma_{\rm scan}$\\
\midrule
40$\times$40&$1.3$&9&$-0.046308434$&$2.82\times10^{-5}$&$7.56\times10^{-5}$\\
40$\times$40&$1.4$&7&$-0.046628988$&$9.54\times10^{-5}$&$6.82\times10^{-5}$\\
50$\times$50&$1.3$&9&$-0.066945364$&$3.04\times10^{-5}$&$7.67\times10^{-5}$\\
50$\times$50&$1.4$&7&$-0.067209861$&$1.25\times10^{-4}$&$6.63\times10^{-5}$\\
\bottomrule
\end{tabular}
\caption[Large-lattice results after continuation at the highest computed order]{Continued large-lattice results at the highest retained order of each run. $\sigma_{\rm MC}$ is obtained from the full coefficient covariance, and $\sigma_{\rm scan}$ is the dispersion over the fixed twelve-point parameter window.}
\label{c4:large:endpoints}
\end{table}

At \mbox{$40\times40$}, the highest-order estimates differ by
$3.21\times10^{-4}$, but this comparison changes both shift and retained
order. At common order seven, the central values are $-0.046408$ and
$-0.046629$, separated by $2.21\times10^{-4}$. The $\alpha=1.3$ sequence
is comparatively flat from orders seven to nine, although
$\sigma_{\rm scan}$ increases at order nine.

At \mbox{$50\times50$}, the highest-order central values are $-0.066945$ and
$-0.067210$. At common order seven they are $-0.067029$ and $-0.067210$,
separated by $1.81\times10^{-4}$. In both geometries the shift dependence
exceeds the quoted Monte Carlo uncertainty and gives an empirical
indication of finite-order sensitivity.

Table~\ref{c4:large:endpoints} gives logarithmic moments per physical
site. At $\alpha=1.3$, the two central values correspond to total
entropies $S_2(A)\simeq74.09$ and $167.36$. Their Monte Carlo errors are
about $0.045$ and $0.076$, and their parameter-window dispersions about
$0.121$ and $0.192$, respectively. These are finite-temperature entropies
of different regions, so they do not on their own determine a
zero-temperature area-law or Fermi-surface logarithmic coefficient.

\subsection{Memory and construction cost}

The seam removes general translation invariance from the
replicated kernel. At fixed time resolution the table is indexed by absolute site pairs. With two replicas there are
\begin{equation}
 N_{\rm pair}=4N_s^2:
 \qquad 10.24\times10^6\quad(40\times40),\qquad
 25.0\times10^6\quad(50\times50).
 \label{c4:large:pairs}
\end{equation}
Their ratio is $(2500/1600)^2\simeq2.44$. At \mbox{$25\times25$} time
resolution, one spin's two-branch twisted table contains $12.8\times10^9$
real coefficients on \mbox{$40\times40$} and $31.25\times10^9$ on \mbox{$50\times50$}.
At eight bytes per coefficient, these arrays alone occupy $102.4$ and
$250$ GB (decimal units), before corner arrays, other tables and
construction workspace are included.

The dense preparation includes cubic operations for diagonalization,
sector solves and basis transforms (Section~\ref{c4:tables:section}).
I build the tables in parallel over blocks of time nodes and site pairs,
which limits memory per job, and hold the completed tables in memory
during sampling.

With the tables in memory, the connected-subset calculation on sampled
vertices dominates the order-dependent arithmetic. Increasing lattice
size mainly increases table construction and storage costs; reaching
higher diagram orders increases the sampling and continuation costs.

\subsection{Comparison with volume-driven sign degradation}

For an ordinary reweighted configuration-space calculation,
the average sign can be written
\begin{equation}
 \langle s\rangle_{|w|}
 =\frac{Z}{Z_{|w|}}
 =e^{-\beta N_s\Delta f}.
 \label{c4:large:sign}
\end{equation}
When $\Delta f$ remains nonzero, obtaining fixed relative
precision by direct sign reweighting requires exponentially growing effort \cite{TroyerWiese2005}. Replica observables can also
introduce poor overlap between sewn and ordinary ensembles
\cite{WangTroyer2014,BroeckerTrebst2014}.

I sum connected contractions at fixed vertex sets before sampling the
normalized replica difference, so each sampled coefficient already
includes these cancellations. The remaining costs arise from coefficient
variance, the order needed at the physical coupling, continuation and
the free tables. At $\beta=8$, the calculations reach tenth order at
$U=2$ on $3\times3$ and ninth order at $U=3$ on $40\times40$ and
$50\times50$. Lower temperatures or stronger coupling can require
higher order and wider arithmetic.

These costs depend differently on temperature, coupling and geometry
from the sign and overlap costs of auxiliary-field methods. A quantitative
comparison therefore requires a common Hamiltonian, temperature,
ensemble, region and total error. The results here establish accessible
orders and table costs, with endpoint accuracy assessed by ED on the
small cluster and by order and shift dependence on the larger lattices.

\subsection{Relation to experiment and the preceding chapters}

Replica moments and subsystem purities are natural targets
of two-copy interference and randomized-measurement protocols
\cite{Pichler2013,Islam2015,Elben2018}.
A numerical comparison must use the same observable,
temperature, filling, lattice and region definition.
At finite temperature, subtracting a thermal contribution or
forming a mutual-information combination requires additional
calculations and their covariance.

As in Chapter~\ref{ch:benchmarks}, the benchmark compares the same
normalized observable in a regime accessible to both calculation and
experiment. The modified Gaussian boundary kernel makes the
connected-subset formulation of Chapter~3 applicable in principle to this
nonlocal observable. I evaluated the coefficients reported here with CDet.

\subsection{Accuracy of the continued entropy}

On the $3\times3$ cluster, the deterministic first three coefficients
agree with ED to better than $10^{-8}$, while the continuations at
$\alpha=0.80$ and $0.85$ give incompatible entropy branches. The
third-order Monte Carlo estimate at $\alpha=0.80$ differs from the
deterministic result by $2.755\sigma$ (Table~\ref{c4:bench:mc-low});
this is a separate coefficient-sampling discrepancy. On the larger
lattices, the two shifts differ at common order seven by
$2.21\times10^{-4}$ and $1.81\times10^{-4}$. The reported endpoints
therefore require their order and shift comparisons in addition to the
propagated Monte Carlo errors.

\chapter{Conclusions and Outlook}
\label{ch:conclusions}

\section{Main results}
\label{c5:v7:results}

\subsection{Inverse-flavour thermodynamics}

The collective $G$--$\Sigma$ formulation in Chapter~\ref{c2:chapter}
organizes the thermodynamics of interacting lattice fermions around a
Hartree reference. Screening enters through its Gaussian fluctuations,
and the remaining terms form a hierarchy in $1/N_f$. The first
non-Gaussian correction consists of a quartic trace vertex and a pair
of cubic trace vertices, which give five connected topological
representatives. Their weights and momentum constraints specify the
Monte Carlo integrals. Using a discrete Lehmann representation makes
selected internal-time integrations analytic and reduces the number of
variables left for sampling.

The density calculation covers 31 state points of the six-flavour
square-lattice Hubbard model on $70\times70$ sites at $T/t=0.15$
(Table~\ref{tab:table3-author-data}; Ref.~\cite{ShiLargeNf2025}). The
Gaussian increment is positive at every point. The central value of the
first non-Gaussian increment is also positive and is smaller than the
Gaussian increment, with their ratio ranging from $0.063$ to $0.298$.
For example, at $U/t=8$ and $\mu/t=1.746251$, the saddle-plus-RPA density
is $0.81310$ and the cumulative density is $0.866(4)$, giving a central
increment of $0.05290$. These contributions quantify the first correction beyond Gaussian
fluctuations. Since the saddle and RPA terms are deterministic, their
subtraction leaves the cumulative Monte Carlo uncertainty unchanged.
Its statistical uncertainty is $0.004$; numerical representation and flavour truncation
remain separate sources of error (Section~\ref{c2:results:precision}).

The comparison in Fig.~\ref{c3:results:sun}(d) uses a different
observable: the three-flavour partition logarithm at $U/t=4.8$,
$\mu/t=1.5$ and $\beta t=5$. The reported inverse-flavour truncation is
$\psi=13.0571(95)$ and lies close to the selected CoS continuation.
The two expansions start from different references and omit different
sequences of terms, making this a comparison between independently
organized approximations. It uses the three-flavour truncation through
second non-Gaussian order; the six-flavour density results above retain
the first non-Gaussian correction.

\subsection{Connected summation and determinant polynomials}

Chapter~\ref{c3:chapter} develops the fixed-coordinate calculations
needed before diagrammatic integration. In CoS++, the head of an open
fermion path carries the flavour information required to close it with
the correct propagator or counterterm. I derived recurrences for
flavour-dependent inputs and described circuit reductions that combine
equivalent remaining computations. For the uniform and two-spin
constructions, terminal readouts give connected values for interaction
subsets while preserving the internal states
needed to continue to higher order. The all-subset interface can therefore
reuse partial products when a sampler considers several retained sets
within one envelope.

The determinant-polynomial sieve of Section~\ref{c3:fast:section}
provides another evaluation of the same connected quantities. It obtains
interaction moments by selecting degree two in each occupied interaction
variable, then removes disconnected contributions with an anchored
recursion. Three flavours require a three-point local filter; higher
flavour numbers also require a projection of the excess degree. The
arithmetic bounds are $O(3^n)$ for three flavours and $O(n3^n)$ for a fixed
$N_f\geq4$. At order nine, the measured C++ speedups over the specified
CoS baselines are $25.18$ and $24.03$ for uniform and non-uniform
three-flavour inputs. The corresponding six-flavour factors are $1.84$
and $1.65$ (Table~\ref{c3:fast:timings}). These measurements cover complete
fixed-coordinate evaluations on dense random inputs; the cost of
sampling physical configurations is discussed separately below.

The thermodynamic examples apply connected summation on a two-spin
honeycomb lattice with $2048$ sites and a three-flavour square lattice
with $32\times32$ sites. For the honeycomb calculation, which uses CDet,
adding the eighth order narrows the spread of the selected Pad\'e
approximants for double occupancy at $U/t=5$, $\mu/t=1.5$ and
$\beta t=9$ (Fig.~\ref{c3:results:honey}). The three-flavour calculation
extends the inputs to unequal chemical potentials and flavour-resolved
reference shifts, while also providing the partition-logarithm comparison
with the inverse-flavour expansion described above.

\subsection{Fermionic subsystem entropy}

In Chapter~\ref{c4:chapter}, I derived the replica trace for the second
R\'enyi entropy of a spatial subsystem, including the signs from fermionic
mode reordering and coherent-state closure. The spatial cut changes the
Gaussian boundary condition and produces a free propagator depending on
two absolute imaginary times. Scaled matrix formulas evaluate this
propagator without forming the exponentially large intermediate factors
that occur in a direct calculation. The interacting coefficients are
then evaluated with CDet, using common coordinates for the twisted and
ordinary terms and propagating their covariance through the normalized
estimator.

The calculations reach tenth order on a $3\times3$ cluster and ninth
order on the larger lattices. On the small cluster, deterministic
integration of the first three coefficients agrees with exact
diagonalization to better than $10^{-8}$
(Table~\ref{c4:bench:lowtable}). On $50\times50$ sites, with a subsystem
of $676$ sites and $(U/t,\mu/t,\beta t)=(3,1,8)$, the ninth-order result
at shift $\alpha=1.3$ is $f_A=-S_2(A)/N_s=-0.0669454$, with a propagated
Monte Carlo error of $3.0\times10^{-5}$
(Table~\ref{c4:large:endpoints}). The continuation and reference-shift
comparisons are essential to interpreting this statistical precision.

\section{Computational capabilities established}
\label{c5:v7:capabilities}

The inverse-flavour calculation connects a chosen approximation to a
specified numerical integral: flavour counting selects its topology
sectors, and momentum routing, contact contractions and internal
integrations determine the remaining measure. This connection makes
individual corrections accessible to calculation. In particular, the
density results resolve how much the Gaussian and first non-Gaussian
terms each change the Hartree prediction, instead of reporting only
the final total.

At fixed interaction coordinates, several algorithms can evaluate the
same connected polynomial. Their agreement at low order checks signs,
flavour factors, counterterms and connectivity independently of Monte
Carlo noise. The sieve timing tests add a quantitative comparison of
arithmetic cost. All-subset values also make it possible to replace a
random choice of one contribution by its conditional average over the
available subsets, as derived in Section~\ref{c3:mc:section}. The resulting
statistical performance depends on both this estimator and the chain
that generates the envelopes.

The replica construction demonstrates how much of the connected-sum
machinery can be reused for a different observable. Subsystem purity
changes the fermionic trace and the Gaussian boundary condition, so its
additional work is concentrated in the two-time kernel and the joint
normalization. Once those are supplied, the interacting coefficients
again follow from connected diagrams. Changing the temperature, region
or replica boundary condition consequently changes the preparation of
the Gaussian problem, even when the connected evaluator is reused.

\section{Implications for reference calculations}
\label{c5:v7:references}

Cold-atom experiments remain a direct motivation for the thermodynamic
calculations. Density and double occupancy enter equation-of-state and
thermometry comparisons, while correlations provide further tests at
fixed fitted parameters. The present lattice results give reference
points at specified Hamiltonians, sizes and temperatures. Applying them
to a trapped cloud requires the corresponding spatial averaging and
readout model. A broader parameter grid would allow the numerical
uncertainties to be propagated through a temperature or coupling fit,
rather than attached only to an isolated calculated value.

The same type of comparison is relevant to other hardware that implements
the target fermionic Hamiltonian and ensemble
(Section~\ref{c1:v7:hardware}). A calculation of selected observables can
remain practical after storing a complete many-body state becomes
prohibitive. Density or energy can be used to constrain state parameters,
and an additional observable can then test the prediction with those
parameters held fixed. A subsystem purity offers a further comparison
when local averages already agree. Such applications require reference
data for the device's actual task, including its geometry and accessible
measurements.

Classical approximate solvers are another use of these results. The
$1/N_f$ hierarchy quantifies corrections to the Hartree and Gaussian
approximations, and its comparison with CoS tests an independently
organized series. Reference data over a common parameter range would
also allow changes in cluster size, bath construction or variational
capacity to be assessed against an external answer. Section~\ref{c1:v7:approximation}
discusses these tasks. 

The partition logarithm has additional uses described in
Section~\ref{c1:v7:free-energy}. It can anchor a thermodynamic integration
and set the equilibrium target for a variational thermal-state
calculation. Thermal entropy surfaces can constrain an initial entropy
budget and an ideal isentropic loading endpoint. These applications need
consistent ensembles and reliable parameter dependence. Assessing a
candidate state's distance from a Gibbs state through its free-energy
gap additionally requires that state's energy and von Neumann entropy.
For subsystem measurements, the second R\'enyi calculation instead gives
the purity associated with a specified cut and temperature. Finite-temperature
purity contains thermal mixedness as well as correlations, so its
interpretation follows the observable and protocol used in the comparison.

\section{Accuracy and present limitations}
\label{c5:v7:accuracy}

Accuracy depends on each stage of the calculation. The exact examples
and random-matrix comparisons in Chapter~\ref{c3:chapter} test the
fixed-coordinate sum, after which integration introduces sampling and
propagator-representation errors. Reconstructing an observable from finitely
many coefficients adds truncation and continuation uncertainty. For
example, the reduction from Gaussian to non-Gaussian density corrections
in Chapter~\ref{c2:chapter} describes the orders actually computed; extending
that hierarchy or applying a justified reconstruction would estimate the
remaining contribution.

The $3\times3$ calculation provides an exact-diagonalization reference,
$f_A=-0.2752498138$. At shift $\alpha=0.90$, continuation through
tenth order gives $-0.2750946$ with a propagated Monte Carlo error of
$6.5\times10^{-6}$. The difference from ED is $1.55\times10^{-4}$,
about 24 times that statistical error. At $\alpha=0.80$, two
continuation branches lie on opposite sides of the ED result and
differ by about $0.00616$. These comparisons identify a reconstruction
error that more precise estimates of the retained coefficients alone
will not resolve.

On the larger lattices, the two replica expansions differ at common order
seven by $2.21\times10^{-4}$ on $40\times40$ sites and
$1.81\times10^{-4}$ on $50\times50$ sites. The corresponding differences
at the respective highest computed orders are approximately
$3.2\times10^{-4}$ and $2.6\times10^{-4}$.
These comparisons vary either the reference alone at common order or
both the reference and the computed order. They should accompany the
individual endpoints and their propagated errors.

The replica error analysis uses 128 chain means per run and fixed
continuation windows. Its statistical errors describe between-chain
dispersion under the equilibration assumptions of Chapter~\ref{c4:chapter}.
Time-resolved measurements would allow within-chain drift and
autocorrelation to be assessed, and a separate window-selection study
would quantify the variability associated with choosing the continuation.
For the exact finite systems considered here, the Borel--Leroy transform
is entire under the stated assumptions. The conformal parameter therefore
reorganizes the truncated series, with its numerical choice assessed
against independent endpoints.

For the density calculation, a further uncertainty comes from the lattice
size. The state near $U/t=8$, $\mu/t=1.746251$ has reported finite-size
sensitivity \cite{ShiLargeNf2025}. A size scan is needed before using
that density as a bulk reference. The replica calculation also has
substantial preparation costs: its tables depend on pairs of sites and
two time coordinates, and the dense construction contains cubic matrix
operations. Increasing the lattice size and increasing the diagram order
therefore place demands on different parts of the calculation. Accuracy
checks must include interpolation, stored precision and the conditioning
of the scaled linear systems.

\section{Next calculations and method extensions}
\label{c5:v7:next}

\subsection{Common thermodynamic grids}

A first extension is to compute the partition logarithm, density,
interaction derivative and thermal entropy on common temperature and
chemical-potential grids. Derivative estimators or finite differences
formed on shared configurations would retain correlations between the
estimated quantities. The covariance should be delivered with the
reference values so that parameter fits and thermodynamic integrations
can propagate it. Overlapping inverse-flavour and CoS calculations would
provide checks at selected points before interpolation is used between
them. Additional lattice sizes are particularly relevant near the
six-flavour state points with known size sensitivity.

\subsection{From evaluator timings to sampling cost}

The fixed-coordinate comparison has already been performed in
Section~\ref{c3:fast:section}. The next comparison should integrate the
same coefficient to a specified uncertainty with each evaluator. It
should include propagator lookup, all requested subset outputs, envelope
updates and measurements, alongside the cost of preparing the input-independent
structures. Proposal rules, shifts and order weights can first be held
fixed to isolate the effect of changing the evaluator. A subsequent
comparison can allow each method to optimize its sampler. Run time,
memory and autocorrelation are all needed to decide whether the
fixed-coordinate gains yield a cheaper physical result.

\subsection{Continuation tests with withheld answers}

The small-cluster data allow reconstruction rules to be specified before
consulting the exact entropy. Applying each rule to series truncated at
successive orders, with the measured coefficient covariance, would test
both its endpoint and its uncertainty estimate. Additional coefficients
could also be withheld and used to assess predictive accuracy. This
would give a stricter test of window selection than inspecting the
flatness of one continuation plot. Larger-system studies can then apply
the same rules and report their remaining order and shift sensitivity.

\subsection{Temperature, geometry and replica observables}

Temperature scans and several subsystem shapes would make the entropy
results more directly usable as reference data. At fixed region area,
changing the boundary tests spatial contributions; changing temperature
helps identify the thermal contribution. Any mutual-information
combination needs the relevant subsystem entropies, their covariance and
an interpretation appropriate to the R\'enyi index. A larger range of
regions also motivates compression of the two-time tables, with errors
checked in the final observable. Higher replica moments or fermionic
negativity require their own operator and boundary-condition derivations.
They are subsequent extensions of the present construction.

\section{Concluding remarks}
\label{c5:v7:closing}

The three developments address successive parts of a reference
calculation: choosing an expansion that resolves interaction corrections,
evaluating its connected coefficients efficiently, and extending the
observable from thermodynamics to subsystem purity. Their numerical
examples show both the precision attainable in the retained terms and
the importance of the omitted orders and continuation. Wider parameter
grids, additional lattice sizes and the reconstruction tests proposed
above would make these calculations more broadly usable in the
comparisons of Chapter~\ref{ch:benchmarks}, from cold-atom thermodynamics
and matching quantum-simulation tasks to approximate solvers and
subsystem-purity measurements.

\clearpage\phantomsection\addcontentsline{toc}{chapter}{Bibliography}
\renewcommand{\bibname}{Bibliography}

\end{document}